\documentclass[12pt]{iopart}

\usepackage{citesort,latexsym,ifthen,graphics,color}

\newcounter{Diagrams}
\Alph{Diagrams}
\newcounter{IllDiagrams}
\Alph{IllDiagrams}
\newcounter{TLDiagrams}
\Alph{TLDiagrams}
\newcounter{cancellation}
\addtocounter{cancellation}{1}

\newtheorem{Diag}{}[Diagrams]
\newtheorem{DC}[Diag]{\{}
\newtheorem{DC-B}[Diag]{$|$}
\newtheorem{cancel}{Cancellation}
\newtheorem{Icancel}{Illustrative Cancellation}

\newtheorem{IllDiag}{}[IllDiagrams]
\newtheorem{IDC}[IllDiag]{\{}

\newtheorem{TLDiag}{}[TLDiagrams]
\newtheorem{TLDC}[TLDiag]{\{}

\newtheorem{D-ID}{Diagrammatic Identity}[section]

\newcommand{\ie}{{\it i.e.}\ }
\newcommand{\cf}{{\it cf.}\ }
\newcommand{\eg}{{\it e.g.}\ }
\newcommand{\aka}{{\it a.k.a.}\ }
\newcommand{\etc}{{\it etc.}\ }
\newcommand{\viz}{{\it viz.}\ }

\newcommand{\wrt}{with respect to}
\newcommand{\CC}{charge conjugation}

\newcommand{\lhs}{left-hand side}
\newcommand{\rhs}{right-hand side}

\newcommand{\role}{role}

\newcommand{\be}{\begin{equation}}
\newcommand{\ee}{\end{equation}}
\newcommand{\bea}{\begin{eqnarray}}
\newcommand{\eea}{\end{eqnarray}}
\newcommand{\beas}{\begin{eqnarray*}}
\newcommand{\eeas}{\end{eqnarray*}}

\newcommand{\bear}{\begin{array}{l}}
\newcommand{\eear}{\end{array}}

\newcommand{\bcf}{\begin{center}\begin{figure}}
\newcommand{\ecf}{\end{figure}\end{center}}

\newcommand{\bct}{\begin{center}\begin{table}}
\newcommand{\ect}{\end{table}\end{center}}

\newcommand{\ds}{\displaystyle}

\def\eq#1{(\ref{#1})}

\def\eqs#1#2{(\ref{#1},\ref{#2})}

\def\sec#1{section~\ref{#1}}
\def\Sec#1{Section~\ref{#1}}

\def\fig#1{figure~\ref{#1}}
\def\Fig#1{Figure~\ref{#1}}
\def\figs#1#2{figures~\ref{#1} and~\ref{#2}}
\def\figrange#1#2{figures~\ref{#1}--\ref{#2}}
\def\tab#1{table~\ref{#1}}
\def\app#1{\ref{#1}}

\newcommand{\xor}{
	\ensuremath{
		\mathsf{\mathbf{xor}} \
	}
}

\def\phi{ \varphi }
\def\A{{\cal A}}
\def\C{{\cal C}}
\def\D{{\cal D}}

\def\G{{\cal G}}
\def\H{{\cal H}}

\newcommand{\nCr}[2]{ \left.\right.^{#1}\!C_{#2}\ }

\newcommand{\bca}[3][1]{
	\left[
		\begin{array}{c}
		\vspace{#1ex}
			\ds #2
		\\
			\ds #3
		\end{array}
	\right]
}

\newcommand{\Tower}[1]{
	\begin{array}{c}
	\vspace{1ex}
		#1
	\\
	\vspace{1ex}
		\TopVertex
	\\
		\SumVertex
	\end{array}
}

\newcommand{\InvTower}[1]{
	\begin{array}{c}
	\vspace{1ex}
		\TopVertex
	\\
	\vspace{1ex}
		\SumVertex
	\\
		#1
	\end{array}
}

\def\hS{\hat{S}}

\def\one{\hbox{1\kern-.8mm l}}
\def\str{\mathrm{str}}

\newcommand{\Op}[1]{\Or (p^{#1})}

\newcommand{\Oepz}{\ensuremath{\Or (\epsilon^0)}}
\newcommand{\OepPow}[1]{\ensuremath{\Or \left(\epsilon^{#1}\right)}}

\newcommand{\Pep}{p^{-2\epsilon}}
\newcommand{\NTE}[1]{\not{\mathrm{T}}(#1)}
\newcommand{\BigNTE}[1]{\not{\! \mathrm{T}}(#1)}
\newcommand{\NTEs}{\not{\mathrm{T}}_{M} (p)}
\newcommand{\BigNTEs}{\not{\! \mathrm{T}}_{M} (p)}

\newlength{\epminusone}
\newlength{\epzero}
\newcommand{\expectation}[1]{\left\langle #1 \right\rangle}

\newcommand{\pder}[2]{\ensuremath{\frac{\partial #1}{\partial #2}}}

\newcommand{\order}[1]{\Or \left( #1 \right)}
\newcommand{\hf}{\frac{1}{2}}

\newcommand{\match}{\eta}

\newlength{\PFheight}
\newcommand{\stats}[1]{
	\left[ #1 \right]
}

\newcommand{\DummyKernel}{\ensuremath{\stackrel{\bullet}{\mbox{\rule{1cm}{.2mm}}}}}
\newcommand{\UnDecKernel}{\ensuremath{\stackrel{\odot}{\mbox{\rule{1cm}{.2mm}}}}}
\newcommand{\DecKernel}{\ensuremath{\stackrel{\circ}{\mbox{\rule{1cm}{.2mm}}}}}
\newcommand{\ExtGR}{\ensuremath{\cdeps{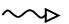}}}
\newcommand{\socket}{\ensuremath{\sqcup}}

\newcommand{\norm}{\ensuremath{\Upsilon}}

\newcommand{\bigdot}[1]{\stackrel{\bullet}{#1}}
\newcommand{\DEP}{\stackrel{\odot}{\Delta}}
\newcommand{\PEP}{\stackrel{\sim}{\Delta}}
\newcommand{\DGRkpr}{\stackrel{\bullet}{\GRkpr}}
\newcommand{\combo}{\ensuremath{\star}}

\newlength{\VertexWidth}

\newcommand{\Vertex}[1]{
\ensuremath{
	\begin{array}{c}
	\settowidth{\VertexWidth}{$#1$}
	\setlength{\unitlength}{1.2\VertexWidth}
	\begin{picture}(1,1)(-0.5,-0.5)
		\put(0,0){\circle{1}}
		\put(-0.4,-0.1){$#1$}
	\end{picture}
	\end{array}
}
}

\newcommand{\Dal}{
	\ensuremath{
		\begin{array}{c}
			\settowidth{\VertexWidth}{\scriptsize $\alpha$}
			\setlength{\unitlength}{1.6\VertexWidth}
			\begin{picture}(1,1)(-0.5,-0.5)
				\put(0,0){\circle{1}}
				\put(-0.45,-0.25){$\alpha$}
			\end{picture}
		\end{array}
	}
}

\newcommand{\SumVertex}{
	\Vertex{n_s, j}
}

\newcommand{\TopVertex}{
	\Vertex{v^{j_+;R}}
}

\newcommand{\Displaced}[1]{
	\begin{array}{c}
	\vspace{1ex}
		\blankstrut
	\\
		#1
	\end{array}
}

\newcommand{\DisplacedVertex}[1]{
	\Displaced{\Vertex{#1}}
}

\newcommand{\flow}{\Lambda \partial_\Lambda}
\newcommand{\flowConstAl}{\Lambda \partial_\Lambda|_\alpha}

\newcommand{\dec}[3][0]{\ensuremath{\left[ #2 \hspace{#1em} \right]^{#3}}}
\newcommand{\decp}[3][0]{\ensuremath{\left[ #2 \hspace{#1em} \right]^{#3}_{p^2}}}
\newcommand{\decpb}[3][0]{\ensuremath{\left. \! #2 \hspace{#1em} \right|^{#3}_{p^2}}}

\newcommand{\decNTEs}[3][0]{\ensuremath{\left[ #2 \hspace{#1em} \right]^{#3}_{\NTEs}}}
\newcommand{\decNTEsb}[3][0]{\ensuremath{\left. \! #2 \hspace{#1em} \right|^{#3}_{\NTEs}}}

\newcommand{\decGR}[3][0]{\ensuremath{\left\{ #2 \hspace{#1em} \right\}^{#3}}}

\newcommand{\Ds}{\mathcal{E}}

\newcommand{\nLDl}[1]{\Ds_{\!n \,\mu \nu}^{#1  1 \, 1}}

\newlength{\tag}
\newlength{\looporder}

\newcommand{\nLV}[4]{
	\settowidth{\looporder}{$\scriptstyle #2$}
	\settowidth{\tag}{$\scriptstyle #4$}
	\ifthenelse{\lengthtest{\looporder > \tag}}%
		{\addtolength{\looporder}{-\tag}
			\ensuremath{	
				{#1}_{#2 #3}^{#4   \hspace{\looporder} 1  1}
			}
		}
		{\addtolength{\tag}{-\looporder}
			\ensuremath{	
				{#1}_{#2 \hspace{\tag} #3}^{#4 1  1}
			}
		}
}

\newcommand{\AGRO}[1]{\left[ #1 \right]_{ES}}

\newcommand{\nothing}{}

\newcommand{\GR}{\cdeps{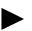}}
\newcommand{\GRk}{\rhd}
\newcommand{\GRkpr}{>}
\newcommand{\DiagDot}{\scriptstyle \bullet}

\newlength{\LabLength}
\newlength{\ProcessRefLength}
\newlength{\ProcessLength}
\newlength{\CancelRefLength}
\newlength{\CancelRefLengthB}
\newlength{\CancelLength}
\newlength{\scriptboldcurlybracket}
\newlength{\strutheight}

\newcommand{\blankstrut}{\rule{0em}{\strutheight}}

\newcommand{\cdeps}[1]{\ensuremath{\begin{array}{c}\includegraphics{./eps/#1.eps} \end{array}}}

\newcommand{\sco}[3][0]{
	\begin{array}{c}
	\vspace{#1ex}
		#2 
	\\
		#3
	\end{array}
}

\newcommand{\LD}[1]{
	\settowidth{\LabLength}{\scriptsize \textbf{\ref{#1}}}
	\addtolength{\LabLength}{0.8em}
	\begin{minipage}{\LabLength}
		\scriptsize
		\begin{Diag}\label{#1}\end{Diag}
	\end{minipage}
}

\newcommand{\LID}[1]{
	\settowidth{\LabLength}{\scriptsize \textbf{\ref{#1}}}
	\addtolength{\LabLength}{0.8em}
	\begin{minipage}{\LabLength}
		\scriptsize
		\begin{IllDiag}\label{#1}\end{IllDiag}
	\end{minipage}
}

\newcommand{\CID}[2]{
	\settowidth{\LabLength}{\scriptsize\textbf{\ref{#1}}}
	\settowidth{\CancelRefLength}{\scriptsize$\ref{#2}$}
	\addtolength{\LabLength}{\CancelRefLength}
	\addtolength{\LabLength}{\CancelLength}
	\begin{minipage}{\LabLength}
		\scriptsize
		\begin{IDC}\label{#1} \ref{#2} \textbf{\}} \end{IDC}
	\end{minipage}
}

\newcommand{\LO}[3][1]{
	\begin{array}{c}
		\LD{#3}
	\\[#1ex]
		#2
	\end{array}
}

\newcommand{\LIO}[3][1]{
	\begin{array}{c}
		\LID{#3}
	\\[#1ex]
		#2
	\end{array}
}

\newcommand{\LDi}[3][1]{\LO[#1]{\ensuremath{\begin{array}{c}\input{pstex/#2.pstex_t} \end{array}}}{#3}}

\newcommand{\LIDi}[3][1]{\LIO[#1]{\ensuremath{\begin{array}{c}\input{pstex/#2.pstex_t} \end{array}}}{#3}}

\newcommand{\LHlabs}[3][0]{
\begin{array}{l}
\vspace{#1ex}
	#2\\
	#3		
\end{array}
}

\newcommand{\LDLD}[3][-1]{
	\LHlabs[#1]{\LD{#2}}{\LD{#3}}
}

\newcommand{\LIDLID}[3][-0.5]{
	\LHlabs[#1]{\LID{#2}}{\LID{#3}}
}

\newcommand{\LOLO}[4][1]{
	\begin{array}{c}
		\LDLD{#3}{#4}
	\\[#1ex]
		#2
	\end{array}
}

\newcommand{\LLDi}[4][1]{
	\LOLO[#1]{\ensuremath{\begin{array}{c}\input{pstex/#2.pstex_t} \end{array}}}{#3}{#4}
}

\newcommand{\ICancel}[2]{
\begin{Icancel}
	Diagram~\ref{#1}  exactly cancels diagram~\ref{#2}.
	\label{Icancel:#1}
\end{Icancel}
}

\newcommand{\ICancelCom}[3]{
\begin{Icancel}
Diagram~\ref{#1}  exactly cancels diagram~\ref{#2}#3
\label{Icancel:#1}
\end{Icancel}
}

\newcommand{\jhep}[3]{{JHEP} #1 (#2) #3}
\newcommand{\NuclPhys}[4]{{Nucl.\ Phys.\ }\textbf{#1 #2} (#3) #4}
\newcommand{\PhysRev}[4]{{Phys.\ Rev.\ }\textbf{#1 #2} (#3) #4}
\newcommand{\IntJModPhys}[4]{{Int.\ J.\ Mod.\ Phys.\ }\textbf{#1 #2} (#3) #4}
\newcommand{\PhysRep}[4]{{Phys.\ Rep.\ }\textbf{#1 #2} (#3) #4}
\newcommand{\PhysRept}[3]{{Phys.\ Rept.\ }\textbf{#1} (#2) #3}

\newcommand{\PhysLett}[4]{{Phys.\ Lett.\ }\textbf{#1 #2} (#3) #4}
\newcommand{\ProgTheorPhys}[3]{{Prog.\ Theor.\ Phys.\ }\textbf{#1} (#2) #3}
\newcommand{\ProgTheorPhysS}[3]{{Prog.\ Theor.\ Phys.\ Suppl.\ }\textbf{#1} (#2) #3}

\newcommand{\EurPhysJ}[4]{{Eur.\ Phys.\ J.\ }\textbf{#1 #2} (#3) #4}
\newcommand{\CEurJPhys}[3]{{Central Eur.\ J.\ Phys.\ }\textbf{#1} (#2) #3}
\newcommand{\ZPhys}[4]{{Z.\ Phys.\ }\textbf{#1 #2} (#3) #4}
\newcommand{\PhysRevLett}[3]{{Phys.\ Rev.\ Lett.\ }\textbf{#1} (#2) #3}

\newcommand{\arxiv}[1]{[arXiv:#1]}
\newcommand{\hepth}[1]{hep-th/#1}
\newcommand{\hepph}[1]{hep-ph/#1}

\newcommand{\condmat}[1]{cond-mat/#1}

\newcommand{\RevModPhys}[3]{{Rev.\ Mod.\ Phys.\ }\textbf{#1} (#2) #3}
\newcommand{\jphysa}[3]{J.\ Phys.\ {\bf A}: Math.\ Gen.\ #1 (#2) #3}
\newcommand{\Physics}[3]{Physics \textbf{#1} (#2) #3}

\begin{document}

\title{A Manifestly Gauge Invariant and Universal Calculus
for $SU(N)$ Yang-Mills}

\author{
	Oliver J.~Rosten
}

\address{School of Physics and Astronomy,  University of Southampton,
	Highfield, Southampton SO17 1BJ, U.K.}
\ead{O.J.Rosten@soton.ac.uk}

\begin{abstract}
	Within the framework of the Exact Renormalization Group,
	a manifestly gauge invariant calculus is constructed for
	$SU(N)$ Yang-Mills. The methodology is comprehensively
	illustrated with a proof, to all orders in perturbation
	theory, that the $\beta$ function has no explicit dependence 
	on either the seed action
	or details of the covariantization of the cutoff. The cancellation
	of these non-universal contributions is done in an entirely
	diagrammatic fashion.
\end{abstract}

\vspace{-60ex}
\hfill SHEP 06-09
\vspace{61ex}

\maketitle

\tableofcontents
\markboth{A Manifestly Gauge Invariant and Universal Calculus
for $SU(N)$ Yang-Mills}{A Manifestly Gauge Invariant and Universal Calculus
for $SU(N)$ Yang-Mills}

\section{Introduction}
\label{sec:Intro}

Of all the problems in theoretical physics,
surely one of the most pressing is a better understanding
of Yang-Mills theories, particularly in the
non-perturbative domain. A promising framework
for addressing this issue
is the Exact Renormalization Group 
(ERG)~\cite{Wil,W&H,Pol}, the continuum version of Wilson's RG.

The essential physical idea behind this approach is
that of integrating out degrees of freedom
between the bare scale of the quantum theory and
some effective scale, $\Lambda$. The effects
of these modes are encoded in the Wilsonian effective action, 
$S_\Lambda$, which describes the physics of the theory
in terms of parameters relevant to the effective scale.
Central to this methodology is the ERG equation
which determines how the Wilsonian effective action
changes under infinitesimal changes of the scale.
By relating physics at different
scales, this equation provides access to the
low energy dynamics of Yang-Mills theories.
Indeed, more generally, 
the ERG has proven itself to be a flexible
and powerful tool for studying 
both perturbative and non-perturbative problems
in a range of field theories 
(see~\cite{Fisher:1998kv,Morris:1998da,Aoki:2000wm,Litim:1998nf,Berges:2000ew,Bagnuls:2000ae,Polonyi:2001se,Salmhofer:2001tr,Delamotte:2003dw} 
for reviews). A particular advantage conferred by the ERG is
that renormalization is built in:
solutions to the flow equation (in pretty much any approximation scheme),
from which physics can be extracted,
can be naturally phrased directly in terms of renormalized parameters.

Given that the notion of a momentum cutoff
is fundamental to the entire ERG construction,
it is clear that a regulator
is required which incorporates this feature. 
Immediately,
this presents a problem for Yang-Mills theories,
since the direct implementation of a momentum 
cutoff breaks non-Abelian gauge invariance. 
Traditional (gauge fixed) approaches,  which can be broadly divided
into those which employ the background field 
method~\cite{Reuter:1993kw,Reuter:1997gx,Bergerhoff:1997cv,Litim:2002ce,Bonini:2001,Freire:2000bq,Gies:2002af,Litim:2002hj}
and those which do 
not~\cite{Becchi:1993,Bonini:1993sj,Bonini:1994kp,Ellwanger:1994iz,Ellwanger:1995qf,Ellwanger:1996wy,Ellwanger:1997tp,Ellwanger:1997wv,Ellwanger:1999vc,Ellwanger:2002xa,D'Attanasio:1996jd,Litim:1998qi,Simionato:2000,Simionato:2000-B,Panza:2000,Pawlowski:2003hq,Fischer:2004uk}, 
accept this breaking,
recovering the physical symmetry in 
the limit that all quantum fluctuations
have been integrated out.\footnote{
For an approach based on the geometric 
effective action see~\cite{Veneziano,Pawlowski-VDW} and for
a summary of the various approaches see~\cite{Pawlowski:2005xe}.}

Nevertheless, in~\cite{SU(N|N)}, a regulator was constructed
for $SU(N)$ Yang-Mills based on a real, gauge invariant cutoff.
This is achieved by embedding the physical gauge theory
into a spontaneously broken $SU(N|N)$ supergauge theory
which is itself regularized by covariant higher 
derivatives. Combining this with earlier work~\cite{ym,ymi,ymii}
allowed the construction of
an ERG for Yang-Mills which respects 
gauge invariance at all scales~\cite{aprop}.  In addition to
the obvious advantages this has over the alternative approaches,
there is a major additional benefit: the gauge invariance
is in fact \emph{manifest}, no gauge fixing being required
at any stage in the computation of the Wilsonian effective
action. Whilst having considerably novelty value, manifest
gauge invariance also provides powerful technical simplifications:
the gauge field is protected from field strength renormalization and
the Ward identities take a particularly simple form~\cite{ymi},
since the Wilsonian effective action is built only from
gauge invariant combinations of the covariant derivative,
even at the quantum level. In the non-perturbative domain, not only are
Gribov copies~\cite{Gribov} entirely avoided, but it should
possible to make statements
about \eg confinement in
a completely gauge independent way. Furthermore, 
such a framework
has the ability to underlie gauge-fixed ERG formalisms
as it should, in principle, be possible to derive all
results obtained by the latter by gauge fixing
at an appropriate stage.

The key to constructing the manifestly gauge invariant 
scheme of this paper
resides in the immense
freedom inherent in  the  ERG~\cite{jose1,jose2,mgierg1}: of the infinite
number suitable for $SU(N)$ Yang-Mills, an infinite subset
allow the Wilsonian effective action to be computed without
fixing the gauge. 
Of these manifestly gauge invariant
ERGs, we further specialize to those
which allow convenient renormalization
to any loop order~\cite{mgierg1,mgierg2,Thesis}. (By this we mean
that the flow equation is of a suitably general form to
treat the physical gauge field and an unphysical copy, 
which
is part of the regularizing structure, asymmetrically. Since
each of these fields comes with their own coupling, which
renormalize separately, allowing the flow equation to distinguish
between them facilitates convenient
computation.) 

Despite these restrictions, there are still an infinite
number of admissible ERGs, the differences between them
amounting to the following non-universal details. The first two 
relate to the
implementation of a gauge invariant cutoff:
the exact forms of both the cutoff
functions and their covariantizations amount to non-universal
choices. The final source of non-universal differences between
the ERGs with which we work is
the `seed action', $\hS$~\cite{Thesis,mgierg1,mgierg2,scalar1,scalar2,aprop,giqed}:
a functional which respects the same symmetries as the Wilsonian effective 
action, $S$, and has the same structure. However, whereas our
aim is to solve the flow for $S$, $\hS$ acts as an input.
Physically, the seed action can be thought of as (partially)
parameterizing a general Kadanoff blocking~\cite{Kadanoff} in the 
continuum~\cite{jose1,jose2,mgierg1}. Crucially, these non-universal
details
need never
be explicitly specified, instead just satisfying general
constraints to ensure that the flow equation is well defined.
We turn this residual freedom
in the construction to our advantage
by recognizing that since all non-universal
details must cancel out in the computation of a universal
quantity, they can be efficiently absorbed into diagrammatic
rules. This observation formed the basis for the manifestly gauge invariant
and universal calculus proposed in~\cite{aprop}, in which
a scheme was developed whereby these non-universal contributions
can be iteratively cancelled out, in perturbative calculations.

In the original work~\cite{aprop},
a small subset of the diagrammatic rules now known
to exist were uncovered and were used in the initial
stages of a manifestly gauge invariant
computation of the one-loop $\beta$
function. In a series of 
works since then~\cite{scalar2,mgierg1,mgierg2,Thesis,giqed},
these diagrammatic rules were extended,
allowing both the one and two-loop $\beta$ functions
(in a variety of Quantum Field Theories (QFTs)) to be reduced
to manifestly universal diagrammatic expressions,
from which the correct numerical coefficients 
were directly extracted. In this paper, 
working in $SU(N)$ Yang-Mills,
we bring together and complete
the overlapping sets of ideas from this collection of works,
developing the diagrammatic calculus to a stage where it
is applicable at any number of loops.  The calculus
is then comprehensively illustrated by 
deriving an expression for the $n$-loop $\beta$ function, $\beta_n$,
which has no explicit dependence on
either the seed action or the details of the
covariantization of the cutoff.

This result is, in itself, interesting for several 
reasons. First, let us recall the standard argument as to why
the coefficients $\beta_1$ and $\beta_2$, but not $\beta_{\geq 3}$,
are guaranteed to
agree between certain renormalization schemes~\cite{Weinberg}.
It is important to recognize that each choice of the non-universal
details within our ERG corresponds, in general, to a different renormalization
scheme. Focusing on one such scheme, we take the coupling
of the physical $SU(N)$ gauge field to be $g(\Lambda)$.\footnote{Beyond this
discussion, we will take $g(\Lambda)$ to represent the coupling
for all renormalization schemes implicitly defined by our approach.}
Now consider a second scheme---either corresponding to some other
choice of non-universal details within our ERG or to an entirely
independent scheme such as $\overline{MS}$---with coupling $\tilde{g}(\mu \mapsto \Lambda)$.
Given the dimensionless
coefficient, $\match$, we can perturbatively
match the two couplings:
\be
\label{eq:match}
	\frac{1}{\tilde{g}^2} = \frac{1}{g^2} + \match + \Or (g^2).
\ee
Using the usual definition for the $\beta$ function of $g$
\[
	\beta \equiv \Lambda \partial_\Lambda g = \sum_{i=1}^{\infty} g^{2i+1} \beta_i,
\]
and a similar definition for $\tilde{\beta} (\tilde{g})$, we can differentiate~\eq{eq:match}
\wrt\ $\Lambda$ to obtain
\begin{equation}
	\tilde{\beta}_1 + g^2 \tilde{\beta}_2 = \beta_1 + g^2 \beta_2 + \flow \match + \Or (g^4).
\label{eq:Universal-beta}
\end{equation}
Therefore, 
if $\flow \match = 0$ (or, at any rate, does not 
contribute until $\mathcal{O}(g^4)$), 
we will obtain $\tilde{\beta}_1 = \beta_1$
and $\tilde{\beta}_2 = \beta_2$. We can expect this 
agreement to be spoilt, however, if there exist
running, dimensionless couplings, besides $g$
(which is equivalent to the introduction of
additional mass scales).

Indeed, within our setup there 
generically exist dimensionless couplings
which run even at tree level
level~\cite{aprop}, spoiling
agreement between $\beta$-function coefficients
at one loop. Of course, 
this is not a sign
of a sick formalism, just a sign that
$\beta_1$ and $\beta_2$ are not physically
observable and can be scheme dependent. Nonetheless, it is possible
to recover the universal values of
$\beta_1$ and $\beta_2$ by suitably tailoring the
setup. The running of
$\match$ has two sources.
The first is $\hS$, which contributes
to the running of $\match$ at all orders, including
tree level. However, through an implicit choice
of $\hS$, this running can be removed, not just
at one-loop, but actually to all orders~\cite{Thesis,mgierg2}.
Indeed, we now assume that this has been done, and promote
this choice to a requirement, necessary in the specification
of the subset of manifestly gauge invariant
ERGs with which we choose to work.

The second contribution to $\flow \match$ comes from
an unphysical, dimensionless coupling, $g_2$,
associated with the regularizing 
structure~\cite{SU(N|N),Thesis,mgierg1,mgierg2}.
The running of this coupling,
which occurs from the one-loop level onwards,
cannot be removed through
a choice of $\hS$; the solution is to tune it to
zero at the end of a calculation. In this manner,
$\beta_2$ can be arranged to coincide with its
standard value~\cite{Thesis,mgierg2}. 
For convenience, we work not with $g_2$ directly
but with
\be
	\alpha := g^2_2/g^2
\label{eq:alpha-defn}
\ee
and so it is $\alpha$, in practise, which is tuned to zero.

Beyond two loops, it is apparent from~\eq{eq:Universal-beta}
that there is no reason to expect any agreement
between $\beta$ function coefficients. In light of
this, it is surprising not only that all contributions
to $\flow \match$ coming from $\hS$ can be removed to all
orders but also that all dependence of $\beta_n$ on
the seed action and details of the covariantization
of the cutoff cancels out, to all orders. (For speculations
on whether it may be possible to push the removal
of non-universal details further still, see~\cite{RG2005}.)
That we do find such
cancellations is perhaps indicative that there
is a more direct framework for performing calculations
in QFTs which retains the advantages of the ERG approach
whilst removing some or all of the redundancy inherent in
the approach. In particular, these observations
could inspire a manifestly 
gauge invariant formalism where the seed action and
details of the covariantization of the cutoff
are relegated to a background role. 

The second point to make about the
derivation of an expression for $\beta_n$
which has no explicit dependence on
either the seed action or the details of the
covariantization of the cutoff is that this is
a huge step forward in
turning this formalism into a practical computational
scheme.
The calculation of $\beta_1$ and $\beta_2$
in this ERG approach was, up until now, an arduous
task; just getting to the expression from which the universal
value can be extracted was extremely difficult. Now, however,
this step is trivial as we simply specialize the new
formula for $\beta_n$ to the appropriate loop order!

To appreciate the various elements of the diagrammatic
calculus, it is worth reviewing the 
procedure for computing $\beta$ function coefficients employed
in~\cite{Primer}.
To compute $\beta_{1,2}$, we start by using the flow
equation to compute the flow of the two-point vertex
corresponding to the physical $SU(N)$ gauge field,
$A^1_\mu$, which we suppose carries momentum, $p$. Next, we specialize to the
appropriate loop order and work at $\Op{2}$; this
latter step constrains the equation by allowing the renormalization
condition for the physical coupling to feed in. At this
point, the equation for $\beta_{1,2}$ contains a small
number of diagrams, each of which contains explicit
dependence on both the seed action and the details
of the covariantization of the cutoff.

Central to the diagrammatic calculus
is the `effective propagator 
relation'~\cite{Thesis,mgierg1,mgierg2,Primer,scalar1,scalar2,aprop,giqed}.
This convenient computational device
arises as a consequence of a choice
of seed action we are free to make: 
we choose the classical, two-point seed action vertices equal to 
their Wilsonian effective action counterparts.
In turn,
this ensures that 
for each independent classical two-point vertex (that cannot be consistently
set to zero~\cite{Thesis}) there exists
an `effective propagator', denoted by $\Delta$, 
which is the inverse of the
given vertex, up to a `gauge remainder'. 
Denoting the classical two-point vertex corresponding to
the fields $X$ and $Y$, which carry indices $R$ and $S$ and
momenta $p$ and $-p$, respectively, 
by $S_{0 R S}^{\ X Y}(p)$
the effective propagator relationship in the $A^1$ sector reads
\[
	S^{\ A^1 A^1}_{0 \mu \ \; \alpha} (p) \Delta^{A^1 A^1}_{\alpha \ \, \nu} (p) = \delta_{\mu\nu} - \frac{p_\mu p_\nu}{p^2}.
\]
Thus, $p_\mu p_\nu / p^2$ is a gauge remainder
and it appears as 
a consequence of the manifest gauge invariance: the effective
propagators are inverses of the classical, two-point vertices
only in the transverse space.
As we will see later, it is convenient to split this gauge remainder 
into two components,
$p_\nu$ and $p_\mu/p^2$.

To proceed, we recognize that certain diagrams generated
by the flow comprise exclusively Wilsonian
effective action vertices joined together by $\bigdot{\Delta}$,
where we define
\be
\label{eq:dot}
	\bigdot{X} \equiv -\flowConstAl X.
\ee
These terms are processed by moving the $-\flowConstAl$
from the effective propagator to the diagram as a whole,
minus correction terms in which $-\flowConstAl$ strikes
the vertices. The former diagrams are called $\Lambda$-derivative
terms; the latter can be processed using the flow equation
and the resulting set of diagrams simplified, using
a set of primary diagrammatic identities, of which the
effective propagator relation is one.
At this point, we are able to identify
cancellations of non-universal contributions, at the
diagrammatic level.
There is, however, a complication to this diagrammatic procedure:
particular classes of sub-diagrams can have two distinct diagrammatic
representations. The equivalence of these representations
is encoded in the secondary diagrammatic identities.

Iterating the diagrammatic procedure,
the expressions for $\beta_{1,2}$ ultimately
reduce to the following sets of diagrams:
\begin{enumerate}
	\item 	$\Lambda$-derivative terms, which are built out
			of Wilsonian effective action vertices, effective
			propagators and (components of) gauge remainders;

	\item 	`$\alpha$-terms', consisting of diagrams
			built out of the same elements as the 
			$\Lambda$-derivative terms but
			struck by $\partial / \partial \alpha$
			(there are no $\alpha$-terms at one loop~\cite{Thesis,mgierg1});

	\item 	`$\Op{2}$-terms', which contain an $\Op{2}$ stub
			\ie a diagrammatic component which is manifestly
			$\Op{2}$.
\end{enumerate}

The $\Op{2}$-terms can be manipulated. In the calculation
of $\beta_1$, at any rate, the structure attaching to the stub
can be directly Taylor expanded to zeroth order in $p$---which 
can once again be done diagrammatically. The above diagrammatic
procedure is then iterated.
At two loops (and beyond), 
this procedure is not so straightforward, since
na\"ive Taylor expansion can generate spurious infra-red (IR)
divergences~\cite{Thesis,mgierg1,mgierg2}. The
solution is to isolate the components which
cannot be Taylor expanded using the 
`subtraction techniques' of~\cite{Thesis,mgierg2}.
Now the $\Op{2}$ can
be processed, and so $\beta_{1,2}$ can be reduced to
just $\Lambda$-derivative and $\alpha$-terms.
From these terms, the universal values of
$\beta_{1,2}$ can be extracted~\cite{aprop,mgierg2,Thesis}.

Given the diagrammatic representation of the flow equation, 
the diagrammatic calculus comprises the following elements:
\begin{enumerate}
	\item	An operator which implements the flow \ie $-\flow$;

	\item	A diagrammatic rule reflecting the \CC\ invariance
			of the theory;

	\item	The primary diagrammatic identities which allow
			the direct manipulation of diagrams possessing particular
			elements;

	\item	The secondary diagrammatic identities which encode
			the equivalence of distinct diagrammatic representations
			of particular classes of sub-diagrams;

	\item	The subtraction techniques.
\end{enumerate}

The primary diagrammatic identities further decompose into three sets. 
Those of the first type are defined
without any reference to perturbation theory, and follow from
general constraints such as gauge invariance. The single primary
diagrammatic identity of the second type is the effective propagator
relation, which we recall follows from the classical flow equation,
given a choice we impose on the seed action. Those of the third type
follow directly from those of the first and second types. However, it
is convenient to give them in their own right, as they are not
necessarily obvious and are heavily used in this paper.

The secondary diagrammatic identities decomposes into two families,
one of which is applicable only to those diagrams which possess
an $\Op{2}$ stub and the other of which is more generally
applicable.

It is apparent that, 
of all the elements of the diagrammatic calculus, 
only one is reliant on perturbation
theory.  An obvious
question, to which we return in the conclusion,
is whether the calculus has a non-perturbative
extension. We emphasise
that, irrespective of whether or not
this is the case, the framework admits 
standard ERG analyses in the non-perturbative domain, 
but with the benefits of manifest gauge invariance.

To derive an expression for arbitrary $\beta_n$,
we could pursue the above strategy (see \cite{Thesis} for an attempt
in this direction). However, there is a much more
efficient way to proceed, motivated by
the form of the $\Lambda$-derivative terms
at one and two loops. We start by constructing
a set of $n$-loop functions which depend only on
Wilsonian effective action vertices, effective
propagators and (components of) gauge remainders.
Considering the $\Op{2}$ parts of these diagrams
(strictly, up to functions not polynomial  in $p$),
we compute their flow. Amongst the generated
terms is $\beta_n$ multiplied by a
coefficient
which is universal as a consequence of specializing
to $\Op{2}$. Using the various diagrammatic identities,
we then demonstrate that all dependence on
the seed action and details of the covariantization
cancels out between the remaining terms. 
By taking the $\Lambda$-derivative terms
as the starting point for the calculation, we
avoid having to iteratively construct them, which
reaps better and better dividends with each loop order.

This paper is organized as follows. In \sec{sec:Review},
we review the aspects of the manifestly gauge
invariant ERG we require for this paper. Following
a brief exposition of $SU(N|N)$ gauge theory, the flow
equation is introduced via its diagrammatic representation.
Some properties of the various diagrammatic
elements of the flow equation are discussed
and the primary diagrammatic identities of the first type
are stated. The section concludes with the examination
of the
form taken by the flow equation in the perturbative regime;
we describe the effective propagator relation
and derive the primary diagrammatic identities of the third type.

\Sec{sec:Further}, the basis of which comes from~\cite{Thesis}, 
is devoted to developing
the diagrammatic techniques to the level
sufficient for computing $\beta_n$. First,
we revisit the gauge remainders, discussing
the type of diagrams they can generate
in the perturbative domain. Next, we
state and prove the secondary diagrammatic
identities and finally we describe
the subtraction techniques necessary for
manipulating the $\Op{2}$ terms. This completes
the description of the calculus and concludes
the first part of the paper.

The second part of the paper is devoted to
a detailed illustration of the calculus.
\Sec{sec:Preliminary}
begins with a description of some further notation
which facilitates the practical
application of the calculus, by allowing us to
represent sets (with a potentially huge
number) of $n$-loop diagrams in an extremely compact
manner. Using this notation,
we introduce a set of diagrammatic functions
which will play a central role in the treatment
of $\beta_n$ to follow. Various important properties of
these functions are analysed, which allows an
illustration of many of the diagrammatic techniques
in their most natural setting.
In \sec{sec:beta_n} we derive an expression for $\beta_n$
which is independent of the seed action and the
details of the covariantization of the cutoff
and we conclude in \sec{sec:Conc}.
The primary and secondary diagrammatic identities are
collected together in \app{app:D-ID} for easy reference.

\section{Review}
\label{sec:Review}

\subsection{Elements of $SU(N|N)$ Gauge Theory}
\label{sec:elements}

Throughout this paper, we work in Euclidean dimension, D.
We regularize $SU(N)$ Yang-Mills by embedding
it in spontaneously broken $SU(N|N)$ Yang-Mills,
which is itself regularized by covariant higher derivatives~\cite{SU(N|N)}. 
The supergauge field, $\A_\mu$, is valued in
the Lie superalgebra and, using the defining representation,
can be written as a Hermitian supertraceless supermatrix:
\[
	\A_\mu = 
	\left(
		\begin{array}{cc}
			A_\mu^1 	& B_\mu 
		\\
			\bar{B}_\mu & A_\mu^2
		\end{array} 
	\right) + \A_\mu^0 \one.
\]
Here, $A^1_\mu(x)\equiv A^1_{a\mu}\tau^a_1$ is the
physical $SU(N)$ gauge field, $\tau^a_1$ being the $SU(N)$
generators orthonormalized to
$\tr(\tau^a_1\tau^b_1)=\delta^{ab}/2$, while $A^2_\mu(x)\equiv
A^2_{a\mu}\tau^a_2$ is a second unphysical $SU(N)$ gauge field.
When labelling \eg vertex coefficient functions,
we often abbreviate $A^{1,2}$ to just 1,2.
The $B$ fields are fermionic gauge fields which will gain a mass
of order $\Lambda$ from the spontaneous symmetry breaking; they play the
role of gauge invariant Pauli-Villars (PV) fields, furnishing the
necessary extra regularization to supplement the covariant
higher derivatives. In order
to unambiguously define contributions which are finite
only by virtue of the PV regularization, a pre-regulator must
be used in $D=4$~\cite{SU(N|N)}. For the time being, we will use
dimensional regularization, emphasising that
this makes sense
non-perturbatively, since 
it is not being used to renormalize the theory, but rather
as a prescription for discarding surface terms in loop
integrals~\cite{SU(N|N)}.

The theory is
subject to the local invariance:
\be
\label{Agauged}
\delta\A_\mu = [\nabla_\mu,\Omega(x)] +\lambda_\mu(x) \one.
\ee
The first term, in which $\nabla_\mu = \partial_\mu -i\A_\mu$, 
generates supergauge transformations. Note that the coupling, $g$,
which we now take to represent the coupling for all renormalization
schemes implictly defined by our approach,
has been scaled out of this definition. It is worth doing
this: since we do not gauge fix, the exact preservation of~\eq{Agauged}
means that none of the fields suffer wavefunction
renormalization, even in the broken phase~\cite{aprop}.

The second term in~\eq{Agauged} divides out the centre of the algebra.
This `no $\A^0$ shift symmetry' ensures that nothing depends on $\A^0$
and that $\A^0$ has no degrees of freedom. We adopt a
prescription whereby we can effectively ignore the field $\A^0$,  altogether, 
using it  to map us into a particular diagrammatic
picture~\cite{Thesis,mgierg1}.

The spontaneous symmetry breaking is carried by a superscalar
field
\[
\C =
	\left(
		\begin{array}{cc}
			C^1		& D
		\\
			\bar{D}	& C^2
		\end{array}
	\right),
\]
which transforms homogeneously:
\be
\label{Cgauged}
\delta\C = -i\,[\C,\Omega].
\ee

It can be shown that, at the classical level, the spontaneous
symmetry
breaking scale (effectively the mass of $B$) tracks the covariant
higher derivative effective cutoff scale, $\Lambda$, if $\C$ is
made dimensionless (by using powers of $\Lambda$) and  $\hS$ has
the minimum of its effective potential at:
\be
\label{sigma}
\sigma \equiv \pmatrix{\one & 0\cr 0 & -\one}.
\ee
In this case the classical action $S_0$ also has a minimum 
at~\eq{sigma}. At the quantum level this can be imposed as a
constraint on $S$ by taking $\expectation{C} = \sigma $ as a renormalization
condition. This ensures that the Wilsonian effective action
does not possess any one-point vertices, which can be
translated into a constraint on
$\hS$~\cite{aprop,Thesis}. In the broken phase, $D$ is
a super-Goldstone mode (eaten by $B$ in the unitary gauge) whilst the
$C^i$ are Higgs bosons and can be given a running mass of order
$\Lambda$~\cite{ym,SU(N|N),aprop}. Working in our manifestly
gauge invariant formalism, $B$ and $D$ gauge transform into each
other; in recognition of this, we define the fields
\numparts
\bea
\label{eq:F}
	F_M & = & (B_\mu, D),
\\
\label{eq:Fbar}
	\bar{F}_N & = & (\bar{B}_\nu, -\bar{D}),
\eea
\endnumparts
where $M$, $N$ are
five-indices~\cite{Thesis,mgierg1}.\footnote{The summation
convention for these indices is that we take each product of
components to contribute with unit weight.} 

The couplings $g$ and $\alpha$ (recall~\eq{eq:alpha-defn})
are defined through their
renormalization conditions:
\bea
\label{defg}
	S[\A=A^1, \C=\sigma]	& =	& {1\over2g^2}\,\str\!\int\!\!d^D\!x\,
									\left(F^1_{\mu\nu}\right)^2+\cdots,
\\	
\label{defg2}
	S[\A=A^2, \C=\sigma] 	& =	& {1\over2 \alpha g^2}\,\str\!\int\!\!d^D\!x\,
									\left(F^2_{\mu\nu}\right)^2+\cdots,
\eea
where the ellipses stand for higher dimension operators and the
ignored vacuum energy. The field strength tensors in the $A^1$ and
$A^2$ sectors, $F^1_{\mu\nu}$ and $F^2_{\mu\nu}$, should really
be embedded in the top left / bottom right entries of a supermatrix,
in order for the supertraces in~\eqs{defg}{defg2} 
to make sense. We will frequently employ
this minor abuse of notation, for convenience.

\subsection{Diagrammatics for the Flow Equation}

\subsubsection{The Exact Flow Equation}
\label{sec:Flow}

The diagrammatic representation of the flow
equation is shown in \fig{fig:Flow}~\cite{Thesis,mgierg1}.
(For a comprehensive description of the diagrammatics see~\cite{Thesis,mgierg1}.)
\bcf[h]
	\beas
	\ds
	-\flow 
	\dec{
		\ensuremath{\begin{array}{c}\begin{picture}(0,0)%
\includegraphics{pstex/Vertex-S.pstex}%
\end{picture}%
\setlength{\unitlength}{3947sp}%
\begingroup\makeatletter\ifx\SetFigFont\undefined%
\gdef\SetFigFont#1#2#3#4#5{%
  \reset@font\fontsize{#1}{#2pt}%
  \fontfamily{#3}\fontseries{#4}\fontshape{#5}%
  \selectfont}%
\fi\endgroup%
\begin{picture}(320,318)(2180,-963)
\put(2291,-859){\makebox(0,0)[lb]{\smash{\SetFigFont{11}{13.2}{\rmdefault}{\mddefault}{\updefault}{\color[rgb]{0,0,0}$S$}%
}}}
\end{picture}
 \end{array}}
	}{\{f\}}
	& = & a_0[S,\Sigma_g]^{\{f\}} - a_1[\Sigma_g]^{\{f\}}
\\
	& = &
	\ds
	\frac{1}{2}
	\dec{
		\ensuremath{\begin{array}{c}\input{pstex/Dumbbell-S-Sigma_g.pstex_t} \end{array}} - \ensuremath{\begin{array}{c}\input{pstex/Padlock-Sigma_g.pstex_t} \end{array}} - \ensuremath{\begin{array}{c}\input{pstex/WBT-Sigma_g.pstex_t} \end{array}}
	}{\{f\}}
	\eeas
\caption{The diagrammatic form of the flow equation.}
\label{fig:Flow}
\ecf

The \lhs\ depicts the flow of all independent Wilsonian effective action
vertex \emph{coefficient functions}, 
which correspond to the set of broken phase
fields, $\{f\}$. Each coefficient function has associated
with it an implied supertrace structure (and symmetry factor which,
as one would want, does not appear in the diagrammatics).
For example,
\be
\label{eq:Vertex-C1C1}
	\dec{
		\ensuremath{\begin{array}{c}\begin{picture}(0,0)%
\includegraphics{pstex/Vertex-S.pstex}%
\end{picture}%
\setlength{\unitlength}{3947sp}%
\begingroup\makeatletter\ifx\SetFigFont\undefined%
\gdef\SetFigFont#1#2#3#4#5{%
  \reset@font\fontsize{#1}{#2pt}%
  \fontfamily{#3}\fontseries{#4}\fontshape{#5}%
  \selectfont}%
\fi\endgroup%
\begin{picture}(320,318)(2180,-963)
\put(2291,-859){\makebox(0,0)[lb]{\smash{\SetFigFont{11}{13.2}{\rmdefault}{\mddefault}{\updefault}{\color[rgb]{0,0,0}$S$}%
}}}
\end{picture}
 \end{array}}
	}{C^1C^1}
\ee
represents both the coefficient functions $S^{C^1 C^1}$ and
$S^{C^1,C^1}$ which, respectively, are associated with the
supertrace structures $\str C^1 C^1$ and $\str C^1 \str C^1$.

The objects on the \rhs\ of \fig{fig:Flow}
have two different types of component. The lobes
represent vertices of action functionals,
where $\Sigma_g \equiv g^2S - 2 \hat{S}$ (recall that $\hS$ is the seed action). 
The object attaching
to the various lobes, \DummyKernel,  is
the sum over vertices of the covariantized ERG kernels~\cite{ymi,aprop}
and, like the action vertices, can be decorated by fields belonging to $\{f\}$.
The appearance of the symbol $\scriptstyle \bullet$ is not accidental, 
meaning $-\flowConstAl$, as in~\eq{eq:dot}.
The dumbbell-like
term, which corresponds to the bilinear functional $a_0$, is referred
to as the classical term. The next two terms, which are both
generated by $a_1$, are referred to as quantum terms. The second
of these contains a kernel which `bites its own tail'. This
diagram is not properly 
UV regularized by the $SU(N|N)$ regularization 
and, in the past, it has been argued that
it can and should be discarded, through an appropriate
constraint on the covariantization~\cite{mgierg1,ym,ymi,aprop}.\footnote{
These diagrams are artefacts of the flow equation. The $SU(N)$
gauge theory \emph{is} fully regularized by the $SU(N|N)$ scheme. However,
regularization of the flow equation does not trivially follow from
the regularization of the underlying theory.}
Here, though, we will keep these diagrams for as
long as possible: as recognized in~\cite{Thesis},
in any calculation $\beta$ function coefficients, 
the explicit ultraviolet divergences in kernel-biting-their-tail
diagrams, which can be dimensionally regularized,
should be cancelled by divergences hidden in other
terms. In this paper, we demonstrate that this
is indeed the case: all explicit dependence
of $\beta_n$ on kernel-biting-their-tail
diagrams cancels out. Nonetheless, there is an implicit
dependence left behind and it seems inevitable that, to further
proceed, the covariantization 
must be suitably constrained,
after all.

It is worth drawing attention to the fact that dimensional
regularization is thus being used for two entirely independent
purposes. On the one hand, it is used as a temporary
measure, at intermediate stages of our calculations, 
to properly define
diagrams possessing a kernel which bites its own tail. 
On the
other hand, it is being employed as a pre-regulator. 
As we will argue in the conclusions, there is evidence
to suggest the existence of a purely diagrammatic
pre-regulator, which would make sense in $D=4$.

Embedded within the diagrammatic rules is a prescription for evaluating the
group theory factors. 
Suppose that we wish to focus on the flow of a particular
vertex coefficient function, which necessarily has a unique
supertrace structure. 
For example, we might be interested in just
the $S^{C^1 C^1}$ component of~\eq{eq:Vertex-C1C1}.

On the \rhs\ of the flow equation, we must
focus on the components of each diagram
with precisely the same (implied) 
supertrace structure as the \lhs,
noting that the kernel, like the vertices,
has multi-supertrace contributions (for more
details see~\cite{Thesis,mgierg1}).
In this more explicit diagrammatic picture,
the kernel is to be considered a double
sided object.
Thus, whilst the dumbbell like term of \fig{fig:Flow}
has at least one associated supertrace, the next diagram
has  at least two, on a account of the loop
(this is strictly true only in the
case that kernel attaches to fields on the same
supertrace). If a closed
circuit formed by a kernel is devoid
of fields then
it contributes 
a factor of $\pm N$, depending on
the flavours of the fields to which the kernel forming
the loop attaches. This is most easily appreciated by
defining the projectors
\be
\label{eq:Projectors}
	\sigma_{+} \equiv
	\left(
		\begin{array}{cc}
			\one	&	0
		\\
				0	&	0
		\end{array}
	\right), \ \ 
		\sigma_{-} \equiv
	\left(
		\begin{array}{cc}
			0	&	0
		\\
			0	&	\one
		\end{array}
	\right)
\ee
and noting that $\str \sigma_\pm = \pm N$. 
In the counterclockwise sense, a $\sigma_+$
can always be inserted for free after an $A^1$, $C^1$ or $\bar{F}$,
whereas a $\sigma_-$
can always be inserted for free after an $A^2$, $C^2$ or $F$.

The rules thus described receive $1/N$ corrections in 
the $A^1$ and $A^2$ sectors. If a kernel
attaches to an $A^1$ or $A^2$, it comprises a direct
attachment and an indirect attachment, as shown 
in \fig{fig:Attach} (see~\cite{Thesis,mgierg1} for more detail).
\bcf[h]
	\[
		\ensuremath{\begin{array}{c}\begin{picture}(0,0)%
\includegraphics{pstex/Direct.pstex}%
\end{picture}%
\setlength{\unitlength}{3947sp}%
\begingroup\makeatletter\ifx\SetFigFont\undefined%
\gdef\SetFigFont#1#2#3#4#5{%
  \reset@font\fontsize{#1}{#2pt}%
  \fontfamily{#3}\fontseries{#4}\fontshape{#5}%
  \selectfont}%
\fi\endgroup%
\begin{picture}(624,477)(2089,-976)
\end{picture}
 \end{array}} \rightarrow \left. \ensuremath{\begin{array}{c}\begin{picture}(0,0)%
\includegraphics{pstex/Direct.pstex}%
\end{picture}%
\setlength{\unitlength}{3947sp}%
\begingroup\makeatletter\ifx\SetFigFont\undefined%
\gdef\SetFigFont#1#2#3#4#5{%
  \reset@font\fontsize{#1}{#2pt}%
  \fontfamily{#3}\fontseries{#4}\fontshape{#5}%
  \selectfont}%
\fi\endgroup%
\begin{picture}(624,477)(2089,-976)
\end{picture}
 \end{array}} \right|_{\mbox{direct}} + \frac{1}{N} \left[ \ensuremath{\begin{array}{c}\begin{picture}(0,0)%
\includegraphics{pstex/Indirect-2.pstex}%
\end{picture}%
\setlength{\unitlength}{3947sp}%
\begingroup\makeatletter\ifx\SetFigFont\undefined%
\gdef\SetFigFont#1#2#3#4#5{%
  \reset@font\fontsize{#1}{#2pt}%
  \fontfamily{#3}\fontseries{#4}\fontshape{#5}%
  \selectfont}%
\fi\endgroup%
\begin{picture}(624,809)(2089,-1279)
\put(2355,-554){\makebox(0,0)[lb]{\smash{\SetFigFont{8}{9.6}{\rmdefault}{\mddefault}{\updefault}{\color[rgb]{0,0,0}$A^2$}%
}}}
\end{picture}
 \end{array}} - \ensuremath{\begin{array}{c}\begin{picture}(0,0)%
\includegraphics{pstex/Indirect-1.pstex}%
\end{picture}%
\setlength{\unitlength}{3947sp}%
\begingroup\makeatletter\ifx\SetFigFont\undefined%
\gdef\SetFigFont#1#2#3#4#5{%
  \reset@font\fontsize{#1}{#2pt}%
  \fontfamily{#3}\fontseries{#4}\fontshape{#5}%
  \selectfont}%
\fi\endgroup%
\begin{picture}(624,809)(2089,-1279)
\put(2355,-554){\makebox(0,0)[lb]{\smash{\SetFigFont{8}{9.6}{\rmdefault}{\mddefault}{\updefault}{\color[rgb]{0,0,0}$A^1$}%
}}}
\end{picture}
 \end{array}} \right]
	\]
\caption{The $1/N$ corrections to the group theory factors.}
\label{fig:Attach}
\ecf 

We can thus consider the diagram on the \lhs\ as having been unpackaged,
to give the terms on the \rhs. The dotted lines in the diagrams with indirect
attachments serve to remind us where the loose end of the kernel attaches
in the parent diagram.

\subsubsection{The Ward Identities}
\label{sec:WIDs}

All vertices, whether they belong to either
of the actions or to the covariantized kernels
are subject to Ward identities which, due to the
manifest gauge invariance, take a 
particularly simple form:
\be
\label{eq:WID-A}
	\ensuremath{\begin{array}{c}\input{pstex/WID-contract.pstex_t} \end{array}} = \ensuremath{\begin{array}{c}\input{pstex/WID-PF.pstex_t} \end{array}} + \ensuremath{\begin{array}{c}\input{pstex/WID-PFb.pstex_t} \end{array}} - \ensuremath{\begin{array}{c}\input{pstex/WID-PB.pstex_t} \end{array}} - \ensuremath{\begin{array}{c}\input{pstex/WID-PBb.pstex_t} \end{array}} + \cdots
\ee

Equation~\eq{eq:WID-A} is the first primary diagrammatic identity
of the first type. On the \lhs, we contract a vertex with the momentum of
the field which carries $p$. This field---which we will
call the active field---can be either
$A^1_\rho$, $A^2_\rho$, $F_R$ or $\bar{F}_R$.
In the first two cases, the triangle $\GRk$ represents
$p_\rho$ whereas, in the latter two cases, it represents
$p_R \equiv (p_\rho,2)$. (Given that we often sum over
all possible fields, we can take the Feynman rule for
$\GRk$ in the $C$-sector to be null.)
In all cases,  $\GRk$ is independent of $\Lambda$ and $\alpha$, which is
encoded in the  second of the primary diagrammatic identities of the 
first type:
\numparts
\bea
\label{eq:LdL-GRk-Pert-a}
	\hspace{0.8em} \stackrel{\bullet}{\GRk} & =  & 0,
\\
\label{eq:dalpha-GRk-a}
	\begin{array}{c}
		\Dal
	\\[-1.5ex]
		\GRk
	\end{array} & = & 0,
\eea
\endnumparts
where $ \Dal\ \equiv \partial / \partial{\alpha}$.

On the \rhs\ of~\eq{eq:WID-A}, we push the contracted momentum forward onto 
the field which directly follows the active field, in the counterclockwise
sense, and pull back (with a minus sign) onto
the field which directly precedes the active field. 
Since our diagrammatics is permutation symmetric, the struck field---which
we will call the target field---can
be either $X$, $Y$ or any of the un-drawn fields represented
by the ellipsis.
Any field(s) besides the active field and the
target field will be called spectators.
Note that we can take
$X$ and / or $Y$ to represent the end of a kernel.
In this case, the struck field is determined to be unambiguously
on one side of the (double sided) kernel; 
the contributions
in which the struck field is on the other side are included
in the ellipsis. 
This highlights the point that allowing the active
field to strike another field necessarily involves a partial
specification of the supertrace structure: it must be the case that
the struck field either directly followed or preceded the active
field. In turn, this means that the Feynman rule for particular
choices of the active and target fields can be zero. For example,
an $F$ can follow, but never precede an $A^1_\mu$, and so the 
pull back of an $A^1_\mu$ onto an $F$ should be assigned a value
of zero. 
The momentum routing follows in an obvious manner: for example,
in the first diagram on the \rhs, momenta $q+p$ and $r$ now flow into
the vertex. In the case that the active field is fermionic,
the field pushed forward / pulled back onto is transformed
into its opposite statistic partner. 
There are some signs 
associated with this in the $C$ and $D$-sectors, 
for which we refer the reader to~\cite{mgierg1,Thesis}. 

The half arrow which terminates the pushed forward / pulled back
active field is of no significance and can go on either side
of the active field line. It is necessary to
keep the active field line---even though the active field
is no longer part of the vertex---in order that
we can unambiguously deduce flavour changes 
and momentum routing, without reference to the parent diagram.

We illustrate~\eq{eq:WID-A} by considering contracting
$\GRk$ into the Wilsonian the effective action
two-point vertex:
\be
\label{eq:GR-TP}
	\ensuremath{\begin{array}{c}\begin{picture}(0,0)%
\includegraphics{pstex/GR-TP.pstex}%
\end{picture}%
\setlength{\unitlength}{3947sp}%
\begingroup\makeatletter\ifx\SetFigFont\undefined%
\gdef\SetFigFont#1#2#3#4#5{%
  \reset@font\fontsize{#1}{#2pt}%
  \fontfamily{#3}\fontseries{#4}\fontshape{#5}%
  \selectfont}%
\fi\endgroup%
\begin{picture}(757,318)(1880,-963)
\put(2278,-861){\makebox(0,0)[lb]{\smash{{\SetFigFont{11}{13.2}{\rmdefault}{\mddefault}{\updefault}{\color[rgb]{0,0,0}$S$}%
}}}}
\end{picture}%
 \end{array}} =  \ensuremath{\begin{array}{c}\begin{picture}(0,0)%
\includegraphics{pstex/GR-TP-PF.pstex}%
\end{picture}%
\setlength{\unitlength}{3947sp}%
\begingroup\makeatletter\ifx\SetFigFont\undefined%
\gdef\SetFigFont#1#2#3#4#5{%
  \reset@font\fontsize{#1}{#2pt}%
  \fontfamily{#3}\fontseries{#4}\fontshape{#5}%
  \selectfont}%
\fi\endgroup%
\begin{picture}(457,414)(2180,-1059)
\put(2278,-861){\makebox(0,0)[lb]{\smash{{\SetFigFont{11}{13.2}{\rmdefault}{\mddefault}{\updefault}{\color[rgb]{0,0,0}$S$}%
}}}}
\end{picture}%
 \end{array}} - \ensuremath{\begin{array}{c}\begin{picture}(0,0)%
\includegraphics{pstex/GR-TP-PB.pstex}%
\end{picture}%
\setlength{\unitlength}{3947sp}%
\begingroup\makeatletter\ifx\SetFigFont\undefined%
\gdef\SetFigFont#1#2#3#4#5{%
  \reset@font\fontsize{#1}{#2pt}%
  \fontfamily{#3}\fontseries{#4}\fontshape{#5}%
  \selectfont}%
\fi\endgroup%
\begin{picture}(457,414)(2180,-963)
\put(2278,-861){\makebox(0,0)[lb]{\smash{{\SetFigFont{11}{13.2}{\rmdefault}{\mddefault}{\updefault}{\color[rgb]{0,0,0}$S$}%
}}}}
\end{picture}%
 \end{array}}.
\ee
Given that $\GRk$ is null in the $C^i$ sector,
the fields decorating the two-point
vertex on the \rhs\ can be either both $A^i$s
or both fermionic. In the former case, 
\eq{eq:GR-TP} reads:
\[
	p_\mu S^{A^i A^i}_{\mu \ \, \nu}(p) = S^{A^i}_{\nu}(0) - S^{A^i}_{\nu}(0) = 0
\]
where we note that $S^{A^i}_{\nu}$ is in fact zero by itself,
as follows by both Lorentz invariance and gauge invariance.
In the latter case, \eq{eq:GR-TP} reads:
\[
	p_M S^{\bar{F} \; F}_{M N}(p) = \left[S^{C^2}(0) - S^{C^1}(0)\right] \delta_{5N},
\]
where we have used~\eq{eq:F} and have discarded
contributions which go like $S^{A^i}_{\nu}(0)$.
However, the $S^{C^i}(0)$ must vanish. This
follows from demanding
that the minimum of the superhiggs potential is
not shifted by quantum corrections~\cite{aprop}.
Therefore, we arrive at the diagrammatic identity
\be
\label{eq:D-ID-GR-TP-A}
	\ensuremath{\begin{array}{c}\begin{picture}(0,0)%
\includegraphics{pstex/GR-TP.pstex}%
\end{picture}%
\setlength{\unitlength}{3947sp}%
\begingroup\makeatletter\ifx\SetFigFont\undefined%
\gdef\SetFigFont#1#2#3#4#5{%
  \reset@font\fontsize{#1}{#2pt}%
  \fontfamily{#3}\fontseries{#4}\fontshape{#5}%
  \selectfont}%
\fi\endgroup%
\begin{picture}(757,318)(1880,-963)
\put(2278,-861){\makebox(0,0)[lb]{\smash{{\SetFigFont{11}{13.2}{\rmdefault}{\mddefault}{\updefault}{\color[rgb]{0,0,0}$S$}%
}}}}
\end{picture}%
 \end{array}} = 0.
\ee

\subsubsection{Taylor Expansion of Vertices}
\label{sec:Taylor}

For the formalism to be properly defined,
it must be the case that all vertices
are Taylor expandable to all orders
in momenta~\cite{ym,ymi,ymii}.
Consider a vertex which is part
of a complete diagram, decorated by some set of internal
fields and by a single external $A^1$ (or $A^2$).
The diagrammatic representation for the zeroth order expansion
in the momentum of the external field is all that is required
for this paper~\cite{Thesis,mgierg1}:
\be
\label{eq:Taylor-A}
	\ensuremath{\begin{array}{c}\input{pstex/Taylor-Parent.pstex_t} \end{array}} = \cdeps{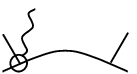} + \cdeps{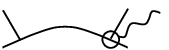} - \cdeps{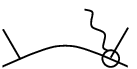} - \cdeps{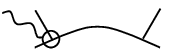} +\cdots
\ee
This is the final primary diagrammatic identity of the first type;
note the similarity to~\eq{eq:WID-A}.

The interpretation of the diagrammatics is as follows. In the first diagram
on the \rhs, the vertex is differentiated \wrt\ the momentum carried
by the field $X$, whilst holding the momentum of the preceding field fixed
(we assume for the time being that both $X$ and the
preceding field carry non-zero momentum).
Of course, using our current diagrammatic notation,
this latter field can be any of those
which decorate the vertex, and so we sum over all possibilities. Thus,
each cyclically ordered push forward like term has a partner,
cyclically ordered pull back like term, such that
the pair can be interpreted as
\be
	\left( \left. \partial^r_\mu \right|_s - \left. \partial^s_\mu \right|_r \right) \mathrm{Vertex},
\label{eq:Momderivs}
\ee
where $r$ and $s$ are momenta entering the vertex. 
In the case that $r=-s$, we can and will
drop either the push forward like term or pull back like term, since
the combination can be expressed as $\partial^r_\mu$; we
interpret the diagrammatic notation appropriately.
If any of the fields decorating the vertex carry
zero momentum (besides the explicitly drawn $A^i$),
then they are transparent to this entire procedure.
Thus, they are never differentiated and, if they precede
a field which is, we must look to the first field carrying
non-zero momentum to figure
out which of the vertex's momenta is held constant.
Just as in~\eq{eq:WID-A}, 
the fields $X$ and or $Y$ can be interpreted as the end
of a kernel. In this case, we introduce some new
notation, since it proves confusing in complete diagrams
to actually locate the derivative
symbol at the end of such an object. The notation for the
derivative \wrt\ the momentum entering the end of a kernel 
is introduced in figure~\ref{fig:Taylor-EndofKernel}.
\bcf[h]
	\[
		\ensuremath{\begin{array}{c}\begin{picture}(0,0)%
\includegraphics{pstex/DifferentiatedKernel-A.pstex}%
\end{picture}%
\setlength{\unitlength}{3947sp}%
\begingroup\makeatletter\ifx\SetFigFont\undefined%
\gdef\SetFigFont#1#2#3#4#5{%
  \reset@font\fontsize{#1}{#2pt}%
  \fontfamily{#3}\fontseries{#4}\fontshape{#5}%
  \selectfont}%
\fi\endgroup%
\begin{picture}(745,453)(2040,-831)
\put(2387,-786){\makebox(0,0)[lb]{\smash{\SetFigFont{11}{13.2}{\rmdefault}{\mddefault}{\updefault}{\color[rgb]{0,0,0}$\DiagDot$}%
}}}
\end{picture}
 \end{array}}
	\]
\caption{Notation for the derivative \wrt\ the momentum entering an
undecorated kernel.}
\label{fig:Taylor-EndofKernel}
\ecf

Recalling that a 
kernel, whose fields are explicitly
cyclically ordered, is a two-sided object, we note
that the field whose momentum we have expanded in is sat on the
top-side of the vertex.
The derivative is taken to be \wrt\
the momentum which flows \emph{into} the end of the vertex which
follows the derivative, in the sense indicated by the
arrow on the derivative symbol.
It is clear that the direction of the
arrow on the derivative symbol can be reversed 
at the expense of a minus sign.

\subsubsection{Charge Conjugation Invariance}
\label{sec:CC}

Charge conjugation invariance can be used to simplify the
diagrammatics, by allowing us to discard certain terms
and to combine others. The diagrammatic prescription for
replacing a diagram which possesses exclusively
bosonic external fields with its charge conjugate is~\cite{Thesis,mgierg1}
to reflect the diagram, picking up a sign for each
\begin{enumerate}
	\item external $A^i$,

	\item \emph{performed} gauge remainder,

	\item momentum derivative symbol (note
			that the direction of the arrow accompanying such
			symbols is reversed by the reflection of the diagram).
\end{enumerate}

\subsection{The Weak Coupling Expansion}
\label{sec:Weak}

\subsubsection{Perturbative Diagrammatics}
\label{sec:WeakDiag}

In the perturbative domain, we have the following
weak coupling expansions~\cite{ymi,aprop,Thesis,mgierg1}.
The Wilsonian effective action is given by
\be
	S = \sum_{i=0}^\infty \left( g^2 \right)^{i-1} S_i = \frac{1}{g^2}S_0 + S_1 + \cdots,
\label{eq:Weak-S}
\ee
where $S_0$ is the classical effective action and the $S_{i>0}$
the $i$th-loop corrections. The seed action has a similar expansion:
\be
	\hat{S} = \sum_{i=0}^\infty  g^{2i}\hat{S}_i.
\label{eq:Weak-hS}
\ee
Recalling~\eq{eq:alpha-defn} we have:
\bea
	\beta \equiv \flow g 		& = & \sum_{i=1}^\infty  g^{2i+1} \beta_i(\alpha)
\label{eq:beta}
\\[1ex]
	\gamma \equiv \flow \alpha 	& = & \sum_{i=1}^{\infty}  g^{2i} \gamma_i(\alpha).
\label{eq:gamma}
\eea

Defining $\Sigma_i = S_i - 2\hS_i$, the weak coupling flow equations
follow from substituting~\eq{eq:Weak-S}--\eq{eq:gamma}
into the flow equation, as shown in 
\fig{fig:WeakCouplingFE}~\cite{Thesis,mgierg1}.
\bcf[h]
	\be
		\dec{
			\ensuremath{\begin{array}{c}\input{pstex/Vertex-n-LdL.pstex_t} \end{array}} 
		}{\{f\}}
		= 
		\dec{
			\begin{array}{c}
				\ds
				\sum_{r=1}^n \left[2\left(n_r -1 \right) \beta_r +\gamma_r \pder{}{\alpha} \right]\ensuremath{\begin{array}{c}\begin{picture}(0,0)%
\includegraphics{pstex/Vertex-n_r-B.pstex}%
\end{picture}%
\setlength{\unitlength}{3947sp}%
\begingroup\makeatletter\ifx\SetFigFont\undefined%
\gdef\SetFigFont#1#2#3#4#5{%
  \reset@font\fontsize{#1}{#2pt}%
  \fontfamily{#3}\fontseries{#4}\fontshape{#5}%
  \selectfont}%
\fi\endgroup%
\begin{picture}(320,318)(1776,-676)
\put(1857,-545){\makebox(0,0)[lb]{\smash{\SetFigFont{11}{13.2}{\rmdefault}{\mddefault}{\updefault}{\color[rgb]{0,0,0}$n_r$}%
}}}
\end{picture}
 \end{array}} 
			\\[4ex]
				\ds
				+ \frac{1}{2} 
				\left( 
					\sum_{r=0}^n \ensuremath{\begin{array}{c}\input{pstex/Dumbbell-n_r-r.pstex_t} \end{array}} - \ensuremath{\begin{array}{c}\input{pstex/Vertex-Sigma_n_-B.pstex_t} \end{array}} - \ensuremath{\begin{array}{c}\input{pstex/WBT-Sigma_n-.pstex_t} \end{array}}
				\right)
			\end{array}
		}{\{f\}}
	\label{eq:WeakFlow}
	\ee
\caption{The weak coupling flow equations.}
\label{fig:WeakCouplingFE}
\ecf

We refer to the first two terms on the \rhs\ of~\eq{eq:WeakFlow} as
$\beta$ and $\alpha$-terms, respectively.
The symbol $\bullet$, as in equation~\eq{eq:dot}, means
$-\flowConstAl$. A vertex whose argument is an unadorned letter, say $n$,
represents $S_n$. We define $n_r \equiv n-r$ and $n_\pm \equiv n \pm 1$. The
bar notation of the dumbbell term is defined as follows:
\be
\label{eq:bar}
	a_0[\bar{S}_{n-r}, \bar{S}_r] 	\equiv 	a_0[S_{n-r}, S_r] - a_0[S_{n-r}, \hat{S}_r] - a_0[\hat{S}_{n-r}, S_r].
\ee

The renormalization condition
for $g$, equation~\eq{defg}, constrains the two-point  vertex of
the physical field $S_{\mu \ \; \nu}^{A^1 A^1}(p)$ as follows:
\bea
\label{eq:S_0-11}
	S_{0 \mu  \nu}^{\ 1 \, 1}(p) & = & 2 (p^2 \delta_{\mu\nu} - p_\mu p_\mu) + \Op{4}
	\equiv 2\Box_{\mu \nu}(p) + \Op{4}
\\
\label{eq:S_>0-11}
	S_{n>0 \mu  \nu}^{\ \ \ \ 1 \, 1}(p) & = & \Op{4},
\eea
where we have abbreviated $A^1$ by just `1'.

\subsubsection{The Effective Propagator Relation}
\label{sec:EP-Reln}

The effective propagator
relation (which we recall is the sole primary
diagrammatic identity of the second type) arises
from examining the flow of all two-point, tree level vertices.
This is done by setting $n=0$ in~\eq{eq:WeakFlow}
and specializing $\{f\}$ to contain two fields, 
as shown in \fig{fig:TLTPs}.
We note that we can and do choose
all such vertices to be single supertrace terms~\cite{Thesis,mgierg1}.
\bcf[h]
	\be
		\ensuremath{\begin{array}{c}\input{pstex/Vertex-TLTP-LdL.pstex_t} \end{array}} = \ensuremath{\begin{array}{c}\input{pstex/Dumbbell-S_0-S_0.pstex_t} \end{array}} - \ensuremath{\begin{array}{c}\input{pstex/Dumbbell-S_0-hS_0.pstex_t} \end{array}} - \ensuremath{\begin{array}{c}\input{pstex/Dumbbell-hS_0-S_0.pstex_t} \end{array}}
	\label{eq:TLTP-flow}
	\ee
\caption{Flow of all possible two-point, tree level vertices.}
\label{fig:TLTPs}
\ecf

Following~\cite{ym,ymi,ymii,aprop,Thesis,mgierg1,scalar2} 
we use the freedom inherent in $\hat{S}$ by choosing the two-point, tree
level seed action vertices equal to the corresponding Wilsonian effective
action vertices. Equation~\eq{eq:TLTP-flow} now simplifies.
Rearranging, integrating \wrt\ $\Lambda$ and choosing the appropriate
integration constants~\cite{Thesis,mgierg1}, we arrive at the following
relationship between the integrated ERG kernels---\aka the
effective propagators---and the two-point,
tree level vertices. 
\be
		\ensuremath{\begin{array}{c}\input{pstex/EffPropReln.pstex_t} \end{array}}	= \ensuremath{\begin{array}{c}\begin{picture}(0,0)%
\includegraphics{pstex/K-Delta.pstex}%
\end{picture}%
\setlength{\unitlength}{3947sp}%
\begingroup\makeatletter\ifx\SetFigFont\undefined%
\gdef\SetFigFont#1#2#3#4#5{%
  \reset@font\fontsize{#1}{#2pt}%
  \fontfamily{#3}\fontseries{#4}\fontshape{#5}%
  \selectfont}%
\fi\endgroup%
\begin{picture}(374,395)(1791,-1006)
\put(1791,-843){\makebox(0,0)[lb]{\smash{{\SetFigFont{8}{9.6}{\rmdefault}{\mddefault}{\updefault}{\color[rgb]{0,0,0}$M$}%
}}}}
\end{picture}%
 \end{array}} - \ensuremath{\begin{array}{c}\begin{picture}(0,0)%
\includegraphics{pstex/FullGaugeRemainder.pstex}%
\end{picture}%
\setlength{\unitlength}{3947sp}%
\begingroup\makeatletter\ifx\SetFigFont\undefined%
\gdef\SetFigFont#1#2#3#4#5{%
  \reset@font\fontsize{#1}{#2pt}%
  \fontfamily{#3}\fontseries{#4}\fontshape{#5}%
  \selectfont}%
\fi\endgroup%
\begin{picture}(424,395)(2053,-930)
\put(2053,-773){\makebox(0,0)[lb]{\smash{{\SetFigFont{8}{9.6}{\rmdefault}{\mddefault}{\updefault}{\color[rgb]{0,0,0}$M$}%
}}}}
\end{picture}%
 \end{array}}
							= \ensuremath{\begin{array}{c}\begin{picture}(0,0)%
\includegraphics{pstex/K-Delta.pstex}%
\end{picture}%
\setlength{\unitlength}{3947sp}%
\begingroup\makeatletter\ifx\SetFigFont\undefined%
\gdef\SetFigFont#1#2#3#4#5{%
  \reset@font\fontsize{#1}{#2pt}%
  \fontfamily{#3}\fontseries{#4}\fontshape{#5}%
  \selectfont}%
\fi\endgroup%
\begin{picture}(374,395)(1791,-1006)
\put(1791,-843){\makebox(0,0)[lb]{\smash{{\SetFigFont{8}{9.6}{\rmdefault}{\mddefault}{\updefault}{\color[rgb]{0,0,0}$M$}%
}}}}
\end{picture}%
 \end{array}} - \ensuremath{\begin{array}{c}\begin{picture}(0,0)%
\includegraphics{pstex/DecomposedGR.pstex}%
\end{picture}%
\setlength{\unitlength}{3947sp}%
\begingroup\makeatletter\ifx\SetFigFont\undefined%
\gdef\SetFigFont#1#2#3#4#5{%
  \reset@font\fontsize{#1}{#2pt}%
  \fontfamily{#3}\fontseries{#4}\fontshape{#5}%
  \selectfont}%
\fi\endgroup%
\begin{picture}(540,395)(1936,-925)
\put(1936,-776){\makebox(0,0)[lb]{\smash{{\SetFigFont{8}{9.6}{\rmdefault}{\mddefault}{\updefault}{\color[rgb]{0,0,0}$M$}%
}}}}
\end{picture}%
 \end{array}}
	\label{eq:EPReln-A}
\ee

Note
that we have attached the effective propagator, which only
ever appears as an internal line, to an arbitrary structure.
The field labelled by $M$ can be any of the broken phase
fields. The object $\GR \!\! \equiv \; \GRkpr \!\!\! \GRk$ is a gauge remainder.
The individual components of
$\GRkpr \!\! \GRk$ will often be loosely
referred to as gauge remainders; where it is necessary to
 unambiguously refer to the composite structure, we will use
the terminology `full gauge remainder'.

The various components on the \rhs\ of~\eq{eq:EPReln-A}
can be interpreted, in the different sectors,
according to table~\ref{tab:NFE:k,k'}, where we take
the gauge remainder to carry momentum $p$ (as denoted
by the subscripts carried by the gauge remainder components).
\renewcommand{\arraystretch}{1.5}
\begin{center}
\begin{table}[h]
	\[
	\begin{array}{c|ccc}
					& \delta_{MN}		& \GRkpr_p						& \GRk_p
	\\ \hline 
		F,\bar{F}	& \delta_{MN}		& (f_p p_\mu / \Lambda^2, g_p)	& (p_\nu, 2)
	\\
		A^i			& \delta_{\mu \nu}	& p_\mu / p^2					& p_\nu
	\\
		C^i			& \one				& \mbox{---}					& \mbox{---}
	\end{array}
	\]
\caption{Prescription for interpreting~\eq{eq:EPReln-A}.}
\label{tab:NFE:k,k'}
\end{table}
\end{center}
\renewcommand{\arraystretch}{1}

From~\eq{eq:S_0-11} and \tab{tab:NFE:k,k'}, it follows
that
\be
\label{eq:EP-leading}
	\Delta^{11}_{\rho \sigma}(p) = \frac{\delta_{\rho \sigma}}{2 p^2} + \Op{0},
\ee
which we will need later. 
The functions $f(p^2/\Lambda^2)$ and $g(p^2/\Lambda^2)$ 
need never be exactly determined (though for a concrete
realization, see~\cite{Thesis});
rather, they must satisfy general constraints enforced
by the requirements of
proper UV regularization of the physical $SU(N)$ theory
and gauge invariance. We will see the effects of the
latter shortly.

Generally speaking, when we encounter
components of gauge remainders in complete diagrams,
we will either sum over flavours, or the flavour
will be obvious, from the context. However, there
is one place where we will find it useful to 
explicitly indicate the flavour (and momenta) 
of both $\GRk$ and $\GRkpr$.
Thus we introduce the notation
\be
\label{GR:flavour}
	\GRk^{i}_{p}, \qquad \GRkpr^{j}_q.
\ee
The superscript indices take the values 0 or 1, denoting
a (non-null) bosonic or fermionic gauge remainder, 
respectively. Later, we will find
it useful to form logical expressions with
these indices.

\subsubsection{Primary Diagrammatic Identities of the Third Type}

We recall that the primary diagrammatic identities
of the third type follow directly from
those of the first and second types but are given
explicitly nonetheless due to the central role
they will play in this paper. The first of
these identities
is simply the classical part of~\eq{eq:D-ID-GR-TP-A}:
\be
	\ensuremath{\begin{array}{c}\begin{picture}(0,0)%
\includegraphics{pstex/GR-TLTP.pstex}%
\end{picture}%
\setlength{\unitlength}{3947sp}%
\begingroup\makeatletter\ifx\SetFigFont\undefined%
\gdef\SetFigFont#1#2#3#4#5{%
  \reset@font\fontsize{#1}{#2pt}%
  \fontfamily{#3}\fontseries{#4}\fontshape{#5}%
  \selectfont}%
\fi\endgroup%
\begin{picture}(757,318)(1880,-963)
\put(2296,-857){\makebox(0,0)[lb]{\smash{{\SetFigFont{11}{13.2}{\rmdefault}{\mddefault}{\updefault}{\color[rgb]{0,0,0}$0$}%
}}}}
\end{picture}%
 \end{array}} = 0.
\label{eq:GR-TLTP-A}
\ee

From the effective propagator relation and~\eq{eq:GR-TLTP-A},
two further diagrammatic identities follow.
First, consider attaching
an effective propagator to the right-hand field in~\eq{eq:GR-TLTP-A}
and applying
the effective propagator before $\GRk$ has acted. Diagrammatically,
this gives
\[
	\ensuremath{\begin{array}{c}\begin{picture}(0,0)%
\includegraphics{pstex/GR-TLTP-EP.pstex}%
\end{picture}%
\setlength{\unitlength}{3947sp}%
\begingroup\makeatletter\ifx\SetFigFont\undefined%
\gdef\SetFigFont#1#2#3#4#5{%
  \reset@font\fontsize{#1}{#2pt}%
  \fontfamily{#3}\fontseries{#4}\fontshape{#5}%
  \selectfont}%
\fi\endgroup%
\begin{picture}(1081,306)(2490,-1356)
\put(2776,-1250){\makebox(0,0)[lb]{\smash{\SetFigFont{11}{13.2}{\rmdefault}{\mddefault}{\updefault}{\color[rgb]{0,0,0}0}%
}}}
\end{picture}
 \end{array}} = 0 = \ensuremath{\begin{array}{c}\begin{picture}(0,0)%
\includegraphics{pstex/k.pstex}%
\end{picture}%
\setlength{\unitlength}{3947sp}%
\begingroup\makeatletter\ifx\SetFigFont\undefined%
\gdef\SetFigFont#1#2#3#4#5{%
  \reset@font\fontsize{#1}{#2pt}%
  \fontfamily{#3}\fontseries{#4}\fontshape{#5}%
  \selectfont}%
\fi\endgroup%
\begin{picture}(174,174)(2239,-821)
\end{picture}
 \end{array}} - \ensuremath{\begin{array}{c}\begin{picture}(0,0)%
\includegraphics{pstex/kkprk.pstex}%
\end{picture}%
\setlength{\unitlength}{3947sp}%
\begingroup\makeatletter\ifx\SetFigFont\undefined%
\gdef\SetFigFont#1#2#3#4#5{%
  \reset@font\fontsize{#1}{#2pt}%
  \fontfamily{#3}\fontseries{#4}\fontshape{#5}%
  \selectfont}%
\fi\endgroup%
\begin{picture}(499,174)(2239,-820)
\end{picture}
 \end{array}},
\]
which implies the following diagrammatic identity:
\be
	\ensuremath{\begin{array}{c}\begin{picture}(0,0)%
\includegraphics{pstex/GR-relation.pstex}%
\end{picture}%
\setlength{\unitlength}{3947sp}%
\begingroup\makeatletter\ifx\SetFigFont\undefined%
\gdef\SetFigFont#1#2#3#4#5{%
  \reset@font\fontsize{#1}{#2pt}%
  \fontfamily{#3}\fontseries{#4}\fontshape{#5}%
  \selectfont}%
\fi\endgroup%
\begin{picture}(314,174)(2239,-821)
\end{picture}
 \end{array}} = 1.
\label{eq:GR-relation-A}
\ee
Notice that this means that the $f$ and $g$ of
\tab{tab:NFE:k,k'} are related by
\be
\label{eq:xf+2g}
	x_p f_p + 2g_p = 1,
\ee
where we have defined $x_p = p^2 / \Lambda^2$.

The effective propagator relation, together
with~\eq{eq:GR-relation-A}, implies that
\[
	\ensuremath{\begin{array}{c}\begin{picture}(0,0)%
\includegraphics{pstex/TLTP-EP-GR.pstex}%
\end{picture}%
\setlength{\unitlength}{3947sp}%
\begingroup\makeatletter\ifx\SetFigFont\undefined%
\gdef\SetFigFont#1#2#3#4#5{%
  \reset@font\fontsize{#1}{#2pt}%
  \fontfamily{#3}\fontseries{#4}\fontshape{#5}%
  \selectfont}%
\fi\endgroup%
\begin{picture}(1059,306)(2512,-1356)
\put(2776,-1250){\makebox(0,0)[lb]{\smash{\SetFigFont{11}{13.2}{\rmdefault}{\mddefault}{\updefault}{\color[rgb]{0,0,0}0}%
}}}
\end{picture}
 \end{array}} = \ensuremath{\begin{array}{c}\begin{picture}(0,0)%
\includegraphics{pstex/kpr.pstex}%
\end{picture}%
\setlength{\unitlength}{3947sp}%
\begingroup\makeatletter\ifx\SetFigFont\undefined%
\gdef\SetFigFont#1#2#3#4#5{%
  \reset@font\fontsize{#1}{#2pt}%
  \fontfamily{#3}\fontseries{#4}\fontshape{#5}%
  \selectfont}%
\fi\endgroup%
\begin{picture}(174,174)(2379,-821)
\end{picture}
 \end{array}} - \ensuremath{\begin{array}{c}\begin{picture}(0,0)%
\includegraphics{pstex/kprkkpr.pstex}%
\end{picture}%
\setlength{\unitlength}{3947sp}%
\begingroup\makeatletter\ifx\SetFigFont\undefined%
\gdef\SetFigFont#1#2#3#4#5{%
  \reset@font\fontsize{#1}{#2pt}%
  \fontfamily{#3}\fontseries{#4}\fontshape{#5}%
  \selectfont}%
\fi\endgroup%
\begin{picture}(508,174)(2379,-817)
\end{picture}
 \end{array}} = 0.
\]
In other words, the (non-zero) structure $\ensuremath{\begin{array}{c}\begin{picture}(0,0)%
\includegraphics{pstex/EP-GR.pstex}%
\end{picture}%
\setlength{\unitlength}{3947sp}%
\begingroup\makeatletter\ifx\SetFigFont\undefined%
\gdef\SetFigFont#1#2#3#4#5{%
  \reset@font\fontsize{#1}{#2pt}%
  \fontfamily{#3}\fontseries{#4}\fontshape{#5}%
  \selectfont}%
\fi\endgroup%
\begin{picture}(406,118)(3165,-1234)
\end{picture}
 \end{array}}$ kills
a classical, two-point vertex. But, by~\eq{eq:GR-TLTP-A}, 
this suggests that the structure $\ensuremath{\begin{array}{c}\begin{picture}(0,0)%
\includegraphics{pstex/EP-GR.pstex}%
\end{picture}%
\setlength{\unitlength}{3947sp}%
\begingroup\makeatletter\ifx\SetFigFont\undefined%
\gdef\SetFigFont#1#2#3#4#5{%
  \reset@font\fontsize{#1}{#2pt}%
  \fontfamily{#3}\fontseries{#4}\fontshape{#5}%
  \selectfont}%
\fi\endgroup%
\begin{picture}(406,118)(3165,-1234)
\end{picture}
 \end{array}}$
must be equal, up to some factor, to $\lhd$. Hence, 
the last of
the primary diagrammatic identities is
\be
	\ensuremath{\begin{array}{c}\begin{picture}(0,0)%
\includegraphics{pstex/EP-GRpr.pstex}%
\end{picture}%
\setlength{\unitlength}{3947sp}%
\begingroup\makeatletter\ifx\SetFigFont\undefined%
\gdef\SetFigFont#1#2#3#4#5{%
  \reset@font\fontsize{#1}{#2pt}%
  \fontfamily{#3}\fontseries{#4}\fontshape{#5}%
  \selectfont}%
\fi\endgroup%
\begin{picture}(628,174)(1926,-821)
\end{picture}
 \end{array}} \equiv \ensuremath{\begin{array}{c}\begin{picture}(0,0)%
\includegraphics{pstex/GR-PEP.pstex}%
\end{picture}%
\setlength{\unitlength}{3947sp}%
\begingroup\makeatletter\ifx\SetFigFont\undefined%
\gdef\SetFigFont#1#2#3#4#5{%
  \reset@font\fontsize{#1}{#2pt}%
  \fontfamily{#3}\fontseries{#4}\fontshape{#5}%
  \selectfont}%
\fi\endgroup%
\begin{picture}(786,174)(2089,-821)
\end{picture}
 \end{array}},
\label{eq:PseudoEP-A}
\ee
where the dot-dash line represents the pseudo effective propagators 
of~\cite{Thesis,mgierg1}, also denoted by $\PEP$.

\section{Further Diagrammatics}
\label{sec:Further}

\subsection{Gauge Remainders, Again}
\label{sec:GRs-2}

The various additional properties of gauge remainders
which we require are most readily introduced by example.
To this end, consider the three diagrams shown
in \fig{fig:GR-Ex}.
\bcf[h]
	\[
	\dec{
		\begin{array}{ccc}
			\vspace{1ex}
			\LID{v-W-GR}&		& \LID{v-WGR}
		\\
			\ensuremath{\begin{array}{c}\input{pstex/v-W-GR.pstex_t} \end{array}} & \ds+ 	& \ensuremath{\begin{array}{c}\input{pstex/v-WGR.pstex_t} \end{array}} 
		\end{array}
	}{\{f\}}
	-\hf
	\dec{
		\LIDi[1]{v-W-GR-w}{v-W-GR-w}
	}{\{f\}\Delta}
	\]
\caption{Examples of diagrams possessing a gauge remainder.}
\label{fig:GR-Ex}
\ecf

There are a number of things to note. First, since these
diagrams are used for illustrative purposes, they
are labelled \textbf{I.\#}; for the $\beta_n$ diagrammatics,
the labels will be \textbf{D.\#}. Secondly, $v$ and $w$ are just
vertex arguments, and so take the values of non-negative
integers. Since these arguments do not carry a hat or a bar
(see~\ref{eq:bar}), the vertices belong to the Wilsonian effective action.
Thirdly, the decorations of diagram~\ref{v-W-GR-w}
include not just the set of
fields, $\{f\}$, but also an effective propagator, $\Delta$, which
can be in any sector. The rules for decorating 
diagram~\ref{v-W-GR-w}
with the effective propagator are simple: if each end
attaches to a different object, then the combinatoric factor
of the attachment is two, recognizing that the effective propagator
can attach either way around. If the two ends attach to the same
object, then the combinatoric factor is unity, recognizing that
each vertex represents a sum over all permutations of the fields arranged
into all possible supertrace structures. Whilst the effective
propagator and fields, $\{f\}$, are not explicitly drawn on the
diagrams, they will be referred to as implicit, or unrealized decorations.

Let us now consider what happens when
the gauge remainders in each
of the diagrams~\ref{v-W-GR}--\ref{v-W-GR-w} act. 
Since no other gauge remainders have acted and since
no Taylor expansions have been performed, we can
use \CC\ to collect together the push forward and
the pull back  (see \sec{sec:CC}).
In diagram~\ref{v-W-GR}, the gauge remainder can strike one
of two things: either the field to which the kernel
attaches or one of the fields, $\{f\}$.
It proves to be technically
very convenient not to specify precisely which field
the gauge remainder
hits in the latter case, only that it hits something.
Consequently, we take the gauge remainder to strike a `socket',
which we suppose can be filled by any of the fields, $\{f\}$.
In diagram~\ref{v-WGR}, the gauge remainder can strike
a socket, the top end of the kernel or the bottom end 
of the kernel. In diagram~\ref{v-W-GR-w}, the gauge remainder
can only strike a socket, though we note
that this socket can be filled not only by $\{f\}$ but
also by an end of the effective propagator. \Fig{fig:GR-Ex-P} shows the result
of processing the gauge remainders of 
diagrams~\ref{v-W-GR}--\ref{v-W-GR-w}.
\bcf[h]
	\[	
	\begin{array}{rcl}
	\vspace{2ex}
		\dec{
			\ensuremath{\begin{array}{c}\input{pstex/v-W-GR.pstex_t} \end{array}}	 
		}{\{f\}}
		& =2 & 
		\dec{
			\begin{array}{ccc}
			\vspace{1ex}
				\CID{v-W-GR-a}{v-W-GR-ai}	&	& \CID{v-W-GR-b}{v-W-GR-b2}
			\\
				\ensuremath{\begin{array}{c}\input{pstex/v-W-GR-a.pstex_t} \end{array}}				&-	& \ensuremath{\begin{array}{c}\input{pstex/v-W-GR-b.pstex_t} \end{array}}
			\end{array}
		}{\{f\}}
	\\
	\vspace{2ex}
		\dec{
			\ensuremath{\begin{array}{c}\input{pstex/v-WGR.pstex_t} \end{array}}	 
		}{\{f\}}
		& =-2 &
		\dec{
			\begin{array}{ccccc}
			\vspace{1ex}
				\CID{v-W-GR-ai}{v-W-GR-a}	&	& \CID{v-WGR-b}{v-WGR-b2}	&	& \LID{v-WGR-c}
				\\
				\ensuremath{\begin{array}{c}\input{pstex/v-W-GR-a.pstex_t} \end{array}}				&+	& \ensuremath{\begin{array}{c}\input{pstex/v-WGR-b.pstex_t} \end{array}}				&-	& \ensuremath{\begin{array}{c}\input{pstex/v-WGR-c.pstex_t} \end{array}}
			\end{array} 
		}{\{f\}}
	\\
		-\ds \hf
		\dec{
			\ensuremath{\begin{array}{c}\input{pstex/v-W-GR-w.pstex_t} \end{array}}
		}{\{f\}\Delta}
		& = &
		\dec{
			\LIDi[1]{v-W-GR-w-b}{v-W-GR-w-b}
		}{\{f\}\Delta}
	\end{array}
	\]
\caption{The result of allowing the gauge remainders
of diagrams~\ref{v-W-GR}--\ref{v-W-GR-w} to act. In all
diagrams with the same (opposite) sign to the parent, the
push forward and pull back have been collected into
twice the push forward (pull back).}
\label{fig:GR-Ex-P}
\ecf

Notice that, rather than terminating the pushed forward / 
pulled back field-line with a half arrow (\cf \eq{eq:WID-A}), 
we just utilize the fact
that the corresponding field line already ends 
in a $\GRkpr$ and use this to indicate the field hit.\footnote{
Note that we will not always have a $\GRkpr$ at our disposal.
In particular, we will encounter gauge remainders involving
pseudo effective propagators terminating in just $\GRk$. In
this case, the notation of \sec{sec:WIDs} must be used.
}
Immediately, we find a cancellation, which we indicate by
enclosing the reference number of the cancelled diagram
in curly brackets, together with the reference number
diagram against
which it cancels.

\ICancelCom{v-W-GR-ai}{v-W-GR-a}{. Although these diagrams
look exactly the same, one might worry that they are
different: in diagram~\ref{v-W-GR-ai}, the gauge remainder
pulls back along the kernel, whereas in diagram~\ref{v-W-GR-a}
the gauge remainder has nothing to do with the kernel, instead
pushing forward around the vertex. However, gauge invariance
ensures that these two diagrams are indeed 
equivalent~\cite{Thesis,mgierg1}, and so they cancel.}

Returning to \fig{fig:GR-Ex-P}, consider next
diagram~\ref{v-WGR-c}. The line segment
which joins the top
of the kernel to the $\GRkpr$---thereby forming a `hook'---performs 
no role other 
than to make this join. In other words, it is neither a section of 
the kernel nor an effective propagator. We could imagine
deforming this line segment so that the hook becomes arbitrarily large. 
Despite
appearances, we must always remember 
that this line segment simply performs the role
of a Kronecker delta. When part of a complete diagram, this line
segment can always be distinguished from an effective propagator,
to which it can be made to look identical, by the context. This
follows because hooks in which the line segment is a Kronecker~$\delta$
only ever attach to effective propagators or kernels, whereas 
hook-like structures
made out of an effective propagator only ever attach to vertices
(this will be particularly clear from the perspective
of \sec{sec:beta_n}).
When viewed in isolation, we will always take the hook structure to
comprise just a line segment and so will draw the hook as tightly as
possible.

To conclude our discussion of the hook, we give its algebraic
form. First, we note from table~\ref{tab:NFE:k,k'} that
the gauge
remainder that forms the hook must be in the $D$ $(\bar{D})$ sector,
else the loop integral over its momenta is odd and vanishes. 
Consequently, 
the field to which the hook attaches must be
in the  $C^1$ ($C^2$) sector; a conclusion which could
also have been drawn from consideration of \CC\ invariance.
Thus, in the former (latter)
case, the gauge remainder of the parent, which is in the
$F$ ($\bar{F}$) sector, strikes the $F$ ($\bar{F}$) at
the top of the kernel, turning it into a $C^1$ ($-C^2$).
The sign in the latter case essentially arises from
the minus sign in fifth component of $\bar{F}_N$ (see
\eqs{eq:F}{eq:Fbar}) (see~\cite{Thesis,mgierg1} for 
painstaking detail). Our prescription is to absorb
any signs associated with the end of the kernel
into the definition of the hook.

The final ingredient we need to obtain the algebraic form for the hook is to
realize that the inside of the hook constitutes an empty loop and so
gives a group theory factor of $\pm N$, depending on flavour. Thus we have:
\numparts
\bea
\label{eq:hook-R-1}
	\ensuremath{\begin{array}{c}\begin{picture}(0,0)%
\includegraphics{pstex/hook-R-1.pstex}%
\end{picture}%
\setlength{\unitlength}{3947sp}%
\begingroup\makeatletter\ifx\SetFigFont\undefined%
\gdef\SetFigFont#1#2#3#4#5{%
  \reset@font\fontsize{#1}{#2pt}%
  \fontfamily{#3}\fontseries{#4}\fontshape{#5}%
  \selectfont}%
\fi\endgroup%
\begin{picture}(153,283)(159,346)
\put(159,346){\makebox(0,0)[lb]{\smash{\SetFigFont{8}{9.6}{\rmdefault}{\mddefault}{\updefault}{\color[rgb]{0,0,0}1}%
}}}
\end{picture}
 \end{array}} & = & +(-N) \int_l g_l 
\\
\label{eq:hook-R-2}
	\ensuremath{\begin{array}{c}\begin{picture}(0,0)%
\includegraphics{pstex/hook-R-2.pstex}%
\end{picture}%
\setlength{\unitlength}{3947sp}%
\begingroup\makeatletter\ifx\SetFigFont\undefined%
\gdef\SetFigFont#1#2#3#4#5{%
  \reset@font\fontsize{#1}{#2pt}%
  \fontfamily{#3}\fontseries{#4}\fontshape{#5}%
  \selectfont}%
\fi\endgroup%
\begin{picture}(153,283)(159,346)
\put(159,346){\makebox(0,0)[lb]{\smash{\SetFigFont{8}{9.6}{\rmdefault}{\mddefault}{\updefault}{\color[rgb]{0,0,0}2}%
}}}
\end{picture}
 \end{array}} & = & - (N) \int_l g_l.
\eea
\endnumparts
The numbers at the base of the hook in the above equations indicate the sector
of the $C^i$ field to which the hooks attach.

We now see a further advantage to having absorbed signs into the definition of
the hook: we can trivially extend the diagrammatic effect of \CC\ to cover
its action on a disconnected hook.
Quite simply, we have that
\begin{equation}
	\ensuremath{\begin{array}{c}\begin{picture}(0,0)%
\includegraphics{pstex/hook-R.pstex}%
\end{picture}%
\setlength{\unitlength}{3947sp}%
\begingroup\makeatletter\ifx\SetFigFont\undefined%
\gdef\SetFigFont#1#2#3#4#5{%
  \reset@font\fontsize{#1}{#2pt}%
  \fontfamily{#3}\fontseries{#4}\fontshape{#5}%
  \selectfont}%
\fi\endgroup%
\begin{picture}(130,178)(182,451)
\end{picture}
 \end{array}} = - \ensuremath{\begin{array}{c}\begin{picture}(0,0)%
\includegraphics{pstex/hook-L.pstex}%
\end{picture}%
\setlength{\unitlength}{3947sp}%
\begingroup\makeatletter\ifx\SetFigFont\undefined%
\gdef\SetFigFont#1#2#3#4#5{%
  \reset@font\fontsize{#1}{#2pt}%
  \fontfamily{#3}\fontseries{#4}\fontshape{#5}%
  \selectfont}%
\fi\endgroup%
\begin{picture}(130,178)(288,451)
\end{picture}
 \end{array}},
\label{eq:hookCC}
\end{equation}
consistent with our previous definition that we take the mirror image, picking
up a sign for every performed gauge remainder.

Now we focus on diagram~\ref{v-W-GR-w-b}. This diagram,
like~\ref{v-W-GR-b} and~\ref{v-WGR-b}, possesses a socket;
we will now see why it is so useful to leave this socket
empty. To process this diagram further, we suppose that
the vertex with argument $w$ possesses a classical, two-point
component. Of course, in this case, this means that
$w$ must be a classical vertex. However, in the computation
of $\beta_n$, we will generally be summing over vertex arguments,
such that $w$ could take all values between zero and some
positive integer. In preparation for this, we  leave the
vertex argument as $w$ rather than explicitly writing it
as zero. To separate off the classical, two-point component
of $w$, we define the reduction of a vertex to be
the full vertex minus its classical, two-point component
(should this component exist). The reduction is denoted
by appending the appropriate vertex argument with a superscript $R$
\viz $w^R$. This decomposition of diagram~\ref{v-W-GR-w-b}
is shown in \fig{fig:Decompose}.
\bcf[h]
	\[
	\dec{
		\ensuremath{\begin{array}{c}\input{pstex/v-W-GR-w-b.pstex_t} \end{array}}
	}{\{f\}\Delta}
	=
	\dec{
		\begin{array}{ccc}
		\vspace{1ex}
			\LID{v-W-GR-w-bR}	&	& \LID{v-W-GR-w-b02}
		\\
			\ensuremath{\begin{array}{c}\input{pstex/v-W-GR-w-bR.pstex_t} \end{array}}	& + & \ensuremath{\begin{array}{c}\input{pstex/v-W-GR-w-b02.pstex_t} \end{array}}
		\end{array}
	}{\{f\}\Delta}
	\]
\caption{Decomposing the vertex of diagram~\ref{v-W-GR-w-b}
(which in this case is implicitly a classical vertex)
into a reduced part and a classical, two-point part.}
\label{fig:Decompose}
\ecf

The top vertex of diagram~\ref{v-W-GR-w-bR}
is reduced. The vertex argument $0^2$ of 
diagram~\ref{v-W-GR-w-b02} tells us that this
vertex must not only be classical but must also
possess precisely two decorations. Thus in addition
to the socket, this vertex must have one and only
one additional decoration. Clearly, this additional
decoration can either be one of the fields $\{f\}$
or one of the ends of the effective propagator. 
The
former case will not concern us here, so we focus
on the latter case. Given that one end of the effective
propagator decorates the classical, two-point vertex,
there are three locations to which the loose end can attach:
\begin{enumerate}
	\item	the socket;
	\label{it:socket}

	\item	the vertex with argument $v$.
	\label{it:vertex}

	\item	the kernel;
	\label{it:kernel}
\end{enumerate}

The first case vanishes by \CC\ invariance. 
To see this,
consider the parent of diagram~\ref{v-W-GR-w-b},
diagram~\ref{v-W-GR-w}. To generate the diagram
corresponding to~\ref{it:socket}, we should
consider the component of diagram~\ref{v-W-GR-w}
in which
the top vertex is a three-point vertex, 
with one of the fields corresponding to the gauge
remainder and the other two corresponding to the
two ends of the effective propagator. Now,
since net fermionic vertices vanish, the gauge
remainder must be bosonic. Furthermore, by \CC\ invariance,
the gauge remainder is forced to be in the $C^{1,2}$
sector, in which it is null (see \sec{sec:CC}).

In both cases~\ref{it:vertex} and~\ref{it:kernel},
we can employ the effective propagator relation.
This is shown in \fig{fig:EP-Ex}. We note that
following. First,
using the effective propagator to join the
classical, two-point vertex to some other object
yields a combinatoric factor of two. Secondly,
when we henceforth talk
of decorating a classical, two-point
vertex we always mean that we are explicitly
drawing on the second decoration of the vertex,
the socket being the first decoration. If we
mean to fill the socket, this will be explicitly
stated.
\bcf[h]
	\beas
	\dec{
		\ensuremath{\begin{array}{c}\input{pstex/v-W-GR-w-b02.pstex_t} \end{array}}
	}{\{f\} \Delta}
	& = &
	2
	\dec{
		\begin{array}{ccc}
		\vspace{1ex}
			\LID{v-W-GR-w-b02-v}&	& \LID{v-W-GR-w-b02-k}
		\\
			\ensuremath{\begin{array}{c}\input{pstex/v-W-GR-w-b02-v.pstex_t} \end{array}}	&+	& \ensuremath{\begin{array}{c}\input{pstex/v-W-GR-w-b02-k.pstex_t} \end{array}}
		\end{array}
	}{\{f\}} + \cdots
	\\
	& = &
	2
	\dec{
		\begin{array}{ccc}
		\vspace{1ex}
			\CID{v-W-GR-b2}{v-W-GR-b}	&	& \LID{v-W-nGR}
		\\
			\ensuremath{\begin{array}{c}\input{pstex/v-W-GR-b.pstex_t} \end{array}}				& - & \ensuremath{\begin{array}{c}\input{pstex/v-W-nGR.pstex_t} \end{array}}
		\\
		\vspace{1ex}
			\CID{v-WGR-b2}{v-WGR-b}		&	& \LID{v-WnGR}
		\\
			\ensuremath{\begin{array}{c}\input{pstex/v-WGR-b.pstex_t} \end{array}}				& - & \ensuremath{\begin{array}{c}\input{pstex/v-WnGR.pstex_t} \end{array}}
		\end{array}
	}{\{f\}} + \cdots
	\eeas
\caption{The decoration of the classical, two-point vertex
of diagram~\ref{v-W-GR-w-b02} by the effective propagator,
together with the subsequent application of the
effective propagator relation. The ellipses
denote diagrams in which the classical, two-point
vertex is decorated by one of the fields, $\{f\}$, rather
than by the effective propagator.}
\label{fig:EP-Ex}
\ecf

\ICancel{v-W-GR-b2}{v-W-GR-b}

\ICancel{v-WGR-b2}{v-WGR-b}	

There are some comments worth making about
diagrams~\ref{v-W-nGR} and~\ref{v-WnGR}.
First, 
the $\GRkpr$
part of the full gauge remainder plays the role of
the socket in the parent diagrams. 
Secondly, these diagrams
have a very similar
structures to diagrams~\ref{v-W-GR} and~\ref{v-WGR}.
The only difference is that the full gauge remainder
in the new diagrams is nested~\cite{Thesis}, meaning
that it does not attach directly to the kernel, but
instead is hit by the gauge remainder component which
ends the kernel.
Processing nested gauge remainders
is much the same as processing un-nested gauge
remainders, but \CC\ invariance cannot generally be used
again to collect together the 
nested push forward and pull back~\cite{Thesis,Primer}:
we must count them separately. However, everything
else goes through as before and so we will find that
cancellation~\ref{Icancel:v-W-GR-ai} is repeated
twice (twice because we count the nested push forward
separately from the nested pull back).

If there were more vertices available, we could imagine
iterating the above procedure and thus generating
arbitrarily nested gauge remainders.
It would then be useful for us to know if there is 
any simple way of
keeping track of which gauge remainders have pushed 
forward and which have pulled back.
With sufficient thought, it is always 
possible to stare at a complicated
diagram and deduce the pattern of pushes 
forward and pulls back
(up to the ambiguity associated with illustrative 
cancellation~\ref{Icancel:v-W-GR-ai}). However, there is
an easier way. Note in diagrams~\ref{v-W-nGR} and~\ref{v-WnGR}
that the $\GRkpr$ part of the full gauge remainder 
is bitten on its right-hand edge by the $\GRkpr$ which 
terminates the kernel. (We define the right-hand edge
to be the edge on our right as we traverse an imaginary
line running into the socket at the apex \viz 
$\left.^{\scriptscriptstyle \mathrm{L}}_{\scriptscriptstyle \mathrm{R}} \ensuremath{\begin{array}{c}\begin{picture}(0,0)%
\includegraphics{pstex/EP-GR.pstex}%
\end{picture}%
\setlength{\unitlength}{3947sp}%
\begingroup\makeatletter\ifx\SetFigFont\undefined%
\gdef\SetFigFont#1#2#3#4#5{%
  \reset@font\fontsize{#1}{#2pt}%
  \fontfamily{#3}\fontseries{#4}\fontshape{#5}%
  \selectfont}%
\fi\endgroup%
\begin{picture}(406,118)(3165,-1234)
\end{picture}
 \end{array}}\right.$;
likewise for sockets on vertices / kernels.)
Had we pushed forward, rather than pulled 
back with the initial
gauge remainder, then this bite would have been to the left. 
With this in mind, consider a string
of gauge remainders which bite a socket 
decorating either a kernel or a vertex; the latter
case is shown in \fig{fig:GRstring}.
\bcf[h]
	\[
	\LIDi{v-GRstring}{v-GRstring}
	\]
\caption{An arbitrarily nested gauge remainder bites a socket on a vertex.}
\label{fig:GRstring}
\ecf

We know from the discussion in illustrative 
cancellation~\ref{Icancel:v-W-GR-ai} that,
were we to fill the socket of diagram~\ref{v-GRstring}
with the end of a kernel (the other end of which
attaches to the gauge remainder at the beginning
of the string), then the sense in which the socket
is bitten can be interpreted in two ways:
either as a push forward around the vertex or
as a pull back along the kernel. Given a diagram like~\ref{v-GRstring},
we will always
interpret the gauge remainder at the end of the string of
gauge remainders as having
bitten the vertex, using this to determine the sense
in which the gauge remainder acts.

To determine
in which sense the nested gauge remainders are bitten
we can equate bites
on the left with pushes forward and bites on the right with pulls back,
the only exception being when a gauge remainder bites
the end of kernel to form a hook: the gauge remainder
that pushed forward to form
the un-nested hook of diagram~\ref{v-WGR-c}
can be thought of as biting itself on the right.
Thus, moving on to consider
the arbitrarily nested hook shown in \fig{fig:nestedhook},
the number of pushes forward is equal to the number of 
bites on the left, plus one, and the number of pulls
back is equal to the number of bites on the right, minus one.
\bcf[h]
	\[
	\ensuremath{\begin{array}{c}\begin{picture}(0,0)%
\includegraphics{pstex/Struc-GR-ring.pstex}%
\end{picture}%
\setlength{\unitlength}{3947sp}%
\begingroup\makeatletter\ifx\SetFigFont\undefined%
\gdef\SetFigFont#1#2#3#4#5{%
  \reset@font\fontsize{#1}{#2pt}%
  \fontfamily{#3}\fontseries{#4}\fontshape{#5}%
  \selectfont}%
\fi\endgroup%
\begin{picture}(353,319)(48,44)
\end{picture}
 \end{array}}
	\]
\caption{An arbitrarily nested version of the hook.}
\label{fig:nestedhook}
\ecf

Structures like those in \fig{fig:GRstring}
have a compact representation, shown below:
\[
	\frac{1}{m!}\ensuremath{\begin{array}{c}\input{pstex/compact.pstex_t} \end{array}} \ .
\]

The notation $\GRkpr^m$ stands for $m$ instances of $\GRkpr$.
These gauge remainder
components form a string which bites the socket. We sum over
all possible ways in which each gauge remainder can bite,
and so diagram~\ref{v-GRstring} is just one component included
our compact diagrammatic representation. When going from
the compact diagrammatic representation to its explicit 
components, we note that any of the $m$ gauge remainders
can bite the socket, and this gauge remainder can, in turn,
be bitten by any of the remaining $m-1$ gauge remainders.
This explains the normalization factor of $1/m!$.
Denoting the number of bites to the left / right by L / R,
the sign of each explicitly drawn diagram is just $(-1)^R$.

Similarly, there is a compact diagrammatic representation	
of the nested hook---or ring---of \fig{fig:nestedhook}, 
which is simply
\[
	\frac{1}{(m-1)!} \decGR{ \ }{\GRkpr^m}.
\] 
The rule to go from this compact
representation to explicit diagrams is as follows:
we draw the set of rings corresponding to
all independent sequences bites to the left / right.
The normalization factor follows 
from noting that a ring is invariant
under a cyclic permutations of the gauge remainders.
The sign of each explicitly drawn ring is $(-1)^{R-1}$.


We conclude this discussion of the gauge 
remainders by examining diagrams possessing
two active gauge remainders. The techniques we use to process such terms are
exactly the same as  the ones detailed already in this section. There is,
however, a qualitatively new type of diagram that arises, 
together with a source
of possible confusion. To investigate both of these issues, we focus on an
example
in which two full gauge remainders bite a kernel\footnote{
Double gauge remainder diagrams are not restricted to those in which both gauge
remainders bite a kernel; one or both of the gauge remainders can bite a vertex.},
as shown in \fig{fig:DoubleGR}.
\bcf[h]
	\[
	\hf
	\LIDi[1]{Struc-WGRx2-B}{Diags-WGRx2-B}
	\]
\caption{An example of a double gauge remainder diagram.}
\label{fig:DoubleGR}
\ecf

We proceed by allowing first one gauge remainder to act and then, if possible,
allowing the second gauge remainder to act. The qualitatively new type of diagram
arises because one of the
effects of the first gauge remainder can be to `trap' the second gauge remainder,
by biting the field on the kernel to which the second gauge remainder attaches.
The remaining full gauge remainder now does 
not have the same momentum flowing through
it as the field it is trying to bite, and so it cannot act.
The other diagrams generated are those in
which the processed gauge remainder bites one of the
ends of the kernel. The result of allowing the first gauge remainder to
act is shown in \fig{fig:DoubleGR-A}, where we have collected pulls
back and pushes forward, as usual.
\bcf[h]
	\[
	\begin{array}{ccccc}
		\LID{Diags-WGRx2-Top}	&	& \LID{Diags-WGRx2-Trap}	&	& \LID{Diags-WGRx2-Bottom}
	\\[1ex]
		\ensuremath{\begin{array}{c}\begin{picture}(0,0)%
\includegraphics{pstex/Struc-WGRx2-Top.pstex}%
\end{picture}%
\setlength{\unitlength}{3947sp}%
\begingroup\makeatletter\ifx\SetFigFont\undefined%
\gdef\SetFigFont#1#2#3#4#5{%
  \reset@font\fontsize{#1}{#2pt}%
  \fontfamily{#3}\fontseries{#4}\fontshape{#5}%
  \selectfont}%
\fi\endgroup%
\begin{picture}(296,662)(501,545)
\end{picture}
 \end{array}}	& -	& \ensuremath{\begin{array}{c}\begin{picture}(0,0)%
\includegraphics{pstex/Struc-WGRx2-Trap.pstex}%
\end{picture}%
\setlength{\unitlength}{3947sp}%
\begingroup\makeatletter\ifx\SetFigFont\undefined%
\gdef\SetFigFont#1#2#3#4#5{%
  \reset@font\fontsize{#1}{#2pt}%
  \fontfamily{#3}\fontseries{#4}\fontshape{#5}%
  \selectfont}%
\fi\endgroup%
\begin{picture}(415,801)(-354,391)
\end{picture}
 \end{array}}	& +	& \ensuremath{\begin{array}{c}\begin{picture}(0,0)%
\includegraphics{pstex/Struc-WGRx2-Bottom.pstex}%
\end{picture}%
\setlength{\unitlength}{3947sp}%
\begingroup\makeatletter\ifx\SetFigFont\undefined%
\gdef\SetFigFont#1#2#3#4#5{%
  \reset@font\fontsize{#1}{#2pt}%
  \fontfamily{#3}\fontseries{#4}\fontshape{#5}%
  \selectfont}%
\fi\endgroup%
\begin{picture}(540,739)(370,442)
\end{picture}
 \end{array}}
	\end{array}
	\]
\caption{Result of allowing one gauge remainder in diagram~\ref{Diags-WGRx2-B} to act.}
\label{fig:DoubleGR-A}
\ecf

Diagram~\ref{Diags-WGRx2-Trap} possesses the 
trapped gauge remainder. In diagram~\ref{Diags-WGRx2-Bottom}
we have recognized that the kernel ends where it attaches 
to the active gauge remainder
and so this is where the processed gauge remainder bites. 
Note that
we can trivially redraw this diagram, as shown in \fig{fig:DoubleGR-B}.
\bcf[h]
	\[
	\LIDi[1]{Struc-WGRx2-Bottom-B}{Diags-WGRx2-Bottom-B}
	\]
\caption{A trivial redrawing of diagram~\ref{Diags-WGRx2-Bottom}.}
\label{fig:DoubleGR-B}
\ecf

Diagram~\ref{Diags-WGRx2-Bottom-B} highlights how one must be 
very careful when drawing
which end of an active gauge remainder a 
processed gauge remainder bites; diagrams~\ref{Diags-WGRx2-Bottom-B}
and~\ref{Diags-WGRx2-Trap} are clearly different. Notice, however, that if
the active gauge remainder were \emph{absent}, then
diagram~\ref{Diags-WGRx2-Bottom}
would cancel diagram~\ref{Diags-WGRx2-Trap}; we will exploit this
later.

Finally, consider allowing the active gauge remainder to 
act in diagrams~\ref{Diags-WGRx2-Top}
and~\ref{Diags-WGRx2-Bottom-B}. We are not 
interested in all the contributions. Rather,
we just want to focus on
\begin{enumerate}
	\item	the term produced by diagram~\ref{Diags-WGRx2-Top} where the
			gauge remainder pulls back along the kernel, to the same end as the hook;

	\item	the term produced by diagram~\ref{Diags-WGRx2-Bottom-B} in which the
			gauge remainder pulls back to the bottom of the kernel.
\end{enumerate}
Reflecting the former diagram about a horizontal line, we arrive at the two diagrams
of \fig{fig:DoubleGR-C}. (The reflection does not yield a net sign as
we pick up one for each of the processed gauge remainders.) 
\bcf[h]
	\[
	\begin{array}{cccc}
			& \LID{Diags-WGRx2-L-Bottom}&	& \LID{Diags-WGR-N1-Bottom-PF}
	\\[1ex]
		-	& \ensuremath{\begin{array}{c}\begin{picture}(0,0)%
\includegraphics{pstex/Struc-WGRx2-L-Bottom.pstex}%
\end{picture}%
\setlength{\unitlength}{3947sp}%
\begingroup\makeatletter\ifx\SetFigFont\undefined%
\gdef\SetFigFont#1#2#3#4#5{%
  \reset@font\fontsize{#1}{#2pt}%
  \fontfamily{#3}\fontseries{#4}\fontshape{#5}%
  \selectfont}%
\fi\endgroup%
\begin{picture}(302,549)(236,111)
\end{picture}
 \end{array}}	& -	& \ensuremath{\begin{array}{c}\begin{picture}(0,0)%
\includegraphics{pstex/Struc-WGR-N1-Bottom-PF.pstex}%
\end{picture}%
\setlength{\unitlength}{3947sp}%
\begingroup\makeatletter\ifx\SetFigFont\undefined%
\gdef\SetFigFont#1#2#3#4#5{%
  \reset@font\fontsize{#1}{#2pt}%
  \fontfamily{#3}\fontseries{#4}\fontshape{#5}%
  \selectfont}%
\fi\endgroup%
\begin{picture}(362,718)(428,481)
\end{picture}
 \end{array}}
	\end{array}
	\]
\caption{Two of the terms produced by processing diagrams~\ref{Diags-WGRx2-Top}
and~\ref{Diags-WGRx2-Bottom-B}.}
\label{fig:DoubleGR-C}
\ecf

From the way in which these two diagrams have been drawn, it is clear that
they are distinct and that they must be treated as such. We can, however, redraw
diagram~\ref{Diags-WGR-N1-Bottom-PF} by 
sliding the outer gauge remainder round the hook
to where the inner gauge remainder bites the kernel. This is shown in 
\fig{fig:DoubleGR-D}.
\bcf[h]
	\[
	-\LIDi[1]{Struc-WGR-N1-Bottom-PF-B}{Diags-WGR-N1-Bottom-PF-B}
	\]
\caption{A trivial redrawing of diagram~\ref{Diags-WGR-N1-Bottom-PF}.}
\label{fig:DoubleGR-D}
\ecf

Diagram~\ref{Diags-WGR-N1-Bottom-PF-B} is, of course, still the same as
diagram~\ref{Diags-WGR-N1-Bottom-PF} but it is starting to look very similar
to diagram~\ref{Diags-WGRx2-L-Bottom}. Indeed, we must make sure that we never
slide the gauge remainder so far round the hook that it appears to bite the kernel
at the same point as the gauge remainder which forms the hook. If this
were to happen, then such a diagram would be ambiguous.

\subsection{The Secondary Diagrammatic Identities}
\label{sec:Secondary}

There are two families of secondary diagrammatic identities.
Members of the first family are
generally applicable, whereas
 members of
the second family are
applicable only to diagrams which have
been manipulated at $\Op{2}$.

\subsubsection{The First Family}
\label{sec:D-ID-Secondary-I}

The first diagrammatic identity in this
family is trivial:
\be
\label{D-ID-Trivial-A}
	\cdeps{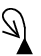} = 0.
\ee
This follows directly from \CC\ invariance:
the field attaching to the hook must
be in the $C^i$ sector, but in this sector
the gauge remainder contracted into the
hook is null. Before moving on, notice
that the diagram contains one empty socket,
corresponding to the $\GRkpr$ part of the
full gauge remainder.

There is a second trivial identity
involving a diagram possessing a single
socket, which we have seen already. We
re-draw it here in what will become a
suggestive form:
\be
\label{eq:D-ID-Trivial-PEP}
	\ensuremath{\begin{array}{c}\begin{picture}(0,0)%
\includegraphics{pstex/TLTP-Soc-EP-GR.pstex}%
\end{picture}%
\setlength{\unitlength}{3947sp}%
\begingroup\makeatletter\ifx\SetFigFont\undefined%
\gdef\SetFigFont#1#2#3#4#5{%
  \reset@font\fontsize{#1}{#2pt}%
  \fontfamily{#3}\fontseries{#4}\fontshape{#5}%
  \selectfont}%
\fi\endgroup%
\begin{picture}(474,497)(1489,-1334)
\put(1567,-1183){\makebox(0,0)[lb]{\smash{{\SetFigFont{11}{13.2}{\rmdefault}{\mddefault}{\updefault}{\color[rgb]{0,0,0}$0^2$}%
}}}}
\end{picture}%
 \end{array}} = 0,
\ee
recalling that this identity follows as
a consequence of the effective propagator
relation and~\eq{eq:GR-relation}. We
explicitly indicate that the vertex is
a two-point vertex, since such structures
will generically 
occur as sub-diagrams in diagrams
with unrealized decorations.

To uncover the next diagrammatic identity,
which is the first we will encounter involving two
sockets, 
consider the pair of diagrams in
\fig{fig:D-ID-Trivial-b}.
\bcf[h]
	\[
	\begin{array}{ccc}
	\vspace{1ex}
		\LID{Trivial-b-A}	&	& \LID{Trivial-b-A2}	
	\\
		\ensuremath{\begin{array}{c}\begin{picture}(0,0)%
\includegraphics{pstex/Trivial-b-A.pstex}%
\end{picture}%
\setlength{\unitlength}{3947sp}%
\begingroup\makeatletter\ifx\SetFigFont\undefined%
\gdef\SetFigFont#1#2#3#4#5{%
  \reset@font\fontsize{#1}{#2pt}%
  \fontfamily{#3}\fontseries{#4}\fontshape{#5}%
  \selectfont}%
\fi\endgroup%
\begin{picture}(689,483)(254,547)
\put(254,881){\makebox(0,0)[lb]{\smash{{\SetFigFont{8}{9.6}{\rmdefault}{\mddefault}{\updefault}{\color[rgb]{0,0,0}$X$}%
}}}}
\end{picture}%
 \end{array}} 	& -	& \ensuremath{\begin{array}{c}\begin{picture}(0,0)%
\includegraphics{pstex/Trivial-b-A2.pstex}%
\end{picture}%
\setlength{\unitlength}{3947sp}%
\begingroup\makeatletter\ifx\SetFigFont\undefined%
\gdef\SetFigFont#1#2#3#4#5{%
  \reset@font\fontsize{#1}{#2pt}%
  \fontfamily{#3}\fontseries{#4}\fontshape{#5}%
  \selectfont}%
\fi\endgroup%
\begin{picture}(793,586)(354,497)
\put(932,528){\makebox(0,0)[lb]{\smash{{\SetFigFont{8}{9.6}{\rmdefault}{\mddefault}{\updefault}{\color[rgb]{0,0,0}$X$}%
}}}}
\end{picture}%
 \end{array}} 	
	\end{array}
	\]
\caption{A pair of diagrams whose sum vanishes.}
\label{fig:D-ID-Trivial-b}
\ecf
These
diagrams naturally arise from 
a common parent: the differences between
the two diagrams come from the sense in
which the full gauge remainder of
the parent, of which 
the component labelled $X$ is left behind, acts.
The point now is that diagrams~\ref{Trivial-b-A}
and~\ref{Trivial-b-A2} cancel. By construction,
the unspecified structure at the bottom of both
diagrams is common. Furthermore, the gauge
remainder labelled $X$ and its attachment to
the common structure is the same between the
two diagrams. From~\eqs{eq:hook-R-1}{eq:hook-R-2},
the two hooks yield the same algebraic contribution.
The only difference between the two diagrams is
the sense in which the gauge remainder labelled $X$
acts. However, effectively
all that this gauge remainder does is determine
whether the hook is of type~\eq{eq:hook-R-1} 
or~\eq{eq:hook-R-2}; and we know that the hook is
just a common algebraic factor between the two diagrams.
Therefore, diagrams~\ref{Trivial-b-A}
and~\ref{Trivial-b-A2} are in fact identical up to a
sign and so cancel.

This leads us to the following diagrammatic
identity:
\be
\label{eq:D-ID-Bitten-hook-A}
	\cdeps{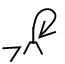} - \cdeps{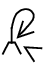} \equiv 0,
\ee
where we understand that this diagram should be viewed
as part of some larger diagram, attached via internal
fields. 
Note that in this diagrammatic
identity, and all that follow, we can strip off
the common $\GRkpr$ which attaches to some
other part of a larger diagram, replacing it with
the half arrow notation of~\eq{eq:WID-A}.

The non-trivial diagrammatic identities which now follow can
be thought of as arising from versions 
of~\eq{D-ID-Trivial-A} and~\eq{eq:D-ID-Bitten-hook-A}
in which the hook is nested and versions of~\eq{eq:D-ID-Trivial-PEP}
in which the gauge remainder is nested. To be
specific, we will find relationships between
the nested versions of~\eq{D-ID-Trivial-A}--\eq{eq:D-ID-Bitten-hook-A}
for which the numbers of sockets are equal; equivalently,
these relationships are between diagrams with an
equal number of \emph{performed} gauge remainders.

The key to proving the non-trivial diagrammatic
identities resides in the following relationship
between gauge remainder components. Using the notation
described around~\eq{GR:flavour}, and taking $j_2 = j_1 \ \xor j_3$
we have:
\be
\label{eq:Template}
	1 - \GRk^{j_1}_{l-m_1-k} \GRkpr^{j_2}_{l-m_1} - \GRk^{j_3}_k \GRkpr^{j_2}_{l-m_1} = 0.
\ee	
To demonstrate this, we simply substitute for the
gauge remainder components using \tab{tab:NFE:k,k'}
and employ~\eq{eq:xf+2g}, as shown in \tab{tab:template}.
\renewcommand{\arraystretch}{1.5}
\bct[h]
	\[
	\begin{array}{cc|c|rcl}
		j_1	& j_3	& j_2	& 	
	\\ \hline
		0	& 0		& 0		& 1 - \frac{(l-m_1-k)\cdot(l-m_1)}{(l-m_1)^2} - \frac{k \cdot (l-m_1)}{(l-m_1)^2} & = & 0
	\\
		0	& 1		& 1		& 1 - f_{l-m_1} \frac{(l-m_1-k)\cdot(l-m_1)}{\Lambda^2} - f_{l-m_1} \frac{k\cdot(l-m_1)}{\Lambda^2} - 2g_{l-m_1} & = & 0
	\\
		1	& 0		& 1		& 1 - f_{l-m_1} \frac{(l-m_1-k)\cdot(l-m_1)}{\Lambda^2} - 2g_{l-m_1} - f_{l-m_1} \frac{k\cdot(l-m_1)}{\Lambda^2} & = & 0
	\\
		1	& 1		& 0		& 1 - \frac{(l-m_1-k)\cdot(l-m_1)}{(l-m_1)^2} - \frac{k \cdot (l-m_1)}{(l-m_1)^2} & = & 0
	\end{array}
	\]
\label{tab:template}
\caption{Break down of~\eq{eq:Template} into the different
sectors.}
\ect
\renewcommand{\arraystretch}{1}

The first non-trivial relationship involves versions 
of~\eq{D-ID-Trivial-A}--\eq{eq:D-ID-Bitten-hook-A} with
two sockets. This identity is, in some sense, still a
special case, since the two-socket version 
of~\eq{eq:D-ID-Bitten-hook-A} is of course trivial.
Consequently, we treat the two-socket case---which is
shown in \fig{fig-D-ID-2}---separately from the rest. 
\bcf[h]
	\be
	\label{eq:D-ID-2}
		\begin{array}{ccccccccc}
		\vspace{1ex}
				& \LID{TLTP-EP-LL}	&	& \LID{GR-ring-RR}	&	& \LID{TLTP-EP-RR}	&	& \LID{GR-ring-LL}	&
		\\
		\vspace{1ex}
				& \ensuremath{\begin{array}{c}\begin{picture}(0,0)%
\includegraphics{pstex/TLTP-EP-LL.pstex}%
\end{picture}%
\setlength{\unitlength}{3947sp}%
\begingroup\makeatletter\ifx\SetFigFont\undefined%
\gdef\SetFigFont#1#2#3#4#5{%
  \reset@font\fontsize{#1}{#2pt}%
  \fontfamily{#3}\fontseries{#4}\fontshape{#5}%
  \selectfont}%
\fi\endgroup%
\begin{picture}(474,558)(1489,-1395)
\put(1567,-1183){\makebox(0,0)[lb]{\smash{{\SetFigFont{11}{13.2}{\rmdefault}{\mddefault}{\updefault}{\color[rgb]{0,0,0}$0^2$}%
}}}}
\end{picture}%
 \end{array}} 	& -	& \ensuremath{\begin{array}{c}\begin{picture}(0,0)%
\includegraphics{pstex/GR-ring-RR.pstex}%
\end{picture}%
\setlength{\unitlength}{3947sp}%
\begingroup\makeatletter\ifx\SetFigFont\undefined%
\gdef\SetFigFont#1#2#3#4#5{%
  \reset@font\fontsize{#1}{#2pt}%
  \fontfamily{#3}\fontseries{#4}\fontshape{#5}%
  \selectfont}%
\fi\endgroup%
\begin{picture}(169,260)(925,791)
\end{picture}%
 \end{array}}	& +	& \ensuremath{\begin{array}{c}\begin{picture}(0,0)%
\includegraphics{pstex/TLTP-EP-RR.pstex}%
\end{picture}%
\setlength{\unitlength}{3947sp}%
\begingroup\makeatletter\ifx\SetFigFont\undefined%
\gdef\SetFigFont#1#2#3#4#5{%
  \reset@font\fontsize{#1}{#2pt}%
  \fontfamily{#3}\fontseries{#4}\fontshape{#5}%
  \selectfont}%
\fi\endgroup%
\begin{picture}(623,558)(1475,-1414)
\put(1714,-1222){\makebox(0,0)[lb]{\smash{{\SetFigFont{11}{13.2}{\rmdefault}{\mddefault}{\updefault}{\color[rgb]{0,0,0}$0^2$}%
}}}}
\end{picture}%
 \end{array}}	& -	& \ensuremath{\begin{array}{c}\begin{picture}(0,0)%
\includegraphics{pstex/GR-ring-LL.pstex}%
\end{picture}%
\setlength{\unitlength}{3947sp}%
\begingroup\makeatletter\ifx\SetFigFont\undefined%
\gdef\SetFigFont#1#2#3#4#5{%
  \reset@font\fontsize{#1}{#2pt}%
  \fontfamily{#3}\fontseries{#4}\fontshape{#5}%
  \selectfont}%
\fi\endgroup%
\begin{picture}(169,260)(780,791)
\end{picture}%
 \end{array}}	&
		\\
		\vspace{1ex}
				&\LID{TLTP-EP-RL}	&	& \LID{GR-ring-RL}	&	& \LID{TLTP-EP-LR}	&	& \LID{GR-ring-LR}	&
		\\
			-	&\ensuremath{\begin{array}{c}\begin{picture}(0,0)%
\includegraphics{pstex/TLTP-EP-RL.pstex}%
\end{picture}%
\setlength{\unitlength}{3947sp}%
\begingroup\makeatletter\ifx\SetFigFont\undefined%
\gdef\SetFigFont#1#2#3#4#5{%
  \reset@font\fontsize{#1}{#2pt}%
  \fontfamily{#3}\fontseries{#4}\fontshape{#5}%
  \selectfont}%
\fi\endgroup%
\begin{picture}(474,497)(1489,-1334)
\put(1567,-1183){\makebox(0,0)[lb]{\smash{{\SetFigFont{11}{13.2}{\rmdefault}{\mddefault}{\updefault}{\color[rgb]{0,0,0}$0^2$}%
}}}}
\end{picture}%
 \end{array}}	& + &\ensuremath{\begin{array}{c}\begin{picture}(0,0)%
\includegraphics{pstex/GR-ring-RL.pstex}%
\end{picture}%
\setlength{\unitlength}{3947sp}%
\begingroup\makeatletter\ifx\SetFigFont\undefined%
\gdef\SetFigFont#1#2#3#4#5{%
  \reset@font\fontsize{#1}{#2pt}%
  \fontfamily{#3}\fontseries{#4}\fontshape{#5}%
  \selectfont}%
\fi\endgroup%
\begin{picture}(247,243)(925,791)
\end{picture}%
 \end{array}} 	& -	& \ensuremath{\begin{array}{c}\begin{picture}(0,0)%
\includegraphics{pstex/TLTP-EP-LR.pstex}%
\end{picture}%
\setlength{\unitlength}{3947sp}%
\begingroup\makeatletter\ifx\SetFigFont\undefined%
\gdef\SetFigFont#1#2#3#4#5{%
  \reset@font\fontsize{#1}{#2pt}%
  \fontfamily{#3}\fontseries{#4}\fontshape{#5}%
  \selectfont}%
\fi\endgroup%
\begin{picture}(623,497)(1017,-1334)
\put(1256,-1192){\makebox(0,0)[lb]{\smash{{\SetFigFont{11}{13.2}{\rmdefault}{\mddefault}{\updefault}{\color[rgb]{0,0,0}$0^2$}%
}}}}
\end{picture}%
 \end{array}}	& +	& \ensuremath{\begin{array}{c}\begin{picture}(0,0)%
\includegraphics{pstex/GR-ring-LR.pstex}%
\end{picture}%
\setlength{\unitlength}{3947sp}%
\begingroup\makeatletter\ifx\SetFigFont\undefined%
\gdef\SetFigFont#1#2#3#4#5{%
  \reset@font\fontsize{#1}{#2pt}%
  \fontfamily{#3}\fontseries{#4}\fontshape{#5}%
  \selectfont}%
\fi\endgroup%
\begin{picture}(247,243)(1148,791)
\end{picture}%
 \end{array}} 	& = 0
		\end{array}
	\ee
\caption{Two sets of diagrams, each of which will be shown
to sum individually to zero.}
\label{fig-D-ID-2}
\ecf

Diagrams~\ref{TLTP-EP-RR}, \ref{TLTP-EP-RL}
and~\ref{TLTP-EP-LR} can each be obtained
from diagram~\ref{TLTP-EP-LL} by changing
the senses in which the gauge remainders
bite (the remaining four diagrams are similarly
related).
There are actually some stronger relationships than
the one given above. For example, the first four diagrams
only ever cancel amongst themselves. However,
for our analysis of $\beta_n$, only the
statement~\eq{eq:D-ID-2} is required.

To prove \eq{eq:D-ID-2},
we begin by focusing on the first
pair of diagrams, which we redraw in
\fig{fig-D-ID-2b}. Notice that we
have used 
the effective propagator relation in
diagram~\ref{TLTP-EP-LL},
have filled the sockets 
with fields of a specific flavour 
and have chosen a convenient momentum
routing for each of the diagrams.
\bcf[h]
	\[
	\begin{array}{ccccc}
	\vspace{1ex}
		\LID{TLTP-EP-LL-Kd2}	&	&\LID{TLTP-EP-LL-GR}	&	&\LID{GR-ring-RR-2}
	\\
		\ensuremath{\begin{array}{c}\input{pstex/TLTP-EP-LL-Kd2.pstex_t} \end{array}}		& -	& \ensuremath{\begin{array}{c}\input{pstex/TLTP-EP-LL-GR.pstex_t} \end{array}}	& -	& \ensuremath{\begin{array}{c}\input{pstex/GR-ring-RR-2.pstex_t} \end{array}}
	\end{array}
	\]
\caption{Re-expression of diagrams~\ref{TLTP-EP-LL} and~\ref{GR-ring-RR}.}
\label{fig-D-ID-2b}
\ecf

The letters $U$--$Z$ denote field flavour, 
and the subscripts
$T$ and $R$ represent indices. It is straightforward to
verify, and intuitively obvious, that $W_R$ and $X_T$
must be either both bosonic or both fermionic. In the former
case, the internal fields $U, \ V,\ Y$ and $Z$ are all
of the same flavour, being either all bosonic or
all fermionic. In the case that the internal
fields are all bosonic, diagrams~\ref{TLTP-EP-LL-Kd2}--\ref{GR-ring-RR-2}
reduce to a common factor multiplied by
\be
\label{eq:D-ID-2-Alg-a}
	1 - \GRk_l^0 \GRkpr_{l+k}^0 - \GRk_k^0 \GRkpr^0_{l+k},
\ee
whereas, in the case that the internal fields
are all fermionic, they reduce to some
other common factor multiplied by
\be
\label{eq:D-ID-2-Alg-b}
	1 - \GRk_l^1 \GRkpr_{l+k}^1 - \GRk_k^0 \GRkpr^1_{l+k}.
\ee
It is immediately
apparent from~\eq{eq:Template}
that both~\eq{eq:D-ID-2-Alg-a} and~\eq{eq:D-ID-2-Alg-b}
yield zero, upon identifying $-m_1$ with $k$.
In exactly the same way, diagrams~\ref{TLTP-EP-RR} and
\ref{GR-ring-LL}, \ref{TLTP-EP-RL} and~\ref{GR-ring-RL},
\ref{TLTP-EP-LR} and~\ref{GR-ring-LR} cancel
when the external fields are bosonic.

Let us now consider filling the
sockets of diagrams~\ref{TLTP-EP-LL}--\ref{GR-ring-LR}
with fermionic fields. The first thing
to realize is that the latter four diagrams
do not exist with these decorations, since
such decoration is incompatible with the group
theory structure~\cite{Thesis}. In \fig{fig:D-ID-2-F2},
we collect together all surviving terms
for which $W_R = \bar{F}$ and $X_T = F$
(and identical analysis can be done for 
$F \leftrightarrow \bar{F}$), and sum
over all possible flavours of the internal fields.
\bcf[h]
\numparts
	\bea
		\begin{array}{ccccccc}
		\vspace{1ex}			
			\LID{TLTP-EP-LL-F2a}&	& \LID{GR-ring-RR-F2a}	&	& \LID{TLTP-EP-RR-F2a}	&	& \LID{GR-ring-LL-F2a}
		\\
			\ensuremath{\begin{array}{c}\input{pstex/TLTP-EP-LL-F2a.pstex_t} \end{array}}	& -	& \ensuremath{\begin{array}{c}\input{pstex/GR-ring-RR-F2a.pstex_t} \end{array}}	& +	& \ensuremath{\begin{array}{c}\input{pstex/TLTP-EP-RR-F2a.pstex_t} \end{array}}	& -	& \ensuremath{\begin{array}{c}\input{pstex/GR-ring-LL-F2a.pstex_t} \end{array}}		
		\end{array}
	\\
		\begin{array}{ccccccc}
		\vspace{1ex}
			\LID{TLTP-EP-LL-F2b}&	& \LID{GR-ring-RR-F2b}	&	& \LID{TLTP-EP-RR-F2b}	&	& \LID{GR-ring-LL-F2b}
		\\
			\ensuremath{\begin{array}{c}\input{pstex/TLTP-EP-LL-F2b.pstex_t} \end{array}}	& -	& \ensuremath{\begin{array}{c}\input{pstex/GR-ring-RR-F2b.pstex_t} \end{array}}	& +	& \ensuremath{\begin{array}{c}\input{pstex/TLTP-EP-RR-F2b.pstex_t} \end{array}}	& -	& \ensuremath{\begin{array}{c}\input{pstex/GR-ring-LL-F2b.pstex_t} \end{array}}		
		\end{array}
	\eea
\endnumparts
\caption{Field decomposition of diagrams~\ref{TLTP-EP-LL}--\ref{GR-ring-LL}
in the case that
the external fields are fermionic.}
\label{fig:D-ID-2-F2}
\ecf

On the basis of~\eqs{eq:D-ID-2-Alg-a}{eq:D-ID-2-Alg-b},
we might expect diagram~\ref{TLTP-EP-LL-F2a}
to cancel diagram~\ref{GR-ring-RR-F2a} \etc However,
this expectation is not borne out; rather we find
that the Kronecker~$\delta$ part of~\ref{TLTP-EP-LL-F2a} (\ref{TLTP-EP-RR-F2a})
combines with the gauge remainder part of~\ref{TLTP-EP-RR-F2a} (\ref{TLTP-EP-LL-F2a})
and~\ref{GR-ring-LL-F2a} (\ref{GR-ring-RR-F2a}), to yield a common factor
multiplied by a term of the form~\eq{eq:Template}.
In this way, we thus find that each row of diagrams
in \fig{fig:D-ID-2-F2} vanishes.

This completes the discussion of diagrammatic
identity~\eq{eq:D-ID-2}
and so now we move on to the general case,
which relates versions 
of~\eq{D-ID-Trivial-A}--\eq{eq:D-ID-Bitten-hook-A}
with any number of sockets~$\geq 3$ to each other. To begin, we consider
by looking at those diagrams with the specific
patterns of bites on the left / right shown in
\fig{fig:D-ID-G-a}.
\bcf[h]
	\bea
		\begin{array}{ccccc}
		\vspace{1ex}
			\LID{TLTP-EP-G-Kd}	&					& \LID{TLTP-EP-G-GR}				&	& \LID{GR-ring-G}
		\\
			\ensuremath{\begin{array}{c}\input{pstex/TLTP-EP-G-Kd.pstex_t} \end{array}} 	& \hspace{-2em} - 	& \hspace{-1em} \ensuremath{\begin{array}{c}\input{pstex/TLTP-EP-G-GR.pstex_t} \end{array}} 	& -	& \ensuremath{\begin{array}{c}\input{pstex/GR-ring-G.pstex_t} \end{array}}
		\end{array}
	\nonumber \\
	\label{eq:D-ID-G-R}
		\begin{array}{ccccc}
		\vspace{1ex}
				& \LID{GR-ring-G-c}		&	& \LID{GR-ring-G-b}	&
		\\
			- 	& \ensuremath{\begin{array}{c}\input{pstex/GR-ring-G-c.pstex_t} \end{array}}		& +	& \ensuremath{\begin{array}{c}\input{pstex/GR-ring-G-b.pstex_t} \end{array}} 	& = 0 
		\end{array}
	\eea
\caption{A generalized diagrammatic identity.}
\label{fig:D-ID-G-a}
\ecf

There are a number of things to note about 
diagrams~\ref{TLTP-EP-G-Kd}--\ref{GR-ring-G-b}.
First, each of the diagrams possesses the fields
$W$, $X$ and $Z^1, \cdots, Z^r$ with indices
$R$, $T$ and $S_1, \cdots S_r$, respectively.
The cyclic ordering of these fields is the same
in each diagram.
In the case that each diagram possesses only
three external fields, we should identify
$Z^1$ with $Z^r$. 
Secondly, 
for each of the fields $X$,
$Z^1, \cdots Z^r$ there is, in each
diagram, an associated $\GRkpr$. The 
flavour of this $\GRkpr$, which may
of course differ from that of the 
associated external
field, is denoted by  $X'$, ${Z^1}', \cdots {Z^r}'$,
as appropriate. Notice that, in diagram~\ref{TLTP-EP-G-GR},
the field line $X$ is associated with not one
but two internal objects, the flavours of which
are denoted by $X'$ and $X''$, as indicated. Diagram~\ref{GR-ring-G}
is the only one in which the field line $W$
is associated not only with an external $\GRkpr$
(the flavour of which, $W'$, is just the same as
that of $W$) but also an internal $\GRkpr$, the flavour
of which is denoted by $W''$.
Finally, we comment on the momentum routing, which is
taken to be the same each diagram. The external field
$W$ is taken to carry momentum $k$. The external
fields ${Z^1}, \cdots, {Z^r}$ carry momenta $m_1, \cdots m_r$,
respectively; momentum conservation now fixes the momentum
carried by $X$. The loop momentum is $l$, and is routed
such that it is carried by ${Z^1}'$. This now uniquely
determines the momenta carried by all other internal fields.

With these points in mind, we now wish to show that the sum
of the five diagrams vanishes. Immediately, we notice
that, whatever the flavours of the internal and external
fields, diagrams~\ref{TLTP-EP-G-Kd} and~\ref{GR-ring-G-c}
cancel. The only difference between these two diagrams
is the order in which $W$ and $Z^r$ bite the field
line of $X$, which affects neither the flavours
of the internal fields or the group theory factors
(this conclusion will change when we consider
versions of these diagrams with different patterns
of bites to the left / right).

Now let us consider diagrams~\ref{TLTP-EP-G-GR},
\ref{GR-ring-G} and~\ref{GR-ring-G-b}. The first
thing that we must do to show that these diagrams
cancel is demonstrate that the flavours of
the external fields and all common internal fields
can be arranged
to be the same in all three cases. Clearly, we 
can always choose the external fields to be
the same. 
Assigning a value of 0 (1) to the bosonic (fermionic) 
fields, as usual, and denoting 
the statistics of the field $V$ by $\stats{V}$,
we see that
\beas
	\stats{X'} & = & \stats{X} \ \xor \stats{Z^{r'}},
\\
	\stats{{Z^1}'} & = & \stats{Z^1} \ \xor \stats{X'} \ \xor \stats{W'},
\\
	\stats{{Z^{(r-i)}}'} & = & \stats{Z^{r-i}} \ \xor \stats{Z^{r-i-1}} \ \xor \cdots \ \xor \stats{{Z^1}'},
\eeas
in all three diagrams, where $i$ is in the range $\{0,r-2\}$.
Thus, if in addition to choosing all 
external fields to be the same in
diagrams~\ref{TLTP-EP-G-GR},
\ref{GR-ring-G} and~\ref{GR-ring-G-b}
we also choose $Z^{r'}$ to be the same,
it follows that all flavours 
$W', \ X'$ and ${Z^1}', \cdots {Z^r}'$ agree.
Consequently, we can write the sum
of the three diagrams as:
\[
	\int_l F(k,m_i,l,j_1,j_2,j_3)
	\left[
		- \GRk^{j_1}_{l-m_1-k} \GRkpr^{j_2}_{l-m_1} - \GRk^{j_3}_k \GRkpr^{j_2}_{l-m_1} + 1
	\right],
\]
where $j_2 = j_1 \ \xor j_3$ and $F(k,m_i,l,j_1,j_2,j_3)$ is common to all
three terms. This term vanishes, courtesy of~\eq{eq:Template}.

Our next task is to show that the versions of~\eq{eq:D-ID-G-R}
with different patterns of bites to the left / right
also vanish. The first thing to notice is that
changing the senses in which the gauge remainder
components carrying the labels ${Z^2}', \cdots {Z^{r-1}}'$
are bitten induces a change which is common to all
five diagrams, and so the cancellations detailed above
go through, just the same. In the case that there
are at least four external fields---in which case
we do not identify $Z^r$ with $Z^1$---the sense in
which 
${Z^r}'$ is bitten factors out as well.

Let us now consider alternatively nested versions
of diagrams~\ref{TLTP-EP-G-Kd} and~\ref{GR-ring-G-c},
in which we consider changing the sense in which
one and only one gauge reminder is bitten / bites.
Changing the sense in which ${Z^1}'$ is bitten
induces the same change in both diagrams, and so they
still cancel. Next, consider changing the sense in which ${Z^r}'$
\emph{bites}. 
This does not induce the same change
in each diagram, on account of the differing orders
in which ${Z^r}'$ and $W'$ act in each of the diagrams.
Moreover, in diagram~\ref{TLTP-EP-G-Kd}
we can change the sense in which $W'$ bites. We cannot
do this in diagram~\ref{GR-ring-G-c} since it has,
in a sense, been done already: diagrams~\ref{GR-ring-G-c}
and~\ref{GR-ring-G-b} form a pair of diagrams, essentially
differing only in the sense in which $W'$ acts.
(This is particularly clear when we consider these
diagrams as part of a larger diagram, in which case
all indices simply become dummy indices which are summed
over).

Similarly, consider alternatively nested versions
of diagrams~\ref{TLTP-EP-G-GR}, \ref{GR-ring-G}
and~\ref{GR-ring-G-b}. Changing the sense in which
$X'$ is bitten induces the same change in all three
diagrams, and so they still cancel. However,
changing the sense in which ${Z^1}'$ is bitten
does not induce the same change  in the
final diagram as in the first two. 
Moreover, in diagram~\ref{TLTP-EP-G-GR} we can
change the sense in which $W'$ bites, and in 
diagram~\ref{GR-ring-G} we can change the sense
in which $W''$ is bitten. Again, neither of
these terms is obviously cancelled.

We collect together the diagrams which no longer
cancel after changing the sense in which one
of the gauge remainders acts
in \fig{fig:D-ID-G-b}. 
\bcf[h]
	\[
	\begin{array}{l}
	\vspace{1ex}
		\begin{array}{cccccc}
		\vspace{1ex}	
				& \LID{TLTP-EP-G-Kd-B}	&	& \LID{TLTP-EP-G-GR-B}	&	& \LID{GR-ring-G-B}
		\\
			-	& \ensuremath{\begin{array}{c}\input{pstex/TLTP-EP-G-Kd-B.pstex_t} \end{array}}	& +	& \ensuremath{\begin{array}{c}\input{pstex/TLTP-EP-G-GR-B.pstex_t} \end{array}}	& +	& \ensuremath{\begin{array}{c}\input{pstex/GR-ring-G-B-Mac.pstex_t} \end{array}}
		\end{array}
	\\
	\vspace{1ex}
		\begin{array}{cccccc}
		\vspace{1ex}	
				& \LID{TLTP-EP-G-Kd-C}	&	& \LID{TLTP-EP-G-GR-C}	&	& \LID{GR-ring-G-C}
		\\
			-	& \ensuremath{\begin{array}{c}\input{pstex/TLTP-EP-G-Kd-C.pstex_t} \end{array}}	& +	& \ensuremath{\begin{array}{c}\input{pstex/TLTP-EP-G-GR-C.pstex_t} \end{array}}	& +	& \ensuremath{\begin{array}{c}\input{pstex/GR-ring-G-C-Mac.pstex_t} \end{array}}
		\end{array}
	\\
		\begin{array}{cccc}
		\vspace{1ex}
				& \LID{GR-ring-G-c-B}		&	& \LID{GR-ring-G-b-B}
		\\
			+	& \ensuremath{\begin{array}{c}\input{pstex/GR-ring-G-c-B.pstex_t} \end{array}}		& -	& \ensuremath{\begin{array}{c}\input{pstex/GR-ring-G-b-B.pstex_t} \end{array}}
		\end{array}
	\end{array}
	\]
\caption{The set of diagrams, obtained from 
changing the sense in which precisely
one of the gauge remainders of
diagrams~\ref{TLTP-EP-G-Kd}--\ref{GR-ring-G-b}
bites, which do not cancel
in the same way as the diagrams of \fig{fig:D-ID-G-a}.}
\label{fig:D-ID-G-b}
\ecf

Notice that, in diagram~\ref{TLTP-EP-G-Kd-B},
we have changed the labelling of
the fields, compared to diagram~\ref{TLTP-EP-G-Kd}.
With the fields labelled in this way,
we can repeat the analysis of 
diagrams~\ref{TLTP-EP-G-GR},
\ref{GR-ring-G} and~\ref{GR-ring-G-b}
and thus prove that, given a set of common external
fields, all internal
fields can be chosen to be the same in 
diagrams~\ref{TLTP-EP-G-Kd-B}--\ref{GR-ring-G-B}.
Furthermore, we see that the group theory structure
of these diagrams is identical: in each case
the field $Z^1_{S_1}$ is inside the loop,
with the remaining field lines having a
cyclic order, common to the three diagrams, on the outside.
Consequently, we can write the sum
of the three diagrams as:
\[
	\int_l F'(k,m_i,l,j_1,j_2,j_3)
	\left[
		1 - \GRk^{j_1}_{l-m_1-k} \GRkpr^{j_2}_{l-m_1} - \GRk^{j_3}_k \GRkpr^{j_2}_{l-m_1}
	\right],
\]
where once again $j_2 = j_1 \ \xor j_3$ and $F'(k,m_i,l,j_1,j_2,j_3)$ is 
common to all
three terms. Thus, in accord with~\eq{eq:Template},
 the sum of diagrams~\ref{TLTP-EP-G-Kd-B}--\ref{GR-ring-G-B}
vanishes. In exactly the same fashion, we can
show that diagrams~\ref{TLTP-EP-G-Kd-C}--\ref{GR-ring-G-C}
also sum to zero.

This still leaves us with diagrams~\ref{GR-ring-G-c-B}
and~\ref{GR-ring-G-b-B}. Let us put these
to one side for a moment and consider instead
different patterns of bites to the left / right
in diagrams~\ref{TLTP-EP-G-Kd-B}--\ref{GR-ring-G-C}.
Let us start with the first three diagrams.
The only changes we can make which do not
lead to diagrams which cancel in exactly the
same way as the parents
(or take us back to the diagrams of \fig{fig:D-ID-G-a})
are as follows:
\begin{enumerate}
	\item	changing the sense in which $W'$ bites
			in diagrams~\ref{TLTP-EP-G-Kd-B} and~\ref{TLTP-EP-G-GR-B};

	\item	changing the sense in which $X'$ bites in
			diagram~\ref{GR-ring-G-B}.
\end{enumerate}
The resulting diagrams are shown in \fig{fig:D-ID-G-c}.
\bcf[h]
	\[
	\begin{array}{ccccc}
	\vspace{1ex}
		\LID{TLTP-EP-G-Kd-B2}	&	& \LID{TLTP-EP-G-GR-C2}	&	& \LID{GR-ring-G-C2}
	\\
		\ensuremath{\begin{array}{c}\input{pstex/TLTP-EP-G-Kd-B2.pstex_t} \end{array}}	& -	& \ensuremath{\begin{array}{c}\input{pstex/TLTP-EP-G-GR-C2.pstex_t} \end{array}}	& -	& \ensuremath{\begin{array}{c}\input{pstex/GR-ring-G-C2.pstex_t} \end{array}}
	\end{array}
	\]
\caption{The set of diagrams, obtained from 
changing the sense in which precisely
one of the gauge remainders of
diagrams~\ref{TLTP-EP-G-Kd-B}--\ref{GR-ring-G-C}
acts, which do not cancel in exactly
the same way as the parent diagrams or reproduce already
drawn diagrams.}
\label{fig:D-ID-G-c}
\ecf

Now consider making similar changes to 
diagrams~\ref{TLTP-EP-G-Kd-C}--\ref{GR-ring-G-C}.
The point is, that the set of diagrams
we would draw simply reproduces the
pattern of bites to the
left / right that we see in 
diagrams~\ref{TLTP-EP-G-Kd-B2}--\ref{GR-ring-G-C2}.
In other words, we have taken account of
these terms already and so we have no further
diagrams to draw. Now, diagram~\ref{TLTP-EP-G-Kd-B2}
cancels diagram~\ref{GR-ring-G-b-B} and
diagrams~\ref{TLTP-EP-G-GR-C2} and~\ref{GR-ring-G-C2}
cancel diagram~\ref{GR-ring-G-c-B}. Thus everything closes.
Any further changes to the pattern of bites to
the left / right will simply replicate cancellations
we have seen already.

We have therefore demonstrated that, in addition
to the diagrams of \fig{fig:D-ID-G-a} summing to
zero, so too do appropriate combinations
of their alternatively nested partners.
We can use this fact, together with
identity~\eq{eq:D-ID-2},
to derive a relationship between all
diagrams of the basic structure just discussed,
possessing between two and $m$ empty sockets.
The relationship, which is crucial in our treatment
of $\beta_n$, is shown in \fig{fig:D-ID-App}.
\bcf[h]
	\be
	\label{eq:D-ID-G-Pre}
	\begin{array}{ll}
	\vspace{1ex}
		&
		\AGRO{
			\LIDi[1]{Struc-LXL}{D-ID-App-LXL} \hspace{0.225em}
			+
			\left[
				\begin{array}{llll}
					& \LIDi[1]{Struc-LLXL}{D-ID-App-LLXL} 	& \hspace{0.5em} + & \cdots
				\\
					- & \LIDi[1]{Struc-LXRL}{D-ID-App-LXLL} 	& \hspace{0.5em} + & \cdots
				\\ 
					& & & \vdots
				\end{array}
			\hspace{5em}
			\right]
		}
	\\
		= &
		\AGRO{
			\begin{array}{lll}
			\vspace{0.2in}
				\LIDi[1]{Struc-GR-RR}{D-ID-App-GR-LL} & - & \hspace{2.3em} \LIDi[1]{Struc-GR-RRR}{D-ID-App-GR-LLL} \hspace{1.6em} +\cdots
			\\
				& - &
				\left[
					\begin{array}{lll}
						\hspace{1em} \LIDi[1]{Struc-PF-TLTP-EP-RR}{D-ID-App-PB-TLTP-EP-LL} 	
						& \hspace{0.5em} -  \LIDi[1]{Struc-PF-TLTP-EP-RRR}{D-ID-App-PB-TLTP-EP-LLL}  
						& +\cdots 
					\\																
						& 
						\hspace{0.5em} - \LIDi[1]{Struc-R-PF-TLTP-EP-RR}{D-ID-App-L-PB-TLTP-EP-LL}
						& +\cdots
					\end{array}
				\right]
			\end{array}
		}
	\end{array}
	\ee
\caption{A relationship between certain diagrams
possessing between two and $m$ empty sockets.}
\label{fig:D-ID-App}
\ecf

The notation $\AGRO{\ }$ tells us to sum over Every Sense
in which all the gauge remainders can bite. Thus, 
diagrams~\ref{D-ID-App-LXL} and~\ref{D-ID-App-GR-LL}
have thee partners, diagrams~\ref{D-ID-App-LLXL}, 
\ref{D-ID-App-LXLL}, \ref{D-ID-App-GR-LLL} and~\ref{D-ID-App-PB-TLTP-EP-LL} 	
seven partners \etc 

It is worth clarifying the diagrams represented by the various ellipses.
The ellipsis following diagram~\ref{D-ID-App-LLXL} denotes additionally
nested diagrams, where the classical, two-point, vertex is still
joined to the innermost gauge remainder. The ellipsis after
diagram~\ref{D-ID-App-LXLL} denotes additionally nested diagrams,
where the classical, two-point vertex joins to the innermost
but one gauge  remainder. The vertical dots on the next line represent additionally
nested diagrams in which the classical, two-point vertex joins to
gauge remainders successively further away from the innermost one.
Note, though, that we never make the join to the outermost gauge remainder.
Thus, we have arranged things such that all diagram in a given
column possess the same number of sockets.
The relationship between the top set of diagrams and
the bottom set holds for any number of columns, so 
long as this number is the same for both sets of
diagrams.

We now expound on how the diagrammatic identities of this
section combine to yield~\eq{eq:D-ID-G-Pre}.
Diagram~\ref{D-ID-App-LXL} and its partners
are clearly
equal to~\ref{D-ID-App-GR-LL} and its partners
by diagrammatic identity~\eq{eq:D-ID-2}.
Applying the effective propagator relationship to
diagram~\ref{D-ID-App-PB-TLTP-EP-LL} and its partners, 
the Kronecker~$\delta$
part combines with diagram~\ref{D-ID-App-GR-LLL}
to give diagram~\ref{D-ID-App-LXLL}, via
diagrammatic identity~\eq{eq:D-ID-G-R}
(and its alternatively nested versions). On the other
hand, the gauge remainder term and its partners
are just equal to
diagram~\ref{D-ID-App-LLXL} 
and its partners courtesy of 
diagrammatic identity~\eq{eq:D-ID-2}.

Using  the
notation described under \fig{fig:nestedhook}, we can represent~\eq{eq:D-ID-G-Pre}
in the compact form shown in \fig{fig:D-ID-G-Compact}.
\bcf[h]
	\bea
	\nonumber
		\sum_{m'=1}^{m-1} \frac{1}{m'!} \ensuremath{\begin{array}{c}\input{pstex/TLTP-EP-ArbGRs.pstex_t} \end{array}} \hspace{2em}
		& = &  \sum_{m'=1}^{m-1} \frac{1}{(m'+1)!} \ensuremath{\begin{array}{c}\begin{picture}(0,0)%
\includegraphics{pstex/GR-ArbGRs.pstex}%
\end{picture}%
\setlength{\unitlength}{3947sp}%
\begingroup\makeatletter\ifx\SetFigFont\undefined%
\gdef\SetFigFont#1#2#3#4#5{%
  \reset@font\fontsize{#1}{#2pt}%
  \fontfamily{#3}\fontseries{#4}\fontshape{#5}%
  \selectfont}%
\fi\endgroup%
\begin{picture}(164,296)(785,791)
\put(785,955){\makebox(0,0)[lb]{\smash{{\SetFigFont{11}{13.2}{\rmdefault}{\mddefault}{\updefault}{\color[rgb]{0,0,0}$\decGR{ \ }{\GRkpr^{m'+1}}$}%
}}}}
\end{picture}%
 \end{array}} \hspace{2em}
	\\ 
	\label{eq:D-ID-G-A}
		&& - \sum_{m'=2}^{m-1} \sum_{m''=0}^{m'-2} \frac{1}{(m'-m'')!m''!} 
		\ensuremath{\begin{array}{c}\input{pstex/ArbGRs-TLTP-EP-ArbGRs.pstex_t} \end{array}}
	\eea
\caption{A compact form of~\eq{eq:D-ID-G-Pre}.}
\label{fig:D-ID-G-Compact}
\ecf

The loose end of the effective propagator
in the first diagram can attach to any
of the $m'$ gauge remainder components. We need not
strictly prohibit it
from attaching to the socket on the vertex since
such a diagram vanishes by \CC\ invariance, anyway.
The relationship~\eq{eq:D-ID-G-A} is, of course, valid separately
for every integer value of $m$ from one to infinity.
The explicitly drawn  $\GRkpr$ in the first and third diagrams
can bite in either sense. Notice that,
since the explicitly drawn $\GRkpr$ is common
to each diagram (in the second diagram it just sits
at the bottom end of the full gauge remainder),
we could strip it off completely, 
replacing it where necessary 
with the 
half arrow notation of~\eq{eq:WID-A}
(\ie we could remove the $\GRkpr$ completely from the second
diagram, where it does not push forward or pull back
on to another field, but would have to replace it
with a half arrow in the first and third diagrams).

\subsubsection{The Second Family}
\label{sec:D-ID-Secondary-II}

The first diagrammatic identity
in this family
follows from the two equivalent
way of processing the diagram on
the \lhs\
of \fig{fig:D-ID-dGR}: our first operation can
either be to process the gauge remainder
or to Taylor expand the vertex (we assume that
the socket is decorated by an $A^i$ sector field).
In the former case, although we do not explicitly
indicate it at this stage, it is understood that
the socket in the daughter diagrams is filled
by an $A^i$ sector field carrying zero momentum,
just as in the parent.
In the latter case, we can move the momentum derivative
from the vertex to the $\GRk$ which bites the vertex,
at the expense of a minus sign, courtesy of diagrammatic
identity~\eq{eq:GR-TLTP}.
\bcf[h]
	\[
	\begin{array}{lcccrr}
		\ensuremath{\begin{array}{c}\input{pstex/D-ID-dGR-A.pstex_t} \end{array}}	& =	& 								 	& =	& \ds \frac{1}{m} \ensuremath{\begin{array}{c}\begin{picture}(0,0)%
\includegraphics{pstex/D-ID-dGR-B.pstex}%
\end{picture}%
\setlength{\unitlength}{3947sp}%
\begingroup\makeatletter\ifx\SetFigFont\undefined%
\gdef\SetFigFont#1#2#3#4#5{%
  \reset@font\fontsize{#1}{#2pt}%
  \fontfamily{#3}\fontseries{#4}\fontshape{#5}%
  \selectfont}%
\fi\endgroup%
\begin{picture}(521,280)(1768,-427)
\put(1911,-279){\makebox(0,0)[lb]{\smash{{\SetFigFont{11}{13.2}{\rmdefault}{\mddefault}{\updefault}{\color[rgb]{0,0,0}$\decGR{ \ }{\GRkpr^{m}}$}%
}}}}
\end{picture}%
 \end{array}} 	& \ds - \frac{1}{m} \ensuremath{\begin{array}{c}\begin{picture}(0,0)%
\includegraphics{pstex/D-ID-dGR-C.pstex}%
\end{picture}%
\setlength{\unitlength}{3947sp}%
\begingroup\makeatletter\ifx\SetFigFont\undefined%
\gdef\SetFigFont#1#2#3#4#5{%
  \reset@font\fontsize{#1}{#2pt}%
  \fontfamily{#3}\fontseries{#4}\fontshape{#5}%
  \selectfont}%
\fi\endgroup%
\begin{picture}(521,327)(1768,-425)
\put(1905,-230){\makebox(0,0)[lb]{\smash{{\SetFigFont{11}{13.2}{\rmdefault}{\mddefault}{\updefault}{\color[rgb]{0,0,0}$\decGR{ \ }{\GRkpr^{m}}$}%
}}}}
\end{picture}%
 \end{array}}
	\\[1ex]
						& = & - 2 \ensuremath{\begin{array}{c}\input{pstex/D-ID-dGR-D.pstex_t} \end{array}} \hspace{1em} 	& =	& -2 \ensuremath{\begin{array}{c}\begin{picture}(0,0)%
\includegraphics{pstex/D-ID-dGR-E.pstex}%
\end{picture}%
\setlength{\unitlength}{3947sp}%
\begingroup\makeatletter\ifx\SetFigFont\undefined%
\gdef\SetFigFont#1#2#3#4#5{%
  \reset@font\fontsize{#1}{#2pt}%
  \fontfamily{#3}\fontseries{#4}\fontshape{#5}%
  \selectfont}%
\fi\endgroup%
\begin{picture}(521,508)(1811,-425)
\put(1916,-49){\makebox(0,0)[lb]{\smash{{\SetFigFont{11}{13.2}{\rmdefault}{\mddefault}{\updefault}{\color[rgb]{0,0,0}$\decGR{ \ }{\GRkpr^{m-1}}$}%
}}}}
\end{picture}%
 \end{array}} 				& + 2 \ensuremath{\begin{array}{c}\begin{picture}(0,0)%
\includegraphics{pstex/D-ID-dGR-F.pstex}%
\end{picture}%
\setlength{\unitlength}{3947sp}%
\begingroup\makeatletter\ifx\SetFigFont\undefined%
\gdef\SetFigFont#1#2#3#4#5{%
  \reset@font\fontsize{#1}{#2pt}%
  \fontfamily{#3}\fontseries{#4}\fontshape{#5}%
  \selectfont}%
\fi\endgroup%
\begin{picture}(521,592)(1826,-509)
\put(1916,-49){\makebox(0,0)[lb]{\smash{{\SetFigFont{11}{13.2}{\rmdefault}{\mddefault}{\updefault}{\color[rgb]{0,0,0}$\decGR{ \ }{\GRkpr^{m-1}}$}%
}}}}
\end{picture}%
 \end{array}}
	\end{array}
	\]
\caption{Deducing a secondary diagrammatic identity; 
the socket must be decorated by an $A^i$ sector
field which carries zero momentum.}
\label{fig:D-ID-dGR}
\ecf

The factor of two on the second line arises from using \CC\ invariance
to collect together momentum derivative terms; we have
not used \CC\ to collect together pushes forward and
pulls back on the first line. By equating the two expressions
with each other, we are lead to the following diagrammatic
identity:
\be
\label{eq:D-ID-dGRk-GT}
	\frac{1}{m} \ensuremath{\begin{array}{c}\begin{picture}(0,0)%
\includegraphics{pstex/D-ID-dGR-Bi.pstex}%
\end{picture}%
\setlength{\unitlength}{3947sp}%
\begingroup\makeatletter\ifx\SetFigFont\undefined%
\gdef\SetFigFont#1#2#3#4#5{%
  \reset@font\fontsize{#1}{#2pt}%
  \fontfamily{#3}\fontseries{#4}\fontshape{#5}%
  \selectfont}%
\fi\endgroup%
\begin{picture}(153,280)(1911,-427)
\put(1911,-279){\makebox(0,0)[lb]{\smash{{\SetFigFont{11}{13.2}{\rmdefault}{\mddefault}{\updefault}{\color[rgb]{0,0,0}$\decGR{ \ }{\GRkpr^{m}}$}%
}}}}
\end{picture}%
 \end{array}} \hspace{1em} = -2 \ensuremath{\begin{array}{c}\begin{picture}(0,0)%
\includegraphics{pstex/D-ID-dGR-Ei.pstex}%
\end{picture}%
\setlength{\unitlength}{3947sp}%
\begingroup\makeatletter\ifx\SetFigFont\undefined%
\gdef\SetFigFont#1#2#3#4#5{%
  \reset@font\fontsize{#1}{#2pt}%
  \fontfamily{#3}\fontseries{#4}\fontshape{#5}%
  \selectfont}%
\fi\endgroup%
\begin{picture}(341,480)(1916,-397)
\put(1916,-49){\makebox(0,0)[lb]{\smash{{\SetFigFont{11}{13.2}{\rmdefault}{\mddefault}{\updefault}{\color[rgb]{0,0,0}$\decGR{ \ }{\GRkpr^{m-1}}$}%
}}}}
\end{picture}%
 \end{array}}
\ee
where, in both diagrams, the socket must be decorated by
an $A^i$ sector carrying zero momentum. In the
special case that the socket of the
first diagram can be identified with the sockets of one
of the $m$ gauge remainders~\eq{eq:D-ID-dGRk-GT} becomes:
\be
\label{eq:D-ID-dGRk-GT-ring-A}
	2m(m-1) \ensuremath{\begin{array}{c}\begin{picture}(0,0)%
\includegraphics{pstex/GRs-dGRk.pstex}%
\end{picture}%
\setlength{\unitlength}{3947sp}%
\begingroup\makeatletter\ifx\SetFigFont\undefined%
\gdef\SetFigFont#1#2#3#4#5{%
  \reset@font\fontsize{#1}{#2pt}%
  \fontfamily{#3}\fontseries{#4}\fontshape{#5}%
  \selectfont}%
\fi\endgroup%
\begin{picture}(506,344)(1446,2019)
\put(1446,2069){\makebox(0,0)[lb]{\smash{{\SetFigFont{11}{13.2}{\rmdefault}{\mddefault}{\updefault}{\color[rgb]{0,0,0}$\decGR{ \hspace{3em} }{\GRkpr^{m-2}}$}%
}}}}
\end{picture}%
 \end{array}} \hspace{3em} = - \decGR{ \ }{\socket \GRkpr^m}.
\ee
Notice that, in the second diagram, 
even though the socket is identified with one of
the gauge remainders, we have kept it
to remind us that the relationship~\eq{eq:D-ID-dGRk-GT-ring-A}
holds only if the sockets on each side of the equation
are filled with an $A^i$ field carrying zero momentum.


The second diagrammatic identity in
this family follows as a consequence of
diagrammatic identities~\eq{eq:WID} 
and~\eq{eq:GR-TLTP}. Defining
\be
\label{eq:Combo}
	\combo \equiv \ensuremath{\begin{array}{c}\begin{picture}(0,0)%
\includegraphics{pstex/Combo.pstex}%
\end{picture}%
\setlength{\unitlength}{3947sp}%
\begingroup\makeatletter\ifx\SetFigFont\undefined%
\gdef\SetFigFont#1#2#3#4#5{%
  \reset@font\fontsize{#1}{#2pt}%
  \fontfamily{#3}\fontseries{#4}\fontshape{#5}%
  \selectfont}%
\fi\endgroup%
\begin{picture}(240,629)(1350,1164)
\end{picture}%
 \end{array}} \equiv \ensuremath{\begin{array}{c}\begin{picture}(0,0)%
\includegraphics{pstex/EP-dGR.pstex}%
\end{picture}%
\setlength{\unitlength}{3947sp}%
\begingroup\makeatletter\ifx\SetFigFont\undefined%
\gdef\SetFigFont#1#2#3#4#5{%
  \reset@font\fontsize{#1}{#2pt}%
  \fontfamily{#3}\fontseries{#4}\fontshape{#5}%
  \selectfont}%
\fi\endgroup%
\begin{picture}(314,693)(1432,1119)
\end{picture}%
 \end{array}} -\hf \ensuremath{\begin{array}{c}\begin{picture}(0,0)%
\includegraphics{pstex/dPEP-GR.pstex}%
\end{picture}%
\setlength{\unitlength}{3947sp}%
\begingroup\makeatletter\ifx\SetFigFont\undefined%
\gdef\SetFigFont#1#2#3#4#5{%
  \reset@font\fontsize{#1}{#2pt}%
  \fontfamily{#3}\fontseries{#4}\fontshape{#5}%
  \selectfont}%
\fi\endgroup%
\begin{picture}(245,630)(969,1222)
\end{picture}%
 \end{array}}
\ee
we have
\be
\label{eq:D-ID-Trapped}
	\ensuremath{\begin{array}{c}\begin{picture}(0,0)%
\includegraphics{pstex/TLTP-Trapped-PF.pstex}%
\end{picture}%
\setlength{\unitlength}{3947sp}%
\begingroup\makeatletter\ifx\SetFigFont\undefined%
\gdef\SetFigFont#1#2#3#4#5{%
  \reset@font\fontsize{#1}{#2pt}%
  \fontfamily{#3}\fontseries{#4}\fontshape{#5}%
  \selectfont}%
\fi\endgroup%
\begin{picture}(563,988)(2025,-1380)
\put(2088,-1063){\makebox(0,0)[lb]{\smash{{\SetFigFont{11}{13.2}{\rmdefault}{\mddefault}{\updefault}{\color[rgb]{0,0,0}$0^2$}%
}}}}
\end{picture}%
 \end{array}} = \ensuremath{\begin{array}{c}\begin{picture}(0,0)%
\includegraphics{pstex/TLThP-GRx2.pstex}%
\end{picture}%
\setlength{\unitlength}{3947sp}%
\begingroup\makeatletter\ifx\SetFigFont\undefined%
\gdef\SetFigFont#1#2#3#4#5{%
  \reset@font\fontsize{#1}{#2pt}%
  \fontfamily{#3}\fontseries{#4}\fontshape{#5}%
  \selectfont}%
\fi\endgroup%
\begin{picture}(554,1048)(2012,-1380)
\put(2090,-1069){\makebox(0,0)[lb]{\smash{{\SetFigFont{11}{13.2}{\rmdefault}{\mddefault}{\updefault}{\color[rgb]{0,0,0}$0^2$}%
}}}}
\end{picture}%
 \end{array}} = - \ensuremath{\begin{array}{c}\begin{picture}(0,0)%
\includegraphics{pstex/TLTP-Trapped-PB.pstex}%
\end{picture}%
\setlength{\unitlength}{3947sp}%
\begingroup\makeatletter\ifx\SetFigFont\undefined%
\gdef\SetFigFont#1#2#3#4#5{%
  \reset@font\fontsize{#1}{#2pt}%
  \fontfamily{#3}\fontseries{#4}\fontshape{#5}%
  \selectfont}%
\fi\endgroup%
\begin{picture}(367,1110)(2025,-1380)
\put(2141,-1037){\makebox(0,0)[lb]{\smash{{\SetFigFont{11}{13.2}{\rmdefault}{\mddefault}{\updefault}{\color[rgb]{0,0,0}$0$}%
}}}}
\end{picture}%
 \end{array}}.
\ee

A number of comments are in order.
Strictly speaking, the presence of $\combo$ (as opposed
to some other internal line involving a momentum derivative) is not necessary
for either this diagrammatic identity or those that follow. However,
in practise, it is a $\combo$ that most commonly occurs in such
scenarios.
Note
that each diagram effectively possesses two external $A^1$s (after
decoration of the socket associated with the $\combo$), which 
we suppose to be the only external fields. 
We understand that the socket
possessed by the $\GRkpr$ and the loose end of the $\combo$
are internal fields, which are somehow tied up into a complete
diagram. 
In the second diagram, if the right-most
gauge remainder pushes forward onto the internal
field, we recover the parent. If this gauge remainder
were instead to pull back onto the internal field,
then the supertrace structure of the diagram
would be uniquely determined to be 
$\str A^1_\mu \str A^1_\nu = 0$.
If either gauge remainder strikes the external field, then the 
other gauge remainder kills the diagram,
courtesy 
of~\eq{eq:GR-TLTP}. To go from the second diagram to the
third, we allow the left-most gauge remainder to act;
the only surviving contribution is the pull back onto
the internal field.

The next diagrammatic identity
is the first of this family which necessitates
consideration of the algebraic form of gauge
remainders, and is shown in \fig{fig:D-ID-Op2-C}.
Notice that, in the first diagram, both the $\combo$
and the $\GRkpr$ bite the socket.
\bcf[h]
	\be
	\label{eq:D-ID-Op2-C}
		\begin{array}{cccccc}
			\LID{D-ID-CTP-E-Cb}	&	& \LID{D-ID-CTP-E-Ab}	&	& \LID{D-ID-CTP-E-Bb}	&
		\\[1ex]
			\ensuremath{\begin{array}{c}\begin{picture}(0,0)%
\includegraphics{pstex/D-ID-CTP-E-Cb.pstex}%
\end{picture}%
\setlength{\unitlength}{3947sp}%
\begingroup\makeatletter\ifx\SetFigFont\undefined%
\gdef\SetFigFont#1#2#3#4#5{%
  \reset@font\fontsize{#1}{#2pt}%
  \fontfamily{#3}\fontseries{#4}\fontshape{#5}%
  \selectfont}%
\fi\endgroup%
\begin{picture}(744,734)(719,714)
\put(801,1067){\makebox(0,0)[lb]{\smash{{\SetFigFont{11}{13.2}{\rmdefault}{\mddefault}{\updefault}{\color[rgb]{0,0,0}$0^2$}%
}}}}
\end{picture}%
 \end{array}} 	& -	& \ensuremath{\begin{array}{c}\begin{picture}(0,0)%
\includegraphics{pstex/D-ID-CTP-E-Ab.pstex}%
\end{picture}%
\setlength{\unitlength}{3947sp}%
\begingroup\makeatletter\ifx\SetFigFont\undefined%
\gdef\SetFigFont#1#2#3#4#5{%
  \reset@font\fontsize{#1}{#2pt}%
  \fontfamily{#3}\fontseries{#4}\fontshape{#5}%
  \selectfont}%
\fi\endgroup%
\begin{picture}(826,789)(719,714)
\put(801,1067){\makebox(0,0)[lb]{\smash{{\SetFigFont{11}{13.2}{\rmdefault}{\mddefault}{\updefault}{\color[rgb]{0,0,0}$0^2$}%
}}}}
\end{picture}%
 \end{array}}	& -	& \ensuremath{\begin{array}{c}\begin{picture}(0,0)%
\includegraphics{pstex/D-ID-CTP-E-Bb.pstex}%
\end{picture}%
\setlength{\unitlength}{3947sp}%
\begingroup\makeatletter\ifx\SetFigFont\undefined%
\gdef\SetFigFont#1#2#3#4#5{%
  \reset@font\fontsize{#1}{#2pt}%
  \fontfamily{#3}\fontseries{#4}\fontshape{#5}%
  \selectfont}%
\fi\endgroup%
\begin{picture}(824,1129)(719,714)
\put(801,1067){\makebox(0,0)[lb]{\smash{{\SetFigFont{11}{13.2}{\rmdefault}{\mddefault}{\updefault}{\color[rgb]{0,0,0}$0^2$}%
}}}}
\end{picture}%
 \end{array}} 	& = 0
		\end{array}
	\ee
\caption{The first diagrammatic identity of the second
family which relies on~\eq{eq:Template}.}
\label{fig:D-ID-Op2-C}
\ecf

This diagrammatic identity follows directly
from~\eq{eq:Template}.
Although we have given 
both~\eqs{eq:D-ID-Trapped}{eq:D-ID-Op2-C}
in un-nested form, it is clear that
they both hold in appropriately nested
cases. Indeed, we now combine
the nested versions of these
diagrammatic identities to give
the final diagrammatic
identity. This is shown in \fig{fig:D-ID-Op2-D}
and is, in some sense, the second
family analogue of diagrammatic
identity~\eq{eq:D-ID-G-A}. Wherever a $\combo$
and $\GRkpr$ bite the same thing, the $\combo$
is taken to act first, as in diagram~\ref{D-ID-CTP-E-Cb}.
\bcf[h]
	\bea
	\lefteqn{
		\sum_{m'=0}^m 
		\left[
			\ensuremath{\begin{array}{c}\input{pstex/CTP-E-GRs-Combo-GRk-GR.pstex_t} \end{array}} 
			-\ensuremath{\begin{array}{c}\input{pstex/CTP-E-GRs-Combo-GR.pstex_t} \end{array}}
		\right]	
	}
	\nonumber
	\\ & &
		- \sum_{m'=0}^m \!\! \nCr{m'}{m''} \!\! \sum_{m''=0}^{m'} 
		\left[
			\ensuremath{\begin{array}{c}\input{pstex/CTP-E-GRs-Combo-CTP-EP-GR.pstex_t} \end{array}}
			\hspace{1.8em}
			-\ensuremath{\begin{array}{c}\input{pstex/CTP-E-GRs-Combo-GRs-CTP-EP.pstex_t} \end{array}}
			\hspace{1.8em}
		\right]
		= 0
	\label{eq:D-ID-Op2-G-A}
	\eea
\caption{The final diagrammatic identity belonging to the second
family.}
\label{fig:D-ID-Op2-D}
\ecf

\subsection{Subtraction Techniques}
\label{sec:Subtraction}

\subsubsection{Introduction}
\label{sec:Subtractions-I}

As discussed in the introduction, and as
we will see shortly, the $n$-loop $\beta$ function
is computed by evaluating a set of diagrams at
$\Op{2}$:
\[
	\beta_n \Box_{\mu \nu}(p) = \left.\mathrm{Diagrams}\right|_{p^2}.
\]
A subset of the diagrams contributing to $\beta_n$ possess
a stub which is
manifestly $\Op{2}$ and so we would like to Taylor expand
the remaining components of these diagrams to zeroth order in $p$.
The problem is that individual diagrams may
possess components which are not Taylor expandable to
$\Op{2}$.
We could proceed simply by ignoring these
components since we know that
the sum of all diagrams contributing to $\beta_n$
must be Taylor expandable to $\Op{2}$. However, for
the purposes of this paper we adopt a higher level of
rigour and so keep them. 
As an example,
consider the first diagram of \fig{fig:TLM-M:TE-Ex-B}
where, in anticipation of what follows, we have added and subtracted
a term.
\bcf[h]
	\[
		\begin{array}{ccc}
			\LID{d:111b.11}		&		& \LIDLID{d:111b.11-s}{d:111b.11-a}
		\\[1ex]
			\ensuremath{\begin{array}{c}\input{pstex/Diagram111b.11.pstex_t} \end{array}} & \mp 	& \left[\ensuremath{\begin{array}{c}\input{pstex/Diagram111b.11-s.pstex_t} \end{array}}\right]_{p^2}
		\end{array}
	\]
\caption{A two-loop diagram with an $\Op{2}$ stub, components
of which cannot be Taylor
expanded to $\Op{2}$, and a term constructed to isolate
these components.}
\label{fig:TLM-M:TE-Ex-B}	
\ecf

We begin by focusing on diagram~\ref{d:111b.11}.
Since we can always Taylor expand vertices in momenta, let us suppose that
we take a power of $l$ from the top-most vertex 
(we cannot take any powers of $p$, at $\Op{2}$) and
let us choose to take a power of $k$ from the other vertex. 
Using \tab{tab:NFE:k,k'} and~\eq{eq:EP-leading}, we see that 
the leading IR behaviour
of the $l$-integral is given by
\[
	\int_l \frac{1}{l^2 (l-p)^2},
\]
which is not Taylor expandable to $\Op{2}$. Note that had we taken a power of $l$ from
the right-hand vertex, rather than a power of $k$, then the extra power of $l$
in the integrand would render the diagram Taylor expandable in $p$ (to the
required order).

Next, let us consider the pair of 
diagrams~\ref{d:111b.11-s} and~\ref{d:111b.11-a},
where the first of these---the subtraction---comes
with the minus sign and the corresponding
addition comes with the plus sign. The tag $p^2$
tells us that only the stub carries momentum
$p$ \ie these diagrams are constructed
by setting $p=0$ in diagram~\ref{d:111b.11}
everywhere but the stub.

The effect of the subtraction on the parent is to cancel
all those components which are Taylor expandable to $\Op{2}$. This immediately tells us
the following about any surviving contributions to diagram~\ref{d:111b.11}:
\begin{enumerate}
	\item	all fields carrying momentum $l$ must be in the $A^1$-sector;

	\item	we must take $\mathcal{O}(l^0)$ from the $k$-integral (note that
			the $k$-integral is Taylor expandable in $l$);

	\item	we must discard any remaining contributions to the $l$
			integral which are Taylor expandable to $\Op{0}$.
\end{enumerate}
The contributions to diagram~\ref{d:111b.11} not removed by
 diagram~\ref{d:111b.11-s} are shown on the \lhs\ of \fig{fig:TLM-M:TE-Ex-C}.
\bcf[h]
	\[
		\left[
			\LIDi{Diagram111b.11-Rem}{d:111b.11-Rem}
		\right]_{\NTE{p}}
		\rightarrow
		\LIO{
			\left[
				\ensuremath{\begin{array}{c}\input{pstex/CTP-E-EP-Thpt-EP-GR.pstex_t} \end{array}}
			\right]_{\NTE{p}}
			\Delta^{11}_{\alpha \beta}(p)
			\decp{
				\ensuremath{\begin{array}{c}\input{pstex/CTP-E-EP-Thpt-K-GR.pstex_t} \end{array}}
			}{}
		}{d:111b.11-Rem-b}
	\]
\caption{The contribution to diagram~\ref{d:111b.11} not removed by its subtraction.}
\label{fig:TLM-M:TE-Ex-C}
\ecf

Focusing first on diagram~\ref{d:111b.11-Rem},
we see that,
as required, we have taken the $\mathcal{O}(l^0)$ from the $k$-integral. 
The tag $\BigNTE{p}$
demands that we retain only the component of the diagram
which, whilst possessing an overall $\Op{2}$ contribution,
is not Taylor expandable in $p$ (in dimensional
regularization, this contribution would go
as $\Op{2} \Pep$, for this diagram).
The bar sitting just before the
$\GRkpr$ which bites the classical,
two-point vertex indicates a discontinuity in
momenta: the $\GRkpr$ which bites
the classical, two-point vertex carries momentum $l$,
but the loop to its right carries only momentum
$k$ along all lines.
We can neaten this diagram up (getting rid of
the momentum discontinuity) 
to obtain diagram~\ref{d:111b.11-Rem-b}.
To do this, we first 
notice 
that the sub-diagram of~\ref{d:111b.11-Rem}
which is independent of $l$
carries two indices, say $\rho$ and $\tau$.
By Lorentz invariance, this sub-diagram must go
as $\delta_{\rho \tau}$, which allows us to tie
up the effective propagator attaching to the
momentum derivative to the gauge remainder
which follows the bar. This procedure yields
the $\BigNTE{p}$ component of diagram~\ref{d:111b.11-Rem-b},
but with indices $\mu$ and $\nu$, rather than
$\mu$ and $\alpha$. We now re-express
\beas
	\Box_{\mu \nu}(p)	& =	& \Box_{\mu \alpha}(p) \frac{\delta_{\alpha \beta}}{2 p^2} 2 \Box_{\beta \nu}(p)
\eeas
which allows us to re-write diagram~\ref{d:111b.11-Rem}
in the form~\ref{d:111b.11-Rem-b}, up to $\Op{4}$ corrections
(\cf~\eq{eq:EP-leading}).

Before discussing the addition, we note that we
define the subtraction to be the diagram which removes
the components of the parent diagram which are Taylor
expandable to $\Op{2}$, not by its sign.
We could always move  
the momentum derivatives in diagrams~\ref{d:111b.11-s}
from one side of the vertex they strike to
the other, at the expense of a minus sign
(this
would amount to replacing $\partial_\nu^{-l}$
with $\partial_\nu^{l}$); were we to do so,
the subtraction would now come with a plus sign
and the addition with a minus. 
 
Additions like our example~\ref{d:111b.11-a} can
be manipulated by using the effective
propagator relation and diagrammatic
identity~\eq{eq:PseudoEP}. For the two structures,
$A$ and $B$ we have:
\be
\label{eq:EP-dCTP-EP}
	\ensuremath{\begin{array}{c}\input{pstex/A-EP-dCTP-EP-B.pstex_t} \end{array}} = 
	\begin{array}{cccccc}
			& \LID{A-dEP-B}	&	& \LID{A-GR-combo-B}	&	& \LID{A-combo-GR-B}
	\\[1ex]
		-	& \ensuremath{\begin{array}{c}\input{pstex/A-dEP-B.pstex_t} \end{array}}	& -	& \ensuremath{\begin{array}{c}\input{pstex/A-GR-combo-B.pstex_t} \end{array}}		& -	& \ensuremath{\begin{array}{c}\input{pstex/A-combo-GR-B.pstex_t} \end{array}},
	\end{array}
\ee
where
\[
	\ensuremath{\begin{array}{c}\begin{picture}(0,0)%
\includegraphics{pstex/Combo-GR-b.pstex}%
\end{picture}%
\setlength{\unitlength}{3947sp}%
\begingroup\makeatletter\ifx\SetFigFont\undefined%
\gdef\SetFigFont#1#2#3#4#5{%
  \reset@font\fontsize{#1}{#2pt}%
  \fontfamily{#3}\fontseries{#4}\fontshape{#5}%
  \selectfont}%
\fi\endgroup%
\begin{picture}(249,641)(996,1828)
\end{picture}%
 \end{array}} \equiv \ensuremath{\begin{array}{c}\begin{picture}(0,0)%
\includegraphics{pstex/EP-dGR-GR-b.pstex}%
\end{picture}%
\setlength{\unitlength}{3947sp}%
\begingroup\makeatletter\ifx\SetFigFont\undefined%
\gdef\SetFigFont#1#2#3#4#5{%
  \reset@font\fontsize{#1}{#2pt}%
  \fontfamily{#3}\fontseries{#4}\fontshape{#5}%
  \selectfont}%
\fi\endgroup%
\begin{picture}(322,700)(956,1830)
\end{picture}%
 \end{array}} -\hf \ensuremath{\begin{array}{c}\begin{picture}(0,0)%
\includegraphics{pstex/dPEP-GR-GR.pstex}%
\end{picture}%
\setlength{\unitlength}{3947sp}%
\begingroup\makeatletter\ifx\SetFigFont\undefined%
\gdef\SetFigFont#1#2#3#4#5{%
  \reset@font\fontsize{#1}{#2pt}%
  \fontfamily{#3}\fontseries{#4}\fontshape{#5}%
  \selectfont}%
\fi\endgroup%
\begin{picture}(243,694)(1001,1764)
\end{picture}%
 \end{array}} 
		\qquad \mathrm{and} \qquad
	\ensuremath{\begin{array}{c}\begin{picture}(0,0)%
\includegraphics{pstex/Combo-GR.pstex}%
\end{picture}%
\setlength{\unitlength}{3947sp}%
\begingroup\makeatletter\ifx\SetFigFont\undefined%
\gdef\SetFigFont#1#2#3#4#5{%
  \reset@font\fontsize{#1}{#2pt}%
  \fontfamily{#3}\fontseries{#4}\fontshape{#5}%
  \selectfont}%
\fi\endgroup%
\begin{picture}(248,632)(996,1766)
\end{picture}%
 \end{array}} \equiv \ensuremath{\begin{array}{c}\begin{picture}(0,0)%
\includegraphics{pstex/EP-dGR-GR.pstex}%
\end{picture}%
\setlength{\unitlength}{3947sp}%
\begingroup\makeatletter\ifx\SetFigFont\undefined%
\gdef\SetFigFont#1#2#3#4#5{%
  \reset@font\fontsize{#1}{#2pt}%
  \fontfamily{#3}\fontseries{#4}\fontshape{#5}%
  \selectfont}%
\fi\endgroup%
\begin{picture}(322,685)(956,1713)
\end{picture}%
 \end{array}} -\hf \ensuremath{\begin{array}{c}\begin{picture}(0,0)%
\includegraphics{pstex/dPEP-GR-GR.pstex}%
\end{picture}%
\setlength{\unitlength}{3947sp}%
\begingroup\makeatletter\ifx\SetFigFont\undefined%
\gdef\SetFigFont#1#2#3#4#5{%
  \reset@font\fontsize{#1}{#2pt}%
  \fontfamily{#3}\fontseries{#4}\fontshape{#5}%
  \selectfont}%
\fi\endgroup%
\begin{picture}(243,694)(1001,1764)
\end{picture}%
 \end{array}}.
\]
It is now apparent why the definition~\eq{eq:Combo} is
useful.

As a consequence of the decomposition~\eq{eq:Combo},
it is clear that diagrams~\ref{A-GR-combo-B} and~\ref{A-combo-GR-B}
have components where an active gauge
remainder strikes both structures $A$ and
$B$. However, in practical calculations,
this turns out to be a red-herring: it
is most efficient never to perform
this decomposition and so we only ever 
process the explicitly drawn gauge remainders
in~\eq{eq:EP-dCTP-EP}.

It is useful to consider the special case
where the realizations of $A$ and $B$ 
are summed over, as we can 
use \CC\ invariance
to combine terms, in this case. 
Specifically, can we identify the structure $A$ $(B)$
of diagram~\ref{A-combo-GR-B} with the structure
$B$ $(A)$ of diagram~\ref{A-GR-combo-B}.
This is shown
in \fig{fig:Op2-CC} where, for the sake
of generality, we suppose that the
active gauge remainders in~\eq{eq:EP-dCTP-EP}
can be arbitrarily nested.
\bcf[h]
	\[
	\begin{array}{ccc}
		\LID{A-GR-combo-A}	&				& \LID{A-combo-GR-A}
	\\[1ex]
		\ensuremath{\begin{array}{c}\begin{picture}(0,0)%
\includegraphics{pstex/A-GR-combo-A.pstex}%
\end{picture}%
\setlength{\unitlength}{3947sp}%
\begingroup\makeatletter\ifx\SetFigFont\undefined%
\gdef\SetFigFont#1#2#3#4#5{%
  \reset@font\fontsize{#1}{#2pt}%
  \fontfamily{#3}\fontseries{#4}\fontshape{#5}%
  \selectfont}%
\fi\endgroup%
\begin{picture}(555,893)(1575,1598)
\put(1712,2202){\makebox(0,0)[lb]{\smash{{\SetFigFont{11}{13.2}{\rmdefault}{\mddefault}{\updefault}{\color[rgb]{0,0,0}$\decGR{ \ }{\GRkpr^{m}}$}%
}}}}
\end{picture}%
 \end{array}}	& \hspace{1em} +& \ensuremath{\begin{array}{c}\begin{picture}(0,0)%
\includegraphics{pstex/A-combo-GR-A.pstex}%
\end{picture}%
\setlength{\unitlength}{3947sp}%
\begingroup\makeatletter\ifx\SetFigFont\undefined%
\gdef\SetFigFont#1#2#3#4#5{%
  \reset@font\fontsize{#1}{#2pt}%
  \fontfamily{#3}\fontseries{#4}\fontshape{#5}%
  \selectfont}%
\fi\endgroup%
\begin{picture}(554,883)(1575,1598)
\put(1712,2202){\makebox(0,0)[lb]{\smash{{\SetFigFont{11}{13.2}{\rmdefault}{\mddefault}{\updefault}{\color[rgb]{0,0,0}$\decGR{ \ }{\GRkpr^{m}}$}%
}}}}
\end{picture}%
 \end{array}}
	\end{array}
	\]
\caption{Generalized versions of the second and third
diagrams of~\eq{eq:EP-dCTP-EP}, where $A$ and $B$
have been identified with each other.}
\label{fig:Op2-CC}
\ecf

Let us now consider a specific arrangement of
bites to the left / right in diagrams~\ref{A-GR-combo-A} and~\ref{A-combo-GR-A}.
To be precise, we suppose that there are a total of $G$ such operations,
where $G = L + R = L' + R'$, the (un)primed variable
corresponding to diagram (\ref{A-GR-combo-A}) \ref{A-combo-GR-A}.
The signs of diagrams~\ref{A-GR-combo-A} and~\ref{A-combo-GR-A} 
are, according to the rules of \sec{sec:GRs-2}, $(-1)^R$ and $(-1)^{R'}$,
respectively.

We now choose to focus on pairs of terms for which
$R' = L$ and consider taking the charge conjugate
of diagram~\ref{A-GR-combo-A} 
which, we recall, amounts to reflecting it, picking up a sign
for each of the $G$ performed gauge remainders
and the momentum derivative. We remove this
latter sign by reversing
the direction of the momentum derivative symbol 
 so that its sense is once again
counterclockwise. By doing this, we have
arranged for the reflected version of diagram~\ref{A-GR-combo-A}
to look exactly the same as diagram~\ref{A-combo-GR-A}.
The associated sign of the
reflected diagram
is $(-1)^{G+R} = (-1)^{R'}$
which is, of course, just the sign associated with
diagram~\ref{A-combo-GR-A}. Therefore 
diagrams~\ref{A-GR-combo-A}
and~\ref{A-combo-GR-A} can be combined.

At the beginning of this section,
we stated that the effect of
subtractions is to isolate the
components of the parent which
are not Taylor expandable to $\Op{2}$
and that, by construction, the
additions are Taylor expandable to the desired
order. Strictly speaking, this latter
statement requires a caveat. 
The point is that we can encounter
diagrams like~\ref{d:111b.11}
as a sub-diagram of some larger,
factorizable diagram. By factorizable,
we mean that the complete diagram 
has (at least one) internal line which
carries just $p$. Thus, we could imagine
the top-most external field of
diagram~\ref{d:111b.11} instead
being an effective propagator, carrying $p$,
attaching to some other sub-diagram
(which would be decorated by the spare
external field).
This latter sub-diagram, which we note
cannot possess a kernel (only one is allowed
per complete diagram),
may contain
components which are not Taylor expandable
to $\Op{2}$. The strategy is to simply leave 
such sub-diagrams alone, but proceed
as usual with the other part of
the complete diagram \ie the sub-diagram
of the form~\ref{d:111b.11}. The rule that
we will effectively obey is that we only ever
construct subtractions for (sub) diagrams
which contain a kernel. The corresponding
additions, whilst always possessing a sub-diagram
which is Taylor expandable in $p$, may also
possess a sub-diagram which is not. This does
not worry us: we simply manipulate the
Taylor expandable sub-diagram and effectively
ignore the other one.

\subsubsection{General Analysis}
\label{sec:Subtractions-G}

In this section, we analyse the construction
and effects of subtractions in complete generality.
In \sec{sec:bn-Op2}, we will discover
that we only have to construct subtractions for
diagrams with two basic templates, which
are shown in \fig{fig:Subs-Templates}.
\bcf[h]
	\[
	\begin{array}{ccc}
		\LID{CTP-E-DK}	&			& \LID{CTP-E-GRs-K}
	\\[1ex]
		\ensuremath{\begin{array}{c}\input{pstex/CTP-E-DK.pstex_t} \end{array}}, 	& \qquad	& \ensuremath{\begin{array}{c}\input{pstex/CTP-E-GRs-K.pstex_t} \end{array}}
	\end{array}
	\]
\caption{The template for diagrams for which we will construct
subtractions.}
\label{fig:Subs-Templates}
\ecf

Clearly, we have suppressed all other diagrammatic
ingredients which can include vertices, effective
propagators and $\GRkpr$s. Our aim now is to start
fleshing out the sub-diagrams~\ref{CTP-E-DK}
and~\ref{CTP-E-GRs-K} to see if we can find
any components which are not Taylor expandable to
$\Op{2}$. To stand any chance of doing so,
we must find effective propagators or gauge remainders
carrying $l-p$
(where $l$ is a loop momentum) since we recall
that, in the $A^i$ sector, these objects go as
$ 1/ (l-p)^2$ and $(l-p)_\alpha / (l-p)^2$, respectively.
To this end, consider decorating diagram~\ref{CTP-E-DK}
with a single socket, and consider the complete set
of momentum routings, as shown in \fig{fig:Subs-MomRoute}.
\bcf[h]
	\[
	\begin{array}{ccccc}
		\LID{CTP-E-DK-Soc-A}	&			& \LID{CTP-E-DK-Soc-B}	&			& \LID{CTP-E-DK-Soc-C}
	\\[1ex]
		\ensuremath{\begin{array}{c}\input{pstex/CTP-E-DK-Soc-A.pstex_t} \end{array}}, 	& \qquad	& \ensuremath{\begin{array}{c}\input{pstex/CTP-E-DK-Soc-B.pstex_t} \end{array}}, 	& \qquad	& \ensuremath{\begin{array}{c}\input{pstex/CTP-E-DK-Soc-C.pstex_t} \end{array}}
	\end{array}
	\]
\caption{Momentum routings for diagram~\ref{CTP-E-DK}.}
\label{fig:Subs-MomRoute}
\ecf

First, consider diagram~\ref{CTP-E-DK-Soc-A}.
There is no possibility of having an effective
propagator carrying $l-p$ and so the (sub) diagram
derived from fully fleshing out \ref{CTP-E-DK-Soc-A}
is Taylor expandable to $\Op{2}$ (we re-iterate
that the field carrying momentum $p$ could attach
to a factorizable sub-diagram which is not Taylor
expandable to the desired order in $p$).

Now we move on to diagram~\ref{CTP-E-DK-Soc-B}.
Notice that there is no need to consider also
a diagram with the momentum arguments
interchanged because such a diagram identical,
by momentum re-routing invariance. Since $l-p$
is carried by the kernel (rather than by an
effective propagator), there is no chance
of generating a factor of $1/(l-p)^2$ from
the structures drawn. However, the $p$
could, in principle, flow into an effective
propagator or gauge remainder such that their
arguments are \eg $k-p$, as illustrated
in \fig{fig:Subs-MomRoute-b}.
\bcf[h]
	\[
	\begin{array}{ccccc}
		\LID{CTP-E-DK-Soc-Ba}	&			& \LID{CTP-E-DK-Soc-Bc}		&			& \LID{CTP-E-DK-Soc-Bb}
	\\[1ex]
		\ensuremath{\begin{array}{c}\input{pstex/CTP-E-DK-Soc-Ba.pstex_t} \end{array}}, 	& \qquad	& \ensuremath{\begin{array}{c}\input{pstex/CTP-E-DK-Soc-Bc.pstex_t} \end{array}}, 	& \qquad	& \ensuremath{\begin{array}{c}\input{pstex/CTP-E-DK-Soc-Bb.pstex_t} \end{array}}
	\end{array}
	\]
\caption{Fleshings out of diagram~\ref{CTP-E-DK-Soc-B} such that
either an effective propagator or gauge remainder
carries $k-p$.}
\label{fig:Subs-MomRoute-b}
\ecf

At first sight, diagram~\ref{CTP-E-DK-Soc-Ba} seems to possess
the momentum dependence we are looking for, so long as we put
the $k$-dependent effective propagators in the $A^i$ sector, which
we do for all that follows. (
Group theory considerations in fact force them to be in
the $A^1$ sector.)
In this case,
the $k$-dependent effective propagators
\[
	\sim \frac{1}{k^2 (k-p)^2},
\]
which is certainly not Taylor expandable to zeroth order in $p$.
However, we must also consider the vertices. Specifically,
by \CC\ invariance, the field filling the socket
of the topmost vertex must be in the $A^1$ sector. 
Taylor expanding this vertex to zeroth order in $p$
(which we can always do) must yield at least one power of $k$ by
Lorentz invariance. Furthermore, Lorentz invariance
and the requirement that (up to powers of $p^{-2 \epsilon}$)
the diagram is $\Op{2}$
forces us to pick up a second power of $k$ from the
four-point vertex. It is thus quite clear, despite
first appearances, that the diagram is Taylor expandable
to $\Op{2}$.

A similar analysis of diagrams~\ref{CTP-E-DK-Soc-Bc}
and~\ref{CTP-E-DK-Soc-Bb}
leads to the conclusion that the worst momentum
dependence is
\[
	\frac{\order{l^2, l.k}}{(k-p)^2 (l-k)^2 l^2},
\]
in both cases. Consequently, both diagrams are Taylor expandable to
$\Op{2}$.

We now attempt to worsen the behaviour of 
diagram~\ref{CTP-E-DK-Soc-Ba} by adding
additional diagrammatic elements. Clearly, 
simply adding more effective propagators
will do nothing. What we can try, however,
is splitting open one of the existing
effective propagators, as shown in 
\fig{fig:Subs-MomRoute-b2}.
\bcf[h]
	\[
	\LIDi{CTP-E-DK-Soc-Ba-i}{CTP-E-DK-Soc-Ba-i}
	\]
\caption{Attempting to worsen the IR behaviour of diagram~\ref{CTP-E-DK-Soc-Ba}.}
\label{fig:Subs-MomRoute-b2}
\ecf

This seems to have done the trick: rather than having
a single effective propagator giving a factor
of $1/k^2$, we now have two (attached to an 
arbitrary structure), and the diagram
appears not to be Taylor expandable to $\Op{2}$.
However, in all the cases we encounter, we will
find that when we sum over all
possible diagrams which can be represented by
the arbitrary structure, the arbitrary
structure turns out to be transverse in its
external momentum (when the legs are in the $A^i$
sector). Once again we are thwarted: diagrams
of the type~\ref{CTP-E-DK-Soc-Ba-i} are, in practise
Taylor expandable to $\Op{2}$.
Similarly, we cannot modify diagrams~\ref{CTP-E-DK-Soc-Bc}	
and~\ref{CTP-E-DK-Soc-Bb} such that they are
no longer Taylor expandable to $\Op{2}$.

We can repeat precisely the same arguments with 
diagram~\ref{CTP-E-DK-Soc-C} 
and are thus lead to the conclusion
that all diagrams with~\ref{CTP-E-DK} as
a template are Taylor expandable to $\Op{2}$.
We can of course still construct subtractions,
but they exactly cancel the parent (leaving
over the additions).

Diagrams built up from the template~\ref{CTP-E-GRs-K}
are, however, a different kettle of fish.
Indeed, the example with
which we began \sec{sec:Subtractions-I} is of
this type. From diagram~\ref{CTP-E-GRs-K},
there are two ways in which we can construct
diagrams which are not Taylor expandable to $\Op{2}$,
as shown in \fig{fig:Subs-NTE}.
\bcf[h]
	\[
	\begin{array}{ccc}
		\LID{CTP-E-NTE-a}	&			& \LID{CTP-E-NTE-b}
	\\[1ex]
		\ensuremath{\begin{array}{c}\input{pstex/CTP-E-NTE-a.pstex_t} \end{array}},	& \qquad	& \ensuremath{\begin{array}{c}\input{pstex/CTP-E-NTE-b.pstex_t} \end{array}}
	\end{array}
	\]
\caption{The two types of diagram derived from~\ref{CTP-E-GRs-K}
which are not Taylor expandable to $\Op{2}$.}
\label{fig:Subs-NTE}
\ecf

Excluding those terms in which the loose ends
of the effective propagator and kernel of
diagram~\ref{CTP-E-NTE-a} attach to something
transverse in $l$ (in the $A^i$ sector),
both diagrams have a component which goes as
\[
	\frac{1}{l^2 (l-p)^2},
\]
and so both diagrams possess
a component which is not Taylor expandable to
order $p^2$. We know from the example
at the beginning of \sec{sec:Subtractions-I}
that further progress can be made by combining
such diagrams with their subtractions. 
To isolate the components which survive, 
the strategy
is as follows:
\begin{enumerate}
	\item	identify all loop integrals over momenta other than
			$l$ but which carry dependence on $l$;

	\item	determine which, if any, of these loop integrals
			are factorizable from the others;

	\item	Focus on the loop integral / non-factorizable 
			loop integrals for which the kernel is in the
			integrand and either
			\begin{enumerate}
				\item 	Taylor expand to zeroth order in $l$,
						directly;

				\item	construct subtractions if appropriate
						and iterate the procedure.
			\end{enumerate}
\end{enumerate}

An example of this is shown in \fig{fig:Op2-NTE-Ex},
where we recognize that the first diagram
is a generalization of~\ref{d:111b.11}.
\bcf[h]
	\[
	\begin{array}{ccc}
			\LID{CTP-E-NTE-EX}	&		& \LIDLID{CTP-E-NTE-EX-s}{CTP-E-NTE-EX-a}
		\\[1ex]
			\ensuremath{\begin{array}{c}\input{pstex/CTP-E-NTE-EX.pstex_t} \end{array}} 	& \mp 	& \decp{\ensuremath{\begin{array}{c}\input{pstex/CTP-E-NTE-EX-s.pstex_t} \end{array}}}{}
		\end{array}
	\]
\caption{A diagram possessing a component which is not Taylor
expandable to $\Op{2}$, together with its subtraction and addition.}
\label{fig:Op2-NTE-Ex}
\ecf

Following the above recipe, we recognize that
the integrals over $k, \ m_1$ and $m_2$ all contain
dependence on $l$. Given that the $m_2$~integral contains
the kernel and that the $m_1$~integral has dependence
on $m_2$ but the $k$~integral does not, we can
factorize out the integral over $k$. 
The parent
and subtraction combine to yield the first diagram
of \fig{fig:Op2-NTE-Ex-b}, for which
we have constructed a subtraction designed to isolate
the part of the $m_1, \ m_2$ dependent sub-diagram
which is not Taylor expandable in $l$.
\bcf[h]
	\[
	\begin{array}{l}
	\vspace{2ex}
		\begin{array}{ccc}
				\LID{CTP-E-NTE-EX-b}						&		& \LIDLID{CTP-E-NTE-EX-ss}{CTP-E-NTE-EX-sa}
			\\[1ex]
				\left[\ensuremath{\begin{array}{c}\input{pstex/CTP-E-NTE-EX.pstex_t} \end{array}}\right]_{\NTE{p}} 	& \mp 	& \left[\ensuremath{\begin{array}{c}\input{pstex/CTP-E-NTE-EX-ss.pstex_t} \end{array}}\right]_{\NTE{p}}
			\end{array}
	\\
	\vspace{2ex}
		\rightarrow
		\dec{
			\LIO{
				\left[
					\ensuremath{\begin{array}{c}\input{pstex/CTP-E-NTE-EX-Rem.pstex_t} \end{array}} \hspace{1em}
				\right]_{\NTE{p}}
				\Delta^{11}_{\alpha \beta}(p)
				\decp{
					\ensuremath{\begin{array}{c}\input{pstex/CTP-E-EP-Thpt-K-GR.pstex_t} \end{array}}
				}{}
			}{CTP-E-NTE-EX-Rem}
		}{}
	\\
		\qquad
		+
		\dec{
			\LIO{
				\left[
					\ensuremath{\begin{array}{c}\input{pstex/CTP-E-NTE-EX-a.pstex_t} \end{array}}
				\right]_{\NTE{p}}
				\Delta^{11}_{\alpha \beta}(p)
				\decp{
					\ensuremath{\begin{array}{c}\input{pstex/Diagram111b.11-s-b.pstex_t} \end{array}}
				}{}
			}{CTP-E-NTE-EX-Rem-a}	
		}{}
	\end{array}
	\]
\caption{The result of combining diagrams~\ref{CTP-E-NTE-EX} and~\ref{CTP-E-NTE-EX-s}.}
\label{fig:Op2-NTE-Ex-b}
\ecf

Diagram~\ref{CTP-E-NTE-EX-b} combines
with diagram~\ref{CTP-E-NTE-EX-ss} 
to yield diagram~\ref{CTP-E-NTE-EX-Rem}.
Just as we have utilized Lorentz
invariance to draw
diagram~\ref{CTP-E-NTE-EX-Rem} in an
appealing form,
so too do
we recognize that a similar thing can be
done with the addition, \ref{CTP-E-NTE-EX-sa},
yielding diagram~\ref{CTP-E-NTE-EX-Rem-a}.

Returning to diagram~\ref{CTP-E-NTE-EX-Rem},
notice that we must take the component of the
$m_1$ integral which, up to the non-polynomial
dependence on $l$ goes as $l^0$. Were we to
take the $l^2$ component, instead,
then we could not take the $\BigNTE{p}$ component of
the diagram as a whole. The appearance of the
tag $\BigNTE{l}$ is awkward, since it requires
specification of a loop momentum, which is unnatural
from our diagrammatic view point. However, with a little
thought, we can replace this tag with something
which refers only to the external momentum, $p$.
Diagram~\ref{CTP-E-NTE-EX-Rem} obviously possesses
three loop integrals. However, only two of them (the ones over $l$ and $m_1$)
are tagged such that we must take the $\BigNTE{\mathrm{mom}}$ component,
where `$\mathrm{mom}$' refers to the momentum external \wrt\
the appropriate sub-diagram. There is a simple rule to
determine the loops from which we must take the
$\BigNTE{\mathrm{mom}}$ component (a similar
rule can be derived for diagrams of type~\ref{CTP-E-NTE-a}). In all that
follows, if there are any sub-diagrams which
 attach to
just two effective propagators carrying equal momenta (\cf\
\fig{fig:Subs-MomRoute-b2}), we can
ignore each sub-diagram and one of the
effective propagators to which it attaches.
Starting with a diagram of arbitrary loop order, the
recipe is as follows:
\begin{enumerate}
	\item	Identify the loop momentum associated with the internal
			field which attaches to the classical, two-point vertex
			(in the above examples, this momentum has always been $l$)
			and route momenta such that 
			\begin{enumerate}
				\item the external field attaching to the classical,
						two-point vertex carries momentum $+p$;

				\item  momentum $l+p$ flows 
						into the vertex decorated by the other external field;
			\end{enumerate}

	\item	Cut the diagram in every place that there is a gauge remainder
			carrying $l$ and cut the effective propagator carrying $l$
			at the end out of which $+l$ flows.

	\item	Discard the sub-diagrams for which there is
			a gauge remainder which carries just $l$
			(this includes the sub-diagram carrying $p$
			and $l$, which is supposed to have already been tagged with $\BigNTE{p}$);
	\label{it:discard}

	\item	Identify the remaining sub-diagram as the one that must be
			tagged $\BigNTE{l}$;

	\item	Iterate the procedure, if necessary, taking
			the sub-diagram tagged with $\BigNTE{l}$ as the starting point.
\end{enumerate}

In \fig{fig:Op2-NTE-Ex-c} we apply this procedure to
diagram~\ref{CTP-E-NTE-EX-Rem}, with the red lines
indicating the cuts. The above recipe tells us to
tag the sub-diagram with loop momentum $m_1$, but
not the one with loop momentum $k$, since the latter sub-diagram
includes a $\GRkpr_l$ and so is excluded by~\ref{it:discard}, above.
\bcf[h]
	\[
	\ensuremath{\begin{array}{c}\input{pstex/CTP-E-NTE-EX-Rem-Cut.pstex_t} \end{array}}
	\]
\caption{Algorithmic determination of which loop integrals
are tagged with $\NTE{}$.}
\label{fig:Op2-NTE-Ex-c}
\ecf

Since this recipe works at any number of loops,
there is no need to explicitly tag the appropriate
loop integrals of some diagram
already tagged by $\BigNTE{p}$; rather, we simply tag the diagram as 
a whole by $\BigNTEs$.

\section{Preliminary Diagrammatics}
\label{sec:Preliminary}

\subsection{Additional Notation}
\label{sec:Notation}

\subsubsection{(Pseudo) Effective Propagators}

In \sec{sec:GRs-2}, we encountered a single effective
propagator amongst the implicit decorations 
(see \eg diagram~\ref{v-W-GR-w}). We now
generalize to consider an arbitrary number of such
effective propagators. The rule for explicit decoration
is defined as follows. If we wish to join two objects
(say two vertices) together with $j'$ out of a total
of $j$ effective propagators, then there are
$\nCr{j}{j'} 2^{j'}$ ways to do this. 
Intuitively, the first factor captures the 
notion that,
so long as they are implicit decorations, the effective
propagators are indistinguishable. The factor of
$2^{j'}$ allows for the fact that we can interchange
the two ends of an effective propagator. If these
effective propagators were instead used to form $j'$
loops on a single vertex, then the factor of
$2^{j'}$ would disappear, since the vertices
are defined such that all cyclically independent
arrangement of their decorative fields are summed
over.

Given that we have seen how
fields and effective propagators can usefully appear
as implicit decorations, it is not surprising that,
in our calculation of $\beta_n$, 
gauge remainder components and pseudo effective
propagators also appear in this fashion.
The combinatorics for pseudo effective propagators
is exactly the same as for effective
propagators. When we come to decorate
with pseudo effective propagators, we can make life easy
for ourselves by noting that, in practise, pseudo
effective propagators can be arranged to appear
only in one way and always in combination with another
term, precisely as in~\eq{eq:Combo}.
The combinatoric factor
associated with partitioning $j$ $\combo$s
into two sets of $j'$ and $j-j'$ is just
$\nCr{j}{j'}$. However, unlike effective
propagators, the ends of $\combo$ are not
interchangeable.

\subsubsection{Gauge Remainders}
\label{sec:Not:GRs}

We begin
by giving the rules for converting a set of $m$
gauge remainder components of type $\GRkpr$ from implicit
to explicit decorations. Thus, we are considering
decorations of the form
\be
\label{eq:Implict-GR}
	\dec{ \
	}{\GRkpr^m \cdots},
\ee
where the square brackets could enclose some
diagrammatic structure, but need not. The ellipsis
represents any additional implicit decorations,
so long as they are not further instances of $\GRkpr$.
The superscript notation $\GRkpr^m$ simply tells
us, as before, that there are $m$ instances of $\GRkpr$.

There are only two structures we are allowed to form from
implicit decorations containing $\GRkpr$. The first
of these is a ring, as 
in~\fig{fig:nestedhook}. Noting that, in some
completely fleshed out diagram, the sockets of
all the $\GRkpr$ will be filled, this means that
we can generate a diagram like~\ref{Diags-WGR-N1-Bottom-PF}
from implicit decorations, but
never
a diagram like~\ref{Diags-WGRx2-L-Bottom}.
However, there is no restriction that we
create a single ring structure from our $m$
gauge remainders: we can create up to $m$
independent structures. Furthermore,
we need not promote all of the $\GRkpr$
to explicit decorations at the same time.
Thus, we can imagine
partitioning the $m$ gauge remainders into
$q+1$ sets, with the first $q$ of these sets forming
an independent ring structure and the final
set remaining as implicit decorations
\viz
\be
\label{eq:Implict-GR-b}
	M \dec{
		\decGR{ \
		}{\GRkpr^{m^1}} 
		\decGR{ \
		}{\GRkpr^{m^2}}
		\cdots
		\decGR{ \
		}{\GRkpr^{m^q}}
	}{\GRkpr^{m^{q+1}} \cdots},
\ee
where $M$ a combinatoric factor and the notation,
$\decGR{ \ }{\GRkpr^{m^1}}$, tells us that
these decorative gauge remainders must form a
single ring \ie they cannot be further partitioned,
nor can this ring be added to by the remaining
implicit decorations.
The $m^i$ are positive integers which sum up to $m$.

The combinatoric factor
is easy to compute: if we partition $m$ gauge
remainders into two sets of $m'$ and $m-m'$
gauge remainders, then the combinatoric factor
is defined to be
\[
	\nCr{m}{m'}.
\]
Therefore, we
can rewrite~\eq{eq:Implict-GR-b} as:
\[
	\frac{m!}{m^{q+1}!} 
	\dec{
		\prod_{i=1}^q \frac{1}{m^i!}
		\decGR{ \ }{\GRkpr^{m^i}}
	}{\GRkpr^{m^{q+1}}\cdots}.
\]

The final gauge remainder structure
we can construct out of implicit decorations
containing $\GRkpr$s occurs only in diagrams
possessing an $\Op{2}$ stub. The
contexts in which this structure occurs are
illustrated in \fig{fig:GR-Op2String}. 
\bcf[h]
	\[
	\left[
		\LIDi{Op2-GR-String-Ex}{Op2-GR-String-Ex}
	\right]_{p^2}, \qquad
	\LIDi{Op2-GR-String-Ex-B-Mac}{Op2-GR-String-Ex-B}
	\]
\caption{A gauge remainder structure which only occurs in diagrams
possessing an $\Op{2}$ stub.}
\label{fig:GR-Op2String}
\ecf

Diagrams~\ref{Op2-GR-String-Ex} and~\ref{Op2-GR-String-Ex-B}
both possess a string (not a ring) of $m'$ gauge remainders
(\cf \fig{fig:GRstring}),
the last of which bites the socket on the classical,
two-point vertex (in either sense).
In diagram~\ref{Op2-GR-String-Ex}, the
first of the $m'$ gauge remainders is 
bitten by a \combo\ whereas, in diagram~\ref{Op2-GR-String-Ex-B}
it is bitten by a $\GRkpr$ attached to an effective
propagator. The loose end of the \combo\ / effective propagator 
can attach to any of the $\GRkpr$s or to the
socket on the classical, two-point vertex or
to any (un-drawn) structures (the \combo\ cannot
attach to its own socket, though).

All diagrams possessing a string of gauge remainders
bitten by a \combo\ are tagged $p^2$. This
tells
us that everything attaching to
the $\Op{2}$ stub 
is independent of
$p$ (as in \sec{sec:Subtraction}). 
Diagrams tagged with $p^2$ can possess
more than one instance of the structure
which attaches to the stub in 
diagram~\ref{Op2-GR-String-Ex}: although there
is only one stub, structures involving a $\combo$
can attach to the momentum derivative symbol
of another such structure, as illustrated in
\fig{fig:GR-Op2String-b}.
\bcf[h]
	\[
	\left[
		\ensuremath{\begin{array}{c}\input{pstex/Op2-GR-String-Ex-b.pstex_t} \end{array}} \hspace{1em}
	\right]_{p^2}
	\]
\caption{An example of a diagram possessing multiple $\combo$s.}
\label{fig:GR-Op2String-b}
\ecf

We note the following. First,
it is not legal to join any of the $m'$
gauge remainders to any of the $m''$ gauge remainders.
In other words, no two $\combo$s are permitted to 
carry a common loop momentum. Secondly, the $\combo$s are always
tied up such that the momentum derivative symbols
are on the `outside' of the diagram. In other words, we can draw
a line from any momentum derivative symbol / external field to any other
without crossing an internal line. Diagrams in which this
is not the case vanish by a combination of group theory
and \CC\ invariance considerations~\cite{Thesis,mgierg1}.

\subsubsection{Vertices}

When deriving an expression for $\beta_n$,
we will find it invaluable to compactly
represent all vertices in a diagram
which are yet to be explicitly decorated.
To this end, we introduce a set of vertex arguments, $v^j$,
where the upper roman index acts as a label. Thus, the $v^j$
are integers, denoting the loop orders of some set of vertices.
We denote the reduction of these vertices by $v^{j;R}$, where
we recall that a reduced vertex lacks a classical, two-point 
component.

Next, we introduce the compact notation
\beas
	v^{j,j_+} & \equiv & v^j - v^{j+1},
\\
	v^{j,j_+;R} & \equiv & v^{j;R} - v^{j+1;R}.
\eeas
We use this notation to define
\be
\label{eq:CompactVertices}
\SumVertex \equiv \prod_{i=0}^j \sum_{v^{i_+} = 0}^{v^i} \Vertex{v^{i,i_+;R}},
\ee
where the first argument of the structure on the \lhs, $n_s$, gives
the value of $v^0$. Notice that all other vertex arguments are summed over.
The interpretation of the product symbol is as a generator
of  $j+1$ vertices.

The structure shown in~\eq{eq:CompactVertices} always appears as a part of
diagrams which
possess an additional vertex, which carries the argument $v^{j_+}$ (this
argument need not appear on its own---it could be part of something more complicated
\eg $v^{j_+,k}$). An example, which will play an important \role\ later,
is
\be
\label{eq:VertexTower}
	\left[
		\sco[2]{\TopVertex}{\SumVertex}
	\right] \equiv
	\prod_{i=0}^j \sum_{v^{i_+} = 0}^{v^i}
	\left[
	\sco[2]{\TopVertex}{\Vertex{v^{i,i_+;R}}}
	\right].
\ee
Notice that the sum over all vertex arguments is trivially $n_s$:
\be
\label{eq:VertexSum}
	\sum_{i=0}^j v^{i,i_+} + v^{j_+} = \sum_{i=0}^j \left( v^i - v^{i+1} \right) + v^{j+1} = v^0 = n_s.
\ee

The interpretation that the structure defined by~\eq{eq:CompactVertices}
possesses $j+1$ vertices allows us to usefully define~\eq{eq:VertexTower}
for $j=-1$:
\be
\label{eq:Tower-1}
	\left[
		\sco[2]{\TopVertex}{\SumVertex}
	\right]_{j=-1} \equiv
	\Vertex{n^R_s \hspace{0.25em}},
\ee
where $n_s^R$ is, of course, just an $n-s$ loop, reduced vertex. (Note that
this example illustrates the rule that $v^{j_+}$ is replaced by $v^0$;
this holds irrespective of whether or not $v^{j_+}$ occurs only as part of some
more complicated vertex argument.) We can
even usefully define what we mean by~\eq{eq:VertexTower}
for $j=-2$:
\be
\label{eq:Tower-2}
	\left[
		\sco[2]{\TopVertex}{\SumVertex}
	\right]_{j=-2} \equiv
	\delta(n-s).
\ee
Notice that~\eq{eq:Tower-1} still makes
sense if one and only one of the vertices
is decorated (in practise, we will
always take this to be the top vertex):
if more than one vertex is decorated then this
implies that the number of vertices is at least two,
which leads to a contradiction.
On the other hand, \eq{eq:Tower-2} makes sense
only as is, and not if any of the vertices are
decorated. In the computation of $\beta_n$,
we will find structures like~\eq{eq:VertexTower},
where we sum over $j$. The lower value of this
sum will start out at $-2$. However, as we perform
explicit decorations of the vertices, so we will
need to raise the lower limit on $j$, such
that the diagrams still make sense.

\subsubsection{Decoration of Kernels \etc}
\label{sec:Not-K}

It will prove useful to isolate the decorated
components of a kernel from the undecorated
parts.  To facilitate this separation, we introduce the 
symbol $\odot$ to indicate an undecorated kernel and $\circ$ 
to indicate a decorated kernel,
such that
\be
\label{eq:Kernel-Notation}
	\DummyKernel \ = \ \UnDecKernel + \DecKernel.
\ee
Noting that decoration is not defined for
either $\combo$s or pseudo effective propagators
(or gauge remainder components)
we define
\be
\label{eq:Combo-Decoration}
	\ensuremath{\begin{array}{c}\begin{picture}(0,0)%
\includegraphics{pstex/LdL-Combo.pstex}%
\end{picture}%
\setlength{\unitlength}{3947sp}%
\begingroup\makeatletter\ifx\SetFigFont\undefined%
\gdef\SetFigFont#1#2#3#4#5{%
  \reset@font\fontsize{#1}{#2pt}%
  \fontfamily{#3}\fontseries{#4}\fontshape{#5}%
  \selectfont}%
\fi\endgroup%
\begin{picture}(352,629)(1046,1769)
\put(1046,2041){\makebox(0,0)[lb]{\smash{{\SetFigFont{8}{9.6}{\rmdefault}{\mddefault}{\updefault}{\color[rgb]{0,0,0}$\bullet$}%
}}}}
\end{picture}%
 \end{array}} \equiv \ensuremath{\begin{array}{c}\begin{picture}(0,0)%
\includegraphics{pstex/DEP-dGR.pstex}%
\end{picture}%
\setlength{\unitlength}{3947sp}%
\begingroup\makeatletter\ifx\SetFigFont\undefined%
\gdef\SetFigFont#1#2#3#4#5{%
  \reset@font\fontsize{#1}{#2pt}%
  \fontfamily{#3}\fontseries{#4}\fontshape{#5}%
  \selectfont}%
\fi\endgroup%
\begin{picture}(406,697)(1037,1768)
\put(1037,2142){\makebox(0,0)[lb]{\smash{{\SetFigFont{8}{9.6}{\rmdefault}{\mddefault}{\updefault}{\color[rgb]{0,0,0}$\odot$}%
}}}}
\end{picture}%
 \end{array}} + \ensuremath{\begin{array}{c}\begin{picture}(0,0)%
\includegraphics{pstex/EP-DdGR.pstex}%
\end{picture}%
\setlength{\unitlength}{3947sp}%
\begingroup\makeatletter\ifx\SetFigFont\undefined%
\gdef\SetFigFont#1#2#3#4#5{%
  \reset@font\fontsize{#1}{#2pt}%
  \fontfamily{#3}\fontseries{#4}\fontshape{#5}%
  \selectfont}%
\fi\endgroup%
\begin{picture}(435,697)(1008,1768)
\put(1008,1842){\makebox(0,0)[lb]{\smash{{\SetFigFont{8}{9.6}{\rmdefault}{\mddefault}{\updefault}{\color[rgb]{0,0,0}$\bullet$}%
}}}}
\end{picture}%
 \end{array}} -\hf \ensuremath{\begin{array}{c}\begin{picture}(0,0)%
\includegraphics{pstex/DdPEP-GR.pstex}%
\end{picture}%
\setlength{\unitlength}{3947sp}%
\begingroup\makeatletter\ifx\SetFigFont\undefined%
\gdef\SetFigFont#1#2#3#4#5{%
  \reset@font\fontsize{#1}{#2pt}%
  \fontfamily{#3}\fontseries{#4}\fontshape{#5}%
  \selectfont}%
\fi\endgroup%
\begin{picture}(337,630)(1062,1828)
\put(1062,2098){\makebox(0,0)[lb]{\smash{{\SetFigFont{8}{9.6}{\rmdefault}{\mddefault}{\updefault}{\color[rgb]{0,0,0}$\bullet$}%
}}}}
\end{picture}%
 \end{array}},
\ee
where we have used~\eqs{eq:LdL-GRk-Pert}{eq:Combo}.

\subsection{Diagrammatic Functions}
\label{sec:Prelim-DiagFns}

In this section we introduce a set of
diagrammatic functions,
$\nLV{\D}{n}{\mu\nu}{i}(p)$, which will
prove crucial in our treatment of $\beta_n$
(recall that the $1$s are shorthand for $A^1$s).
By analysing their various
properties, we will be able to illustrate
the use of many of the diagrammatic techniques,
in their natural setting. We will ultimately
be interested in the $\Op{2}$ parts of these
diagrams (up to functions not polynomial in $p$).

\subsubsection{The Function $\protect \nLV{\D}{n}{\mu\nu}{a}(p)$}

We introduce the function
\be
\label{eq:nL-a}
	\nLV{\D}{n}{\mu \nu}{a}(p) \equiv
	2 \sum_{s=1}^n  \sum_{m=0}^{2s} \sum_{j=-2}^{n+s-m-2} 
	\frac{\norm_{j+s+1,j+2}}{m!}
	\dec{
		\sco[1]{\TopVertex}{\SumVertex}
	}{11\Delta^{j+s+1}\GRkpr^m},
\ee
where, for the non-negative integers $a$ and $b$,
we define
\be
\label{eq:norm}
	\norm_{a,b} = \frac{(-1)^{b+1}}{a!b!} \left(\frac{1}{2}\right)^{a+1},
\ee
which we see obeys the following relationships:
\numparts
\bea
\label{eq:norm-b}
	2a \norm_{a,b} 		& = & \norm_{a-1,b},
\\
\label{eq:norm-c}
	b\norm_{a,b} 		& = & -\norm_{a,b-1}.
\eea
\endnumparts
In the case that either $a$ or $b$ are negative, $\norm_{a,b}$
is null. The overall combinatoric factor, $\norm_{j+s+1,j+2}/m!$,
is said to be canonical, since the three arguments, $j+s+1$, $j+2$
and $m$ match, respectively, the number of effective propagators,
the number of vertices and the number of $\GRkpr$s.

There is a simple, intuitive explanation for the relationship
between the number of vertices, the number of effective propagators
and the sum over the vertex arguments.
Consider the components of $\nLV{\D}{n}{\mu\nu}{a}(p)$
for which there is at least one vertex \ie for which $j>-2$ and for which
$m=0$. We
know from~\eq{eq:VertexSum} that the sum of the vertex arguments is $n-s$.
Now, given
$j+2$ vertices, exactly $j+1$ effective propagators
are required to create a connected diagram. This leaves
over $s$ effective propagators, each of which must create
a loop. Therefore, the loop order of the diagram is $n-s+s = n$,
as must be the case.
Now we relax the restriction that $m=0$ and suppose that
we create $m'$ gauge remainder rings (\cf \fig{fig:nestedhook}). 
To ensure that each of these rings is connected to
some other part of the diagram, we must use up 
 $m'$ of our
$s$ `spare' effective propagators. However, this is
precisely counteracted by the contribution of
each of the rings to the loop order of the diagram.
Next, consider the component of the diagram
for which $j=-2$:
\be
\label{eq:Dn-j=-2}
	2 \sum_{m=0}^{2n} \frac{\norm_{n-1,0}}{m!} 
		\dec{
			\hspace{1em}
		}{11\Delta^{n-1} \GRkpr^{m}}.
\ee
To ensure that the diagram is fully connected, $m$ must
take the value $2n$ since,
for any smaller value, there are not enough $\GRkpr$s
for the $2(n-1)+2$ fields which require a socket in
which to reside. The loop order of all diagrams
generated by~\eq{eq:Dn-j=-2} is clearly $n$.

Having discussed how the various diagrammatic elements 
conspire to yield a diagram of loop order $n$, we now
discuss the ranges on the various sums. The maximum
value of $s$ clearly follows from the requirement
that the loop order of the diagram is $\geq 0$. 
The maximum values of the sums over $m$ and $j$
and the minimum value of $s$
follow from
the constraint that all fully fleshed out diagrams
must be connected and must have an $\Op{2}$ contribution. 
Suppose that there were a term contributing to $\nLV{\D}{n}{\mu\nu}{a}(p)$
with $s=0$.
In this case, we would no longer have any
decorative gauge remainder components, being left
with just $j+2$ reduced, Wilsonian effective
action vertices, $j+1$ effective propagators
and two external fields. Recalling that effective
propagators are two-ended objects, we see that
there are $2j+4$ available decorations.
Now, from 
\sec{sec:elements} we know that it is imposed as a constraint that all
one-point, Wilsonian effective action vertices vanish
(though one-point, \emph{seed} action vertices do exist, beyond tree level),
in order that the vacuum expectation value of the
superscalar which breaks the $SU(N|N)$ symmetry
is not shifted by quantum corrections.
We can thus insist that all 
Wilsonian effective action vertices are at 
least two-point. This requires $2(j+2)$
decorations, which is precisely equal to
the number of available decorations
for the diagram
as a whole to be connected. Consequently, 
each vertex
must be precisely two-point. Immediately,
this precludes the existence of any classical
vertices, since reduced vertices
do not possess such a component.
Thus we have a line of vertices with a loop
order of at least one apiece. At each end
of the line is a vertex decorated by one external
field and by one end of an effective propagator;
all other vertices are joined to two other vertices,
in each case by a single effective propagator.
Now, since none of these vertices is classical,
they must each, by~\eq{eq:S_>0-11}, contribute $\Op{4}$.
Although the effective propagators each contain
a $1/p^2$ component this is insufficient to prevent
the diagram as a whole vanishing at $\Op{2}$.

In similar fashion, we can understand the
maximum values of the sums over $m$ and $j$.
Of the $j+2$ vertices, suppose
that $T$ are classical vertices. Since the
vertices are reduced,
a minimum of $3T$ decorations are required for
the classical vertices and so, if there are
$m$ gauge remainders, a minimum of  
\be
\label{eq:required}
	2(j+2-T) + 3T + m = 2j+T+m+4
\ee
decorations are required. 
Noting that the 
number of
available decorations is
$2(j+s+1)+2$, it is clear
that 
\be
\label{eq:T+m}
	T+m \leq 2s.
\ee
It therefore follows that
\be
\label{eq:Maxm}
	m \leq 2s.
\ee

Next, let us deduce the maximum number of vertices \ie
the maximum number taken by $j+2$ for some values of
$s$ and $m$. From~\eq{eq:VertexSum},
we know that
the sum over vertex arguments is $n-s$.
Therefore, we can have at most $n-s$ vertices which
are not tree level and so a total of at most $n-s+T$
vertices. Hence,
\[
	j+2 \leq n+s-m,
\]
where we have used~\eq{eq:T+m}.
Before moving on, let us return to the conclusion~\eq{eq:Maxm}.
In fact, there may be a tighter constraint than this
since, if there are more than $n-s$ vertices,
then $T$ is compelled to be greater than zero.
However, we stick with the limit given by~\eq{eq:Maxm},
discarding any diagrams which
turn out to be disconnected and adopting
the prescription that terms for which the
maximum value of the sum over $j$ is less
than its minimum value do not exist.

This completes our discussion of the form of 
$\nLV{\D}{n}{\mu\nu}{a}(p)$ and we will now
discuss some of its properties. However,
rather than working with $\nLV{\D}{n}{\mu\nu}{a}(p)$
directly, we introduce the auxiliary function
\be
\label{eq:nLaux-a}
	\nLV{\Ds}{n}{\mu \nu}{a}(p) \equiv
	2 \sum_{s=0}^n  \sum_{m=0}^{2s+1} \sum_{j=-2}^{n+s-m-1} 
	\frac{\norm_{j+s+1,j+2}}{m!}
	\dec{
		\sco[1]{\TopVertex}{\SumVertex}
	}{11\Delta^{j+s+1}\GRkpr^m}.
\ee
Comparing $\nLV{\Ds}{n}{\mu\nu}{a}(p)$ with $\nLV{\D}{n}{\mu\nu}{a}(p)$,
we notice 
that the ranges on the sums over $s, \ m$ and $j$
differ between the two expressions.
Thus, we know from our previous arguments that
there are diagrams contained in $\nLV{\Ds}{n}{\mu\nu}{a}(p)$ 
which vanish
at $\Op{2}$ and / or contain one-point, Wilsonian effective
action vertices. The point is that, in our computation of
$\beta_n$, we will consider $\dec{\nLV{\Ds}{n}{\mu\nu}{a}(p)}{\bullet}$.
To process this expression, we will use the weak coupling
flow equations. However, when we do so, there are
generically  elements
of the resulting set of terms which,
though not generated by $\dec{\nLV{\D}{n}{\mu\nu}{a}(p)}{\bullet}$,
do not \emph{individually}
vanish.  We will find that some of these terms cancel against
terms generated by the flow of other diagrams and so
we can view $\nLV{\Ds}{n}{\mu\nu}{a}(p)$ as being 
the same as $\nLV{\D}{n}{\mu\nu}{a}(p)$, at $\Op{2}$,
up to a useful re-expression of zero.

We now demonstrate that 
\be
\label{eq:Transverse+TE}
	\nLV{\Ds}{n}{\mu\nu}{a}(p) \sim \Box_{\mu \nu}(p) + \Op{4}.
\ee
To do this, we first show that $\nLV{\Ds}{n}{\mu\nu}{a}(p)$
is transverse in $p$. This is done by contracting
$\nLV{\Ds}{n}{\mu\nu}{a}(p)$
with the momenta
$p_\mu$ and $(-p)_\nu$ of the external 
fields.\footnote{
To prove the transversality of  $\nLDl{a}(p)$ 
it is necessary only to contract
$\nLDl{a}(p)$ with one of the external momenta,
but it technically easier to do it with both.}
Since we are not computing the flow of any of
the vertices, we can discard all one-point
vertices, and so decrease the upper limits
on the sums over $m$ and $j$ by one each.
Explicitly decorating with one of
the external fields but leaving the other
as an implicit decoration 
yields the diagrams of \fig{fig:bn-Transverse}.
\bcf[h]
	\beas
		p_\mu (-p)_\nu \nLDl{a}(p) & = &
		-2
		\bca{
			\sum_{s=0}^n  \sum_{m=0}^{2s} \sum_{j=-1}^{n+s-m-2} 
			\frac{\norm_{j+s+1,j+1}}{m!}
		}
			{
			\dec{
				\sco[1]{
					\LIDi{vj+R-EGR}{vj+R-EGR}
				}{\SumVertex}
			}{\ExtGR \!\! \Delta^{j+s+1}\GRkpr^m}
		}
	\\[2ex]
		& & +2
		\bca{
			\sum_{s=0}^n  \sum_{m=2}^{2s} \sum_{j=-2}^{n+s-m-2} 
			\frac{\norm_{j+s+1,j+2}}{m!} \sum_{m'=2}^m
		}
			{
			 \!\! \nCr{m}{m'} \!\!
			\dec{
				\Tower{
					\LIDi{GRs-EGR}{GRs-EGR}
				}
			}{\ExtGR \!\! \Delta^{j+s+2}\GRkpr^{m-m'}}
		}
	\eeas
\caption{Contraction of $\nLDl{a}(p)$ with its external
momenta.}
\label{fig:bn-Transverse}
\ecf

Diagram~\ref{vj+R-EGR} comes with a factor of $j-2$,
compared to the parent (which have have absorbed into
$\norm$, using~\eq{eq:norm-c}), recognizing
the indistinguishability of the $j+2$ vertices,
prior to explicit decoration. 
Furthermore, the lower limit on the sum over $j$ is
$-1$ and not $-2$, since diagram~\ref{vj+R-EGR}
must possess at least one vertex.
The combinatorics
for diagram~\ref{GRs-EGR} follows from~\sec{sec:Not:GRs};
we have recognized that for the diagram as a whole to
be connected (or, alternatively, on account of diagrammatic
identity~\eq{D-ID-Trivial}), the explicit gauge remainder
structure must possess a minimum of two $\GRkpr$s.

The reduced vertex of 
diagram~\ref{vj+R-EGR} 
which is struck by the gauge remainder
can
be promoted to a full vertex, since a classical,
two-point vertices contracted into a $\GRk$
dies, courtesy of diagrammatic 
identity~\eq{eq:GR-TLTP}.
Allowing the gauge remainder
to act---whereupon it necessarily strikes 
a socket---we split the resulting vertex
into a reduced part and a classical, two-point
part. In the latter case, we shift $j \rightarrow j+1$.
Since no other gauge remainders have acted (and no Taylor
expansions have been performed), we could use
\CC\ invariance to collect together
pushes forward and pulls back. However, we refrain
from doing this, for reasons that will become
apparent.
The result of processing diagram~\ref{vj+R-EGR} 
is shown in \fig{fig:bn-Transverse-b} where,
for each term,
the explicitly drawn external field can bite the
socket in either sense.
\bcf[h]
	\[
	-2
		\bca{
			\sum_{s=0}^n  \sum_{m=0}^{2s-1} \sum_{j=-1}^{n+s-m-3} 
			\frac{\norm_{j+s+1,j+1}}{m!}
		}
			{
			\dec{
				\sco[1]{
					\LIDi{vj+R-EGR-Socket}{vj+R-EGR-Socket}
				}{\SumVertex}
			}{\ExtGR \!\! \Delta^{j+s+1}\GRkpr^m}
		}
	-2
		\bca{
			\sum_{s=0}^n  \sum_{m=0}^{2s-1} \sum_{j=-2}^{n+s-m-3} 
			\frac{\norm_{j+s+2,j+2}}{m!}
		}
			{
			\dec{
				\Tower{
					\LIDi{CTP-EGR-Socket}{CTP-EGR-Socket}
				}
			}{\ExtGR \!\! \Delta^{j+s+2}\GRkpr^m}
		}
	\]
\caption{The result of processing diagram~\ref{vj+R-EGR}.}
\label{fig:bn-Transverse-b}
\ecf

Notice that the limits on the sums over $m$ and 
$j$ differ from those of the parent. For example, 
in diagram~\ref{vj+R-EGR-Socket}, where the gauge
remainder strikes a socket on a reduced vertex,
we have `used up' a decoration and so can 
decrease the limits on the sums over
$m$ and $j$ by one each.
In diagram~\ref{CTP-EGR-Socket}
similar consideration apply, but here 
notice that the lower limit on the sum over $j$ is now
$-2$, as a consequence of having
shifted $j \rightarrow j+1$.

The next step is to decorate the classical,
two-point vertex of diagram~\ref{CTP-EGR-Socket},
just as we did with diagram~\ref{v-W-GR-w-b02}
(see \figs{fig:Decompose}{fig:EP-Ex}). If we
attach the remaining \ExtGR, the diagram
dies, courtesy of diagrammatic identity~\eq{eq:GR-TLTP}.
Thus, all we  can do is
attach an effective propagator,
the loose end of which joins
either to a reduced Wilsonian effective
action vertex or to a gauge remainder
structure (recall that if the both
ends of the effective propagator attach
to the classical, two-point vertex then the
diagram dies as a consequence of \CC\ invariance). 
In each case, the combinatoric
factor associated with choosing one effective
propagator, either end of which can attach to
the classical, two-point vertex, is
$2(j+s+2)$. In the case that the other end attaches
to one of the $j+2$ reduced vertex, there is an additional
factor of $j+2$.

\begin{Icancel}
\label{Icancel:Tansverse}
	The component of diagram~\ref{CTP-EGR-Socket}
	in which the classical, two-point vertex
	is attached to a reduced, Wilsonian effective
	action vertex cancels diagram~\ref{vj+R-EGR-Socket},
	up to a gauge remainder contribution.
\end{Icancel}

Crucially, the surviving gauge remainder contribution
is a nested version of diagram~\ref{vj+R-EGR}.
Thus, processing the nested gauge remainder
just repeats the above cancellations,
generating successively more nested diagrams.
The process terminates when there are insufficient
decorations for the gauge remainder to strike
a reduced vertex. What, then, remains? At each level of
nesting (including the un-nested case, as discussed
already), we can
attach the classical, two-point vertex to a
gauge remainder structure. Additionally,
from the first level of nesting onwards,
we can attach the classical, two-point vertex
to one of the gauge remainders which is nested
\wrt\ the gauge remainder which bites the vertex. The
complete set of surviving diagrams is collected
together in \fig{fig:bn-Transverse-c}.
\bcf
	\[
	\begin{array}{l}
	\vspace{2ex}
		-2
		\bca{
			\sum_{s=0}^n  \sum_{m'=1}^{2s-1} \sum_{m=0}^{2s-1-m'} \sum_{j=-1}^{n+s-m-m'-3} 
			\frac{\norm_{j+s+1,j+2}}{m!}
		}
			{
			\dec{
				\Tower{
					\LIDi{EGR-GRs-EP-TLTP}{EGR-GRs-EP-TLTP}
				}
			}{\ExtGR \!\! \Delta^{j+s+1}\GRkpr^m}
		}
	\\
		-2
		\bca{
			\sum_{s=0}^n  \sum_{m''=0}^{2s-3}  \sum_{m=2}^{2s-1-m''} \sum_{j=-1}^{n+s-m-m''-3} 
			\frac{\norm_{j+s+1,j+2}}{m!m''!} \sum_{m'=2}^{m}
		}
			{
			\!\! \nCr{m}{m'} \!\!
			\dec{
				\Tower{\LIDi{EGR-Skt-CTP-EP-GRs}{EGR-Skt-CTP-EP-GRs}}	
			}{\ExtGR \!\! \Delta^{j+s+1}\GRkpr^{m-m'}}
		}
	\end{array}
	\]
\caption{All surviving terms generated by the diagrams
of \fig{fig:bn-Transverse-b}.}
\label{fig:bn-Transverse-c}
\ecf

Notice that, in diagram~\ref{EGR-Skt-CTP-EP-GRs},
the sum over $m'$ starts from two, as the $m'=1$
case vanishes on account of diagrammatic 
identity~\eq{eq:D-ID-Bitten-hook}.
There is a very important point to make about
diagrams~\ref{EGR-GRs-EP-TLTP} 
and~\ref{EGR-Skt-CTP-EP-GRs}. Recall that, in
\sec{sec:Not:GRs}, we stated strings of gauge remainders
formed by implicit decorations can only occur
in diagrams with $\Op{2}$ stubs, yet here we
appear to have a counter-example. 
However, the strings of gauge remainders
in diagrams~\ref{EGR-GRs-EP-TLTP} 
and~\ref{EGR-Skt-CTP-EP-GRs} are
formed by the action of nested gauge remainders
and \emph{not} by the original set of implicit decorations.
With these structures formed, it is useful
to sum over nestings, in the usual way,
using the notation $\decGR{ \ }{\GRkpr^{q}}$, as indicated
(remembering to compensate with a
combinatoric factor of $1/q!$).
However, we can not further promote these gauge
remainders to join the implicit decorations of the diagram
as a whole.

We now find a wonderful cancellation.

\begin{Icancel}
	Diagrams~\ref{GRs-EGR}, \ref{EGR-GRs-EP-TLTP}
	and~\ref{EGR-Skt-CTP-EP-GRs} cancel,
	courtesy of diagrammatic identity~\eq{eq:D-ID-G}.
	This is seen most clearly my making the following
	changes of variables:
	\begin{description}
		\item[Diagram~\ref{GRs-EGR}] 
			$m \rightarrow m+1$, $m' \rightarrow m'+1$;
	
		\item[Diagram~\ref{EGR-GRs-EP-TLTP}] 
			$m \rightarrow m-m'$,
			and recognising that
			\[
			\sum_{m'=1}^{2s-1} \sum_{m=m'}^{2s-1} = \sum_{m=1}^{2s-1} \sum_{m'=1}^m;
			\]

		\item[Diagram~\ref{EGR-Skt-CTP-EP-GRs}]
			$m \rightarrow m-m''$, $m' \rightarrow m'-m''$
			and recognising that
			\[
			\sum_{m''=0}^{2s-3} \sum_{m=m''+2}^{2s-1} \sum_{m'=2+m''}^m =
			\sum_{m=2}^{2s-1} \sum_{m'=2}^m \sum_{m''=0}^{m'-2}.
			\]	
	\end{description}
\end{Icancel}

We have thus demonstrated that
\be
\label{eq:Transverse}
	p_\mu (-p)_\nu \nLV{\Ds}{n}{\mu\nu}{a}(p) = 0,
\ee
as we set out to. The patterns of cancellations
involved are the template for all cancellations
we will encounter when we deal with gauge
remainders henceforth.

At first sight, \eq{eq:Transverse+TE} follows
directly from~\eq{eq:Transverse}.
However, consider the components of $\nLV{\Ds}{n}{\mu\nu}{a}(p)$
which are factorizable. The worry is that such diagrams
necessarily contain at least one
effective propagator carrying just
the external momentum, which contributes
a factor of $1/p^2 + \Op{0}$. The resolution to this
problem comes from breaking all such contributions
down into non-factorizable sub-diagrams and
recognizing that these sub-diagrams are
transverse, thereby providing
the necessary powers of $p$ to cancel
those coming from the effective propagators.

To
see this explicitly, we consider the component of
$\nLV{\Ds}{n}{\mu\nu}{a}(p)$ which possesses two non-factorizable
sub-diagrams. To this end, we split up 
\begin{enumerate}
	\item	the two external fields;

	\item	the $j+s+1$ effective
			propagators into three sets of $j+s-s'$, $s'$ and 1
			effective propagator(s) with the single effective
			propagator required to join together the two factorizable
			sub-diagrams;

	\item	the $j+2$ vertices into two sets of $j-j'$
			and $j'+2$ vertices; 

	\item	the $m$ gauge remainders
			into two sets of $m-m'$ and $m'$ gauge remainders;

\end{enumerate}

Noting that we sum over $j'$, $m'$ and $s'$, the combinatoric
factor is
\be
\label{eq:Factorize-combins}
	2 (j+s+1) \nCr{j+s}{j+s-s'} \nCr{j+2}{j'+2} \nCr{m}{m'}.
\ee
Including this
with~\eq{eq:Factorize-combins} and combining the
result
with $\norm_{j+s+1,j+2} / m!$ yields
\[
	-2 \frac{\norm_{j+s-s', j-j'}}{(m-m')!} \frac{\norm_{s',j'+2}}{m'!}.
\]
It is now convenient to shift $j \rightarrow j+j'+2$,
$s' \rightarrow s' + j' +1$ and $m \rightarrow m+m'$,
which yields the contribution shown in \fig{fig:Factorize}.
It is understood that the repeated vertex argument $v^{j+2}$
is summed over.
\bcf[h]
	\[
	\begin{array}{c}
	\vspace{2ex}
		\ds
		-4 	\sum_{s=0}^n  \sum_{m=0}^{2s+1} \sum_{m'=0}^{2s+1-m} \sum_{j=-2}^{n+s-m-m'-1} 
			\sum_{j'=-2}^{n+s-m-m'-j-3} \sum_{s'=-j'-1}^{j+s+1} 	
	\\
		\left[
			\begin{array}{cc}
			\vspace{1ex}
				\ds \frac{\norm_{j'+s'+1,j'+2}}{m'!} 	& 
				\dec{
					\sco[1]{
						\ensuremath{\begin{array}{c}\begin{picture}(0,0)%
\includegraphics{pstex/vj+jpr+3-R.pstex}%
\end{picture}%
\setlength{\unitlength}{3947sp}%
\begingroup\makeatletter\ifx\SetFigFont\undefined%
\gdef\SetFigFont#1#2#3#4#5{%
  \reset@font\fontsize{#1}{#2pt}%
  \fontfamily{#3}\fontseries{#4}\fontshape{#5}%
  \selectfont}%
\fi\endgroup%
\begin{picture}(784,780)(466,1971)
\put(502,2296){\makebox(0,0)[lb]{\smash{{\SetFigFont{11}{13.2}{\rmdefault}{\mddefault}{\updefault}{\color[rgb]{0,0,0}$\nothing {v}^{j+j'+3;R}$}%
}}}}
\end{picture}%
 \end{array}}
					}{\Vertex{v^{j+2},j'}}
				}{1 \Delta^{j'+s'+1} \GRkpr^{m'}}
			\\
			\vspace{1ex}
				& \hspace{-4.5em} \mbox{\rule{.2mm}{1cm}}
			\\
				\ds \frac{\norm_{j+s+1-s',j+2}}{m!} &
				\dec{
					\sco[1]{
						\ensuremath{\begin{array}{c}\begin{picture}(0,0)%
\includegraphics{pstex/vj_+j+2-R.pstex}%
\end{picture}%
\setlength{\unitlength}{3947sp}%
\begingroup\makeatletter\ifx\SetFigFont\undefined%
\gdef\SetFigFont#1#2#3#4#5{%
  \reset@font\fontsize{#1}{#2pt}%
  \fontfamily{#3}\fontseries{#4}\fontshape{#5}%
  \selectfont}%
\fi\endgroup%
\begin{picture}(760,756)(478,1983)
\put(523,2291){\makebox(0,0)[lb]{\smash{{\SetFigFont{11}{13.2}{\rmdefault}{\mddefault}{\updefault}{\color[rgb]{0,0,0}$\nothing {v}^{j_+,j+2;R}$}%
}}}}
\end{picture}%
 \end{array}}
					}{\SumVertex}
				}{1 \Delta^{j+s+1-s'} \GRkpr^{m}}
			\end{array}
		\right]
	\end{array}
	\]
\caption{A factorizable contribution to $\nLDl{a}(p)$. The two sub-diagrams are understood to be non-factorizable.}
\label{fig:Factorize}
\ecf

The crucial point to recognize is that the
combinatoric factor for each of the sub-diagrams
is the canonical one. Immediately we see
that each of these sub-diagrams is transverse.
It is trivial to generalize this analysis to the case
where there are an arbitrary number of non-factorizable
sub-diagrams, from which~\eq{eq:Transverse+TE} follows.
In view of this, it is hardly surprising
that we can split  $\nLV{\Ds}{n}{\mu\nu}{a}(p)$
into Non-Factorizable (NF) components
in an enlightening way.
Indeed, shifting $s \rightarrow s+s'$
and $v^{i\leq j+2} \rightarrow v^i - s'$ it is apparent that
\be
\label{nLDa-factorize}
	\nLV{{\Ds}}{n}{\mu\nu}{a}(p) 
	=
	\nLV{\overline{\Ds}}{n}{\mu\nu}{a}(p)
			- \sum_{n'=1}^{n-1}
			\nLV{\overline{\Ds}}{n-n'}{\mu \rho}{a}(p) 
			\Delta^{1\,1}_{\rho \sigma}(p)
			\nLV{\overline{\Ds}}{n'}{\sigma \nu}{a}(p)
	+\ldots,
\ee
where 
$\nLV{\overline{\Ds}}{n}{\mu\nu}{a}(p)
\equiv \left. \nLV{\Ds}{n}{\mu\nu}{a}(p) \right|_{\mathrm{NF}}$,
and the ellipsis builds up a full geometric series.

\subsubsection{The Function $\protect \nLV{\D}{n}{\mu\nu}{b}(p)$}

We define the function $\nLV{\D}{n}{\mu\nu}{b}(p)$
and two auxiliary functions as shown in \fig{fig:nL-b}.
Notice that the range for $s$ is the same for all three
functions; only the upper limits on $m$ and $j$ differ.
\bcf
\beas
	\nLV{\D}{n}{\mu\nu}{b}(p) \equiv
	\bca{
		4 \sum_{s=1}^n  \sum_{r=1}^s \sum_{m=0}^{2(s-r)} 
		\sum_{j=-2}^{n+s-2r-m-2} 
		\frac{\norm_{j+s+2-r,j+2}}{m!r!}
	}
		{
		\decp{
				\Tower{\ensuremath{\begin{array}{c}\begin{picture}(0,0)%
\includegraphics{pstex/CTP-E.pstex}%
\end{picture}%
\setlength{\unitlength}{3947sp}%
\begingroup\makeatletter\ifx\SetFigFont\undefined%
\gdef\SetFigFont#1#2#3#4#5{%
  \reset@font\fontsize{#1}{#2pt}%
  \fontfamily{#3}\fontseries{#4}\fontshape{#5}%
  \selectfont}%
\fi\endgroup%
\begin{picture}(466,586)(719,713)
\put(801,1067){\makebox(0,0)[lb]{\smash{{\SetFigFont{11}{13.2}{\rmdefault}{\mddefault}{\updefault}{\color[rgb]{0,0,0}$0^2$}%
}}}}
\end{picture}%
 \end{array}}}
		}{1\Delta^{j+s+2-r} \combo^r \GRkpr^m }{}
	}
\\[2ex]
	\nLV{\Ds}{n}{\mu\nu}{b}(p) \equiv
	\bca{ 
		4 \sum_{s=1}^n  \sum_{r=1}^s \sum_{m=0}^{2(s-r)+1} 
		\sum_{j=-2}^{n+s-2r-m-1} 
		\frac{\norm_{j+s+2-r,j+2}}{m!r!}
	}
		{
		\qquad \times
		\decp{
				\Tower{\ensuremath{\begin{array}{c}\begin{picture}(0,0)%
\includegraphics{pstex/CTP-E.pstex}%
\end{picture}%
\setlength{\unitlength}{3947sp}%
\begingroup\makeatletter\ifx\SetFigFont\undefined%
\gdef\SetFigFont#1#2#3#4#5{%
  \reset@font\fontsize{#1}{#2pt}%
  \fontfamily{#3}\fontseries{#4}\fontshape{#5}%
  \selectfont}%
\fi\endgroup%
\begin{picture}(466,586)(719,713)
\put(801,1067){\makebox(0,0)[lb]{\smash{{\SetFigFont{11}{13.2}{\rmdefault}{\mddefault}{\updefault}{\color[rgb]{0,0,0}$0^2$}%
}}}}
\end{picture}%
 \end{array}}}
		}{1\Delta^{j+s+2-r} \combo^r \GRkpr^m }{}		
	}
\\[2ex]
	\nLV{\Ds}{n}{\mu\nu}{b'}(p) \equiv
	\bca{ 
		4 \sum_{s=1}^n  \sum_{r=1}^s \sum_{m=0}^{2(s-r)+1} 
		\sum_{j=-2}^{n+s-2r-m-1} 
		\frac{\norm_{j+s+2-r,j+2}}{m!r!}
	}
		{
		\dec{
				\Tower{\decp{\ensuremath{\begin{array}{c}\begin{picture}(0,0)%
\includegraphics{pstex/CTP-E.pstex}%
\end{picture}%
\setlength{\unitlength}{3947sp}%
\begingroup\makeatletter\ifx\SetFigFont\undefined%
\gdef\SetFigFont#1#2#3#4#5{%
  \reset@font\fontsize{#1}{#2pt}%
  \fontfamily{#3}\fontseries{#4}\fontshape{#5}%
  \selectfont}%
\fi\endgroup%
\begin{picture}(466,586)(719,713)
\put(801,1067){\makebox(0,0)[lb]{\smash{{\SetFigFont{11}{13.2}{\rmdefault}{\mddefault}{\updefault}{\color[rgb]{0,0,0}$0^2$}%
}}}}
\end{picture}%
 \end{array}}}{\combo^r}}
			}{1\Delta^{j+s+2-r}  \GRkpr^m }
	}
\eeas
\caption{The definition of $\nLV{\D}{n}{\mu\nu}{b}(p)$
and two auxiliary functions.}
\label{fig:nL-b}
\ecf

In the case of $\nLV{\Ds}{n}{\mu\nu}{b'}(p)$,
we must be careful to define precisely
what it is we mean by $\decp{\ }{}$, in this context.
The basic notion is that for all diagrammatic 
components enclosed by $\decp{\ }{}$,
all $p$ dependence comes from the $\Op{2}$ stub, alone;
those components not explicitly enclosed
by $\decp{\ }{}$ can either be brought under
its influence or not, as we begin to flesh
out the diagrams. Thus, for the
non-factorizable components of $\nLV{\Ds}{n}{\mu\nu}{b'}(p)$,
all diagrammatic elements can be brought under
the influence of $\decp{\ }{}$, whereas this is
not the case for
the factorizable components. Notice, though, that all $\combo$s
must decorate the part of the diagram under the influence of
$\decp{\ }{}$.
With these points in mind, we can
write
\be
\label{eq:b'-decomp}
	\nLV{\Ds}{n}{\mu\nu}{b'}(p) = \nLV{\Ds}{n}{\mu\nu}{b}(p)
	- \sum_{n'=1}^{n-1} \nLV{\Ds}{n-n'}{\mu\rho}{b}(p)
	\Delta^{11}_{\rho \sigma}(p) \nLV{\Ds}{n'}{\sigma\nu}{a}(p).
\ee

\subsubsection{The Function $\protect \nLV{\D}{n}{\mu\nu}{c}(p)$}

We define the function $\nLV{\D}{n}{\mu\nu}{c}(p)$,
together with an auxiliary function, as follows:
\bea
\label{eq:nL-c}
	\nLV{\D}{n}{\mu\nu}{c}(p) \equiv
	\sum_{s=1}^n \sum_{m=1}^{2s} \sum_{j=-2}^{n+s-m-2} 
	\frac{\norm_{j+s+2,j+2}}{m!}
	\dec{
		\Tower{\ensuremath{\begin{array}{c}\begin{picture}(0,0)%
\includegraphics{pstex/CTP-E.pstex}%
\end{picture}%
\setlength{\unitlength}{3947sp}%
\begingroup\makeatletter\ifx\SetFigFont\undefined%
\gdef\SetFigFont#1#2#3#4#5{%
  \reset@font\fontsize{#1}{#2pt}%
  \fontfamily{#3}\fontseries{#4}\fontshape{#5}%
  \selectfont}%
\fi\endgroup%
\begin{picture}(466,586)(719,713)
\put(801,1067){\makebox(0,0)[lb]{\smash{{\SetFigFont{11}{13.2}{\rmdefault}{\mddefault}{\updefault}{\color[rgb]{0,0,0}$0^2$}%
}}}}
\end{picture}%
 \end{array}}}
	}{1 \Delta^{j+s+2}\GRkpr^m}
\\
\label{eq:nLaux-c}
	\nLV{\Ds}{n}{\mu\nu}{c}(p) \equiv
	\sum_{s=1}^n \sum_{m=1}^{2s+1} \sum_{j=-2}^{n+s-m-1} 
	\frac{\norm_{j+s+2,j+2}}{m!}
	\dec{
		\Tower{\ensuremath{\begin{array}{c}\begin{picture}(0,0)%
\includegraphics{pstex/CTP-E.pstex}%
\end{picture}%
\setlength{\unitlength}{3947sp}%
\begingroup\makeatletter\ifx\SetFigFont\undefined%
\gdef\SetFigFont#1#2#3#4#5{%
  \reset@font\fontsize{#1}{#2pt}%
  \fontfamily{#3}\fontseries{#4}\fontshape{#5}%
  \selectfont}%
\fi\endgroup%
\begin{picture}(466,586)(719,713)
\put(801,1067){\makebox(0,0)[lb]{\smash{{\SetFigFont{11}{13.2}{\rmdefault}{\mddefault}{\updefault}{\color[rgb]{0,0,0}$0^2$}%
}}}}
\end{picture}%
 \end{array}}}
	}{1 \Delta^{j+s+2}\GRkpr^m}.
\eea

As with $\Ds^a$, it is useful to isolate the non-factorizable components
of $\Ds^c$:
\be
\label{nLDc-factorize}
	\nLV{\Ds}{n}{\mu\nu}{c}(p) = \nLV{\overline{\Ds}}{n}{\mu\nu}{c}(p)
	- \sum_{n'=1}^{n-1}
	\nLV{\overline{\Ds}}{n-n'}{\mu\rho}{c}(p)
	\Delta^{1\,1}_{\rho \sigma}(p)
	\nLV{\Ds}{n'}{\sigma\nu}{a}(p).
\ee

We now demonstrate that
\be
\label{eq:nLV-c-Op2}
	2 \decp{\nLV{\overline{\Ds}}{n}{\mu\nu}{c}(p)}{} = \nLV{\Ds}{n}{\mu\nu}{b}(p) + \mathrm{remainders}.
\ee

The presence of the $\decp{\ }{}$ allows us to Taylor
expand all diagrammatic components of $\nLV{\overline{\Ds}}{n}{\mu\nu}{c}(p)$, besides
the $\Op{2}$ stub, to $\Op{0}$. The result
of this procedure is shown in \fig{fig:nLV-c-Properties-a}.
\bcf
	\[
	\begin{array}{l}
	\vspace{2ex}
		-4
		\bca{
			\sum_{s=1}^n  \sum_{m=1}^{2s}  \sum_{j=-1}^{n+s-m-2} 
			\frac{\norm_{j+s+2,j+1}}{m!}
		}
			{
			\decp{
				\begin{array}{c}
				\vspace{1ex}
					\LIDi{dvj+R}{nLV-c-dvj+R}
				\\
				\vspace{1ex}
					\SumVertex
				\\
					\ensuremath{\begin{array}{c}\begin{picture}(0,0)%
\includegraphics{pstex/CTP-E.pstex}%
\end{picture}%
\setlength{\unitlength}{3947sp}%
\begingroup\makeatletter\ifx\SetFigFont\undefined%
\gdef\SetFigFont#1#2#3#4#5{%
  \reset@font\fontsize{#1}{#2pt}%
  \fontfamily{#3}\fontseries{#4}\fontshape{#5}%
  \selectfont}%
\fi\endgroup%
\begin{picture}(466,586)(719,713)
\put(801,1067){\makebox(0,0)[lb]{\smash{{\SetFigFont{11}{13.2}{\rmdefault}{\mddefault}{\updefault}{\color[rgb]{0,0,0}$0^2$}%
}}}}
\end{picture}%
 \end{array}}
				\end{array}
			}{1\Delta^{j+s+2} \GRkpr^m}
		}
		-4
		\bca{
			\sum_{s=1}^n  \sum_{m=1}^{2s}  \sum_{j=-2}^{n+s-m-2} 
			\frac{\norm_{j+s+3,j+2}}{m!}
		}
			{
			\decp{
				\begin{array}{c}
				\vspace{1ex}
					\LIDi{dCTP}{nLV-c-dCTP}
				\\
				\vspace{1ex}
					\TopVertex
					\SumVertex
				\\
					\ensuremath{\begin{array}{c}\begin{picture}(0,0)%
\includegraphics{pstex/CTP-E.pstex}%
\end{picture}%
\setlength{\unitlength}{3947sp}%
\begingroup\makeatletter\ifx\SetFigFont\undefined%
\gdef\SetFigFont#1#2#3#4#5{%
  \reset@font\fontsize{#1}{#2pt}%
  \fontfamily{#3}\fontseries{#4}\fontshape{#5}%
  \selectfont}%
\fi\endgroup%
\begin{picture}(466,586)(719,713)
\put(801,1067){\makebox(0,0)[lb]{\smash{{\SetFigFont{11}{13.2}{\rmdefault}{\mddefault}{\updefault}{\color[rgb]{0,0,0}$0^2$}%
}}}}
\end{picture}%
 \end{array}}
				\end{array}
			}{1 \Delta^{j+s+3} \GRkpr^m}
		}
	\\
	\vspace{2ex}
		+2
		\bca{
			\sum_{s=1}^n  \sum_{m=2}^{2s}  \sum_{j=-2}^{n+s-m-2} 
			\frac{\norm_{j+s+2,j+2}}{m!} \sum_{m'=2}^m \nCr{m}{m'}
		}
			{
			\decp{
				\begin{array}{c}
				\vspace{1ex}
					\LIO{\TopVertex}{nLV-c-CTP-GRs-TE}
				\\
				\vspace{1ex}
					\SumVertex
				\\
					\ensuremath{\begin{array}{c}\input{pstex/CTP-E-TEGRs.pstex_t} \end{array}}
				\end{array}
			}{1\Delta^{j+s+2} \GRkpr^{m-m'}}
		}
	\\
		+2
		\bca{
			\sum_{s=1}^n  \sum_{m=4}^{2s}  \sum_{j=-2}^{n+s-m-2} 
			\frac{\norm_{j+s+2,j+2}}{m!} \sum_{m'=3}^{m-1} \nCr{m}{m'}
		}
			{
			\decp{
				\begin{array}{c}
				\vspace{1ex}
					\LIO{
						\decGR{\ }{\socket \GRkpr^{m'}}
					}{nLV-c-TE-GRs}
				\\
				\vspace{1ex}
					\TopVertex
				\\
				\vspace{1ex}
					\SumVertex
				\\
					\ensuremath{\begin{array}{c}\begin{picture}(0,0)%
\includegraphics{pstex/CTP-E.pstex}%
\end{picture}%
\setlength{\unitlength}{3947sp}%
\begingroup\makeatletter\ifx\SetFigFont\undefined%
\gdef\SetFigFont#1#2#3#4#5{%
  \reset@font\fontsize{#1}{#2pt}%
  \fontfamily{#3}\fontseries{#4}\fontshape{#5}%
  \selectfont}%
\fi\endgroup%
\begin{picture}(466,586)(719,713)
\put(801,1067){\makebox(0,0)[lb]{\smash{{\SetFigFont{11}{13.2}{\rmdefault}{\mddefault}{\updefault}{\color[rgb]{0,0,0}$0^2$}%
}}}}
\end{picture}%
 \end{array}}
				\end{array}
			}{1\Delta^{j+s+2} \GRkpr^{m-m'}}
		}
	\end{array}
	\]
\caption{The diagrams contributing to $2 \protect \decp{\nLV{\overline{\Ds}}{n}{\mu\nu}{c}(p)}{}$.}
\label{fig:nLV-c-Properties-a}
\ecf

There are a number of comments to make about the
diagrams of \fig{fig:nLV-c-Properties-a}. First,
each diagram
has a socket that we demand is filled by the
external field (we will see shortly why
we do not fill the socket).  
In diagram~\ref{nLV-c-CTP-GRs-TE} 
it is understood that, given that the final
gauge remainder in the string is the one
that bites the vertex,
the first gauge remainder in the string
cannot be filled by the external field,
as this generates an illegal diagram
(see \sec{sec:Not:GRs}).

Secondly, the factor of two
that diagrams~\ref{nLV-c-dvj+R}
and~\ref{nLV-c-dCTP}
have acquired relative to the parent diagram
comes from using \CC\ to combine the pull back like
momentum derivative with the push forward like derivative
momentum.
There is no such factor of two in  diagrams~\ref{nLV-c-CTP-GRs-TE}
and~\ref{nLV-c-TE-GRs} since are no momentum
derivatives.

Thirdly, the fact that the diagrams are under the influence
of $\decp{\ }{}$ leads to some novelties. Notice
that in diagram~\ref{nLV-c-TE-GRs},
the explicitly drawn gauge remainder structure
is required to
have at least three gauge remainders. If there
were only a single gauge remainder, whose socket
is necessarily filled by an  $A^1$ sector field, then
the diagram would vanish as a consequence of \CC\ 
invariance---nothing
unusual there. But why is it not possible for there
to be only two gauge remainders? The reason
is that such a structure which, crucially, is
decorated by a $\socket$, would be compelled
to be attached to the rest of the diagram by an 
effective propagator carrying just $p$. 
Such a term
is forbidden by
the influence of $\decp{\ }{}$.

To make progress, we attach two effective propagators
to the classical, two-point vertex of diagram~\ref{nLV-c-dCTP}.
As a consequence of the  $\decp{\ }{}$, these
two effective propagators must carry the same
(loop) momentum and so we can utilise~\eq{eq:EP-dCTP-EP}.
Processing
the gauge remainders, we arrive at the diagrams
of \fig{fig:nLV-c-Properties-b}, where we have 
used \CC\ invariance to collect
terms together, as described under~\eq{eq:EP-dCTP-EP}.
\bcf
	\[
	\begin{array}{l}
	\vspace{2ex}
		2
		\bca{
			\sum_{s=1}^n  \sum_{m=1}^{2s}  \sum_{j=-2}^{n+s-m-2} 
			\frac{\norm_{j+s+1,j+2}}{m!}
		}
			{
			\decp{
				\begin{array}{c}
				\vspace{1ex}
					\LIO{\cdeps{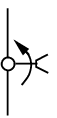}}{nLV-c-dEP}
				\\
					\InvTower{\ensuremath{\begin{array}{c}\begin{picture}(0,0)%
\includegraphics{pstex/CTP-E.pstex}%
\end{picture}%
\setlength{\unitlength}{3947sp}%
\begingroup\makeatletter\ifx\SetFigFont\undefined%
\gdef\SetFigFont#1#2#3#4#5{%
  \reset@font\fontsize{#1}{#2pt}%
  \fontfamily{#3}\fontseries{#4}\fontshape{#5}%
  \selectfont}%
\fi\endgroup%
\begin{picture}(466,586)(719,713)
\put(801,1067){\makebox(0,0)[lb]{\smash{{\SetFigFont{11}{13.2}{\rmdefault}{\mddefault}{\updefault}{\color[rgb]{0,0,0}$0^2$}%
}}}}
\end{picture}%
 \end{array}}}
				\end{array}
			}{1\Delta^{j+s+1} \GRkpr^m}
		}
		-4
		\bca{
			\sum_{s=1}^n  \sum_{m=1}^{2s-1}  \sum_{j=-1}^{n+s-m-2} 
			\frac{\norm_{j+s+1,j+1}}{m!}
		}
			{
			\decp{
				\begin{array}{c}
				\vspace{1ex}
					\LIDi{Combo-Bite-Vertex-nn}{nLV-c-Combo-Bite-Vertex-nn}
				\\
				\vspace{1ex}
					\SumVertex
				\\
					\ensuremath{\begin{array}{c}\begin{picture}(0,0)%
\includegraphics{pstex/CTP-E.pstex}%
\end{picture}%
\setlength{\unitlength}{3947sp}%
\begingroup\makeatletter\ifx\SetFigFont\undefined%
\gdef\SetFigFont#1#2#3#4#5{%
  \reset@font\fontsize{#1}{#2pt}%
  \fontfamily{#3}\fontseries{#4}\fontshape{#5}%
  \selectfont}%
\fi\endgroup%
\begin{picture}(466,586)(719,713)
\put(801,1067){\makebox(0,0)[lb]{\smash{{\SetFigFont{11}{13.2}{\rmdefault}{\mddefault}{\updefault}{\color[rgb]{0,0,0}$0^2$}%
}}}}
\end{picture}%
 \end{array}}
				\end{array}
			}{1\Delta^{j+s+1} \GRkpr^m}
		}
	\\
	\vspace{2ex}
		+4
		\bca{
			\sum_{s=1}^n  \sum_{m=3}^{2s}  \sum_{j=-2}^{n+s-m-2} 
			\frac{\norm_{j+s+1,j+2}}{m!} \sum_{m'=2}^{m-1} \nCr{m}{m'}
		}
			{
			\decp{
				\LIO{
					\ensuremath{\begin{array}{c}\begin{picture}(0,0)%
\includegraphics{pstex/CTP-E.pstex}%
\end{picture}%
\setlength{\unitlength}{3947sp}%
\begingroup\makeatletter\ifx\SetFigFont\undefined%
\gdef\SetFigFont#1#2#3#4#5{%
  \reset@font\fontsize{#1}{#2pt}%
  \fontfamily{#3}\fontseries{#4}\fontshape{#5}%
  \selectfont}%
\fi\endgroup%
\begin{picture}(466,586)(719,713)
\put(801,1067){\makebox(0,0)[lb]{\smash{{\SetFigFont{11}{13.2}{\rmdefault}{\mddefault}{\updefault}{\color[rgb]{0,0,0}$0^2$}%
}}}}
\end{picture}%
 \end{array}} \ensuremath{\begin{array}{c}\begin{picture}(0,0)%
\includegraphics{pstex/Combo-GR-GRs.pstex}%
\end{picture}%
\setlength{\unitlength}{3947sp}%
\begingroup\makeatletter\ifx\SetFigFont\undefined%
\gdef\SetFigFont#1#2#3#4#5{%
  \reset@font\fontsize{#1}{#2pt}%
  \fontfamily{#3}\fontseries{#4}\fontshape{#5}%
  \selectfont}%
\fi\endgroup%
\begin{picture}(252,731)(1226,-626)
\put(1318,-27){\makebox(0,0)[lb]{\smash{{\SetFigFont{11}{13.2}{\rmdefault}{\mddefault}{\updefault}{\color[rgb]{0,0,0}$\decGR{\ }{\GRkpr^{m'}}$}%
}}}}
\end{picture}%
 \end{array}} \hspace{1em}
					\sco[1]{
						\TopVertex
					}{\SumVertex}
				}{nLV-c-Combo-GR-GRs}
			}{1\Delta^{j+s+1} \GRkpr^{m-m'}}
		}
	\\
		+4
		\bca{
			\sum_{s=1}^n  \sum_{m=1}^{2s}  \sum_{j=-2}^{n+s-m-2} 
			\frac{\norm_{j+s+1,j+2}}{m!} \sum_{m'=1}^{m} \nCr{m}{m'} m'
		}
			{
			\decp{
				\Tower{
					\LIDi{CTP-E-GRs-Combo-b}{nLV-c-CTP-E-GRs-Combo-b}
					\hspace{1.5em}
					+
					\LIDi{CTP-E-GRs-Combo-c}{nLV-c-CTP-E-GRs-Combo-c}
				}
			}{1\Delta^{j+s+1} \GRkpr^{m-m'}}
		}
	\end{array}
	\]
\caption{Diagrams spawned by \ref{nLV-c-dCTP}.}
\label{fig:nLV-c-Properties-b}
\ecf

Diagrams~\ref{nLV-c-CTP-E-GRs-Combo-b} and~\ref{nLV-c-CTP-E-GRs-Combo-c}
are related, arising from
isolating particular components of a single diagram. In the
parent diagram from which~\ref{nLV-c-CTP-E-GRs-Combo-b} and~\ref{nLV-c-CTP-E-GRs-Combo-c}
are derived, the classical, two-point vertex is decorated
by $m'$ $\GRkpr$s. The $\GRk$ at the end of the $\combo$
can be contracted into any of these $\GRkpr$s or the
socket on the classical, two-point vertex. However,
it is useful to isolate the component in which
the $\GRk$ is contracted into the $\GRkpr$
at the top of the string \ie diagram~\ref{nLV-c-CTP-E-GRs-Combo-c}. 
In this case alone,
we can use diagrammatic identity~\eq{eq:GR-relation}.
We recognize the resulting term as the $r=1$ component
of $\nLV{\Ds}{n}{\mu\nu}{b}(p)$, demonstrating
that we are well on the way to proving~\eq{eq:nLV-c-Op2}.

In anticipation of what is to follow,
we note that diagrams~\ref{nLV-c-dvj+R}
and~\ref{nLV-c-dEP} naturally form part of a set
in which all diagrammatic elements
are struck by a momentum derivative.
To find a further member of the
set, we process diagram~\ref{nLV-c-Combo-Bite-Vertex-nn}
by isolating the component possessing a classical,
two-point vertex and proceeding as usual.
The reason we have held back with this step,
rather than using our experience of
dealing with such terms to jump straight
to the result of iterating the diagrammatic
procedure until exhaustion, is because
of the presence of $\decp{\ }{}$.

The loose end of the $\combo$
can attach to any available socket in the
diagram, besides its own. Suppose that it attaches to the
vertex with argument $v^{j_+}$, such that
the $\combo$ bites its own base. In this
case, the vertex does not possess a (classical)
two-point component due to the influence of
$\decp{\ }{}$. Given that the vertex
with argument $v^{j_+}$ must be at least
three point, how do we proceed?
The point is that, since the $\combo$ bites
itself, the field on the vertex to 
which the $\combo$ attaches carries zero momentum.
Furthermore, \CC\ invariance forces
this field to be in the $A^i$
sector (group theory considerations then
force it to be in the $A^1$ sector). 
Thus, we \emph{effectively} have a vertex
which has been Taylor expanded to zeroth
order in the momentum of a decorative $A^1$
field, and so we can repeat the above analysis!
The difference is that, rather than the socket
associated with the Taylor expansion being filled
by an external field, it is now filled by a $\combo$
which bites its own base. 

Putting this diagram to one side for the moment,
consider those components of diagram~\ref{nLV-c-Combo-Bite-Vertex-nn}
that we \emph{can} process in the usual way.
Cancelling terms where possible, we
are left with the following diagrams:
\begin{enumerate}
	\item	the component of the nested version of
			diagram~\ref{nLV-c-Combo-Bite-Vertex-nn}
			in which the $\combo$ bites a socket on a classical,
			two-point vertex, the \emph{other} socket of which
			is filled by the other end of the $\combo$;
	\label{it:nLV-c-Combo-Bite-Vertex}

	\item	nested versions of
			diagram~\ref{nLV-c-Combo-Bite-Vertex-nn}
			where the bitten field on the
			vertex with argument $v^{j_+}$ carries
			zero momentum;
	\label{it:nLV-c-combo-kv}

	\item	remainders from cancellations via diagrammatic identity~\eq{eq:D-ID-G};
			the cancellation no longer being exact, as some of the diagrams necessary for
			the cancellation are forbidden when under the influence of
			$\decp{\ }{}$;
	\label{it:nLV-c-combo-GR}

	\item	the component of the nested version of
			diagram~\ref{nLV-c-Combo-Bite-Vertex-nn}
			in which the classical, two-point vertex
			bitten by the $\combo$ is attached
			to the gauge remainder structure or socket
			decorating the $\Op{2}$ stub.
	\label{it:nLV-c-TLTP-EP-GRs}
\end{enumerate}

We begin by analysing the diagrams of item~\ref{it:nLV-c-Combo-Bite-Vertex},
which we have drawn in \fig{fig:nLV-c-Properties-c}.
\bcf[h]
	\[
	-4
	\bca{
		\sum_{s=1}^n  \sum_{m=2}^{2s-1}  \sum_{j=-2}^{n+s-m-3} 
		\frac{\norm_{j+s+2,j+2}}{m!} \sum_{m'=1}^{m-1} \nCr{m}{m'}
	}
		{
		\decp{
			\LIO{
				\ensuremath{\begin{array}{c}\begin{picture}(0,0)%
\includegraphics{pstex/CTP-E.pstex}%
\end{picture}%
\setlength{\unitlength}{3947sp}%
\begingroup\makeatletter\ifx\SetFigFont\undefined%
\gdef\SetFigFont#1#2#3#4#5{%
  \reset@font\fontsize{#1}{#2pt}%
  \fontfamily{#3}\fontseries{#4}\fontshape{#5}%
  \selectfont}%
\fi\endgroup%
\begin{picture}(466,586)(719,713)
\put(801,1067){\makebox(0,0)[lb]{\smash{{\SetFigFont{11}{13.2}{\rmdefault}{\mddefault}{\updefault}{\color[rgb]{0,0,0}$0^2$}%
}}}}
\end{picture}%
 \end{array}} \ensuremath{\begin{array}{c}\input{pstex/TLTP-Combo-GR.pstex_t} \end{array}}
				\sco[1]{
					\TopVertex
				}{\SumVertex}
			}{nLV-c-TLTP-Combo-GR}
		}{1\Delta^{j+s+2} \GRkpr^{m-m'}}
	}
	\]
\caption{One of the surviving terms produced by processing
diagram~\ref{nLV-c-Combo-Bite-Vertex-nn}.}
\label{fig:nLV-c-Properties-c}
\ecf

To proceed, we decompose the $\combo$,  according to~\eq{eq:Combo}.
 The first
term arising from this decomposition can be processed
using the effective propagator relation, whereas the
second term vanishes, courtesy of diagrammatic 
identity~\eq{eq:GR-TLTP}. The result of
combining
the surviving terms with diagrams~\ref{nLV-c-dvj+R}
and~\ref{nLV-c-dEP} is shown in \fig{fig:nLV-c-Properties-d}.
\bcf
	\[
	\begin{array}{l}
	\vspace{2ex}
		4
		\bca{
			\sum_{s=1}^n  \sum_{m=1}^{2s}  \sum_{j=-2}^{n+s-m-2} 
			\frac{\norm_{j+s+2,j+2}}{m!}
		}
			{
			\decp{
				\left[
					\Tower{\LIDi{CTP-E}{nLV-c-MomDer}}
					\cdeps{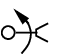} \hspace{-1.8em}
				\right]
				\hspace{1em}
			}{1 \Delta^{j+s+2} \GRkpr^{m}}
		}
	\\
	\vspace{2ex}
		-4
		\bca{
			\sum_{s=1}^n  \sum_{m=1}^{2s}  \sum_{j=-2}^{n+s-m-2} 
			\frac{\norm_{j+s+2,j+2}}{m!} \sum_{m'=1}^{m} \nCr{m}{m'}
		}
			{
			\decp{
				\LIO{
					\ensuremath{\begin{array}{c}\input{pstex/CTP-E-dGRs.pstex_t} \end{array}}
					\sco[1]{
						\TopVertex
					}{\SumVertex}
				}{nLV-c-dGRs}
			}{1 \Delta^{j+s+2} \GRkpr^{m-m'}}
		}
	\\
		-4
		\bca{
			\sum_{s=1}^n  \sum_{m=4}^{2s-1}  \sum_{j=-2}^{n+s-m-3} 
			\frac{\norm_{j+s+2,j+2}}{m!} \sum_{m'=3}^{m-1} \nCr{m}{m'} m'(m'-1)
		}
			{
			\decp{
				\LIO{
					\ensuremath{\begin{array}{c}\begin{picture}(0,0)%
\includegraphics{pstex/CTP-E.pstex}%
\end{picture}%
\setlength{\unitlength}{3947sp}%
\begingroup\makeatletter\ifx\SetFigFont\undefined%
\gdef\SetFigFont#1#2#3#4#5{%
  \reset@font\fontsize{#1}{#2pt}%
  \fontfamily{#3}\fontseries{#4}\fontshape{#5}%
  \selectfont}%
\fi\endgroup%
\begin{picture}(466,586)(719,713)
\put(801,1067){\makebox(0,0)[lb]{\smash{{\SetFigFont{11}{13.2}{\rmdefault}{\mddefault}{\updefault}{\color[rgb]{0,0,0}$0^2$}%
}}}}
\end{picture}%
 \end{array}} \ensuremath{\begin{array}{c}\begin{picture}(0,0)%
\includegraphics{pstex/GRs-Full-dGR.pstex}%
\end{picture}%
\setlength{\unitlength}{3947sp}%
\begingroup\makeatletter\ifx\SetFigFont\undefined%
\gdef\SetFigFont#1#2#3#4#5{%
  \reset@font\fontsize{#1}{#2pt}%
  \fontfamily{#3}\fontseries{#4}\fontshape{#5}%
  \selectfont}%
\fi\endgroup%
\begin{picture}(565,343)(1468,2019)
\put(1468,2069){\makebox(0,0)[lb]{\smash{{\SetFigFont{11}{13.2}{\rmdefault}{\mddefault}{\updefault}{\color[rgb]{0,0,0}$\decGR{ \hspace{3em} }{\GRkpr^{m'-2}}$}%
}}}}
\end{picture}%
 \end{array}} \hspace{2ex}
					\sco[1]{
						\TopVertex
					}{\SumVertex}
				}{nLV-c-GRs-Full-dGR}
			}{1 \Delta^{j+s+2} \GRkpr^{m-m'}}
		}	
	\end{array}
	\]
\caption{The result of
combining diagrams~\ref{nLV-c-dvj+R}
and~\ref{nLV-c-dEP} with the surviving components
of diagram~\ref{nLV-c-TLTP-Combo-GR}.}
\label{fig:nLV-c-Properties-d}
\ecf

Diagram~\ref{nLV-c-MomDer} contains a sum  of
total momentum derivatives \wrt\
the various loop momenta carried by the fully fleshed
out constituent diagrams~\cite{mgierg2}. Thus, diagram~\ref{nLV-c-MomDer}
vanishes as a consequence of the 
dimensional pre-regularisation. We comment on this further
in the conclusions.

\begin{Icancel}

Diagram~\ref{nLV-c-GRs-Full-dGR} exactly cancels
diagram~\ref{nLV-c-TE-GRs}. 
This follows from~\eq{eq:D-ID-dGRk-GT-ring}
upon recognizing
that diagrammatic 
identity~\eq{eq:GR-relation}
allows
the momentum derivative
of diagram~\ref{nLV-c-GRs-Full-dGR} to be transfered from the $\GRkpr$ to the
$\GRk$, at the expense of a minus sign. 

\end{Icancel}

Next, consider the diagrams of item~\ref{it:nLV-c-combo-kv}, above.
From the earlier comments about the
un-nested diagram~\ref{nLV-c-Combo-Bite-Vertex-nn}
it is clear that the diagrams of item~\ref{it:nLV-c-combo-kv}
simply yield versions of
diagram~\ref{nLV-c-dvj+R} and~\ref{nLV-c-dCTP}
but where the socket associated with the momentum
derivative is filled by an external field of a sub-diagram
containing a $\combo$ (this is why we left the socket
unfilled in the first place, rather than decorating
it with the external field). The similar generalization
of diagram~\ref{nLV-c-TE-GRs} arises from the
diagrams of item~\ref{it:nLV-c-combo-GR}, above.
We can manipulate the new version of
diagram~\ref{nLV-c-dCTP} just as we manipulated
the original version and precisely the same pattern
of terms and cancellations will be generated. In particular,
we will generate versions of diagram~\ref{nLV-c-CTP-E-GRs-Combo-c}.
We recognise these as the $r>1$ components of
$\nLV{\Ds}{n}{\mu\nu}{b}(p)$. Therefore,
we have succeeded in demonstrating~\eq{eq:nLV-c-Op2};
now we find an explicit expression for the remainders.

The remainders can be thought of as a set of primitive
diagrams, together with their analogues possessing
implicit $\combo$s. The primitive diagrams are:
the diagrams of item~\ref{it:nLV-c-TLTP-EP-GRs}
and diagrams~\ref{nLV-c-CTP-GRs-TE},
\ref{nLV-c-CTP-E-GRs-Combo-b} and~\ref{nLV-c-dGRs}.

Diagrams of item~\ref{it:nLV-c-TLTP-EP-GRs}
can be redrawn  via diagrammatic identities~\eq{eq:D-ID-Op2-G},
upon which
one of the resultant diagrams
exactly cancels 
diagram~\ref{nLV-c-CTP-E-GRs-Combo-b}.
Diagram~\ref{nLV-c-CTP-GRs-TE} can be
redrawn using diagrammatic identity~\eq{eq:D-ID-dGRk-GT-ring}.
The resultant diagram naturally combines
with diagram~\ref{nLV-c-dGRs} upon recognizing
that
\be
\label{eq:TLTP-Combo}
	\ensuremath{\begin{array}{c}\begin{picture}(0,0)%
\includegraphics{pstex/CTP-Combo.pstex}%
\end{picture}%
\setlength{\unitlength}{3947sp}%
\begingroup\makeatletter\ifx\SetFigFont\undefined%
\gdef\SetFigFont#1#2#3#4#5{%
  \reset@font\fontsize{#1}{#2pt}%
  \fontfamily{#3}\fontseries{#4}\fontshape{#5}%
  \selectfont}%
\fi\endgroup%
\begin{picture}(814,372)(854,1164)
\put(936,1284){\makebox(0,0)[lb]{\smash{{\SetFigFont{11}{13.2}{\rmdefault}{\mddefault}{\updefault}{\color[rgb]{0,0,0}$0^2$}%
}}}}
\end{picture}%
 \end{array}} = \ensuremath{\begin{array}{c}\begin{picture}(0,0)%
\includegraphics{pstex/CTP-EP-dGR.pstex}%
\end{picture}%
\setlength{\unitlength}{3947sp}%
\begingroup\makeatletter\ifx\SetFigFont\undefined%
\gdef\SetFigFont#1#2#3#4#5{%
  \reset@font\fontsize{#1}{#2pt}%
  \fontfamily{#3}\fontseries{#4}\fontshape{#5}%
  \selectfont}%
\fi\endgroup%
\begin{picture}(867,403)(854,1164)
\put(936,1284){\makebox(0,0)[lb]{\smash{{\SetFigFont{11}{13.2}{\rmdefault}{\mddefault}{\updefault}{\color[rgb]{0,0,0}$0^2$}%
}}}}
\end{picture}%
 \end{array}} = \ensuremath{\begin{array}{c}\begin{picture}(0,0)%
\includegraphics{pstex/dGR.pstex}%
\end{picture}%
\setlength{\unitlength}{3947sp}%
\begingroup\makeatletter\ifx\SetFigFont\undefined%
\gdef\SetFigFont#1#2#3#4#5{%
  \reset@font\fontsize{#1}{#2pt}%
  \fontfamily{#3}\fontseries{#4}\fontshape{#5}%
  \selectfont}%
\fi\endgroup%
\begin{picture}(240,300)(1481,1267)
\end{picture}%
 \end{array}} + \ensuremath{\begin{array}{c}\begin{picture}(0,0)%
\includegraphics{pstex/GR-dGRk-GR.pstex}%
\end{picture}%
\setlength{\unitlength}{3947sp}%
\begingroup\makeatletter\ifx\SetFigFont\undefined%
\gdef\SetFigFont#1#2#3#4#5{%
  \reset@font\fontsize{#1}{#2pt}%
  \fontfamily{#3}\fontseries{#4}\fontshape{#5}%
  \selectfont}%
\fi\endgroup%
\begin{picture}(354,300)(1035,1226)
\end{picture}%
 \end{array}}.
\ee
We thus find that
\be
\label{eq:nLV-c-Op2-b}
	2 \decp{\nLV{\overline{\Ds}}{n}{\mu\nu}{c}(p)}{} = \nLV{\Ds}{n}{\mu\nu}{b}(p) 
	+ 2 \nLV{\Ds}{n}{\mu\nu}{c'}(p),
\ee
where $\nLV{\Ds}{n}{\mu\nu}{c'}(p)$
is given in \fig{fig:nLV-c-Properties-e}.
\bcf
	\[
	\begin{array}{l}
	\vspace{2ex}
		\nLV{\Ds}{n}{\mu\nu}{c'}(p) \equiv
	\\
	\vspace{2ex}
		-2
		\bca{
			\sum_{s=1}^n \sum_{r=1}^{s-1} \sum_{m=1}^{2(s-r)+1}
			\sum_{j=-2}^{n+s-m-2r-1} \frac{\norm_{j+s+2-r,j+2}}{m!(r-1!)}
			\sum_{m'=1}^{m} \nCr{m}{m'} \sum_{m''=1}^{m'} \nCr{m'}{m''}
		}
			{
			\decp{
				\LIDi{CTP-E-GRs-Combo-GRs-CTP-EP-b}{nLV-c-CTP-E-GRs-Combo-GRs-CTP-EP}
				\sco[1]{
					\TopVertex
				}{\SumVertex}
			}{1\Delta^{j+s+2-r} \combo^r \GRkpr^{m-m'}}
		}
	\\
	\vspace{2ex}
		+2
		\bca{
			\sum_{s=1}^n \sum_{r=1}^{s-1} \sum_{m=0}^{2(s-r)+1}
			\sum_{j=-2}^{n+s-m-2r-1} \frac{\norm_{j+s+2-r,j+2}}{m!(r-1!)}
			\sum_{m'=0}^{m} \nCr{m}{m'} 
		}
			{
			\decp{
				\LIDi{CTP-E-GRs-Combo-GR}{nLV-c-CTP-E-GRs-Combo-GR}
				\sco[1]{
					\TopVertex
				}{\SumVertex}
			}{1\Delta^{j+s+2-r} \combo^r \GRkpr^{m-m'}}
		}
	\\
		-2
		\bca{
			\sum_{s=1}^n \sum_{r=1}^{s-1} \sum_{m=0}^{2(s-r)+1}
			\sum_{j=-2}^{n+s-m-2r-1} \frac{\norm_{j+s+3-r,j+2}}{m!(r-1!)}
			\sum_{m'=0}^{m} \nCr{m}{m'} \sum_{m''=0}^{m'} \nCr{m'}{m''}
		}
			{
			\decp{
				\LIDi{CTP-E-GRs-Combo-CTP-GRs}{nLV-c-CTP-E-GRs-Combo-CTP-GRs}
				\hspace{1.5em}
				\sco[1]{
						\TopVertex
				}{\SumVertex}
			}{1\Delta^{j+s+3-r} \combo^r \GRkpr^{m-m'}}	
		}
	\end{array}
	\]
\caption{The expression for $\nLV{\Ds}{n}{\mu\nu}{c'}(p)$.}
\label{fig:nLV-c-Properties-e}
\ecf

\section{An Expression for $\beta_n$}
\label{sec:beta_n}

\subsection{Initial Manipulations}

In this section,
we will derive a compact expression
for $\beta_n$ which has no explicit
dependence on either the seed action
or the details of the covariantization 
of the cutoff. Our starting point is
to consider 
\[
	\dec{
		\nLV{\Ds}{n}{\mu\nu}{a}(p) + \nLV{\Ds}{n}{\mu\nu}{b}(p)
	}{\bullet}.
\]
(We work with these auxiliary function
because it is technically easier to do
so than to work directly with
$\D^{a,b}$. At the end of the calculation,
will be able to conveniently trade
$\Ds$s for $\D$s.)

As a precursor to computing the flow of
these functions,
it is clear that we will need to understand
how to compute the flow of a reduced vertex.
Recall that a reduced vertex lacks a classical,
two-point component. Consequently, the flow of
a reduced vertex lacks a component given
by the flow of a classical, two-point vertex,
which we can read off from \fig{fig:TLTPs}.
Now, the classical flow equation involves only
the $a_0$ term; there are no $\beta$, $\alpha$
or $a_1$ terms. Hence, we need only take care
with the dumbbell term generated by the flow
of a reduced vertex, which we will tag with an
$R$. It is straightforward to check that a dumbbell 
term tagged in this way must either possess
at least one reduced vertex or must have a decorated
kernel.

We begin by focusing on the terms
generated by $\dec{\nLV{\Ds}{n}{\mu\nu}{a}(p)}{\bullet}$
and adjust the ranges the sums
over $s$, $m$ and $j$ to remove diagrams which vanish
even after the action of $-\flowConstAl$.
This is trivial in the case where
$-\flowConstAl$ strikes an effective propagator: the limits of the
sums just reduce to those of $\nLV{\D}{n}{\mu\nu}{a}(p)$.
If $-\flowConstAl$ strikes a $\GRkpr$ things are much
the same only now 
the sum over $m$ must, of course, start from
one and not zero.

However, things are more subtle in the case
that $-\flowConstAl$ strikes a vertex. 
Immediately, since we must have at least one vertex,
the lower limit of the sum over $j$ increases from $-2$
to $-1$. Consider
now the resulting $\alpha$ and $\beta$ terms.
Compared to the parent vertex, the loop order
of the daughter vertex is changed, but the number
of legs remains the same. Thus, if
the parent
diagram vanishes at $\Op{2}$, it is not necessarily
the case that the daughter $\alpha$ and $\beta$-terms also 
vanish\footnote{Of course, it must be the case that the sum over
all daughters, including those generated
by $a_0$ and $a_1$, does vanish at $\Op{2}$.}:
it is clear from~\eqs{eq:S_0-11}{eq:S_>0-11}
that a change to the loop order of a two-point
vertex carrying momentum $p$ can
change the order of its leading contribution in $p$
.
However, if the parent diagram vanishes because it
possesses a one-point vertex, then the daughter $\alpha$ and $\beta$-terms
will vanish for this reason also. Consequently, 
for the $\alpha$ and $\beta$-terms
the minimum
value  of $s$ stays the same, but the maximum values of both
$m$ and $j$ are reduced by one. 

With these points in mind, it is worth considering
the $\alpha$ and $\beta$-terms in more detail and
so, to this end, we explicitly give them in \fig{fig:beta+alpha-Prelim}.
The combinatoric factor follows from recognizing
that the $-\flowConstAl$ can strike any of the $j+2$ vertices
and using~\eq{eq:norm-c}.
\bcf[h]
	\[
	-2
	\dec{
		\begin{array}{c}
		\vspace{1ex}
			\ds
			\sum_{u}
			\sum_{s=0}^{n-1}   \sum_{m=0}^{2s} \sum_{j=-1}^{n+s-m-1-u}\frac{\norm_{j+s+1,j+1}}{m!}
			\sum_{v^{k}=1}^{v^{j_+}} \delta(v^k - u)
		\\
			\left[
				\sco[1]{
						\left[
								2 \left( v^{j_+,k}-1 \right) \beta_{v^{k}} 
								+ \gamma_{v^{k}} \pder{}{\alpha}
					\right]
					\Vertex{v^{j_+\!,k}}
				}{\SumVertex}
			\right]
		\end{array}
	}{11\Delta^{j+s+1} \GRkpr^m}
	\]
\caption{The $\alpha$ and $\beta$ terms produced by 
$\dec{\nLV{\Ds}{n}{\mu\nu}{a}(p)}{\bullet}$.}
\label{fig:beta+alpha-Prelim}
\ecf

There are a number of important points to make.
First, it is understood 
that the sum over the vertex argument $v^k$---which is identified with
$v^{j+2}$---is
performed \emph{after}
the sums over $v^1 \cdots v^{j_+}$ buried in the diagrammatics 
(\cf \eq{eq:VertexTower}). Secondly, we notice a
sum over the new variable, $u$, and a $\delta$-function
$\delta(v^k - u)$. This is to ensure that the upper limit
on the sum over $j$ is given correctly, given that
the sum over
vertex arguments is now $n-s-v^k$.
Finally, the upper limit on the sum over $s$ has been reduced
by unity. This follows from the minimum value of $v^k$
being unity ($\beta_0$ does not exist) which, in turn, forces $n-s$
to take a minimum value of unity. The expression for the $\beta$
and $\alpha$ terms is quite ugly in its current form,
but it can be neatened up by a change of variables:
\[
 v^{i>0} \rightarrow v^i + v^k,
\]
to give the diagrams of \fig{fig:beta+alpha-Prelim-b}. Notice that
we have changed the dummy variable $k$ to $n'$ which
now appears as the outer sum and also as the label for
both $\alpha$ and $\beta$.
\bcf[h]
\[
		-2\dec{
			\begin{array}{c}
			\vspace{1ex}
				\ds
				\sum_{n'=1}^{n}
				\sum_{s=0}^{n-n'}   \sum_{m=0}^{2s} \sum_{j=-1}^{n-n'+s-m-1}\frac{\norm_{j+s+1,j+1}}{m!}
			\\
				\left[
					\sco[1]{
						\left[
								2 \left( v^{j_+}-1 \right) \beta_{n'} 	+  \gamma_{n'} \pder{}{\alpha}
						\right]
						\DisplacedVertex{v^{j_+}}
					}{\Vertex{n_{n'+s}, j}}
				\right]
			\end{array}
		}{11\Delta^{j+s+1} \GRkpr^m}
\]
\caption{A re-expression of the $\alpha$ and $\beta$-terms.}
\label{fig:beta+alpha-Prelim-b}
\ecf

We now focus on the $s=0$ components of these diagrams.
Demanding that the resulting terms are connected,
we discard all diagrams possessing one-point (Wilsonian
effective action) vertices. Utilizing~\eqs{eq:S_0-11}{eq:S_>0-11}
it is straightforward to demonstrate that the sole
contribution which survives at $\Op{2}$ is:
\beas
	&
	-2 \norm_{0,0}
	\left(
		-2\beta_n + \gamma_n \pder{}{\alpha}
	\right)
	\ensuremath{\begin{array}{c}\begin{picture}(0,0)%
\includegraphics{pstex/CTP-EE.pstex}%
\end{picture}%
\setlength{\unitlength}{3947sp}%
\begingroup\makeatletter\ifx\SetFigFont\undefined%
\gdef\SetFigFont#1#2#3#4#5{%
  \reset@font\fontsize{#1}{#2pt}%
  \fontfamily{#3}\fontseries{#4}\fontshape{#5}%
  \selectfont}%
\fi\endgroup%
\begin{picture}(466,822)(719,713)
\put(801,1067){\makebox(0,0)[lb]{\smash{{\SetFigFont{11}{13.2}{\rmdefault}{\mddefault}{\updefault}{\color[rgb]{0,0,0}$0^2$}%
}}}}
\end{picture}%
 \end{array}}
\\
	= & -4\beta_n \Box_{\mu \nu}(p) + \Op{4},
\eeas
where we have used~\eq{eq:norm}.

Returning to our consideration of 
$\dec{\nLV{\Ds}{n}{\mu\nu}{a}(p)}{\bullet}$,
we note that
terms generated by the action of $a_0$ are particularly
simple to deal with: the ranges of the sums over 
$s$, $m$ and $j$ just stay the same. This follows 
because $a_0$ converts a Wilsonian effective action
vertex into a dumbbell structure, components of
which possess seed action vertices; seed action
vertices, unlike their Wilsonian effective
action counterparts, exist at the one-point level.
Suppose
that the parent diagram vanishes at $\Op{2}$,
on account of an $\nLV{S}{n'}{\mu\nu}{}(p)$
vertex. This can be converted
into (amongst other terms) a dumbbell possessing
a one-point, $n'$-loop seed action vertex,
joined to a classical, three-point vertex.
This term no longer vanishes at $\Op{2}$.
If the parent diagram vanishes on account
of a one-point Wilsonian effective action
vertex, the daughter
can survive 
since the vanishing vertex of the parent can
be converted into a dumbbell structure
possessing a one-point, seed action vertex
joined to a two-point vertex.

Finally, consider the effect of the action
of $a_1$. Due to the equivalence of each of the vertices
in a given term, we can take the $a_1$ to strike
the vertex with argument $v^{j_+;R}$ (so long as we multiply by $j+2$), causing
the argument to become
$\Sigma_{v^{j_+}-1}$.
The $R$ has been dropped,
since a quantum term is necessarily formed from a 
vertex whose argument is greater than zero. The
lower limit on the sum over $v^{j_+}$ should
be now changed from zero to one, in recognition of the
fact that
$a_1$ does not act on tree level terms. Furthermore,
the sum over all vertex arguments is now $n+s-1$,
rather than $n+s$, and so we should reduce the
upper limit of the sum over $s$ by one. It is
convenient to change variables:
\[
	v^{i>0} \rightarrow v^i + 1
\]
and then to let $s \rightarrow s-1$. This affects
the ranges of all three sums.

The diagrams formed by  $\dec{\nLV{\Ds}{n}{\mu\nu}{a}(p)}{\bullet}$
are shown in \fig{fig:bn-b-P}. It is understood
here, and in all that follows, that the
vertex argument $v^k$ is to be identified with $v^{j+2}$.
If this vertex argument appears in more than one vertex
(as will always be the case) it is understood to be
summed over, in  the usual manner.
\bcf
	\[
	\begin{array}{l}
	\vspace{2.5ex}
		\dec{\nLV{\Ds}{n}{\mu\nu}{a}(p)}{\bullet} + \Op{4} = -4 \beta_n \Box_{\mu\nu}(p) 
	\\
	\vspace{3ex} 
		-2\dec{
			\begin{array}{c}
			\vspace{1ex}
				\ds
				\sum_{n'=1}^{n-1}
				\sum_{s=1}^{n-n'}   \sum_{m=0}^{2s} \sum_{j=-1}^{n-n'+s-m-1}\frac{\norm_{j+s+1,j+1}}{m!}
			\\
				\left[
					\sco[1]{
						\left[
							\begin{array}{ccc}
							\vspace{1ex}	
								\LD{beta-terms}							&								& \LD{alpha-terms}			
							\\	
								2 \left( v^{j_+}-1 \right) \beta_{n'} 	& \hspace{-1em} + \hspace{-1em}	& \gamma_{n'} \pder{}{\alpha}
							\end{array}
						\right]
						\DisplacedVertex{v^{j_+}}
					}{\Vertex{n_{n'+s}, j}}
				\right]
			\end{array}
		}{11\Delta^{j+s+1} \GRkpr^m}
	\\
	\vspace{3ex}
		\ds
		-\sum_{s=0}^n   \sum_{m=0}^{2s+1} \sum_{j=-1}^{n+s-m-1}\frac{\norm_{j+s+1,j+1}}{m!}
		\dec{
			\sco[1]{
				\ds
				\left[
					\LDi{Dumbbell-vj_+kb-vkb}{D-vj_+kb-vkb}
				\right]_R
			}{\Vertex{ n_s, j}}
		}{11\Delta^{j+s+1} \GRkpr^m}
	\\
	\vspace{3ex}
		\ds
		+\sum_{s=1}^n   \sum_{m=0}^{2s-1} \sum_{j=-1}^{n+s-m-2}
		\frac{\norm_{j+s,j+1}}{m!}
		\dec{
			\sco[1]{
				\LDi{Padlock-Sig_vj_+}{P-Sig_vj_+}
				+\LDi{WBT-Sig_vj_+}{WBT-Sig_vj_+}
			}{\Vertex{ n_{s}, j}}
		}{11\Delta^{j+s} \GRkpr^m}
	\\
		\ds
		+
		\left[
			\begin{array}{c}
				\ds
				\sum_{s=1}^n   \sum_{m=0}^{2s} \sum_{j=-2}^{n+s-m-2}
				\frac{\norm_{j+s,j+2}}{m!}
			\\
				\dec{
					\sco[1]{\LO{\Vertex{v^{j_+\!;R}}}{j+2-DEP}}{\Vertex{n_s, j}}
				}{11\Delta^{j+s} \DEP \GRkpr^m}
			\end{array}
		\right]
		+
		\left[
			\begin{array}{c}
				\ds
				2 \sum_{s=1}^n   \sum_{m=1}^{2s} \sum_{j=-2}^{n+s-m-2}
				\frac{\norm_{j+s+1,j+2}}{(m-1)!}
			\\
				\dec{
					\sco[1]{\LO{\Vertex{v^{j_+\!;R}}}{j+2-DGR}}{\Vertex{n_s, j}}
				}{11\Delta^{j+s+1} \GRkpr^{m-1} \DGRkpr}
			\end{array}
		\right]
	\end{array}
	\]
\caption{An expression for $\dec{\nLV{\Ds}{n}{\mu\nu}{a}(p)}{\bullet}$.}
\label{fig:bn-b-P}
\ecf

It is worth considering the diagrams
arising from explicitly decorating~\ref{j+2-DEP}
with the differentiated effective propagator
in more detail. For convenience, these diagrams
are collected together in \fig{fig:app:DEP}.
Explicit decoration with the differentiated effective
propagator has the capacity to further change
the limits on the various sums. For example,
if we attach both ends to the same vertex,
then this will reduce the maximum value of $m$
by one since, in order to form a connected
diagram, the vertex decorated by the
differentiated effective propagator must be at least
three-point. This effectively uses up a decoration
which could otherwise have filled a $\GRkpr$.

 In the case that each end of the differentiated
effective propagator attaches to a different vertex,
we note that such a diagram can only exist for $j>-1$.
Our strategy here is to shift $j \rightarrow j+1$,
so that the sum over $j$ starts from $-1$,
identifying $v^{j+2}$ with $v^k$ as usual.

We now isolate all classical, two-point vertices
in diagrams~\ref{D-vj_+kb-vkb}--\ref{WBT-Sig_vj_+}. 
This step is crucial to
the entire diagrammatic procedure: if a classical,
two-point vertex is attached to an effective propagator,
we can employ the effective propagator relation, whereas
if it is attached to an external field, we can perform
manipulations at $\Op{2}$. Although
the classical, two-point vertices of diagram~\ref{P-Sig_vj_+}
are fully decorated, it will nonetheless prove useful
to isolate them, as this will help elucidate
the structure of the forthcoming cancellations.

Let us consider isolating the classical, two-point
vertices of diagram~\ref{D-vj_+kb-vkb} in more detail.
First, let us take both vertices
of the dumbbell structure to be reduced vertices.
This immediately allows us to reduce the maximum 
value of the sum over $j$ by one,
since the
formation of the dumbbell term causes the
total number of vertices to increase from
$j+2$ to $j+3$ (and the maximum number of reduced
vertices we can decorate in the formation of
a connected diagram remains the same).
Of the terms which remain, consider
those possessing exclusively 
Wilsonian 
effective action vertices. Since we discard
one-point, Wilsonian effective action vertices
in all diagrams in which $\flowConstAl$ has acted,
we can reduce the maximum values of 
$j$ and $m$ and increase
the minimum value of $s$ by one.
The component of the resulting diagram in which
the kernel is undecorated is exactly cancelled
by the component of diagram~\ref{j+2-DEP}
in which the differentiated effective propagator
joins two different vertices (see diagram~\ref{D-vj_+kR-DEP-vkR}). 
Since the surviving
component has a decorated kernel,
the maximum values of both $m$ and $j$ are reduced
by one, again.

Isolating a single classical, two-point vertex
in diagram~\ref{D-vj_+kb-vkb} is straightforward:
taking the argument of the top or bottom vertex
of the dumbbell structure to be a classical, two-point
vertex amounts to the same thing; hence we will choose
to isolate the classical, two-point part of $\overline{v}^k$ 
and multiply by two. When taking the classical
part of $\overline{v}^k$,
the other vertex argument,
$\overline{v}^{j_+,k;R} \equiv \overline{v}^{j_+;R} - \overline{v}^{k;R}$, reduces to 
simply $\overline{v}^{j_+;R}$. 
We now expanded out the barred vertices,
according to~\eq{eq:bar}, noting that certain terms cancel
on account of us having
set the classical, two-point seed action vertices
equal to their Wilsonian effective action counterparts.
There is no need to change the 
limits on any of the sums: compared to the
parent diagram, we have an extra two-point vertex,
but also an extra two decorative fields, corresponding
to the two ends of the kernel.

Taking two classical, two-point vertices in
diagram~\ref{D-vj_+kb-vkb} requires some thought.
First, we note that the resulting dumbbell
structure cannot have been formed by
the flow of a one-point vertex. Since we are
interested only in one-point Wilsonian effective
action vertices if
they have been processed,
we can reduce the maximum values of $m$
and $j$ and increase the
minimum value of $s$ to remove any unwanted diagrams.
If $j=-1$, then there are only two vertices
in total, and so this case should be treated differently
from $j \geq 0$. In the latter case, it will prove
useful to shift variables $j \rightarrow j+1$, so
that the sum over $j$ starts, once again, from $-1$.
We can then recombine terms into a diagram
with a sum that starts from $j=-2$.
The isolation of the
classical, two-point vertices of 
diagram~\ref{D-vj_+kb-vkb} is shown in 
\fig{fig:bn-Ia}.
\bcf
	\[
		\begin{array}{l}
		\vspace{2ex}
			-
			\left[
				\begin{array}{c}
				\vspace{1ex}
					\ds
					\sum_{s=1}^n \sum_{m=0}^{2s-1}  \sum_{j=-1}^{n+s-m-4} 
					\frac{\norm_{j+s+1,j+1}}{m!}
				\\
					\ds
					\dec{
						\LO{\ensuremath{\begin{array}{c}\input{pstex/Dumbbell-vj_+kR-RW-vkR.pstex_t} \end{array}} \Vertex{n_s,j}}{D-vj_+kR-vkR}
					}{11\Delta^{j+s+1} \GRkpr^m}
				\end{array}
			\right]
		\\
		\vspace{2ex}
			+2
			\left[
				\begin{array}{c}
				\vspace{1ex}
					\ds
					\sum_{s=0}^n \sum_{m=0}^{2s+1}  \sum_{j=-1}^{n+s-m-2} 
					\frac{\norm_{j+s+1,j+1}}{m!}
				\\
					\ds
					\dec{
						\LO{\ensuremath{\begin{array}{c}\input{pstex/Dumbbell-vj_+kR-vkhR.pstex_t} \end{array}} \Vertex{n_s,j}}{D-vj_+kR-vkhR}
					}{11\Delta^{j+s+1} \GRkpr^m}
				\end{array}
			\right]
		\\
			+
			\left[
				\begin{array}{c}
				\vspace{1ex}
					\ds
					\sum_{s=1}^n \sum_{m=0}^{2s-1}  \sum_{j=-2}^{n+s-m-3} 
					\frac{\norm_{j+s+2,j+2}}{m!}
				\\
					\ds
					\dec{
						\begin{array}{c}
						\vspace{1ex}
							\ds
							\LDi{Dumbbell-02x2-R}{D-02x2-R-VS}
						\\
						\vspace{1ex}
							\Vertex{v^{j_+\!;R}}
						\\
							\Vertex{n_s, j}
						\end{array}
					}{11\Delta^{j+s+2} \GRkpr^m}
				\end{array}
			\right]
			+2
			\left[
				\begin{array}{c}
				\vspace{1ex}
					\ds
					\sum_{s=0}^n \sum_{m=0}^{2s+1}  \sum_{j=-1}^{n+s-m-1} 
					\frac{\norm_{j+s+1,j+1}}{m!}
				\\
					\ds
					\dec{
						\sco[1]{
							\ds				
							\LDi{Dumbbell-vj_+h-02}{D-vj_+R-02}
						}{\Vertex{n_s, j}}		
					}{11\Delta^{j+s+1} \GRkpr^m}	
				\end{array}
			\right]
		\end{array}
	\]
\caption{Isolation of the classical, two-point vertices
of diagram~\ref{D-vj_+kb-vkb} plus diagram~\ref{D-vj_+kR-DEP-vkR}.}
\label{fig:bn-Ia}
\ecf

It is now easy to isolate the classical, two-point
vertices of diagrams~\ref{P-Sig_vj_+} and~\ref{WBT-Sig_vj_+},
though we note the following. First, on account
of the equality of the Wilsonian effective action,
classical, two-point vertices and their seed action 
counterparts, $\Sigma_{0 RS}^{\ XX}(k) = -S_{0 RS}^{\ XX}(k)$;
in other words,
the classical, two-point component of $\Sigma$ has a minus
sign, relative to the reduced part. Secondly, the
kernel attaching to the classical, two-point
vertex contained in diagram~\ref{P-Sig_vj_+} must be
decorated, in order that a connected diagram
can be formed.
Finally, the component of diagram~\ref{P-Sig_vj_+}
in which all vertices are Wilsonian effective
action vertices and the kernel is undecorated
is cancelled by
 the component of diagram~\ref{j+2-DEP}
in which the differentiated effective propagator
attaches at both ends to the same vertex (see diagram~\ref{vj_+R-DEP}).
The isolation of the
classical, two-point vertices of 
diagrams~\ref{P-Sig_vj_+} and~\ref{WBT-Sig_vj_+} is shown in 
\fig{fig:bn-Ib}.
\bcf
	\[
	\begin{array}{l}
	\vspace{2ex}
		\bca{
			\sum_{s=1}^n   \sum_{m=0}^{2s-1} \sum_{j=-1}^{n+s-m-2}
			\frac{\norm_{j+s,j+1}}{m!}
		}
			{
			\dec{
				\sco[1]{
					
					-2 \LDi{Padlock-hS_vj_+R}{P-hS_vj_+R}
				}{\Vertex{ n_{s}, j}}
			}{11\Delta^{j+s} \GRkpr^m}
		}
		-
		\bca{
			\sum_{s=1}^n   \sum_{m=0}^{2s-1} \sum_{j=-2}^{n+s-m-3}
			\frac{\norm_{j+s,j+1}}{m!}
		}
			{
			\dec{
				\begin{array}{c}
				\vspace{1ex}
					\ds
					\LDi{Padlock-02}{P-02-MV}
				\\
				\vspace{1ex}
					\Vertex{v^{j_+\!;R}}
				\\
					\Vertex{n_s, j}
				\end{array}
			}{11\Delta^{j+s} \GRkpr^m}
		}
	\\
	\vspace{2ex}
		+
		\bca{
			\sum_{s=1}^n   \sum_{m=0}^{2s-1} \sum_{j=-1}^{n+s-m-3}
			\frac{\norm_{j+s,j+1}}{m!}
		}
			{	
			\dec{
				\sco[1]{
					\ds	
					\LDi{Padlock-S_vj_+R}{P-S_vj_+R}			
					+\LDi{WBT-vj_+R}{WBT-vj_+R}
					-2\LDi{WBT-vj_+hR}{WBT-vj_+hR}
				}{\Vertex{n_s, j}}		
			}{11\Delta^{j+s} \GRkpr^m}
		}
	\\
		-
		\bca{
			\sum_{s=1}^n   \sum_{m=0}^{2s-1} \sum_{j=-2}^{n+s-m-3}
			\frac{\norm_{j+s+1,j+2}}{m!}		
		}
			{		
			\dec{
				\begin{array}{c}
				\vspace{1ex}
					\ds
					\LDi{WBT-02}{WBT-02-VS}
				\\
				\vspace{1ex}
					\Vertex{v^{j_+\!;R}}
				\\
					\Vertex{n_s, j}
				\end{array}
			}{11\Delta^{j+s+1} \GRkpr^m}
		}
	\end{array}
	\]
\caption{Isolation of the classical, two-point vertices
of diagrams~\ref{P-Sig_vj_+} and~\ref{WBT-Sig_vj_+} plus
diagram~\ref{vj_+R-DEP}.}
\label{fig:bn-Ib}
\ecf

The next step of the diagrammatic procedure
is to decorate the classical, two-point
vertices of diagrams~\ref{D-02x2-R-VS}, \ref{D-vj_+R-02}
and~\ref{WBT-02-VS} with either
an external field
or an end of an effective propagator. 
In the latter case we must then attach
the loose end of the effective propagator
to an available structure. We refer to the 
primary part of 
a diagram as the component left over after
applying the effective propagator relation
as many times as possible but, each time,
retaining only the 
Kronecker-$\delta$ contribution.

Assuming that
the necessary structures exist, we can do the
following with a diagram possessing a single
classical, two-point vertex:
\begin{enumerate}
	\item	attach an external field;

	\item	attach one end of an effective propagator,
			with the other end attaching to:
			\begin{enumerate}
				\item	the seed action vertex to which the kernel
						attaches;
				\label{EP-SV}

				\item	one of the Wilsonian  effective action vertices;
				\label{EP-DV}

				\item	the kernel;
				\label{EP-k}

				\item	a gauge remainder.
				\label{EP-GR}
			\end{enumerate}
\end{enumerate}

In each of~\ref{EP-SV}--\ref{EP-GR}
the effective propagator relation
can be applied. For the time
being, we will concern ourselves
with just the primary
parts of~\ref{EP-SV}--\ref{EP-k},
which yield a series of cancellations,
returning to both the gauge remainder
contributions and other diagrams later.
First, we analyse the result of
decorating the classical, two-point
vertex of diagram~\ref{D-vj_+R-02}.

\begin{cancel}[Diagram~\ref{D-vj_+kR-vkhR}]
\label{Cancel:D-vj_+kR-vkhR}
Consider the primary part of
diagram~\ref{D-vj_+R-02} corresponding
to~\ref{EP-DV}, above, which we note
exists only for $j>-1$. 
For comparison with diagram~\ref{D-vj_+kR-vkhR},
it is convenient
to change variables $j \rightarrow j+1$,
so that the sum over $j$ once again starts
from $-1$ and
to identify $\hat{v}^{j+2}$ with $\hat{v}^{k}$.
Thus, 
the two-point, tree level vertex of diagram\ref{D-vj_+R-02}
can be joined to any of $j+2$ identical Wilsonian
effective action vertices, using
any of $j+s+2$ effective propagators.
Noting that this effective propagator can
attach either way round,  the
combinatoric factor is 
\[
	2 (j+s+2)(j+2),
\]
which, from \eqs{eq:norm-b}{eq:norm-c}, 
combines with $\norm_{j+s+2,j+2}$
to give $-\norm_{j+s+1,j+1}$. Thus,
the primary part of
diagram~\ref{D-vj_+R-02} corresponding
to~\ref{EP-DV}, above,
precisely cancels
diagram~\ref{D-vj_+kR-vkhR}.
\end{cancel}

\begin{cancel}[Diagram~\ref{P-hS_vj_+R}]
\label{Cancel:P-Sig_vj_+R-hS}

Consider the primary
part of diagram~\ref{D-vj_+R-02} corresponding
to~\ref{EP-SV}, above, which we might
hope cancels diagram~\ref{P-hS_vj_+R}.
At first sight, though, it does not
look like the cancellation will quite work
since, although the combinatorics
are fine, the limits on the sums over $s, \ m$
and $j$ differ between diagrams~\ref{D-vj_+R-02}
and~\ref{P-hS_vj_+R}. However, the formation
of a loop in diagram~\ref{D-vj_+R-02}
allows us to restrict the ranges
over these sums (just as we did for diagram~\ref{vj_+R-DEP}),
and so the two diagrams do exactly cancel.
\end{cancel}

\begin{cancel}[Diagram~\ref{WBT-vj_+hR}]

Consider the primary
part of diagram~\ref{D-vj_+R-02} corresponding
to~\ref{EP-k}, above. Following through
similar arguments to those invoked in 
cancellation~\ref{Cancel:P-Sig_vj_+R-hS}, it can
be shown that the primary
part of diagram~\ref{D-vj_+R-02} corresponding
to~\ref{EP-k} exactly cancels diagram~\ref{WBT-vj_+hR}.

\end{cancel}

This concludes our initial discussion of
diagram~\ref{D-vj_+R-02}.
Rather than immediately analysing
diagram~\ref{WBT-02-VS}, which also possesses
a single classical, two-point vertex,
it is useful to first consider
the decoration of diagram~\ref{D-02x2-R-VS}
which
possesses two such vertices.
At $\Op{2}$, we can discard the case in which
an external field attaches to each of the classical,
two-point vertices. If one of the vertices
is decorated by an external field, then
we can attach an effective propagator to the
other according to~\ref{EP-DV}--\ref{EP-GR}
above. If both vertices are decorated by an
end of an effective propagator, then we can tie up
the associated loose ends in the following
independent ways: 
\begin{enumerate}
	\item	Join one classical, two-point vertex 
			to a Wilsonian effective action vertex 
			and the other to:
			\begin{enumerate}
				\item	the same vertex;
				\label{EPx2-SV}

				\item	a different vertex;
				\label{EPx2-DV}

				\item	the kernel;
				\label{EPx2-V-k}

				\item	a gauge remainder;
			\end{enumerate}

	\item	Join one classical, two-point vertex  to the kernel and the other to:
			\begin{enumerate}
				\item	the kernel;
				\label{EPx2-k-k}
	
				\item	a gauge remainder;
				\label{EPx2-k-GR}
			\end{enumerate}

	\item	Join one classical, two-point vertex  
			to a gauge remainder and the other to a gauge remainder
			\begin{enumerate}
				\item	in the same gauge remainder structure;

				\item	in a different gauge remainder structure;
			\end{enumerate}

	\item	Join the classical, two-point vertices together.
			\label{EP-02x2}
\end{enumerate}

We now consider the results of decorating
diagram~\ref{D-02x2-R-VS} according
to~\ref{EPx2-SV}--\ref{EP-02x2}.

\begin{cancel}[Diagram~\ref{P-S_vj_+R}]

The primary part of~\ref{D-02x2-R-VS}
corresponding to~\ref{EPx2-SV}, above, 
exactly cancels diagram~\ref{P-S_vj_+R}.
\end{cancel}

\begin{cancel}[Diagram~\ref{P-02-MV}]
Diagram~\ref{P-02-MV}
is exactly cancelled by the primary part
of diagram~\ref{D-02x2-R-VS}
corresponding to~\ref{EP-02x2},
above.
\end{cancel}

\begin{cancel}[Diagram~\ref{D-vj_+kR-vkR}]
The primary part of diagram~\ref{D-02x2-R-VS}
corresponding to~\ref{EPx2-DV}, above,
exactly cancels diagram~\ref{D-vj_+kR-vkR}.
\end{cancel}

\begin{cancel}[Components of diagram~\ref{WBT-02-VS}]
The primary part of diagram~\ref{D-02x2-R-VS}
corresponding to~\ref{EPx2-k-k}, above,
exactly cancels the primary part of diagram~\ref{WBT-02-VS}
in which the classical, two-point vertex attaches
to the kernel.
\label{Cancel:WBT-02}
\end{cancel}

Looking at cancellation~\ref{Cancel:WBT-02},
it is tempting to say that the primary
parts of diagram~\ref{WBT-02-VS}
are all
exactly cancelled by a subset of the
primary parts
of diagram~\ref{D-02x2-R-VS}.
However, this is not true: consider
the primary part of diagram~\ref{D-02x2-R-VS}
corresponding to~\ref{EPx2-k-GR}, above.
In this case, because each of the classical, two-point
vertices attaches to something different,
we pick up a factor of two, in recognition
of the fact that either of the vertices
could attach to the kernel / gauge remainder.
However, the primary part of diagram~\ref{WBT-02-VS}
in which the classical, two-point vertex attaches
to a gauge remainder can only be formed in one way;
hence the cancellation is only partially realized.
The key to cancellation~\ref{Cancel:WBT-02}
is that both classical, two-point vertices
of diagram~\ref{D-02x2-R-VS} attach to the same structure,
and so there is no factor of two picked up.

A
diagram generated by the partial cancellation of
a component of
diagram~\ref{WBT-02-VS}
against a component of diagram~\ref{D-02x2-R-VS}
is immediately involved in a cancellation.

\begin{cancel}[Diagram~\ref{WBT-vj_+R}]
\label{Cancel:WBT-vj_+R}
The primary part of diagram~\ref{D-02x2-R-VS}
corresponding to~\ref{EPx2-V-k} above and
the primary part of diagram~\ref{WBT-02-VS}
in which the classical, two-point attaches
to a vertex combine to exactly cancel
diagram~\ref{WBT-vj_+R}.
\end{cancel}

Using cancellations~\ref{Cancel:D-vj_+kR-vkhR}--\ref{Cancel:WBT-vj_+R} 
and partial cancellations of the type just discussed,
we can rewrite $\nLV{\Ds}{n}{\mu\nu}{a}(p)$ as shown in \fig{fig:bn-b-Pb}.
$\nLV{\G}{n}{\mu\nu}{a}(p)$  and $\nLV{\H}{n}{\mu\nu}{a}(p)$
are disjoint sets whose elements are
all spawned by $a_0$ and $a_1$: all such diagrams which 
possess an $\Op{2}$ stub belong to
$\nLV{\H}{n}{\mu\nu}{a}(p)$;
all elements
of  $\nLV{\G}{n}{\mu\nu}{a}(p)$ possess a
full gauge remainder
(arising from the application of the effective propagator relation).
\bcf
	\[
	\begin{array}{l}
	\vspace{3ex}
		\ds \dec{\nLV{\Ds}{n}{\mu\nu}{a}(p)}{\bullet} + \Op{4} = 
		-4 \beta_n \Box_{\mu\nu}(p)  +
		\mbox{\ref{beta-terms}} + \mbox{\ref{alpha-terms}} +
		\nLV{\G}{n}{\mu\nu}{a}(p) + \nLV{\H}{n}{\mu\nu}{a}(p)
	\\
	\vspace{3ex}
		+
		\left[
			\begin{array}{c}
				\ds
				\sum_{s=1}^n \sum_{m=1}^{2s} \sum_{j=-2}^{n+s-m-2}\frac{\norm_{j+s,j+2}}{m!}
				\sum_{m'=1}^{m} \!\! \nCr{m}{m'} 
			\\
				\dec{
				\begin{array}{c}
				\vspace{1ex}
					\ds \sum_{m''=1}^{m'-1}  \!\! \nCr{m'}{m''} \!\! 
					\LDi{GRs-W-GRs}{GRs-W-GRs-MV} \hspace{1.5em}
					+ \LDi{Ubend-W}{Ubend-W-MV} \hspace{1em}
					+ \LDi{WBT-GRs}{WBT-GRs-MV} 
				\\
				\vspace{1ex}
					\TopVertex
				\\
					\SumVertex
				\end{array}
			}{11\Delta^{j+s} \GRkpr^{m-m'}}
			\end{array}
		\right]
	\\
	\vspace{3ex}
		\ds
		+2 \sum_{s=1}^n \sum_{m=1}^{2s} \sum_{j=-2}^{n+s-m-2} 
		\frac{\norm_{j+s+1,j+2}}{(m-1)!}
		\dec{
			\sco[1]{\LO{\Vertex{v^{j_+\!;R}}}{j+2-DGR-b}}{\Vertex{n_s, j}}
		}{11\Delta^{j+s+1} \GRkpr^{m-1} \DGRkpr}
	\\
		+2
		\bca{
			\sum_{s=1}^n   \sum_{m=1}^{2s+1} \sum_{m'=1}^m \sum_{j=-1}^{n+s-m-1}
			\frac{\norm_{j+s,j+1}}{m!}
		}
			{
			\nCr{m}{m'}	
			\dec{
				\sco[1]{\LO{\ensuremath{\begin{array}{c}\input{pstex/v_j+hR-W-GRs.pstex_t} \end{array}}}{v_j+hR-W-GRs}}{\Vertex{n_s, j}}
				-\sco[1]{\LO{\ensuremath{\begin{array}{c}\input{pstex/v_j+R-W-GRs.pstex_t} \end{array}}}{v_j+R-W-GRs}}{\Vertex{n_s, j}}
			}{11\Delta^{j+s} \GRkpr^{m-m'}}
		}
	\end{array}
	\]
\caption{A re-expression of $\dec{\nLV{\Ds}{n}{\mu\nu}{a}(p)}{\bullet}$.}
\label{fig:bn-b-Pb}
\ecf

The formation of diagram~\ref{GRs-W-GRs-MV}
deserves further comment. This diagram
is a combination of the following terms:
\begin{enumerate}
	\item	the primary part
			of~\ref{D-02x2-R-VS} in which both classical,
			two-point vertices attaches to a different
			gauge remainder structure

	\item	the component of diagram~\ref{j+2-DEP} in which both
			ends of the differentiated effective propagator
			attaches to a different
			gauge remainder structure (see diagram~\ref{GRs-DEP-GRs-MV}).
\end{enumerate}

Notice that the limits on the sums over $m$ and $j$,
in diagram~\ref{GRs-W-GRs-MV} are the same as in
diagram~\ref{j+2-DEP} and not~\ref{D-02x2-R-VS}.
Diagram~\ref{D-02x2-R-VS} supplements diagram~\ref{j+2-DEP},
converting a differentiated effective propagator
into a full kernel in all cases where there
are sufficiently few gauge remainders and vertices
to permit decoration of the kernel. 
The values of $m$ and $j$ for which diagram~\ref{j+2-DEP}
exists but diagram~\ref{D-02x2-R-VS} does not are
those for which the differentiated effective
propagator of~\ref{j+2-DEP} can be trivially
replaced by a full kernel, since this kernel
can effectively never be decorated.

Similar considerations apply to diagrams~\ref{Ubend-W-MV}
and~\ref{v_j+R-W-GRs}. In the
latter case, to save space, we have put this diagram
under the same summations as diagram~\ref{v_j+hR-W-GRs}.
Whereas diagram~\ref{v_j+hR-W-GRs} exists for
the maximum values of $m$ and $j$, diagram~\ref{v_j+R-W-GRs}
possesses a Wilsonian effective action, one-point
vertex in these cases and so has no support: the
maximum values of $m$ and $j$ for which diagram~\ref{v_j+R-W-GRs}
exists are $2s$ and $n+s-m-2$, respectively.
In the same fashion, diagram~\ref{WBT-GRs-MV} does 
not in fact exist for the
maximum values of $m$ indicated. Having taken great
care, up until now, to restrict the ranges on
the sums as much as possible for each diagram,
we will now often put diagrams under the same summation
sign to save space, mindful that not all diagrams
necessarily exist for all values of $s$, $m$ and $j$.

\subsection{Gauge Remainders}
\label{sec:bn-GRs}

The terms of $\nLV{\G}{n}{\mu\nu}{a}(p)$ can be usefully
decomposed into five sets comprising:
\begin{enumerate}
\renewcommand{\theenumi}{\Roman{enumi}}
	\item	diagrams possessing a single full gauge remainder, 
			the $\GRkpr$ part of which attaches to a kernel,
			the other end of which attaches to:
			\begin{enumerate}
				\item	itself;
				\label{GR-WBT}
	
				\item	a reduced seed action vertex;
				\label{GR-S}

				\item	a reduced Wilsonian effective action vertex;
				\label{GR-W}

				\item	a gauge remainder structure;
				\label{GR-GRs}
			\end{enumerate}

	\item	diagrams possessing two full gauge remainders.
	\label{GRsx2}
\end{enumerate}

The strategy for dealing with the gauge remainders
is by now familiar:
allowing the full gauge remainder
to act, wherever possible, we isolate any
classical, two-point vertices and proceed as usual. 
We start with the 
gauge remainder terms of type~\ref{GR-WBT},
which we collect together in \fig{fig:bn:GR-WBT}.
\bcf
	\[
	\begin{array}{l}
	\vspace{2ex}
		+
		\bca{
			\sum_{s=1}^n \sum_{m=0}^{2s-1} \sum_{j=-2}^{n+s-m-3} 
		\frac{\norm_{j+s,j+1}}{m!}
		}
			{
			\dec{
				\sco[1]{
					\LDi{WBT-GR-vj_+R}{WBT-GR-vj_+R}
				}{\SumVertex}
			}{11\Delta^{j+s} \GRkpr^m}
		}
		-
		\bca{
			\sum_{s=1}^n \sum_{m=0}^{2s-2} \sum_{j=-2}^{n+s-m-3} 
			\frac{\norm_{j+s,j+2}}{m!}
		}
			{
			\dec{
				\Tower{\LDi{WBT-GR-B}{WBT-GR-B-MV}}
			}{11\Delta^{j+s} \GRkpr^m}
		}	
	\\
		-
		\bca{
			\sum_{s=1}^n \sum_{m=2}^{2s-1} \sum_{j=-2}^{n+s-m-3} 
			\frac{\norm_{j+s,j+2}}{m!} \sum_{m'=2}^m \nCr{m}{m'}
		}
			{
			\dec{
				\Tower{\LDi{WBT-GR-GRs}{WBT-GR-GRs-MV}}
			}{11\Delta^{j+s} \GRkpr^{m-m'}}
		}
	\end{array}
	\]
\caption{Gauge remainders of type~\ref{GR-WBT}.}
\label{fig:bn:GR-WBT}
\ecf

The pattern of cancellations we will find
is very similar to that observed in the
demonstration of the transversality
of $\nLV{\Ds}{n}{\mu\nu}{a}(p)$. Thus, although
diagram~\ref{WBT-GR-GRs-MV}
possesses a trapped gauge remainder, and so cannot
be processed, it has been collected with
the other terms in anticipation of a
cancellation via diagrammatic identity~\eq{eq:D-ID-G}.

Diagram~\ref{WBT-GR-B-MV} 
can be processed by allowing the gauge remainder
to act. This gauge remainder can strike one of three
locations:
\begin{enumerate}
	\item	the top end of the kernel (which itself attaches
			to the middle of the kernel);

	\item	the bottom end of the kernel;

	\item	a socket, which can be filled by one of the
			decorations. 
\end{enumerate}

Summing over diagrams of the first item
yields zero:
because the kernel
bites itself, its top end can be struck in two
independent ways. First, the active gauge remainder can 
push forward (pull back) onto the field on the
kernel which the kernel bites; this field
is identified with the top end of the kernel.
Secondly, the active gauge remainder
can pull back (push forward)  unhindered to the top.
These two independent cases exactly cancel
each other (see also the discussion around
diagrams~\ref{Diags-WGRx2-Trap}--\ref{Diags-WGRx2-Bottom-B}).

Now we consider processing the active
gauge remainder of diagram~\ref{WBT-GR-vj_+R}
and splitting off the classical, two-point
component from the reduced component.
Compared to diagram~\ref{CTP-EGR-Socket},
the classical, two-point vertex can,
in addition to being attached  
to a reduced Wilsonian
effective action vertex or gauge remainder
structure, be attached  
to the kernel or decorated with an 
external field. If it attaches
to the kernel, this just cancels
the component of diagram~\ref{WBT-GR-B-MV} 
where the gauge remainder bites a socket
on the kernel, up to a gauge remainder
contribution. This cancellation,
together with the analogue of
illustrative cancellation~\ref{Icancel:Tansverse}
occurs at each level of nesting.
Of those diagrams which survive, two
cancel against diagram~\ref{WBT-GR-GRs-MV}
via diagrammatic identity~\eq{eq:D-ID-G}.

Up to those diagrams with an $\Op{2}$
stub, the only diagrams which survive
are those spawned by diagram~\ref{WBT-GR-B-MV} 
and its nested partners in which the
gauge remainder strikes the bottom end
of the kernel, as shown in \fig{fig:bn-GR-WBT-Surv}.
\bcf[h]
	\[
		-
		\bca{
			\sum_{s=1}^n \sum_{m'=1}^{2s-1} \sum_{m=0}^{2s-1-m'} \sum_{j=-2}^{n+s-m-m'-3} 	
		}
			{
			\frac{\norm_{j+s,j+2}}{m!m'!}
			\dec{
				\Tower{\LDi{WBT-GRs}{WBT-ring}}
			}{11\Delta^{j+s} \GRkpr^m}
		}
	\]
\caption{Surviving terms generated by the diagrams of \fig{fig:bn:GR-WBT}, 
up to terms with an $\Op{2}$ stub.}
\label{fig:bn-GR-WBT-Surv}
\ecf
The sign and combinatoric factor of diagram~\ref{WBT-ring} 
follow from the discussion
around \fig{fig:nestedhook}.

\begin{cancel}[Diagram~\ref{WBT-GRs-MV}]
	Diagram~\ref{WBT-ring} exactly cancels
	diagram~\ref{WBT-GRs-MV}. This follows
	from first letting $m \rightarrow m-m'$ in~\ref{WBT-ring}
	and recognizing that 
	\[
	\sum_{m'=1}^{2s-1} \sum_{m=m'}^{2s-1} = \sum_{m=1}^{2s-1} \sum_{m'=1}^m,
	\]
	and secondly recalling that the upper limits on the sums over
	$m$ and $j$ in diagram~\ref{WBT-GRs-MV} can be reduced by
	one apiece.
\end{cancel}

We have thus demonstrated how all diagrams
spawned by the type~\ref{GR-WBT}
gauge remainders, up to those with an $\Op{2}$
stub, cancel. 
In preparation for our treatment of the other 
gauge remainders, it is worth reviewing this
how this happens.
If an active gauge remainder
strikes a socket on a reduced vertex
or on a kernel, this diagram will be
cancelled by a diagram in which the gauge
remainder instead strikes a socket on
a classical, two-point vertex, which
is subsequently attached to either
a reduced vertex or a kernel with an 
effective propagator. Such cancellations
leave over diagrams of exactly the same
basic form as the parent diagrams,
but where the gauge remainder is nested.
Allowing the nested gauge remainder
to act then just repeats the cancellations
just described. We are left with following:
\begin{enumerate}
	\item	Diagrams possessing a 
			classical, two-point vertex attached to
			\begin{enumerate}
				\item	an $\Op{2}$ stub;

				\item  a gauge remainder nested \wrt\ to
						the original, active gauge remainder;
				\label{it:nested-wrt-active}

				\item	a gauge remainder structure.
				\label{it:GR-structure}
			\end{enumerate}

	\item	The original type~\ref{GR-WBT} diagram possessing a trapped
			gauge remainder;
	\label{it:trapped}

	\item	Diagrams in which a kernel is bitten at its end by
			a (nested) gauge remainder.
	\label{it:BittenKernel}
\end{enumerate}
Diagrams of type~\ref{it:nested-wrt-active}, \ref{it:GR-structure}
and~\ref{it:trapped} cancel, courtesy of diagrammatic
identity~\ref{eq:D-ID-G}.
Diagrams of type~\ref{it:BittenKernel} cancel
against terms generated elsewhere in the calculation.

We thus see that most of the diagrams spawned by the
type~\ref{GR-WBT} gauge remainders cancel amongst themselves.
The wonderful thing is that this basic pattern of cancellations
is repeated for all the other types of gauge remainders. 
The differences come with the types of terms generated
which cancel against terms generated elsewhere in
the calculation. With this in mind, we now treat
the gauge remainders of types~\ref{GR-S} and~\ref{GR-W}, which
are collected together in \figs{fig:bn-GR-S}{fig:bn-GR-W},
respectively.
\bcf
	\[
	\begin{array}{c}
	\vspace{4ex}
		+2
			\left[
				\begin{array}{c}
				\vspace{1ex}
					\ds
					\sum_{s=0}^n \sum_{m=0}^{2s}  \sum_{j=-1}^{n+s-m-3} 
					\frac{\norm_{j+s+1,j+1}}{m!}
				\\
					\ds
					\dec{
						\LO{\ensuremath{\begin{array}{c}\input{pstex/Dumbbell-vj_+kR-vkhR-GR.pstex_t} \end{array}} \SumVertex}{D-vj_+kR-vkhR-GR}
					}{11\Delta^{j+s+1} \GRkpr^m}
				\end{array}
			\right]
	\\
	\vspace{4ex}
		-2
		\bca{
			\sum_{s=1}^n   \sum_{m=0}^{2s-1} \sum_{j=-1}^{n+s-m-2}
			\frac{\norm_{j+s,j+1}}{m!}
		}
			{
			\dec{
				\sco[1]{
					\LDi{Padlock-hS_vj_+R-GR}{P-hS_vj_+R-GR}
					+\LDi{WBT-vj_+hR-GR}{WBT-vj_+hR-GR}
				}{\SumVertex}
			}{11\Delta^{j+s} \GRkpr^m}
		}
	\\
	-2
		\bca{
			\sum_{s=0}^n \sum_{m=2}^{2s} \sum_{m'=2}^m \sum_{j=-1}^{n+s-m-2} 
			\frac{\norm_{j+s,j+1}}{m!}
		}
			{
			\nCr{m}{m'}
			\dec{
				\sco[1]{
						\LDi{vj_+h-W-GR-GRs}{vj_+h-W-GR-GRs}
				}{\SumVertex}		
			}{11\Delta^{j+s} \GRkpr^{m-m'}}

		}
	\end{array}
	\]
\caption{Gauge remainders of type~\ref{GR-S}.}
\label{fig:bn-GR-S}
\ecf

\bcf
	\[
	\begin{array}{c}
	\vspace{4ex}
		-2
			\left[
				\begin{array}{c}
				\vspace{1ex}
					\ds
					\sum_{s=0}^n \sum_{m=0}^{2s-1}  \sum_{j=-1}^{n+s-m-4} 
					\frac{\norm_{j+s+1,j+1}}{m!}
				\\
					\ds
					\dec{
						\LO{\ensuremath{\begin{array}{c}\input{pstex/Dumbbell-vj_+kR-vkR-DW-GR.pstex_t} \end{array}} \SumVertex}{D-vj_+kR-vkR-DW-GR}
					}{11\Delta^{j+s+1} \GRkpr^m}
				\end{array}
			\right]
	\\
	\vspace{4ex}
	+2
		\bca{
			\sum_{s=1}^n   \sum_{m=0}^{2s-1} \sum_{j=-1}^{n+s-m-2}
			\frac{\norm_{j+s,j+1}}{m!}
		}
			{
			\dec{
				\sco[1]{
					\LDi{Padlock-S_vj_+R-DW-GR}{P-S_vj_+R-DW-GR}
					+\LDi{WBT-vj_+R-DW-GR}{WBT-vj_+R-DW-GR}
				}{\SumVertex}
			}{11\Delta^{j+s} \GRkpr^m}
		}
	\\
	+2
		\bca{
			\sum_{s=0}^n \sum_{m=2}^{2s-1} \sum_{m'=2}^m \sum_{j=-1}^{n+s-m-3} 
			\frac{\norm_{j+s,j+1}}{m!}
		}
			{
			\nCr{m}{m'}
			\dec{
				\sco[1]{
						\LDi{vj_+R-DW-GR-GRs}{vj_+R-DW-GR-GRs}
				}{\SumVertex}		
			}{11\Delta^{j+s} \GRkpr^{m-m'}}

		}
	\end{array}
	\]
\caption{Gauge remainders of type~\ref{GR-W}.}
\label{fig:bn-GR-W}
\ecf

It is apparent that the gauge remainders
of types~\ref{GR-S} and~\ref{GR-W} are
almost exactly the same, which is why
we choose to treat them together.
The differences are first that the gauge
remainders of type~\ref{GR-S} possess
a seed action vertex and secondly that the kernels
of all type~\ref{GR-W} diagrams
must be decorated. In diagrams~\ref{D-vj_+kR-vkhR-GR}
and~\ref{vj_+h-W-GR-GRs}, we note
that we can reduce the maximum values
of the sums over $m$ and $j$ by one more
than expected: if the seed action vertex is
one-point, then the kernel must be decorated,
else the full gauge remainder is forced
to be in the $C^i$ sector---where of course it
is null---by \CC\ invariance.

Processing the gauge remainders, it is
obvious what will happen. 
Diagram~\ref{D-vj_+kR-vkhR-GR} (\ref{D-vj_+kR-vkR-DW-GR})
has two components: one in which the
gauge remainder strikes a socket
on a reduced vertex and one in which
the gauge remainder strikes a socket
on a classical, two-point vertex.
The former case is cancelled, up to a nested 
gauge remainder by the contribution in which
the classical, two-point
vertex is joined to one of the Wilsonian
effective action vertices. If, instead,
the classical, two-point vertex is attached
to a socket on either the seed action vertex or
kernel then, up to a nested gauge remainder, 
this cancels the contributions
from diagrams~\ref{P-hS_vj_+R-GR} 
(\ref{P-S_vj_+R-DW-GR}) and~\ref{WBT-vj_+hR-GR} (\ref{WBT-vj_+R-DW-GR})
in which, respectively, the gauge strikes a socket
on the reduced component of the seed action
vertex and a socket on the kernel.
Up to terms with an $\Op{2}$ stub, the final type of diagram 
spawned by~\ref{D-vj_+kR-vkhR-GR} (\ref{D-vj_+kR-vkR-DW-GR})
is one in which the classical, two-point vertex
formed by the action of the gauge remainder
is attached to a gauge remainder structure.
Putting this to one side for the moment,
we note that  processing the nested gauge remainders
repeats this pattern of cancellations. 

At the first level of nesting we get, as usual, a
new type of diagram: one in which the classical,
two-point vertex generated by~\ref{D-vj_+kR-vkhR-GR} 
(\ref{D-vj_+kR-vkR-DW-GR}) attaches to the nested
gauge remainder. Iterating the
diagrammatic procedure, diagrams of this type, together
with diagrams of the type temporarily put aside,
exactly cancel diagram~\ref{vj_+h-W-GR-GRs} (\ref{vj_+R-DW-GR-GRs}),
courtesy of diagrammatic identity~\eq{eq:D-ID-G}.

Of the diagrams that remain, two are immediately involved
in cancellations.

\begin{cancel}[Diagrams~\ref{v_j+hR-W-GRs} and~\ref{v_j+R-W-GRs}]

Diagram~\ref{v_j+hR-W-GRs} (\ref{v_j+R-W-GRs}) is exactly
cancelled by the component of diagrams~\ref{WBT-vj_+hR-GR}
(\ref{WBT-vj_+R-DW-GR}) and its nested partners in which the
gauge remainder strikes the top end of the kernel.

\end{cancel}

Up to $\Op{2}$ terms, the diagrams which survive are the following terms and
their nested partners.
\begin{enumerate}
	\item	The component of diagram~\ref{P-hS_vj_+R-GR}
			in which the gauge remainder strikes
			\begin{enumerate}
				\item	a socket on a classical, two-point vertex;
						\label{GR-S-CTP}

				\item	the field which attaches to the
						base of the kernel, and for which the
						seed action vertex is:
						\begin{enumerate}
							\item a reduced vertex;
							\label{GR-S-BoK-R}			
							\item a classical, two-point vertex;
							\label{GR-S-BoK-CTP}
						\end{enumerate}
			\end{enumerate}

	\item	The component of diagram~\ref{WBT-vj_+hR-GR} in which 
			the gauge remainder strikes the bottom end of the
			kernel;
			\label{GR-S-K}
	
	\item	The component of diagram~\ref{P-S_vj_+R-DW-GR}
			in which the gauge remainder strikes
			\begin{enumerate}
				\item	a socket on a classical, two-point vertex;
						\label{GR-W-CTP}

				\item	the field which attaches to the
						base of the kernel, and for which the
						seed action vertex is:
						\begin{enumerate}
							\item a reduced vertex;
							\label{GR-W-BoK-R}		
							\item a classical, two-point vertex;
							\label{GR-W-BoK-CTP}
						\end{enumerate}
			\end{enumerate}

	\item	The component of diagram~\ref{WBT-vj_+R-DW-GR} in which 
			the gauge remainder strikes the bottom end of the
			kernel;
			\label{GR-W-K}
\end{enumerate}

Now, \ref{GR-S-BoK-R} and~\ref{GR-S-K} exactly cancel
(\cf illustrative cancellation~\ref{Icancel:v-W-GR-ai}).
However, \ref{GR-W-BoK-R} and~\ref{GR-W-K} do not:
in the former case, the kernel must be decorated
whereas, in the latter case, there is no such restriction.
Thus, a diagram possessing a differentiated effective
propagator is left behind.
The component of diagram~\ref{GR-S-CTP} (\ref{GR-S-BoK-CTP})
in which the kernel is decorated is exactly
cancelled by diagram~\ref{GR-W-CTP} (\ref{GR-W-BoK-CTP}).
Thus, we are left with a whole set of diagrams,
collected together in \fig{fig:TLTP-Survivors},
possessing differentiated effective propagators.

\bcf
	\[
	\begin{array}{l}
	\vspace{2ex}
		-2
		\bca{
			\sum_{s=1}^n \sum_{m=0}^{2s-1}  \sum_{j=-1}^{n+s-m-3} 
			\frac{\norm_{j+s,j+1}}{m!} \sum_{m'=0}^{m}
		}
			{
			\dec{
				\nCr{m}{m'} \!\!
				\sco[1]{
					\LDi{RV-DEP-B-DEP}{RV-DEP-B-DEP} \ 
				}{\SumVertex}
			}{11\Delta^{j+s}\GRkpr^{m-m'}}
		}	
	\\
		-2
		\bca{
			\sum_{s=1}^n \sum_{m=0}^{2s-1} \sum_{j=-2}^{n+s-m-3} 
			\frac{\norm_{j+s+1,j+2}}{m!}  \sum_{m'=0}^{m} \!\! \nCr{m}{m'} \!\!
		}
			{
			\dec{
				\Tower{
					\LDi{TLTP-DEP-B-DEP}{TLTP-DEP-B-DEP-MV} \ 
					+ \LDi{TLTP-DEP-B-Soc}{TLTP-DEP-B-Soc-MV}}
			}{11\Delta^{j+s+1}\GRkpr^{m-m'}}
		}
	\end{array}
	\]
\caption{Surviving terms from the type~\ref{GR-S} and~\ref{GR-W}
gauge remainders, up to those with an $\Op{2}$ stub.}
\label{fig:TLTP-Survivors}
\ecf

Diagram~\ref{TLTP-DEP-B-Soc-MV} can be simplified
by utilizing the primary diagrammatic identities
and the classical flow equation. We illustrate this
by considering the un-nested version; identical 
manipulations can be performed in the nested case.
\bea
	\ensuremath{\begin{array}{c}\input{pstex/TLTP-DEP-GR.pstex_t} \end{array}} \	& =	& \dec{\ensuremath{\begin{array}{c}\begin{picture}(0,0)%
\includegraphics{pstex/TLTP-EP-GR-B.pstex}%
\end{picture}%
\setlength{\unitlength}{3947sp}%
\begingroup\makeatletter\ifx\SetFigFont\undefined%
\gdef\SetFigFont#1#2#3#4#5{%
  \reset@font\fontsize{#1}{#2pt}%
  \fontfamily{#3}\fontseries{#4}\fontshape{#5}%
  \selectfont}%
\fi\endgroup%
\begin{picture}(439,594)(1649,144)
\put(1751,332){\makebox(0,0)[lb]{\smash{\SetFigFont{8}{9.6}{\rmdefault}{\mddefault}{\updefault}{\color[rgb]{0,0,0}$0^2$}%
}}}
\end{picture}
 \end{array}}}{\bullet} - \ensuremath{\begin{array}{c}\input{pstex/TLTP-LdL-EP-GR.pstex_t} \end{array}} - \ensuremath{\begin{array}{c}\input{pstex/TLTP-EP-GR-LdL.pstex_t} \end{array}}	\nonumber
\\[1ex]
						& =	& \dec{\ensuremath{\begin{array}{c}\begin{picture}(0,0)%
\includegraphics{pstex/TLTP-GR-PEP.pstex}%
\end{picture}%
\setlength{\unitlength}{3947sp}%
\begingroup\makeatletter\ifx\SetFigFont\undefined%
\gdef\SetFigFont#1#2#3#4#5{%
  \reset@font\fontsize{#1}{#2pt}%
  \fontfamily{#3}\fontseries{#4}\fontshape{#5}%
  \selectfont}%
\fi\endgroup%
\begin{picture}(438,603)(1649,163)
\put(1751,332){\makebox(0,0)[lb]{\smash{\SetFigFont{8}{9.6}{\rmdefault}{\mddefault}{\updefault}{\color[rgb]{0,0,0}$0^2$}%
}}}
\end{picture}
 \end{array}}}{\bullet} - \ensuremath{\begin{array}{c}\input{pstex/TLTP-LdL-GR-PEP.pstex_t} \end{array}} - \ensuremath{\begin{array}{c}\begin{picture}(0,0)%
\includegraphics{pstex/GR-LdL.pstex}%
\end{picture}%
\setlength{\unitlength}{3947sp}%
\begingroup\makeatletter\ifx\SetFigFont\undefined%
\gdef\SetFigFont#1#2#3#4#5{%
  \reset@font\fontsize{#1}{#2pt}%
  \fontfamily{#3}\fontseries{#4}\fontshape{#5}%
  \selectfont}%
\fi\endgroup%
\begin{picture}(205,238)(1748,126)
\put(1914,157){\makebox(0,0)[lb]{\smash{\SetFigFont{8}{9.6}{\rmdefault}{\mddefault}{\updefault}{\color[rgb]{0,0,0}$\bullet$}%
}}}
\end{picture}
 \end{array}} + \ensuremath{\begin{array}{c}\begin{picture}(0,0)%
\includegraphics{pstex/GR-GR-LdL.pstex}%
\end{picture}%
\setlength{\unitlength}{3947sp}%
\begingroup\makeatletter\ifx\SetFigFont\undefined%
\gdef\SetFigFont#1#2#3#4#5{%
  \reset@font\fontsize{#1}{#2pt}%
  \fontfamily{#3}\fontseries{#4}\fontshape{#5}%
  \selectfont}%
\fi\endgroup%
\begin{picture}(217,349)(1748,126)
\put(1914,157){\makebox(0,0)[lb]{\smash{\SetFigFont{8}{9.6}{\rmdefault}{\mddefault}{\updefault}{\color[rgb]{0,0,0}$\bullet$}%
}}}
\end{picture}
 \end{array}} \nonumber
\\[1ex]
						& =	& - \ensuremath{\begin{array}{c}\begin{picture}(0,0)%
\includegraphics{pstex/GR-LdL.pstex}%
\end{picture}%
\setlength{\unitlength}{3947sp}%
\begingroup\makeatletter\ifx\SetFigFont\undefined%
\gdef\SetFigFont#1#2#3#4#5{%
  \reset@font\fontsize{#1}{#2pt}%
  \fontfamily{#3}\fontseries{#4}\fontshape{#5}%
  \selectfont}%
\fi\endgroup%
\begin{picture}(205,238)(1748,126)
\put(1914,157){\makebox(0,0)[lb]{\smash{\SetFigFont{8}{9.6}{\rmdefault}{\mddefault}{\updefault}{\color[rgb]{0,0,0}$\bullet$}%
}}}
\end{picture}
 \end{array}}
\label{eq:GR-LdL}
\eea

To go from the first line to the second, we have employed diagrammatic 
identity~\eq{eq:PseudoEP} and the effective
propagator relation. On the second line, the
first term vanishes courtesy of diagrammatic
identity~\eq{eq:GR-TLTP}; similarly, the second
term, if we employ~\eq{eq:LdL-GRk-Pert}. 
The final term on the second line
vanishes on account of diagrammatic identities~\eq{eq:GR-relation}
and~\eq{eq:LdL-GRk-Pert}:
\[
	\dec{ \GRk \!\! \GRkpr}{\bullet} = 0 = \stackrel{\bullet}{\GRk} \! \GRkpr + \GRk \! \! \stackrel{\bullet}{\GRkpr} = \GRk \!\!  \stackrel{\bullet}{\GRkpr}.
\]
Thus, the structure in diagram~\ref{TLTP-DEP-B-Soc-MV}
comprising a classical, two-point vertex,
differentiated effective propagator
and $\GRkpr$ reduces to just
a differentiated $\GRkpr$. We now promote
the $\DGRkpr$ to an implicit
decoration noting that, according to the rules
for converting implicit decorations to explicit
decorations, this comes with a minus sign
(see the discussion around \fig{fig:nestedhook}).
Finally, the decorations $\decGR{\ }{\GRkpr^{m'} \DGRkpr}$
can be combined with the overall decorations $\GRkpr^m$.
Upon shifting $m \rightarrow m' -1$ we find that:
\begin{cancel}[Diagram~\ref{j+2-DGR-b}]
	Diagram~\ref{TLTP-DEP-B-Soc-MV} exactly
	cancels diagram~\ref{j+2-DGR-b}. 
\end{cancel}

We now proceed to decorate the classical, two-point
vertex of diagram~\ref{TLTP-DEP-B-DEP-MV}. In
the case that we join this vertex to a Wilsonian 
effective action vertex, we just cancel diagram~\ref{RV-DEP-B-DEP},
up to a gauge remainder. Processing this
gauge remainder, we can think of
the resultant diagram as versions 
of~\ref{RV-DEP-B-DEP} and~\ref{TLTP-DEP-B-DEP-MV} but where the end
of the differentiated effective propagator
which attaches to the two-point vertex
now attaches to one of the nested gauge
remainders, instead. 
Thus, this version
of diagram~\ref{RV-DEP-B-DEP} is cancelled also.
Up to diagrams with an $\Op{2}$ stub, the only terms
that remain are those of the 
form~\ref{TLTP-DEP-B-DEP-MV} 
(but where the differentiated effective propagator
can attach to nested gauge remainders, in addition to
the classical, two-point vertex)
where 
the classical, two-point vertex is attached to
one of the nested gauge remainders or to a gauge remainder
structure. These terms can
be combined, courtesy of diagrammatic
identity~\eq{eq:D-ID-G}, to yield the diagrams of
\fig{fig:TLTP-Survivors-b}.
\bcf
	\[
	\begin{array}{l}
	\vspace{2ex}	
		-2
		\bca{
			\sum_{s=1}^n \sum_{m=1}^{2s-1} \sum_{j=-2}^{n+s-m-3} \sum_{m'=1}^{m}
			\frac{\norm_{j+s,j+2}}{(m-m')!(m'+1)!}  
		}
			{
			\dec{
				\Tower{\LDi{GR-ArbGRs-DEP}{GR-ArbGRs-DEP}} \hspace{2em}
			}{11\Delta^{j+s+1}\GRkpr^{m-m'}}
		}
	\\
		+2
		\bca{
			\sum_{s=1}^n \sum_{m=1}^{2s-1} \sum_{j=-2}^{n+s-m-3}  
			\frac{\norm_{j+s,j+2}}{m!} \sum_{m'=2}^{m} \!\! \nCr{m}{m'} \!\!
		}
			{
			\sum_{m''=0}^{m'-2} \!\! \nCr{m'}{m''} \!\!
			\dec{
				\Tower{\LDi{GRs-TLTP-GRs-DEP}{GRs-TLTP-GRs-DEP}} \hspace{2em}
			}{11\Delta^{j+s}\GRkpr^{m-m'}}
		}
	\end{array}
	\]
\caption{The surviving 
diagrams spawned by the gauge remainders
of types~\ref{GR-S} and~\ref{GR-W}, up to those
with an $\Op{2}$ stub.}
\label{fig:TLTP-Survivors-b}
\ecf

It is worth expanding on the origin 
of diagrams~\ref{GR-ArbGRs-DEP} 
and~\ref{GRs-TLTP-GRs-DEP} a little further,
which is most readily done by returning to~\eq{eq:D-ID-G}.
Essentially, we have combined the following terms:
\begin{enumerate}
	\item	a version of the first term in which a differentiated
			effective propagator attaches at one end to the
			explicitly drawn $\GRkpr$ and at the other end
			to either the socket decorating the classical, two-point
			vertex or to one of the implicit decorations;

	\item	a version of the final term in which a differentiated
			effective propagator attaches at one end to the explicitly
			drawn $\GRkpr$ and at the other end
			to either the socket decorating the classical, two-point
			vertex or to one of the implicit decorations which
			forms the string of gauge remainders biting this socket.
\end{enumerate}
Clearly, we can re-express these diagrams in terms of:
\begin{enumerate}
	\item	the second diagram of~\eq{eq:D-ID-G} in which
			a differentiated
			effective propagator attaches at one end to the
			full gauge remainder and at the other end
			to one of the implicit decorations;

	\item	a version of the final term in which a differentiated
			effective propagator attaches at one end to the explicitly
			drawn $\GRkpr$ and at the other end to the gauge remainder
			components which decorate the top of the diagram.
\end{enumerate}

Having completed the analysis
of the type~\ref{GR-S} and~\ref{GR-W} gauge remainders,  
we now move on to gauge remainders of
type~\ref{GR-GRs}, which we collect together
in \fig{fig:GR-GRs}.
\bcf
	\[
	\begin{array}{c}
	\vspace{2ex}
		2
		\bca{
			\sum_{s=1}^n \sum_{m=1}^{2s-1} \sum_{j=-1}^{n+s-m-3} 
		\frac{\norm_{j+s,j+1}}{m!} \sum_{m'=1}^m \nCr{m}{m'}
		}
			{
			\dec{
				\sco[1]{
					\LDi{GRs-RW-GR-RWV}{GRs-RW-GR-RWV}
				}{\SumVertex}
			}{11\Delta^{j+s} \GRkpr^{m-m'}}
		}
	\\
		-2
		\bca{
			\sum_{s=1}^n \sum_{m=1}^{2s-1} \sum_{j=-2}^{n+s-m-3} 
			\frac{\norm_{j+s,j+2}}{m!} \sum_{m'=1}^m \!\! \nCr{m}{m'} \!\!
		}
			{
			
			\dec{
				\Tower{
					\ds
					\sum_{m''=1}^{m'-1} \!\! \nCr{m'}{m''} \hspace{-1em}
					\LDi{GRs-DW-GR-GRs}{GRs-DW-GR-GRs} \hspace{1.5em}
					+\LDi{Ubend-DW-GR}{Ubend-DW-GR} \hspace{1em}
					+\LDi{GRBT-GRs}{GRBT-GRs} \hspace{0.5em}
				}
			}{11\Delta^{j+s} \GRkpr^{m-m'}}
		}
	\end{array}
	\]
\caption{Gauge remainders of type~\ref{GR-GRs}.}
\label{fig:GR-GRs}
\ecf

Allowing the gauge remainder to act in
diagrams~\ref{GRs-RW-GR-RWV} and~\ref{GRBT-GRs},
we uncover the by now familiar pattern of
cancellations. Up to terms with an $\Op{2}$
stub, we are left with:
\begin{enumerate}
	\item	diagram~\ref{Ubend-DW-GR}, which we notice combines
			with diagram~\ref{GR-ArbGRs-DEP};

	\item	diagrams in which the (nested) gauge remainder of (the nested version of)
			\ref{GRBT-GRs} strikes:
			\begin{enumerate}
				\item	the top end of the kernel;

				\item	the bottom end of the kernel;
			\end{enumerate}

	\item	diagrams in which the classical, two-point vertex of~\ref{GRs-RW-GR-RWV}
			and its nested partners
			is attached to one of the $m'$ gauge remainders,
			at the top of the diagram.
\end{enumerate}

With a little thought,
we see that diagrams of the last item are
of the same structure as diagram~\ref{GRs-TLTP-GRs-DEP}. 
Indeed, these diagrams
combine, to yield a single term with a full kernel.
The resultant term is shown in \fig{fig:GR-GRs-P},
together with the other surviving diagrams
spawned by the gauge remainders of type~\ref{GR-GRs}, up
to those with an $\Op{2}$ stub.
\bcf
	\[
	\begin{array}{l}
	\vspace{2ex}
		-2
		\bca{
			\sum_{s=1}^n \sum_{m=2}^{2s} \sum_{j=-2}^{n+s-m-2} \frac{\norm_{j+s,j+2}}{m!}
			\sum_{m'=2}^m \!\! \nCr{m}{m'} \!\! 
		}
			{
			\dec{
				\Tower{
					\ds
					\sum_{m''=1}^{m'-1} \!\! \nCr{m'}{m''} \!\!
					\LDi{GRs-W-GRs}{GRs-W-GRs} \hspace{1.5em} 
					+\LDi{Ubend-W-GR}{Ubend-W-GR} \hspace{1em}
				}
			}{11\Delta^{j+s} \GRkpr^{m-m'}}
		}
	\\
			+2
		\bca{
			\sum_{s=1}^n \sum_{m=2}^{2s-1} \sum_{j=-2}^{n+s-m-3} \frac{\norm_{j+s,j+2}}{m!}
			\sum_{m'=2}^m \!\! \nCr{m}{m'} \!\! 
		}
			{
			\sum_{m''=0}^{m'-2} \!\! \nCr{m'}{m''} \!\! 
			\dec{
				\Tower{					
					\LDi{K-GRs-B-K-GRs}{K-GRs-B-K-GRs} \hspace{2em}	
					+\LDi{GRs-TLTP-GRs-K}{GRs-TLTP-GRs-K} \hspace{2em}
				} 
			}{11\Delta^{j+s} \GRkpr^{m-m'}}
		}
	\end{array}
	\]
\caption{Surviving diagrams spawned by the 
gauge remainders of type~\ref{GR-GRs} plus 
diagrams~\ref{GR-ArbGRs-DEP} and~\ref{GRs-TLTP-GRs-DEP}.}
\label{fig:GR-GRs-P}
\ecf

We note the following about the diagram of \fig{fig:GR-GRs-P}.
First, we should explain the sign of diagram~\ref{K-GRs-B-K-GRs},
since the parent diagram comes with a minus sign.
Consider the string of $m''+1$ gauge remainders (the $+1$ corresponding
to the explicitly drawn gauge remainder). The
last of these bites the end of the kernel, which
is then attached to one of the $m'-m''$ gauge remainders.
In other words, we can think of the first 
of the $m'-m''$ gauge remainders not as being at the beginning of a string
of $m'-m''$ gauge remainders, but in the middle (or end if $m'-m'' =1$) 
of a string of $m'+1$ gauge remainders. Upon explicit decoration
of diagram~\ref{K-GRs-B-K-GRs} with the $m'$ gauge remainders,
our rule for determining the sign is to associate
associate bites to the left (right)
with pushes forward (pulls back), yielding $(-1)^R$.
However, if the gauge remainder at the end
of the kernel is bitten on the left (right),
it is in fact pulled back (pushed forward)
onto (\cf the discussion around \figs{fig:GRstring}{fig:nestedhook}).
Hence, given our rule for decoration, we must give
the diagram an overall minus sign.

Secondly, notice that diagram~\ref{GRs-W-GRs} is
exactly the same as diagram~\ref{GRs-W-GRs-MV}
(the difference between the lower limits on
the sums over $m$ and $m'$ is just an artefact
of putting diagram~\ref{GRs-W-GRs} under the same
summation signs as diagram~\ref{WBT-GRs-MV}---which does
exist for $m,\ m'=1$),
up to a relative factor of $-2$.
Thus, these two diagrams do not exactly cancel;
to find the missing term that completes the 
cancellation---and also to understand the
roles of diagrams~\ref{K-GRs-B-K-GRs}--\ref{GRs-TLTP-GRs-K}---we 
must conclude  our discussion of
the gauge remainders by analysing the
gauge remainders of type~\ref{GRsx2},
which are collected together in \figs{fig:GRsx2-a}{fig:GRsx2-b}.
\bcf
	\[
	\begin{array}{l}
	\vspace{2ex}
		
			\left[
				\begin{array}{c}
				\vspace{1ex}
					\ds
					\sum_{s=0}^n \sum_{m=0}^{2s-1}  \sum_{j=-1}^{n+s-m-4} 
					\frac{\norm_{j+s+1,j+1}}{m!} 
				\\
					\ds		
					\dec{
						\LO{\ensuremath{\begin{array}{c}\input{pstex/Dumbbell-vj_+kR-vkR-DW-GRx2.pstex_t} \end{array}} \SumVertex}{D-vj_+kR-vkR-DW-GRx2}
					}{11\Delta^{j+s+1} \GRkpr^m}
				\end{array}
			\right]
			+
		\bca{
			\sum_{s=1}^n   \sum_{m=0}^{2s-3} \sum_{j=-2}^{n+s-m-5}
			\frac{\norm_{j+s,j+2}}{m!}		
		}
			{
			\dec{
				\Tower{\LDi{Struc-WGRx2-B}{WGRx2}}
			}{11\Delta^{j+s} \GRkpr^m}
		}
	\\
	-
		\bca{
			\sum_{s=1}^n   \sum_{m=0}^{2s-1} \sum_{j=-1}^{n+s-m-3}
			\frac{\norm_{j+s,j+1}}{m!}
		}
			{
			\dec{
				\sco[1]{
					\LDi{Padlock-S_vj_+R-DW-GRx2}{P-S_vj_+R-DW-GRx2}
					+2\LDi{WBT-vj_+R-DW-GRx2}{WBT-vj_+R-DW-GRx2}
				}{\SumVertex}
			}{11\Delta^{j+s} \GRkpr^m}
		}
	\end{array}
	\]
\caption{Gauge remainders of type~\ref{GRsx2}, part~1.}
\label{fig:GRsx2-a}
\ecf

\bcf
	\[
	\begin{array}{l}
	\vspace{2ex}
		-2
		\bca{
			\sum_{s=0}^n \sum_{m=2}^{2s-1} \sum_{m'=2}^m \sum_{j=-1}^{n+s-m-3} 
			\frac{\norm_{j+s,j+1}}{m!}
		}
			{
			\nCr{m}{m'}
			\dec{
				\sco[1]{
						\LDi{vj_+R-DW-GRx2-GRs}{vj_+R-DW-GRx2-GRs}
				}{\SumVertex}		
			}{11\Delta^{j+s} \GRkpr^{m-m'}}

		}
	\\
		+
		\bca{
			\sum_{s=1}^n \sum_{m=1}^{2s-1} \sum_{j=-2}^{n+s-m-3} 
			\frac{\norm_{j+s,j+2}}{m!} \sum_{m'=1}^m \!\! \nCr{m}{m'} \!\!
		}
			{
			
			\dec{
				\Tower{
					\ds
					\sum_{m''=1}^{m'-1} \!\! \nCr{m'}{m''} 	\hspace{-1em} 
					\LDi{GRs-DW-GRx2-GRs}{GRs-DW-GRx2-GRs} 	\hspace{1.5em}
					+\LDi{Ubend-GRsx2}{Ubend-GRsx2}			\hspace{1em}
					+2\LDi{WBT-GRx2-GRs}{WBT-GRx2-GRs}		\hspace{0.8em}
				}
			}{11\Delta^{j+s} \GRkpr^{m-m'}}
		}
	\end{array}
	\]
\caption{Gauge remainders of type~\ref{GRsx2}, part~2.}
\label{fig:GRsx2-b}
\ecf

The first thing to note about the
gauge remainders of type~\ref{GRsx2} is
that we are guaranteed to generate
trapped gauge remainders in 
diagrams~\ref{WGRx2}
and~\ref{P-S_vj_+R-DW-GRx2} 
(\cf \figs{fig:DoubleGR}{fig:DoubleGR-A}). 
In each of these
cases, we can choose to act with
either gauge remainder first, and one of
the things it will do is bite the field
on the structure to which the tip of
the other gauge remainder is attached. If,
on the other hand, the first action of 
one of the gauge remainders is to bite a socket
or the end of a kernel, then 
we are free to perform the 
other gauge remainder as well. However,
it is inefficient to allow the second
gauge remainder to act immediately, 
as we can identify
cancellations prior to this; we employ this
strategy with diagrams~\ref{D-vj_+kR-vkR-DW-GRx2}
and~\ref{WBT-vj_+R-DW-GRx2}, also.
Note, though, that diagram~\ref{WBT-vj_+R-DW-GRx2} is not
symmetrical \wrt\ the action of its two
gauge remainders; the most efficient way to proceed
is to be diplomatic:
we take one instance of this diagram 
where one gauge remainder
acts first and one instance where the other 
acts first, dividing by two to avoid over-counting.

Amongst the diagrams generated by processing
one of the gauge remainders of 
diagrams~\ref{D-vj_+kR-vkR-DW-GRx2}--\ref{WBT-vj_+R-DW-GRx2}
are versions of the parents, nested \wrt\ the gauge
remainder which acted. The strategy for dealing
with these terms is to process only the nested
gauge remainder, for the time being.

We delay processing diagrams~\ref{vj_+R-DW-GRx2-GRs}
and~\ref{WBT-GRx2-GRs}, which have only
a single active gauge remainder,
until after one of the gauge remainders
in each of 
diagrams~\ref{D-vj_+kR-vkR-DW-GRx2} 
and~\ref{WBT-vj_+R-DW-GRx2}, and their nested partners,
has acted. It is easy to see why we do
this. Consider the component of diagram~\ref{D-vj_+kR-vkR-DW-GRx2},
and its nested partners, in which 
a classical, two-point vertex, created
by the action of one of the gauge remainders,
is joined to either a nested gauge
remainder or a gauge remainder structure. This
diagram partially cancels~\ref{vj_+R-DW-GRx2-GRs},
courtesy of diagrammatic identity~\eq{eq:D-ID-G} and without
the need to process any further gauge remainders.
Similarly, we process the gauge remainders
of diagram~\ref{WBT-vj_+R-DW-GRx2} according to the
strategy outlined above, and diagram~\ref{WBT-GRx2-GRs}
is also partially cancelled. Treating the partially
cancelled diagrams together (and using also 
diagram~\ref{GRs-DW-GRx2-GRs}), it is evident that
diagrams in which the active gauge remainder
strikes a socket cancel in the usual way. Up to $\Op{2}$ terms,
the only
survivors from this chain of cancellations
are the partially cancelled components of 
\begin{enumerate}
 	\item	diagram~\ref{vj_+R-DW-GRx2-GRs} and it nested partners
			in which a classical, two-point vertex, created
			by the action of the gauge remainder, is attached
			to the explicitly specified gauge remainder structure;
	\label{it:CTP-ExplicitGR}

	\item	diagram~\ref{WBT-GRx2-GRs}  and it nested partners
			in which the active gauge
			remainder strikes either end of the kernel.
	\label{it:GR-B-EOK}
\end{enumerate}

Returning to diagrams~\ref{WGRx2} 
and~\ref{WBT-vj_+R-DW-GRx2} and their nested partners,
if the first gauge remainder strikes the end
of a kernel, we then have no choice but to
go ahead and process the second gauge remainder.
Diagrams in which this second
gauge remainder strikes a socket, combined
with diagrams of item~\ref{it:GR-B-EOK}, above,
cancel, in the
usual manner.

Thus, up to diagrams with an $\Op{2}$ stub, all
surviving terms possess either a trapped gauge
remainder, or a kernel which is bitten at
one or both of its ends or a classical,
two-point vertex attached to an explicitly
specified gauge remainder structure 
(\ie \ref{it:CTP-ExplicitGR}, above).

The terms with a trapped gauge remainder,
which are collected together in 
\fig{fig:GRsx2-P}, can be simplified.
Notice that the trapped gauge remainder
diagrams spawned by~\ref{P-S_vj_+R-DW-GRx2}
(see diagrams~\ref{RV-RW-GR-B-RW} and~\ref{TLTP-RW-GR-B-RW})
are very similar to diagrams~\ref{RV-DEP-B-DEP}
and~\ref{TLTP-DEP-B-DEP-MV}.
(There is no analogue
of diagram~\ref{TLTP-DEP-B-Soc-MV}: if
one of the
gauge remainders of
diagram~\ref{P-S_vj_+R-DW-GRx2}
were to strike a socket on a classical,
two-point vertex then this vertex would be
killed by the other gauge
remainder, courtesy
of diagrammatic identity~\eq{eq:GR-TLTP}.)
The essential difference
is the decoration of the kernel and the trapped
gauge remainder. Now, we know
that diagrams~\ref{RV-DEP-B-DEP}
and~\ref{TLTP-DEP-B-DEP-MV} can be processed
to yield diagrams~\ref{GR-ArbGRs-DEP} 
and~\ref{GRs-TLTP-GRs-DEP}. Clearly, processing
the diagrams with trapped gauge remainders
will generate analogues of these diagrams, but
with the differentiated effective propagator
replaced by a decorated kernel, ending in a
trapped gauge remainder. Noting that the overall
factor of these diagrams is half of 
diagrams~\ref{GR-ArbGRs-DEP} 
and~\ref{GRs-TLTP-GRs-DEP} we see that both the
diagrams corresponding to item~\ref{it:CTP-ExplicitGR},
above, and diagram~\ref{Ubend-GRsx2}
are exactly cancelled.

However, there are some new diagrams left over
as a consequence of the kernel admitting
decorations. First, the
classical, two-point vertex of diagram~\ref{TLTP-RW-GR-B-RW},
can be attached to the kernel. The corresponding
primary part is simply cancelled by
diagram~\ref{Trapped-KBK}. The secondary
part yields a version of diagram~\ref{Trapped-KBK} 
with an active gauge remainder inserted between
the end of the vertical line emanating from the
trapped gauge remainder and the kernel. If this
gauge remainder strikes a socket, then
we can think of the resulting diagram as
a version of diagram~\ref{Trapped-KBK} but
where the trapped gauge remainder is contracted
into the gauge remainder which bites the kernel,
rather than the kernel itself.
This diagram is cancelled by a term generated in the following way.
Attach the classical, two-point vertex of diagram~\ref{TLTP-RW-GR-B-RW}
to a vertex and apply the effective propagator relation. 
Focusing on the gauge remainder part, allow
the gauge remainder to act and attach the resulting
classical, two-point vertex to the kernel. The primary part
yields the term we require for our cancellation.
Iterating the diagrammatic
procedure, the only terms that survive are those
in which an active gauge remainder bites one of
the ends of the kernel (or which possess an $\Op{2}$ stub).

We collect together all surviving diagrams
spawned by the gauge remainders of type~\ref{GR-GRs},
up to those with an $\Op{2}$ stub,
in \fig{fig:GRsx2-S}.
\bcf
	\[
	\begin{array}{l}	
	\vspace{0.8ex}
		\bca{
			\sum_{s=1}^n \sum_{m=2}^{2s} \sum_{j=-2}^{n+s-m-2} \frac{\norm_{j+s,j+2}}{m!}
			\sum_{m'=2}^m \!\! \nCr{m}{m'} \!\! 
		}
			{
			\dec{
				
				\Tower{
					\ds	
					\sum_{m''=1}^{m'-1} \!\! \nCr{m'}{m''} \!\!
					\LDi{GRs-W-GRs}{GRs-W-GRs-b} \hspace{1.5em}
					-\LDi{Ubend-W}{Ubend-W-MV-b} \hspace{1em} 
					+ \LDi{Ubend-W-GR}{Ubend-W-GR-b} \hspace{1em}
				}
			}{11\Delta^{j+s} \GRkpr^{m-m'}}
		}
	\\
	\vspace{0.8ex}
		-
		\bca{
			\sum_{s=1}^n \sum_{m=2}^{2s-1} \sum_{j=-2}^{n+s-m-3} \frac{\norm_{j+s,j+2}}{m!}
			\sum_{m'=2}^m \!\! \nCr{m}{m'} \!\! \sum_{m''=1}^{m'-1} \!\! \nCr{m'}{m''} \!\!
		}
			{
			\dec{
				\Tower{
					\LDi{K-GRs-B-K-GRs}{K-GRs-B-K-GRs-b} \hspace{2em}
					+\LDi{K-GRs-B-K-GR-GRs}{K-GRs-B-K-GR-GRs} \hspace{2em}
					+\LDi{GRs-TLTP-GRs-K}{GRs-TLTP-GRs-K-b} \hspace{1em}
				}
			}{11\Delta^{j+s} \GRkpr^{m-m'}}
		}
	\\
		+
		\bca{
			\sum_{s=1}^n \sum_{m=1}^{2s-1} \sum_{j=-2}^{n+s-m-3} \frac{\norm_{j+s,j+2}}{m!}
			\sum_{m'=1}^m \!\! \nCr{m}{m'} \!\!
		}
			{
			\dec{
				\Tower{
					\LDi{K-B-K-TLTP-EP}{K-B-K-TLTP-EP}
				}
			}{11\Delta^{j+s} \GRkpr^{m-m'}}
		}	
	\end{array}
	\]
\caption{Surviving diagrams
spawned by the gauge remainders of type~\ref{GRsx2},
up to those with an $\Op{2}$ stub.}
\label{fig:GRsx2-S}
\ecf

The sign of diagram~\ref{Ubend-W-MV-b}
requires comment. 
This diagram is formed
from diagram~\ref{WGRx2} and its nested partners;
however,
we can understand the sign from the un-nested term, alone.
In this case, the gauge remainder structure is formed
as follows. The first gauge remainder bites the end
of the kernel, just above the second (active)
gauge remainder. This active gauge remainder
then bites the kernel at the end to which
the $\GRkpr$ corresponding to the first gauge
remainder attaches. The point is that each
of the gauge remainders bites the end of a kernel
and so, for this pair of gauge remainders---which
form part of the same gauge remainder structure---bites 
to the left (right) must be interpreted as
pulls back (pushes forward). Thus, the sign of
the diagram is given by  $(-1)^{R-2}$.
Promoting these gauge remainders to an explicitly
specified structure, $\decGR{ \ }{\GRkpr^2}$,
we see from the discussion around \fig{fig:nestedhook} that we must compensate with 
minus sign. For \eg diagram~\ref{GRs-W-GRs-b}
there is no such sign, since the gauge
remainders of diagram~\ref{WGRx2}
and its nested partners have 
formed separate structures.

Finally, we comment on diagram~\ref{K-B-K-TLTP-EP}.
The loose end of the kernel can attach to any
of the $m'$ $\GRkpr$s or to the socket on the classical,
two-point vertex. The primary part of the component
of the diagram in which the loose end of the kernel attaches to the
socket
is spawned by diagram~\ref{WGRx2} and its
nested partners. Indeed, the $m'=1$ case
corresponds to the illustrative 
diagram~\ref{Diags-WGR-N1-Bottom-PF} of \fig{fig:DoubleGR-C}.
The corresponding secondary part 
is spawned by a version of diagram~\ref{Trapped-KBK}
with a (nested) gauge remainder inserted between
the end of the vertical line emanating from the
trapped gauge remainder and the kernel. Both
the primary and secondary parts of the component
of diagram~\ref{K-B-K-TLTP-EP} in which the loose end of the 
kernel attaches to a gauge remainder are spawned
by diagram~\ref{WBT-vj_+R-DW-GRx2} and its nested
partners.

\begin{cancel}[Diagram~\ref{Ubend-W-MV}]

Diagram~\ref{Ubend-W-MV-b} exactly cancels
diagram~\ref{Ubend-W-MV}.

\end{cancel}

\begin{cancel}[Diagram~\ref{GRs-W-GRs-MV}]

Diagrams~\ref{GRs-W-GRs-b} and \ref{GRs-W-GRs}
exactly cancel diagram~\ref{GRs-W-GRs-MV}.

\end{cancel}

Finally, we conclude our treatment of
the gauge remainders with the
following elaborate cancellation.

\begin{cancel}[Diagram~\ref{Ubend-W-GR}]
	Diagrams~\ref{K-GRs-B-K-GRs} and~\ref{GRs-TLTP-GRs-K}
	combine with diagrams~\ref{K-GRs-B-K-GRs-b} 
	and~\ref{GRs-TLTP-GRs-K-b}, reversing their
	signs. The resultant diagrams, when combined with
	diagrams~\ref{K-GRs-B-K-GR-GRs} and~\ref{K-B-K-TLTP-EP},
	can be redrawn, courtesy of diagrammatic identity~\eq{eq:D-ID-G},
	to yield an exact copy of diagram~\ref{Ubend-W-GR-b}.
	Thus the overall factor of diagram~\ref{Ubend-W-GR-b}
	is doubled, providing precisely the necessary
	contribution to exactly cancel diagram~\ref{Ubend-W-GR}.
\end{cancel}

Referring back to \fig{fig:bn-b-P},
we have therefore demonstrated
that
\[
	\dec{\nLV{\Ds}{n}{\mu\nu}{a}(p)}{\bullet} +\Op{4} = -4 \beta_n \Box_{\mu\nu}(p) +\mbox{\ref{beta-terms}} + \mbox{\ref{alpha-terms}} + \cdots,
\]
where the ellipsis stands for terms with an $\Op{2}$ stub, to which we
now turn.

\subsection{Terms with an $\Op{2}$ Stub}
\label{sec:bn-Op2}

Every time a classical, two-point vertex was
generated in the computation of $\dec{\nLV{\Ds}{n}{\mu\nu}{a}(p)}{\bullet}$,
we had the option of decorating it with an external
field. In \figs{fig:bn-Op2-a}{fig:bn-Op2-b} 
we collect together the set of
terms spawned by $a_0$ and $a_1$
which remain after all active gauge remainders have
been processed, all cancellations have been identified
and all diagrams which manifestly vanish at $\Op{2}$
have been discarded.
The treatment of gauge remainders in diagrams with an $\Op{2}$
stub is identical to the analysis just performed, and the
same patterns of cancellations are observed. 
Wherever possible, we have combined terms possessing
a seed action vertex with their counterparts
possessing just Wilsonian effective action vertices by
introducing the notation
\[
	\Pi = S - \hat{S}.
\]
Notice, though, that since the terms possessing
exclusively Wilsonian effective action vertices
necessarily possess a decorated kernel, we are
left over with terms possessing a seed action
vertex attached to a differentiated effective propagator.
\bcf
	\[
	\begin{array}{l}
	\vspace{2ex}
			2
			\bca{
				\sum_{s=1}^n \sum_{m=0}^{2s} \sum_{j=-1}^{n+s-m-2}
				\frac{\norm_{j+s+1,j+1}}{m!}
			}
				{
				\dec{
					\sco[1]{
						\LDi{Dumbbell-CTP-E-DEP-vh_j+R}{D-CTP-E-DEP-vh_j+R}
						-\LDi{Dumbbell-CTP-E-DW-Pi_j+R}{D-CTP-E-DW-Pi_j+R}
					}{\SumVertex}
				}{1\Delta^{j+s+1}\GRkpr^m}
			}
	\\
		+\bca{
			\sum_{s=1}^n \sum_{m=0}^{2s-2} \sum_{j=-2}^{n+s-m-4}
				\frac{\norm_{j+s+1,j+2}}{m!}	
		}
			{
			\dec{
				\Tower{
					\LDi{CTP-E-WBT}{CTP-E-WBT}
				}
			}{1\Delta^{j+s+1}\GRkpr^m}
		}
	\end{array}
	\]
\caption{$\Op{2}$ terms generated by $\dec{\nLV{\Ds}{n}{\mu\nu}{a}(p)}{\bullet}$, part~1.}
\label{fig:bn-Op2-a}
\ecf

\bcf
	\[
	\begin{array}{l}
	\vspace{2ex}
		2
		\bca{
			\sum_{s=1}^n \sum_{m=0}^{2s-1} \sum_{j=-1}^{n+s-m-3}
			\frac{\norm_{j+s+1,j+1}}{m!} \sum_{m'=0}^m \!\! \nCr{m}{m'} \!\!
		}
			{
			\dec{	
				\sco[1]{
					\LDi{vh_j+R-DEP-GRs-Skt-CTP-E}{vh_j+R-DEP-GRs-Skt-CTP-E}
					-\LDi{Pi_j+R-DW-GRs-Skt-CTP-E}{Pi_j+R-DW-GRs-Skt-CTP-E}
				}{\SumVertex}
			}{1\Delta^{j+s+1}\GRkpr^{m-m'}}
		}
	\\
	\vspace{2ex}
		+
		\bca{
			\sum_{s=1}^n \sum_{m=0}^{2s-1} \sum_{j=-2}^{n+s-m-3}
			\frac{\norm_{j+s+1,j+2}}{m!} 
		}
			{
			\sum_{m'=0}^m \!\! \nCr{m}{m'} \!\!
			\dec{
				\sco[1]{
					\LDi{WBT-GRs-Skt-CTP-E}{WBT-GRs-Skt-CTP-E}
					-\LDi{DW-GR-GRs-Skt-CTP-E}{DW-GR-GRs-Skt-CTP-E}
					+2\LDi{DW-GRs-Skt-CTP-E}{DW-GRs-Skt-CTP-E}
				}{\TopVertex \SumVertex}
			}{1\Delta^{j+s+1}\GRkpr^{m-m'}}
		}
	\\
		-2
		\bca{
			\sum_{s=1}^n \sum_{m=2}^{2s-1} \sum_{j=-2}^{n+s-m-3}
			\frac{\norm_{j+s+1,j+2}}{m!} 
		}
			{
			\sum_{m'=2}^m \!\! \nCr{m}{m'} \!\! \sum_{m''=2}^{m'} \!\! \nCr{m'}{m''} \!\!
			\dec{
				\sco[1]{
					\LDi{GRs-dEP-GR-Skt-CTP-E}{GRs-dEP-GR-Skt-CTP-E}
				}{\TopVertex \SumVertex}
			}{1\Delta^{j+s+1}\GRkpr^{m-m'}}
		}
	\end{array}
	\]
\caption{$\Op{2}$ terms generated by $\dec{\nLV{\Ds}{n}{\mu\nu}{a}(p)}{\bullet}$, part~2.}
\label{fig:bn-Op2-b}
\ecf

Before processing diagrams~\ref{D-CTP-E-DEP-vh_j+R}--\ref{GRs-dEP-GR-Skt-CTP-E},
we emphasise a very important observation.
Consider the set of diagrams obtained by promoting
the explicitly drawn external field to an implicit
decoration. This set of diagrams is, of course,
nothing other than
the complete set of terms generated by
$\dec{\nLV{\Ds}{n}{\mu\nu}{a}(p)}{\bullet}$ which possess a single,
classical, 
two-point vertex (with a free socket)
\emph{after} all gauge remainders
have acted. We can use this to immediately deduce
the set of diagrams left over after 
taking $\dec{\nLV{\Ds}{n}{\mu\nu}{b}(p)}{\bullet}$
(see \fig{fig:nL-b})
and iterating the diagrammatic procedure until exhaustion.
First, there will be the $\alpha$ and $\beta$ terms.
Secondly, there will be diagrams in which $-\flowConstAl$
strikes either one of the $\combo$s or one of the elements
of a string of $\GRkpr$s. Finally, there are those diagrams
arising from attaching all classical, two-point vertices
generated at each stage of the calculation either to a
$\combo$ or to one of the elements of a string of
$\GRkpr$s. In the case that there is just a single,
classical, two-point vertex to play with, 
the resulting set of diagrams can be directly
deduced from diagrams~\ref{D-CTP-E-DEP-vh_j+R}--\ref{GRs-dEP-GR-Skt-CTP-E}.
In the case that there are two classical, two-point 
vertices to play with, the resultant diagrams
will arise from an analogue of diagram~\ref{D-02x2-R-VS}.
(Note that the two classical, two-point 
vertices must be attached to elements of
the same 
structure comprising a single
$\combo$ plus string
of $\GRkpr$s plus vertices plus $\GRkpr$s, 
as follows from  \sec{sec:Not:GRs}.)

Diagram~\ref{D-CTP-E-DEP-vh_j+R} is simple
to deal with. First, notice that the differentiated
effective propagator must be in the $A^1$ sector
(which, incidentally, means that the seed action vertex, to which
it attaches at one end, cannot be a one-point vertex).
Now we redraw this diagram by detaching 
the differentiated effective
propagator from the seed action vertex
and
decorating the seed action vertex with an $A^1$ which
carries the same index as the loose end of the
effective propagator, say $\alpha$. Next, we
decorate the diagram with the remaining external
field, which we suppose carries index $\nu$.
Putting the $\Op{2}$ stub and differentiated
effective propagator to one side, consider
contracting the remaining part of
the diagram with $(-p)_\nu p_\alpha$.
It is straightforward to demonstrate
that the final outcome is zero. Therefore,
by Lorentz invariance, the diagram to which
the $\Op{2}$ stub attaches via the 
differentiated effective propagator 
is transverse in $p$ and, just like
$\nLV{\Ds}{n}{\mu\nu}{a}(p)$, in fact
goes as
$\Box_{\mu \nu}(p) + \Op{4}$. 
Consequently,
diagram~\ref{D-CTP-E-DEP-vh_j+R} as a whole
is $\Op{4}$
and so does not contribute to 
$\dec{\nLV{\Ds}{n}{\mu\nu}{a}(p)}{\bullet}$ at $\Op{2}$.

To process diagrams~\ref{D-CTP-E-DW-Pi_j+R}--\ref{DW-GRs-Skt-CTP-E}
we construct subtractions. We begin with diagram~\ref{CTP-E-WBT}
for which, as we know from \sec{sec:Subtractions-G},
the subtractions completely kill the parents. This leaves
behind only the additions, which are collected together in
\fig{fig:CTP-E-WBT-Add}.
\bcf
	\[
	\begin{array}{l}
	\vspace{2ex}
		2
		\bca{
			\sum_{s=1}^n \sum_{m=0}^{2s-3} \sum_{j=-2}^{n+s-m-5}
				\frac{\norm_{j+s+1,j+2}}{m!}	
		}
			{
			\dec{
				\decp{\LDi{CTP-E-WBT-E}{CTP-E-WBT-E}}{}
				\sco[1]{
					\TopVertex
				}{\SumVertex}
			}{1\Delta^{j+s+1}\GRkpr^m}
		}
	\\
	\vspace{2ex}
		-2
		\bca{
			\sum_{s=1}^n \sum_{m=0}^{2s-3} \sum_{j=-1}^{n+s-m-5}
				\frac{\norm_{j+s+1,j+1}}{m!}	
		}
			{
			\dec{
				\decp{
					\sco[1]{
						\LDi{CTP-E-WBT}{CTP-E-WBT-dV}
					}{\ensuremath{\begin{array}{c}\begin{picture}(0,0)%
\includegraphics{pstex/dvj+hR.pstex}%
\end{picture}%
\setlength{\unitlength}{3947sp}%
\begingroup\makeatletter\ifx\SetFigFont\undefined%
\gdef\SetFigFont#1#2#3#4#5{%
  \reset@font\fontsize{#1}{#2pt}%
  \fontfamily{#3}\fontseries{#4}\fontshape{#5}%
  \selectfont}%
\fi\endgroup%
\begin{picture}(586,684)(1081,1705)
\put(1141,1904){\makebox(0,0)[lb]{\smash{{\SetFigFont{11}{13.2}{\rmdefault}{\mddefault}{\updefault}{\color[rgb]{0,0,0}$\hat{v}^{j_+;R}$}%
}}}}
\end{picture}%
 \end{array}}}
				}{} \SumVertex
			}{1\Delta^{j+s+1}\GRkpr^m}
		}
	\\
	\vspace{2ex}
		-2
		\bca{
			\sum_{s=1}^n \sum_{m=0}^{2s-3} \sum_{j=-2}^{n+s-m-5}
				\frac{\norm_{j+s+2,j+2}}{m!}	
		}
			{
			\dec{
				\decp{
					\sco[1]{
						\LDi{CTP-E-WBT}{CTP-E-WBT-dCTP}
					}{\ensuremath{\begin{array}{c}\begin{picture}(0,0)%
\includegraphics{pstex/dCTP.pstex}%
\end{picture}%
\setlength{\unitlength}{3947sp}%
\begingroup\makeatletter\ifx\SetFigFont\undefined%
\gdef\SetFigFont#1#2#3#4#5{%
  \reset@font\fontsize{#1}{#2pt}%
  \fontfamily{#3}\fontseries{#4}\fontshape{#5}%
  \selectfont}%
\fi\endgroup%
\begin{picture}(504,534)(1185,1840)
\put(1267,1960){\makebox(0,0)[lb]{\smash{{\SetFigFont{11}{13.2}{\rmdefault}{\mddefault}{\updefault}{\color[rgb]{0,0,0}$0^2$}%
}}}}
\end{picture}%
 \end{array}}}
				}{}
				\sco[1]{
					\TopVertex
				}{\SumVertex}
			}{1\Delta^{j+s+2}\GRkpr^m}
		}
	\\
		+
		\bca{
			\sum_{s=1}^n \sum_{m=3}^{2s-2} \sum_{j=-2}^{n+s-m-4}
				\frac{\norm_{j+s+1,j+1}}{m!} \sum_{m'=3}^m	\nCr{m}{m'} 
		}
			{
			\dec{
				\decp{
					\sco[1]{
						\LDi{CTP-E-WBT}{CTP-E-WBT-GRs}
					}{\decGR{\ }{\socket \GRkpr^{m'}}}
				}{}
				\sco[1]{
					\TopVertex
				}{\SumVertex}
			}{1\Delta^{j+s+1}\GRkpr^{m-m'}}
		}
	\end{array}
	\]
\caption{Additions for diagram~\ref{CTP-E-WBT}.}
\label{fig:CTP-E-WBT-Add}
\ecf

The first thing to notice about the
diagrams of \fig{fig:CTP-E-WBT-Add}
is that they are very similar
to the diagrams of \fig{fig:nLV-c-Properties-a}
(which we recall arose from considering 
$\decp{\nLV{\overline{\Ds}}{n}{\mu\nu}{c}(p)}{}$). Consequently,
they will be very easy to manipulate.
The differences are essentially three-fold.
First, we must be careful to 
properly take account of the
effects of $\decp{\ }{}$, since it does not
necessarily apply to all elements of a fully
fleshed out diagram.
Consider first the 
non-factorizable components of
diagrams~\ref{CTP-E-WBT-E}--\ref{CTP-E-WBT-dCTP}.
In this case, the socket of each diagram is
filled
by the external field; there can
be no effective propagators carrying just $p$
and all diagrammatic
elements can be brought under the influence
of $\decp{\ }{}$. In other words, 
each of the additions can be thought of
as arising from Taylor expanding the corresponding components
of the parent
diagram to $\Op{2}$.
In the factorizable case, however,
$p$ dependence comes not only from the $\Op{2}$
stub, but also from the effective propagators
carrying just $p$
and the sub-diagrams to which they
attach. These sub-diagrams may 
possess components which are 
not Taylor expandable in $p$ and so we leave them alone.
Only those diagrammatic components which are not
part of a $p$ dependent sub-diagrams can be brought
under the influence of $\decp{\ }{}$, in this case.

The second difference between the diagrams
of \figs{fig:nLV-c-Properties-a}{fig:CTP-E-WBT-Add}
is that the latter diagrams do not possess
a string of gauge remainders decorating their
classical, two-point vertex decorated by
the external field. This makes life easier
and means that we will never generate the analogue
of the diagrams of \fig{fig:nLV-c-Properties-e}.
The third difference, however, makes our life slightly
harder: the presence of a kernel in the diagrams
of \fig{fig:CTP-E-WBT-Add} means there is an additional
structure to which effective propagators can attach.
As usual, this ultimately generates a set of
surviving diagrams in which gauge remainders
which bite the kernel slide down to its base.
Immediately, then, we can deduce the single diagram
left over from iterating the diagrammatic procedure
until exhaustion, which is shown in \fig{fig:CTP-E-WBT-Final}.
\bcf[h]
	\[
		-2
		\bca{
			\sum_{s=2}^n \sum_{r=1}^{s-1} \sum_{m=0}^{2(s-r)-1} \sum_{j=-2}^{n+s-2r-m-3}
				\frac{\norm_{j+s+1-r,j+2}}{m!r!}
		}
			{
			\dec{
				\decp{
					\LDi{CTP-E-WBT}{CTP-E-WBT-Final}
				}{\combo^r }
				\sco[1]{
					\TopVertex
				}{\SumVertex}
			}{1\Delta^{j+s+1-r} \GRkpr^m}
		}
	\]
\caption{The only diagram left from the full treatment
of the additions of \fig{fig:CTP-E-WBT-Add}.}
\label{fig:CTP-E-WBT-Final}
\ecf

What is to be the fate of diagram~\ref{CTP-E-WBT-Final}?
As we will demonstrate shortly, it will be exactly
cancelled by terms arising from processing
\be
\label{eq:b-derivs}
	\dec{\nLV{\Ds}{n}{\mu\nu}{b}(p)}{\bullet}  -
	\sum_{n'=1}^{n-1} \dec{\nLV{\Ds}{n-n'}{\mu\rho}{b}(p)}{\bullet}
	\Delta^{11}_{\rho \sigma}(p) \nLV{\Ds}{n'}{\sigma\nu}{a}(p)
\ee
(see \fig{fig:nL-b} and equation~\eq{eq:b'-decomp}).
Before seeing this explicitly, we will deal with
the remaining diagrams of \figs{fig:bn-Op2-a}{fig:bn-Op2-b}.
First, we deal with diagram~\ref{D-CTP-E-DW-Pi_j+R},
the only term other than~\ref{CTP-E-WBT} 
that is completely killed
by its subtractions. Inspired by the analysis
of~\ref{CTP-E-WBT}, it is obvious that the
final result of processing diagram~\ref{D-CTP-E-DW-Pi_j+R}
is the
diagram of \fig{fig:CTP-E-DW-Pi_j+R-Add}.
\bcf[h]
	\[
	4
	\bca{
		\sum_{s=2}^n \sum_{r=1}^{s-1} \sum_{m=0}^{2(s-r)-1} \sum_{j=-2}^{n+s-2r-m-3}
				\frac{\norm_{j+s+1-r,j+1}}{m!r!}
	}
		{
		\dec{
			\decp{
				\LDi{Dumbbell-CTP-E-DW-Pi_j+R}{D-CTP-E-DW-Pi_j+R-Final}
			}{\combo^r}
			\SumVertex
		}{1\Delta^{j+s+1-r}  \GRkpr^m}
	}
	\]
\caption{The final result of processing diagram~\ref{D-CTP-E-DW-Pi_j+R}.}
\label{fig:CTP-E-DW-Pi_j+R-Add}
\ecf

We must now bite the bullet and deal with those diagrams
which are not completely cancelled by their subtractions,
starting with diagram~\ref{WBT-GRs-Skt-CTP-E}---whose
subtractions and additions are collected together in 
\fig{fig:WBT-GRs-Skt-CTP-E-Add}.
\bcf
	\[
	\begin{array}{l}
	\vspace{0.5ex}
		\mp 2
		\bca{
			\sum_{s=1}^n \sum_{m=0}^{2s-2} \sum_{j=-2}^{n+s-m-4}
			\frac{\norm_{j+s+1,j+2}}{m!} \sum_{m'=0}^m \!\! \nCr{m}{m'} \!\!
		}
			{
			\dec{
				\decp{
					\LLDi{dWBT-GRs-Skt-CTP-E}{dWBT-GRs-Skt-CTP-E-s}{dWBT-GRs-Skt-CTP-E-a}
				}{}
				\sco[1]{
					\TopVertex
				}{\SumVertex}
			}{1\Delta^{j+s+1}\GRkpr^{m-m'}}
		}
	\\
	\vspace{0.5ex}
		\pm 2
		\bca{
			\sum_{s=1}^n \sum_{m=0}^{2s-1} \sum_{j=-1}^{n+s-m-3}
			\frac{\norm_{j+s+1,j+1}}{m!} \sum_{m'=0}^m \!\! \nCr{m}{m'} \!\!
		}
			{
			\dec{
				\decp{
					\LOLO{
						\ensuremath{\begin{array}{c}\input{pstex/WBT-GRs-Skt-CTP-E.pstex_t} \end{array}}  \hspace{0.5em} \ensuremath{\begin{array}{c}\begin{picture}(0,0)%
\includegraphics{pstex/dvj+.pstex}%
\end{picture}%
\setlength{\unitlength}{3947sp}%
\begingroup\makeatletter\ifx\SetFigFont\undefined%
\gdef\SetFigFont#1#2#3#4#5{%
  \reset@font\fontsize{#1}{#2pt}%
  \fontfamily{#3}\fontseries{#4}\fontshape{#5}%
  \selectfont}%
\fi\endgroup%
\begin{picture}(540,594)(1127,1795)
\put(1203,1941){\makebox(0,0)[lb]{\smash{{\SetFigFont{11}{13.2}{\rmdefault}{\mddefault}{\updefault}{\color[rgb]{0,0,0}$\nothing {v}^{j_+}$}%
}}}}
\end{picture}%
 \end{array}}
					}{WBT-GRs-Skt-CTP-E-dvj+-s}{WBT-GRs-Skt-CTP-E-dvj+-a}
				}{}
				\SumVertex
			}{1\Delta^{j+s+1}\GRkpr^{m-m'}}
		}
	\\
	\vspace{0.5ex}
		\mp
		\bca{
			\sum_{s=1}^n \sum_{m=2}^{2s-1} \sum_{j=-1}^{n+s-m-3}
			\frac{\norm_{j+s+1,j+2}}{m!} \sum_{m'=2}^m \!\! \nCr{m}{m'} \!\!
			\sum_{m''=2}^{m'} \!\! \nCr{m'}{m''} \!\!
		}
			{
			\dec{
				\decp{
					\LOLO{
						\ensuremath{\begin{array}{c}\input{pstex/WBT-GRs-Skt-CTP-E-b.pstex_t} \end{array}}
						\hspace{2em}
						\decGR{\ }{\socket \GRkpr^{m''}}
					}{WBT-GRs-Skt-CTP-E--TEGRs-s}{WBT-GRs-Skt-CTP-E--TEGRs-a}
				}{}
				\sco[1]{
					\TopVertex
				}{\SumVertex}
			}{1\Delta^{j+s+1}\GRkpr^{m-m'}}
		}
	\\
		\mp
		\bca{
			\sum_{s=1}^n \sum_{m=1}^{2s-1} \sum_{j=-1}^{n+s-m-3}
			\frac{\norm_{j+s+1,j+2}}{m!} \sum_{m'=1}^m \!\! \nCr{m}{m'} \!\!
		}
			{
			\dec{
				\decp{
					\LLDi{WBT-TEGRs-Skt-CTP-E}{WBT-TEGRs-Skt-CTP-E-s}{WBT-TEGRs-Skt-CTP-E-a}
					\hspace{1em}
				}{}
				\sco[1]{
					\TopVertex
				}{\SumVertex}
			}{1\Delta^{j+s+1}\GRkpr^{m-m'}}
		}
	\end{array}
	\]
\caption{Subtractions and additions for diagram~\ref{WBT-GRs-Skt-CTP-E}.}
\label{fig:WBT-GRs-Skt-CTP-E-Add}
\ecf

We proceed as before by
isolating the classical, two-point component
of the differentiated vertex of 
diagram~\ref{WBT-GRs-Skt-CTP-E-dvj+-a}, decorating it with
two effective propagators and utilizing~\eq{eq:EP-dCTP-EP}.
However, compared to the treatment of diagram~\ref{CTP-E-WBT},
we will encounter four novelties, all associated with
the $m'$ $\GRkpr$s which bite
the classical, two-point vertex decorated by an external field.
First, the active gauge 
remainders associated with~\eq{eq:EP-dCTP-EP}
can 
be contracted into one of these $m'$  $\GRkpr$s or the
socket which they bite.
Secondly, classical two-point vertices
generated by the action of (arbitrarily)
nested gauge remainders
can be similarly attached.
Thirdly, we can no longer exactly combine terms into total
momentum derivatives since, at this
stage of the calculation, those terms involving
momentum derivatives of any of the $m'$ $\GRkpr$s
or the $\GRkpr$ which bites them are missing.
However, it proves
convenient to express the sum of momentum derivative terms
that we do have as a total momentum derivative (which we
throw away) minus the missing terms.
Lastly, we have not yet encountered
an analogue of diagram~\ref{WBT-TEGRs-Skt-CTP-E-a},
and so we put this to one side for a moment.

Iterating the diagrammatic procedure until exhaustion,
we can re-express the additions~\ref{dWBT-GRs-Skt-CTP-E-a}, 
\ref{WBT-GRs-Skt-CTP-E-dvj+-a} and~\ref{WBT-GRs-Skt-CTP-E--TEGRs-a}
as shown in \figs{fig:WBT-GRs-Skt-CTP-E-+Subs}{fig:WBT-GRs-Skt-CTP-E-+Subs-II},
where we have absorbed diagram~\ref{CTP-E-WBT-Final}
into diagram~\ref{CTP-E-GRs-Combo-WBT}, in preparation for what is to follow.
\bcf[h]
	\[
	\begin{array}{l}
	\vspace{2ex}
		-2
		\bca{
			\sum_{s=2}^n \sum_{r=1}^{s-1} \sum_{m=0}^{2(s-r)-1} \sum_{j=-2}^{n+s-2r-m-3}
				\frac{\norm_{j+s+2-r,j+2}}{m!(r-1)!} \sum_{m'=0}^m \!\! \nCr{m}{m'} \!\!
		}
			{
			\dec{
				\decp{
					\LDi{CTP-E-dGRs-WBT-GR}{CTP-E-dGRs-WBT-GR}
				}{\combo^{r-1}}
				\sco[1]{
					\TopVertex
				}{\SumVertex}
			}{1\Delta^{j+s+2-r}  \GRkpr^{m-m'}}
		}
	\\
		-2
		\bca{
			\sum_{s=2}^n \sum_{r=1}^{s-1} \sum_{m=0}^{2(s-r)-1} \sum_{j=-2}^{n+s-2r-m-3}
				\frac{\norm_{j+s+1-r,j+2}}{m!(r-1)!} \sum_{m'=0}^m \!\! \nCr{m}{m'} \!\!
		}
			{
			\dec{
				\decp{
					\LDi{CTP-E-GRs-Combo-WBT}{CTP-E-GRs-Combo-WBT}
				}{\combo^{r-1}}
				\sco[1]{
					\TopVertex
				}{\SumVertex}
			}{1\Delta^{j+s+1-r}  \GRkpr^{m-m'}}
		}
	\end{array}
	\]
\caption{The final result of processing
diagrams~\ref{dWBT-GRs-Skt-CTP-E-a}, 
\ref{WBT-GRs-Skt-CTP-E-dvj+-a} and~\ref{WBT-GRs-Skt-CTP-E--TEGRs-a}
plus
diagram~\ref{CTP-E-WBT-Final}, part~1.}
\label{fig:WBT-GRs-Skt-CTP-E-+Subs}
\ecf
\bcf
	\[
	\begin{array}{l}
	\vspace{2ex}
		+2
		\bca{
			\sum_{s=2}^n \sum_{r=1}^{s-1} \sum_{m=0}^{2(s-r)-1} \sum_{j=-2}^{n+s-2r-m-3}
				\frac{\norm_{j+s+1-r,j+2}}{m!(r-1)!} \sum_{m'=0}^m \!\! \nCr{m}{m'} \!\!
		}
			{
			\dec{
				\decp{
					\LDi{CTP-E-GRs-Combo-GRk-WBT-GR}{CTP-E-GRs-Combo-GRk-WBT-GR}
				}{\combo^{r-1}}
				\sco[1]{
					\TopVertex
				}{\SumVertex}
			}{1\Delta^{j+s+1-r}  \GRkpr^{m-m'}}
		}
	\\
		-2
		\bca{
			\sum_{s=2}^n \sum_{r=1}^{s-1} \sum_{m=0}^{2(s-r)-1} \sum_{j=-2}^{n+s-2r-m-3}
				\frac{\norm_{j+s-r,j+1}}{m!(r-1)!} \sum_{m'=0}^m \!\! \nCr{m}{m'} \!\!
				\sum_{m''=0}^{m'} \!\! \nCr{m'}{m''} \!\!
		}
			{
			\dec{
				\decp{
					\LDi{CTP-E-GRs-Combo-CTP-EP-WBT-GR}{CTP-E-GRs-Combo-CTP-EP-WBT-GR}
					\hspace{1em}
				}{\combo^{r-1}}
				\sco[1]{
					\TopVertex
				}{\SumVertex}
			}{1\Delta^{j+s-r}  \GRkpr^{m-m'}}
		}
	\end{array}
	\]
\caption{The final result of processing
diagrams~\ref{dWBT-GRs-Skt-CTP-E-a}, 
\ref{WBT-GRs-Skt-CTP-E-dvj+-a} and~\ref{WBT-GRs-Skt-CTP-E--TEGRs-a}
plus
diagram~\ref{CTP-E-WBT-Final}, part~2.}
\label{fig:WBT-GRs-Skt-CTP-E-+Subs-II}
\ecf

The cancellation of diagrams~\ref{WBT-TEGRs-Skt-CTP-E-a} 
and~\ref{CTP-E-dGRs-WBT-GR}--\ref{CTP-E-GRs-Combo-CTP-EP-WBT-GR}
follows from processing~\eq{eq:b-derivs}
and
focusing on all surviving terms in which a kernel
bites its own tail. As commented under \figs{fig:bn-Op2-a}{fig:bn-Op2-b},
this set of terms arises from the versions
of diagrams~\ref{CTP-E-WBT} and~\ref{WBT-GRs-Skt-CTP-E}
appropriate to~\eq{eq:b-derivs}  \ie diagrams
for which the classical, two-point vertices of
\ref{CTP-E-WBT} and~\ref{WBT-GRs-Skt-CTP-E}, rather than being decorated
by an external field, are instead attached  either to a
$\combo$ or to one of the elements of a string of
$\GRkpr$s. 
Using~\eq{eq:TLTP-Combo}, we collect together
all terms spawned by 
\be
\label{eq:LdL-b-ba}
	\dec{\nLV{\Ds}{n}{\mu\nu}{b}(p)}{\bullet}
	-
	\sum_{n'=1}^{n-1} \dec{\nLV{\Ds}{n-n'}{\mu\rho}{b}(p)}{\bullet}
	\Delta^{11}_{\rho \sigma}(p) \nLV{\Ds}{n'}{\sigma\nu}{a}(p)
\ee
in which a kernel
bites its own tail
in \figs{fig:nLDlb-KBT}{fig:nLDlb-KBT-2}.

\begin{cancel}[Diagram~\ref{CTP-E-dGRs-WBT-GR}]
Diagrams~\ref{CTP-E-WBT-GR-GRs-dGR-GRs} 
and~\ref{CTP-E-WBT-dGR-GRs} exactly
cancel diagram~\ref{CTP-E-dGRs-WBT-GR}.

\end{cancel}

\begin{cancel}[Diagram~\ref{CTP-E-GRs-Combo-WBT}]

Diagram~\ref{CTP-E-GRs-Combo-WBT-b}
exactly cancels diagram~\ref{CTP-E-GRs-Combo-WBT}.

\end{cancel}

\begin{cancel}[Diagram~\ref{WBT-TEGRs-Skt-CTP-E-a}]

Diagrams~\ref{CTP-E-WBT-GR-GRs-GR-dGR-GRs}
and~\ref{CTP-E-WBT-GR-dGR-GRs}
exactly cancel diagram~\ref{WBT-TEGRs-Skt-CTP-E-a},
courtesy of diagrammatic identity~\eq{eq:D-ID-dGRk-GT-ring}.

\end{cancel}

\begin{cancel}[Diagrams~\ref{CTP-E-GRs-Combo-GRk-WBT-GR} and~\ref{CTP-E-GRs-Combo-CTP-EP-WBT-GR}]

Diagrams~\ref{CTP-E-GRs-Combo-WBT-GR} and~\ref{CTP-E-GRs-Combo-WBT-GRs-CTP-EP}
exactly cancel 
diagrams~\ref{CTP-E-GRs-Combo-GRk-WBT-GR} 
and~\ref{CTP-E-GRs-Combo-CTP-EP-WBT-GR},
courtesy of diagrammatic identity~\eq{eq:D-ID-Op2-G}.

\end{cancel}

This completes the analysis of
the additions of diagram~\ref{WBT-GRs-Skt-CTP-E}.
Note that the only diagrams remaining
in which a kernel bites its own tail
are diagram~\ref{WBT-GRs-Skt-CTP-E}, itself,
and its subtractions. It should therefore
not come as a surprise that we can
now reduce 
\be
\label{eq:al+be+subs}
	\dec{\nLV{\Ds}{n}{\mu\nu}{a}(p)+ \nLV{\Ds}{n}{\mu\nu}{b}(p)}{\bullet}
	-
	\sum_{n'=1}^{n-1} \dec{\nLV{\Ds}{n-n'}{\mu\rho}{b}(p)}{\bullet}
	\Delta^{11}_{\rho \sigma}(p) \nLV{\Ds}{n'}{\sigma\nu}{a}(p)
\ee
to just $\alpha$-terms, $\beta$-terms and
the diagrams of \fig{fig:bn-Op2-b} and their 
subtractions. To see this,
we must explicitly construct subtractions for
diagrams~\ref{vh_j+R-DEP-GRs-Skt-CTP-E},
\ref{Pi_j+R-DW-GRs-Skt-CTP-E},
\ref{DW-GR-GRs-Skt-CTP-E},
\ref{DW-GRs-Skt-CTP-E}
and~\ref{GRs-dEP-GR-Skt-CTP-E},
which are shown in 
\figrange{fig:vh_j+R-DEP-GRs-Skt-CTP-E-Add}{fig:GRs-dEP-GR-Skt-CTP-E-Add-II}.

When processing the additions, the only novelty is
associated with diagram~\ref{dvh_j-DEP-GRs-Skt-CTP-E-a}:
this diagram has a component which
possesses a classical, two-point vertex,
differentiated \wrt\ momentum, that cannot
be manipulated in the usual way, on
account of the \emph{differentiated} effective
propagator, already attached to one of
its sockets. We can proceed by utilizing the
various primary diagrammatic identities to
re-express
\bea
	\ensuremath{\begin{array}{c}\input{pstex/GR-DEP-dCTP-EP.pstex_t} \end{array}}	& =	& \ensuremath{\begin{array}{c}\begin{picture}(0,0)%
\includegraphics{pstex/DGR-dEP.pstex}%
\end{picture}%
\setlength{\unitlength}{3947sp}%
\begingroup\makeatletter\ifx\SetFigFont\undefined%
\gdef\SetFigFont#1#2#3#4#5{%
  \reset@font\fontsize{#1}{#2pt}%
  \fontfamily{#3}\fontseries{#4}\fontshape{#5}%
  \selectfont}%
\fi\endgroup%
\begin{picture}(443,368)(297,923)
\put(306,1207){\makebox(0,0)[lb]{\smash{{\SetFigFont{8}{9.6}{\rmdefault}{\mddefault}{\updefault}{\color[rgb]{0,0,0}$\bullet$}%
}}}}
\end{picture}%
 \end{array}} - \ensuremath{\begin{array}{c}\begin{picture}(0,0)%
\includegraphics{pstex/GR-DEP-dFGR.pstex}%
\end{picture}%
\setlength{\unitlength}{3947sp}%
\begingroup\makeatletter\ifx\SetFigFont\undefined%
\gdef\SetFigFont#1#2#3#4#5{%
  \reset@font\fontsize{#1}{#2pt}%
  \fontfamily{#3}\fontseries{#4}\fontshape{#5}%
  \selectfont}%
\fi\endgroup%
\begin{picture}(639,390)(297,888)
\put(511,1194){\makebox(0,0)[lb]{\smash{{\SetFigFont{8}{9.6}{\rmdefault}{\mddefault}{\updefault}{\color[rgb]{0,0,0}$\odot$}%
}}}}
\end{picture}%
 \end{array}}
\nonumber
\\
						& =	& -\ensuremath{\begin{array}{c}\begin{picture}(0,0)%
\includegraphics{pstex/Dcombo.pstex}%
\end{picture}%
\setlength{\unitlength}{3947sp}%
\begingroup\makeatletter\ifx\SetFigFont\undefined%
\gdef\SetFigFont#1#2#3#4#5{%
  \reset@font\fontsize{#1}{#2pt}%
  \fontfamily{#3}\fontseries{#4}\fontshape{#5}%
  \selectfont}%
\fi\endgroup%
\begin{picture}(442,367)(298,927)
\put(459,1210){\makebox(0,0)[lb]{\smash{{\SetFigFont{8}{9.6}{\rmdefault}{\mddefault}{\updefault}{\color[rgb]{0,0,0}$\bullet$}%
}}}}
\end{picture}%
 \end{array}} - \ensuremath{\begin{array}{c}\begin{picture}(0,0)%
\includegraphics{pstex/DdEP-GR.pstex}%
\end{picture}%
\setlength{\unitlength}{3947sp}%
\begingroup\makeatletter\ifx\SetFigFont\undefined%
\gdef\SetFigFont#1#2#3#4#5{%
  \reset@font\fontsize{#1}{#2pt}%
  \fontfamily{#3}\fontseries{#4}\fontshape{#5}%
  \selectfont}%
\fi\endgroup%
\begin{picture}(443,392)(297,923)
\put(478,1231){\makebox(0,0)[lb]{\smash{{\SetFigFont{8}{9.6}{\rmdefault}{\mddefault}{\updefault}{\color[rgb]{0,0,0}$\odot$}%
}}}}
\end{picture}%
 \end{array}} + \ds \hf \ensuremath{\begin{array}{c}\begin{picture}(0,0)%
\includegraphics{pstex/GR-DdEP-FGR.pstex}%
\end{picture}%
\setlength{\unitlength}{3947sp}%
\begingroup\makeatletter\ifx\SetFigFont\undefined%
\gdef\SetFigFont#1#2#3#4#5{%
  \reset@font\fontsize{#1}{#2pt}%
  \fontfamily{#3}\fontseries{#4}\fontshape{#5}%
  \selectfont}%
\fi\endgroup%
\begin{picture}(512,392)(297,923)
\put(478,1231){\makebox(0,0)[lb]{\smash{{\SetFigFont{8}{9.6}{\rmdefault}{\mddefault}{\updefault}{\color[rgb]{0,0,0}$\odot$}%
}}}}
\end{picture}%
 \end{array}}
\nonumber
\\
						&	& -\hf \ensuremath{\begin{array}{c}\begin{picture}(0,0)%
\includegraphics{pstex/GR-DEP-dGR-GRk.pstex}%
\end{picture}%
\setlength{\unitlength}{3947sp}%
\begingroup\makeatletter\ifx\SetFigFont\undefined%
\gdef\SetFigFont#1#2#3#4#5{%
  \reset@font\fontsize{#1}{#2pt}%
  \fontfamily{#3}\fontseries{#4}\fontshape{#5}%
  \selectfont}%
\fi\endgroup%
\begin{picture}(686,390)(297,888)
\put(511,1194){\makebox(0,0)[lb]{\smash{{\SetFigFont{8}{9.6}{\rmdefault}{\mddefault}{\updefault}{\color[rgb]{0,0,0}$\odot$}%
}}}}
\end{picture}%
 \end{array}} + \hf \ensuremath{\begin{array}{c}\begin{picture}(0,0)%
\includegraphics{pstex/dGR-DEP-FGR.pstex}%
\end{picture}%
\setlength{\unitlength}{3947sp}%
\begingroup\makeatletter\ifx\SetFigFont\undefined%
\gdef\SetFigFont#1#2#3#4#5{%
  \reset@font\fontsize{#1}{#2pt}%
  \fontfamily{#3}\fontseries{#4}\fontshape{#5}%
  \selectfont}%
\fi\endgroup%
\begin{picture}(682,385)(239,892)
\put(558,1193){\makebox(0,0)[lb]{\smash{{\SetFigFont{8}{9.6}{\rmdefault}{\mddefault}{\updefault}{\color[rgb]{0,0,0}$\odot$}%
}}}}
\end{picture}%
 \end{array}}
\label{eq:GR-DEP-dCTP-EP}
\eea
where,
for all elements for which decoration is not defined, we
have traded $\odot$ for $\bullet$.

Including~\eq{eq:GR-DEP-dCTP-EP} amongst our tricks,
we can now iterate the diagrammatic procedure
until exhaustion. There are no further novelties
and it is straightforward to crank the handle and
demonstrate that~\eq{eq:al+be+subs}
does indeed reduce 
to just $\alpha$-terms, $\beta$-terms and
the diagrams of \fig{fig:bn-Op2-b} and their 
subtractions. This latter collection of
terms is remarkably easy to treat,
and we can immediately write down
the result, by using the techniques of
\sec{sec:Subtractions-G}. As an illustration,
we shown the result of combining
diagram~\ref{vh_j+R-DEP-GRs-Skt-CTP-E} with its subtractions, 
\ref{dvh_j-DEP-GRs-Skt-CTP-E-s}, \ref{vh_kR-dvj+kh-s}, 
\ref{vh_j-DEP-TEGRs-Skt-CTP-E-s} and~\ref{vh_j+R-DEP-GRs-Skt-CTP-E-b-s}
in \fig{fig:Subs-P}.
\bcf[h]
	\[
	\begin{array}{c}
	\vspace{1ex}
		\ds
		-2 \sum_{s=1}^n \sum_{m=0}^{2s-1} \sum_{m'=1}^{2s-1-m} \sum_{j=-1}^{n+s-m-m'-3}
		\sum_{j'=-1}^{n+s-m-m'-j-5} \sum_{s'=-j'+1}^{j+s+1}
	\\
	\vspace{2ex}
		\ds
		\Delta^{1\, 1}_{\rho \sigma}(p) \frac{\norm_{j'+s'+2,j'+2}}{m'!} \frac{\norm_{j+s+1-s',j+1}}{m!} \sum_{m''=0}^{m}
	\\
		\decNTEs{
			\begin{array}{c}
			\vspace{2ex}
				\ensuremath{\begin{array}{c}\begin{picture}(0,0)%
\includegraphics{pstex/vj+jpr+3-R.pstex}%
\end{picture}%
\setlength{\unitlength}{3947sp}%
\begingroup\makeatletter\ifx\SetFigFont\undefined%
\gdef\SetFigFont#1#2#3#4#5{%
  \reset@font\fontsize{#1}{#2pt}%
  \fontfamily{#3}\fontseries{#4}\fontshape{#5}%
  \selectfont}%
\fi\endgroup%
\begin{picture}(784,780)(466,1971)
\put(502,2296){\makebox(0,0)[lb]{\smash{{\SetFigFont{11}{13.2}{\rmdefault}{\mddefault}{\updefault}{\color[rgb]{0,0,0}$\nothing {v}^{j+j'+3;R}$}%
}}}}
\end{picture}%
 \end{array}}
			\\
			\vspace{2ex}
				\Vertex{v^{j+2},j'}
			\\
				\ensuremath{\begin{array}{c}\begin{picture}(0,0)%
\includegraphics{pstex/CTP-E.pstex}%
\end{picture}%
\setlength{\unitlength}{3947sp}%
\begingroup\makeatletter\ifx\SetFigFont\undefined%
\gdef\SetFigFont#1#2#3#4#5{%
  \reset@font\fontsize{#1}{#2pt}%
  \fontfamily{#3}\fontseries{#4}\fontshape{#5}%
  \selectfont}%
\fi\endgroup%
\begin{picture}(466,586)(719,713)
\put(801,1067){\makebox(0,0)[lb]{\smash{{\SetFigFont{11}{13.2}{\rmdefault}{\mddefault}{\updefault}{\color[rgb]{0,0,0}$0^2$}%
}}}}
\end{picture}%
 \end{array}}
			\end{array}
		}{1_\rho \Delta^{j'+s'+2} \GRkpr^{m'}}
		
		\dec{
			\decp{
				\ensuremath{\begin{array}{c}\input{pstex/vh_j+R-DEP-GRs-Skt-CTP-E-c.pstex_t} \end{array}} \hspace{0.5em}
			}{}
			\SumVertex
		}{1_\sigma \Delta^{j+s+1-s'} \GRkpr^{m-m''}}
	\end{array}
	\]
\caption{The result of combining
diagram~\ref{vh_j+R-DEP-GRs-Skt-CTP-E} with its subtractions.}
\label{fig:Subs-P}
\ecf

As expected, the contributions factorize into two sub-diagrams
joined by an effective propagator, $\Delta^{1\, 1}_{\rho \sigma}(p)$.
The first sub-diagram is under the influence of $\NTEs$.
The second sub-diagram has exactly the same structure
as the additions for diagram~\ref{vh_j+R-DEP-GRs-Skt-CTP-E}.
If we now include those terms obtained by combining
diagrams~\ref{Pi_j+R-DW-GRs-Skt-CTP-E}--\ref{GRs-dEP-GR-Skt-CTP-E}
with their subtractions we will generate the additions
for these diagrams as factorizable sub-diagrams also. 
The crucial point to recognize is that we already know exactly
what happens when we manipulate this complete set of
additions! We can therefore write:
\bea
	-4\beta_n \Box_{\mu\nu}(p) + \Op{4} = 
	\dec{\nLV{\Ds}{n}{\mu\nu}{a}(p) + \nLV{\Ds}{n}{\mu\nu}{b}(p)}{\bullet}
\nonumber
\\
		-\sum_{n'=1}^{n-1} 
		\dec{
			\Ds_{n-n' \mu \rho}^{b \hspace{1.2em} 1 \, 1}(p)
		}{\bullet}
		\Delta^{1\,1}_{\rho \sigma}(p) 
		\left(
			\nLV{\Ds}{n'}{\sigma\nu}{a}(p) +
			\decNTEs{ \nLV{\overline{\Ds}}{n'}{\sigma\nu}{c}(p)}{}
		\right)
\nonumber
\\
		+
		\sum_{n'=1}^{n-1} \sum_{n''=1}^{n'-1}
		\dec{
			\nLV{\Ds}{n-n'-n''}{\mu\rho}{b}(p) 
		}{\bullet}
		\Delta^{11}_{\rho \kappa}(p)
		\nLV{\Ds}{n''}{\kappa\tau}{a}(p)
		\Delta^{11}_{\tau \sigma}(p)
		\decNTEs{\nLV{\overline{\Ds}}{n'}{\sigma \nu}{c}(p)}{}
\nonumber
\\
		+ \cdots
\label{eq:bn-nearly}
\eea
where the ellipsis represents the $\alpha$ and $\beta$ terms.
There is no need to explicitly draw all of
these, since the complete set can be built up out
of just the
$\alpha$ and $\beta$ terms formed by
$\dec{\nLV{\Ds}{n}{\mu\nu}{a}(p)}{\bullet}$
and $\dec{\nLV{\Ds}{n}{\mu\nu}{b}(p)}{\bullet}$.
These diagrams are collected together in
\fig{fig:bn-al+be}; the overall signs of
the diagrams are appropriate to the \rhs\ 
of~\eq{eq:bn-nearly}.
\bcf
	\[
	\begin{array}{l}
	\vspace{2ex}
		2
		\bca{
			\ds
			\sum_{n'=1}^{n-1}
			\sum_{s=1}^{n-n'}   \sum_{m=0}^{2s} \sum_{j=-1}^{n-n'+s-m-1}\frac{\norm_{j+s+1,j+1}}{m!}
		}
			{
			\dec{
				\sco[1]{
					\left[
						\begin{array}{ccc}
						\vspace{1ex}	
							\LD{beta-terms-a}							&								& \LD{alpha-terms-a}			
						\\	
							2 \left( v^{j_+}-1 \right) \beta_{n'} 	& \hspace{-1em} + \hspace{-1em}	& \gamma_{n'} \pder{}{\alpha}
						\end{array}
					\right]
					\DisplacedVertex{v^{j_+}}
				}{\Vertex{n_{n'+s}, j}}
			}{11\Delta^{j+s+1} \GRkpr^m}
		}
	\\
		4
		\bca{
			\sum_{n'=1}^{n-1} \sum_{s=1}^{n-n'}  \sum_{r=1}^s 
			\sum_{m=0}^{2(s-r)} \sum_{j=-1}^{n-n'+s-2r-m-1}
			\frac{\norm_{j+s+2-r,j+1}}{m!r!}
		}
			{
			\decp{
				\begin{array}{c}
				\vspace{1ex}
					\left[
						\begin{array}{ccc}
						\vspace{1ex}	
							\LD{beta-terms-b}							&								& \LD{alpha-terms-b}			
						\\	
							2 \left( v^{j_+}-1 \right) \beta_{n'} 	& \hspace{-1em} + \hspace{-1em}	& \gamma_{n'} \pder{}{\alpha}
						\end{array}
					\right]
					\DisplacedVertex{v^{j_+}}
				\\
				\vspace{1ex}
					\Vertex{n_{n^{'} +s, j}} 
				\\
					\ensuremath{\begin{array}{c}\begin{picture}(0,0)%
\includegraphics{pstex/CTP-E.pstex}%
\end{picture}%
\setlength{\unitlength}{3947sp}%
\begingroup\makeatletter\ifx\SetFigFont\undefined%
\gdef\SetFigFont#1#2#3#4#5{%
  \reset@font\fontsize{#1}{#2pt}%
  \fontfamily{#3}\fontseries{#4}\fontshape{#5}%
  \selectfont}%
\fi\endgroup%
\begin{picture}(466,586)(719,713)
\put(801,1067){\makebox(0,0)[lb]{\smash{{\SetFigFont{11}{13.2}{\rmdefault}{\mddefault}{\updefault}{\color[rgb]{0,0,0}$0^2$}%
}}}}
\end{picture}%
 \end{array}}
				\end{array}
			}{1 \Delta^{j+s+2-r} \combo^r \GRkpr^m }
		}
	\end{array}
	\]
\caption{The $\alpha$ and $\beta$ terms contributing 
to $-4 \beta_n \Box_{\mu\nu}(p)$
formed by $\dec{\nLV{\Ds}{n}{\mu\nu}{a}(p)}{\bullet}$
and $\dec{\nLV{\Ds}{n}{\mu\nu}{b}(p)}{\bullet}$.}
\label{fig:bn-al+be}
\ecf

Examining these diagrams, and recalling the expressions
for the $\Ds^{i}$, we see that we have succeeded
in deriving an expression for $\beta_n$
which has no explicit dependence on either
the seed action or the details of the covariantization!

\subsection{The $\alpha$ and $\beta$ Terms}
\label{sec:al+be}

We can simplify the diagrams of \fig{fig:bn-al+be}
and do so by starting with the
$\alpha$
terms.
This is done by using the diagrammatic relationship:
\be
\label{eq:EP-dalDTP-EP}
	\ensuremath{\begin{array}{c}\input{pstex/A-EP-dalCTP-EP-B.pstex_t} \end{array}} = -\ensuremath{\begin{array}{c}\input{pstex/A-dalEP-B.pstex_t} \end{array}} - \hf \ensuremath{\begin{array}{c}\input{pstex/A-GR-dalCombo-B.pstex_t} \end{array}} -\hf \ensuremath{\begin{array}{c}\input{pstex/A-dalCombo-GR-B.pstex_t} \end{array}},
\ee
where
\[
	\ensuremath{\begin{array}{c}\begin{picture}(0,0)%
\includegraphics{pstex/dalCombo-GR.pstex}%
\end{picture}%
\setlength{\unitlength}{3947sp}%
\begingroup\makeatletter\ifx\SetFigFont\undefined%
\gdef\SetFigFont#1#2#3#4#5{%
  \reset@font\fontsize{#1}{#2pt}%
  \fontfamily{#3}\fontseries{#4}\fontshape{#5}%
  \selectfont}%
\fi\endgroup%
\begin{picture}(90,632)(1154,1766)
\put(1176,2044){\makebox(0,0)[lb]{\smash{{\SetFigFont{11}{13.2}{\rmdefault}{\mddefault}{\updefault}{\color[rgb]{0,0,0}$\Dal$}%
}}}}
\end{picture}%
 \end{array}} \hspace{1em} \equiv \ensuremath{\begin{array}{c}\begin{picture}(0,0)%
\includegraphics{pstex/EP-dalGR-GR.pstex}%
\end{picture}%
\setlength{\unitlength}{3947sp}%
\begingroup\makeatletter\ifx\SetFigFont\undefined%
\gdef\SetFigFont#1#2#3#4#5{%
  \reset@font\fontsize{#1}{#2pt}%
  \fontfamily{#3}\fontseries{#4}\fontshape{#5}%
  \selectfont}%
\fi\endgroup%
\begin{picture}(100,685)(1146,1713)
\put(1198,1851){\makebox(0,0)[lb]{\smash{{\SetFigFont{11}{13.2}{\rmdefault}{\mddefault}{\updefault}{\color[rgb]{0,0,0}$\Dal$}%
}}}}
\end{picture}%
 \end{array}} \hspace{1em} -\hf \ensuremath{\begin{array}{c}\begin{picture}(0,0)%
\includegraphics{pstex/GR-dalPEP-GR.pstex}%
\end{picture}%
\setlength{\unitlength}{3947sp}%
\begingroup\makeatletter\ifx\SetFigFont\undefined%
\gdef\SetFigFont#1#2#3#4#5{%
  \reset@font\fontsize{#1}{#2pt}%
  \fontfamily{#3}\fontseries{#4}\fontshape{#5}%
  \selectfont}%
\fi\endgroup%
\begin{picture}(90,694)(1154,1764)
\put(1176,2042){\makebox(0,0)[lb]{\smash{{\SetFigFont{11}{13.2}{\rmdefault}{\mddefault}{\updefault}{\color[rgb]{0,0,0}$\Dal$}%
}}}}
\end{picture}%
 \end{array}}
\]
(and we recall that $\Dal \equiv \partial / \partial \alpha$).
Comparing~\eq{eq:EP-dalDTP-EP} with~\eq{eq:EP-dCTP-EP}
it is clear that the analysis of the $\alpha$
terms is very similar to the analysis
of diagrams with momentum derivatives.
Indeed, it is straightforward
to show that the $\alpha$ terms
yield a contribution to
$-4\beta_n \Box_{\mu \nu}(p)$ 
such that for every appearance
of $\dec{X_n}{\bullet}$, there
is a partner term
which
can be generated by making the substitution
\[
\dec{X_n}{\bullet} \rightarrow - \sum_{n'=1}^{n-1} \gamma_{n'} 
	\pder{}{\alpha} X_{n-n'}.
\]
Equivalently, we can explicitly include the
$\alpha$-terms in~\eq{eq:bn-nearly}
by inserting $\sum_n g^{2n+1}$
at the beginning of both sides of the equation
and then
making the replacement
\be
\label{eq:replace}
	-\flowConstAl \rightarrow -\flow.
\ee

The $\beta$ terms are more subtle
than one might expect. Let us begin with
diagram~\ref{beta-terms-a}. Clearly, the way to proceed
is to isolate any diagrams which possess
a classical, two-point vertex and then decorate
this vertex. There are two possible decorations:
either we can attach one external field and one
effective propagator, or we can attach two
effective propagators; either way, we can apply
the effective propagator relation. In the former
case, the primary part simply yields
\be
\label{eq:beta-a-nLV-a}
	-4 \sum_{n'=1}^{n-1} \beta_{n'} 
	\nLV{\Ds}{n-n'}{\mu\nu}{a}(p).
\ee
The secondary part vanishes, since the
associated gauge remainder strikes $\Ds^a$, which 
we know to be transverse.
With this in mind,
we re-express
diagram~\ref{beta-terms-a}  in \fig{fig:bn-beta}.
\bcf[h]
	\[
	\begin{array}{l}
	\vspace{2ex}
		\ds
		-4 \sum_{n'=1}^{n-1} \beta_{n'} \Ds_{n-n' \mu \nu}^{a \hspace{1.2em} 1 \, 1}(p)
	\\
	\vspace{2ex}
		+4
		\bca{
			\sum_{n'=1}^{n-1} \sum_{s=1}^{n-n'}   \sum_{m=0}^{2s} 
			\sum_{j=-1}^{n-n'+s-m-2} \frac{\norm_{j+s+1,j+1}}{m!}
			\beta_{n'}
		}
			{
			\left(
				\begin{array}{ccc}
					\LD{beta-vj+}	&	& \LD{beta-novj+}
				\\[1ex]
					v^{j+}			& - & 1
				\end{array}
			\right)
			\dec{
				\sco[1]{
						\TopVertex
				}{\Vertex{n_{n'+s},j}}
			}{11\Delta^{j+s+1} \GRkpr^m}
		}

	\\
		-2
		\bca{
			\sum_{n'=1}^{n-1} \sum_{s=1}^{n-n'}   \sum_{m=0}^{2s} 
			\sum_{j=-2}^{n-n'+s-m-2} \frac{\norm_{j+s,j+2}}{m!}
			\beta_{n'}
		}
			{
			\dec{
				\begin{array}{c}
				\vspace{1ex}
					\begin{array}{ccc}					
					\vspace{1ex}
						\LD{beta-EP}	& 	& \LD{beta-EP-GR}
					\\
						\ensuremath{\begin{array}{c}\begin{picture}(0,0)%
\includegraphics{pstex/Vert-EP.pstex}%
\end{picture}%
\setlength{\unitlength}{3947sp}%
\begingroup\makeatletter\ifx\SetFigFont\undefined%
\gdef\SetFigFont#1#2#3#4#5{%
  \reset@font\fontsize{#1}{#2pt}%
  \fontfamily{#3}\fontseries{#4}\fontshape{#5}%
  \selectfont}%
\fi\endgroup%
\begin{picture}(24,441)(1424,1811)
\end{picture}%
 \end{array}}	& -	& \ensuremath{\begin{array}{c}\begin{picture}(0,0)%
\includegraphics{pstex/Vert-EP-GR.pstex}%
\end{picture}%
\setlength{\unitlength}{3947sp}%
\begingroup\makeatletter\ifx\SetFigFont\undefined%
\gdef\SetFigFont#1#2#3#4#5{%
  \reset@font\fontsize{#1}{#2pt}%
  \fontfamily{#3}\fontseries{#4}\fontshape{#5}%
  \selectfont}%
\fi\endgroup%
\begin{picture}(90,506)(1390,1746)
\end{picture}%
 \end{array}}
					\end{array}
				\\
				\vspace{1ex}
					\TopVertex
				\\
					\Vertex{n_{n'+s},j}
				\end{array}
			}{11\Delta^{j+s} \GRkpr^m}
		}
	\end{array}
	\]
\caption{A re-expression of 
diagram~\ref{beta-terms-a}.}
\label{fig:bn-beta}
\ecf

We can now see that there are some
novelties associated with the
$\beta$ terms.
First, diagram~\ref{beta-novj+}
does not have a canonical normalization factor:
there are $j+2$ identical vertices, but the
normalization factor $\sim 1/(j+1)!$. 
Secondly, the presence of the multiplicative
factor $v^{j_+}$ in diagram~\ref{beta-vj+}
makes this term different from any we have 
encountered so far. Note, though, that
the combinatoric factor is effectively
canonical, since the factor of $v^{j_+}$
singles out one of the vertices, leaving
behind only $j+1$ identical vertices.
Thirdly, just as diagram~\ref{beta-vj+} has a special
vertex, so diagram~\ref{beta-EP} has a special
effective propagator. This effective
propagator cannot be promoted to an implicit
decoration in the usual way, since then
the combinatorics will be wrong. 
To put
it another way, if we decorate a vertex with
a total of $q+1$ effective propagators,
comprising the special one and $q$
others, the combinatoric factor
associated with choosing the effective propagators
in this way 
will be $\nCr{j+s}{q}$ and \emph{not} $\nCr{j+s+1}{q}$.

Nevertheless, diagrams~\ref{beta-vj+}, \ref{beta-novj+}
and~\ref{beta-EP} combine in a miraculous way.
Starting with diagram~\ref{beta-vj+}, we exploit
the indistinguishability of the $j+2$ vertices by
replacing $v^{j_+}$ with
\[
	\frac{1}{j+2} \left[ \sum_{i=0}^j v^{i,i_+} + v^{j_+} \right] = \frac{n_{n'}-s}{j+2},
\]
where we have used~\eq{eq:VertexSum} (but with $v^0 = n_{n'} -s$).

Next, consider creating some fully fleshed out
diagram from~\ref{beta-EP}. The total of $j+s+1$
effective propagators are to be divided into $q$
sets, each containing $L_i$ effective propagators. 
Since the special effective propagator can reside
in any of these sets, there are $q$ different ways
to make the sets. The overall combinatoric factor
associated with this partitioning
is, therefore,
\[
	\frac{(j+s)!}{\prod_i L_i!} \sum_i L_i = \frac{(j+s+1)!}{\prod_i L_i!},
\]
which is just the combinatoric factor expected
from partitioning $j+s+1$
effective propagators into $q$ sets.
Therefore, we can now promote the special effective propagator
of diagram~\ref{beta-EP}
to an implicit decorations but, counterintuitively, 
the combinatoric
factor of the diagram, $\norm_{j+s,j+2}$, \emph{stays the same}!

With these points in mind, it is a trivial matter to show that 
\be
\label{eq:nLV-d}
	\mbox{\ref{beta-vj+}} + \mbox{\ref{beta-novj+}} 
		+ \mbox{\ref{beta-EP}}
	= -2 \sum_{n'=1}^{n-1} \beta_{n'} (n_{n'}-1) \nLV{\Ds}{n-n'}{\mu \nu}{a}(p)
\ee

%
%

Diagram~\ref{beta-EP-GR} can be processed, yielding:
\be
\label{eq:beta-a-nLV-c}
	-2 \sum_{n'=1}^{n-1} \beta_{n'} 
	\nLV{\Ds}{n-n'}{\mu\nu}{c}(p)
\ee
Putting together~\eq{eq:beta-a-nLV-a}, \eq{eq:nLV-d} and~\eq{eq:beta-a-nLV-c},
diagram~\ref{beta-terms-a} ultimately reduces to 
\be
\label{beta-terms-a-Final}
	-2\sum_{n'=1}^{n-1} \beta_{n'} 
	\left[
		(n_{n'}+1) 	\nLV{\Ds}{n-n'}{\mu\nu}{a}(p) + \nLV{\Ds}{n-n'}{\mu\nu}{c}(p)
	\right].
\ee

The treatment of diagram~\ref{beta-terms-b} follows in 
similar fashion. Now, however, our options for
decorating the classical, two-point vertex
are to attach either two effective propagators or
one effective propagator and one $\combo$
(it is illegal to attach two $\combo$s---see
\sec{sec:Not:GRs}).
In the latter case we proceed by expressing
\[
	\ensuremath{\begin{array}{c}\begin{picture}(0,0)%
\includegraphics{pstex/EP-CTP-Combo.pstex}%
\end{picture}%
\setlength{\unitlength}{3947sp}%
\begingroup\makeatletter\ifx\SetFigFont\undefined%
\gdef\SetFigFont#1#2#3#4#5{%
  \reset@font\fontsize{#1}{#2pt}%
  \fontfamily{#3}\fontseries{#4}\fontshape{#5}%
  \selectfont}%
\fi\endgroup%
\begin{picture}(1259,372)(409,1164)
\put(936,1284){\makebox(0,0)[lb]{\smash{{\SetFigFont{11}{13.2}{\rmdefault}{\mddefault}{\updefault}{\color[rgb]{0,0,0}$0^2$}%
}}}}
\end{picture}%
 \end{array}} = 
	\hf
	\left[
		\cdeps{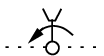} - \cdeps{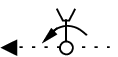} + \cdeps{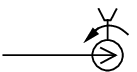} - \cdeps{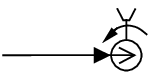}
	\right].
\]
In this way, we can reduce diagram~\ref{beta-terms-b} 
to 
\be
\label{beta-terms-b-Final}
	\sum_{n'=1}^{n-1} \beta_{n'} 
	\left[
		2 \nLV{\Ds}{n-n'}{\mu\nu}{c'}(p) - (2n_{n'}+1) \nLV{\Ds}{n-n'}{\mu\nu}{b}(p) + 2\nLV{\tilde{\Ds}}{n-n'}{\mu\nu}{b}(p)
	\right],
\ee
where
\[
	\nLV{\tilde{\Ds}}{n}{\mu\nu}{b}(p)  \equiv
	\bca{ 
		4 \sum_{s=1}^n  \sum_{r=1}^s \sum_{m=0}^{2(s-r)+1} 
		\sum_{j=-2}^{n+s-2r-m-1} 
		\frac{\norm_{j+s+2-r,j+2}}{m!(r-1)!}
	}
		{
		\qquad
		\decp{
				\Tower{\ensuremath{\begin{array}{c}\begin{picture}(0,0)%
\includegraphics{pstex/CTP-E.pstex}%
\end{picture}%
\setlength{\unitlength}{3947sp}%
\begingroup\makeatletter\ifx\SetFigFont\undefined%
\gdef\SetFigFont#1#2#3#4#5{%
  \reset@font\fontsize{#1}{#2pt}%
  \fontfamily{#3}\fontseries{#4}\fontshape{#5}%
  \selectfont}%
\fi\endgroup%
\begin{picture}(466,586)(719,713)
\put(801,1067){\makebox(0,0)[lb]{\smash{{\SetFigFont{11}{13.2}{\rmdefault}{\mddefault}{\updefault}{\color[rgb]{0,0,0}$0^2$}%
}}}}
\end{picture}%
 \end{array}}}
		}{1\Delta^{j+s+2-r} \combo^r \GRkpr^m }{}		
	}.
\]
Notice that the $r=1$ contribution is the same as
for $\nLV{\Ds}{n}{\mu\nu}{b}(p)$.

By combining~\eq{eq:bn-nearly} with~\eqs{beta-terms-a-Final}{beta-terms-b-Final}
and utilizing~\eq{eq:replace} we can write down the
complete
expression for $\beta_n$:
\be
\label{eq:bn-Complete}
\fl
	\begin{array}{c}
	\vspace{1ex}
		\ds
		-4 \sum_{n} g^{2n+1} \beta_n \Box_{\mu\nu}(p) + \Op{4} = 
	\\
	\vspace{1ex}
			\ds
			\sum_{n} g^{2n+1}
	\\
		\ds
		\left[
			\begin{array}{l}
			\vspace{2ex}
				\ds
				-\flow\dec{\nLV{\D}{n}{\mu\nu}{a}(p) + \nLV{\D}{n}{\mu\nu}{b}(p)}{}
				
			\\
			\vspace{1ex}
				\ds
			  	-2\sum_{n'=1}^{n-1} \beta_{n'}
			\\
			\vspace{2ex}
				\ds
				\qquad
				\times
				\left\{
					\begin{array}{l}
					\vspace{1ex}
						(n_{n'} +1) \nLV{\D}{n-n'}{\mu\nu}{a}(p)
						+\nLV{\D}{n-n'}{\mu\nu}{c}(p)
						-\decpb{\nLV{\overline{\D}}{n-n'}{\mu\nu}{c}(p)}{}
					\\
						+(n_{n'}+1) \nLV{\D}{n-n'}{\mu\nu}{b}(p) -  \nLV{\tilde{\D}}{n-n'}{\mu\nu}{b}(p)
					\end{array}
				\right\}
			\\
			\vspace{2ex}
				\ds
				+
				\sum_{n'=1}^{n-1}
				\left[
					\nLV{\D}{n'}{\mu\rho}{a}(p) +
					\decNTEsb{
						\nLV{\overline{\D}}{n'}{\mu\rho}{c}(p)
					}{}
				\right]
				\Delta^{1\,1}_{\rho \sigma}(p) 
			\\
			\vspace{2ex}
				\ds
				\qquad
				\times
				\left\{
					\begin{array}{l}
					\vspace{1ex}
						\flow
						\dec{
							\nLV{\D}{n-n'}{\sigma\nu}{b}(p)
						}{}						
					\\
						\ds
						-2\sum_{n''=1}^{n'-1} \beta_{n''}
						\left[
							\begin{array}{l}
							\vspace{1ex}
								\decpb{\nLV{\overline{\D}}{n-n'-n''}{\sigma\nu}{c}(p)}{} 
							\\
							\vspace{1ex}
								- (n_{n'+n''}+1)\nLV{\D}{n-n'-n''}{\sigma\nu}{b}(p)
							\\
								+ \nLV{\tilde{\D}}{n-n'-n''}{\sigma\nu}{b}(p)
							\end{array}
						\right]	
					\end{array}
				\right\}
			\\
			\vspace{1ex}
				\ds
				-\sum_{n'=1}^{n-1} \sum_{n''=1}^{n'-1}
				
				\decNTEsb{
						\nLV{\overline{\D}}{n'}{\mu\rho}{c}(p)
					}{}
				\Delta^{11}_{\rho\tau}(p)
				\nLV{\D}{n''}{\tau\kappa}{a}(p)
				\Delta^{11}_{\kappa \sigma}(p)
			\\
				\qquad
				\times
				\left\{
					\begin{array}{l}
					\vspace{1ex}
						\ds
						\flow
						\dec{
							\nLV{\D}{n-n'-n''}{\sigma\nu}{b}(p) 
						}{}						
					\\
						\ds
						-2\sum_{n'''=1}^{n''-1} \beta_{n'''}
						\left[
							\begin{array}{l}
							\vspace{1ex}
								\ds
								\decpb{\nLV{\overline{\D}}{n-n'-n''-n'''}{\sigma\nu}{c}(p)}{} 
							\\
							\vspace{1ex}
								- (n_{n'+n''+n'''}+1)\nLV{\D}{n-n'-n''}{\sigma\nu}{b}(p)
							\\
								+ \nLV{\tilde{\D}}{n-n'-n''}{\sigma\nu}{b}(p)
							\end{array}
						\right]	
					\end{array}
				\right\}
			\end{array}
		\right]
	\end{array}
\ee

Notice that we have substituted for $\Ds^{c'}$
via~\eq{eq:nLV-c-Op2-b} and have traded
the $\Ds^i$ for the $\D^i$, as this is the
correct time to discard all diagrams which
vanish at $\Op{2}$ and / or contain a Wilsonian
effective action one-point vertex.

Up until now, we have been very careful not to
Taylor expand individual diagrams to $\Op{2}$ unless
we are certain that this step is valid. However,
we know that the \emph{sum} 
of diagrams contributing to $\beta_n$ is
Taylor expandable to $\Op{2}$. It must therefore
be the case that all contributions which include
non-polynomial dependence on $p$, at $\Op{2}$,
cancel out of~\eq{eq:bn-Complete}. At two loops
it has already been demonstrated that a subset of
such contributions cancel out diagrammatically~\cite{Thesis,mgierg2}.
It is very likely that this analysis can be extended
to any number of loops, but we leave the investigation
of this for the future. In the meantime,
we note that we can considerably simplify~\eq{eq:bn-Complete}
if we focus just on 
the $\Op{2}$ component of each diagram, 
which yields
the following equation (we have used~\eq{nLDc-factorize}). 
\be
\fl
	\begin{array}{c}
	\vspace{1ex}
		\ds
		-4 \sum_{n} g^{2n+1} \beta_n \Box_{\mu\nu}(p) + \Op{4} = 
	\\
	\vspace{1ex}
			\ds
			\sum_{n} g^{2n+1}
	\\
		\ds
		\left[
			\begin{array}{l}
			\vspace{2ex}
				\ds
				-\flow\dec{\nLV{\D}{n}{\mu\nu}{a}(p) + \nLV{\D}{n}{\mu\nu}{b}(p)}{}
				+\sum_{n'=1}^{n-1}
					\nLV{\D}{n'}{\mu \rho}{a}(p) \Delta^{1\,1}_{\rho \sigma}(p) 
					\flow
					\dec{
						\nLV{\D}{n-n'}{\sigma\nu}{b}(p)
					}{}
			\\
			\vspace{2ex}
				\ds
				\left.
				\rule{0em}{4ex}
			  	-2\sum_{n'=1}^{n-1} \beta_{n'}
				\right\{
						(n_{n'} +1) 
						\left[
							\nLV{\D}{n-n'}{\mu\nu}{a}(p)
							+\nLV{\D}{n-n'}{\mu\nu}{b}(p)
						\right]
						- \nLV{\tilde{\D}}{n-n'}{\mu\nu}{b}(p)
			\\
			\vspace{2ex}
				\ds
				\qquad +
				\left.
				\sum_{n''=1}^{n'-1} 
					\nLV{\D}{n''}{\mu \rho}{a}(p) \Delta^{1\,1}_{\rho \sigma}(p) 
				\left[
					\nLV{\tilde{\D}}{n-n'-n''}{\sigma\nu}{b}(p) - 
					(n_{n'+n''}+1)\nLV{\D}{n-n'-n''}{\sigma\nu}{b}(p)
				\right]
				\rule{0em}{4ex}
				\right\}
			\end{array}
		\right]_{p^2}
	\end{array}
\label{eq:bn-Final}
\ee

Equation~\eq{eq:bn-Final} reduces to particularly simple
forms at one and two loops:
\bea
\label{eq:beta1}
\vspace{2ex}
	-4 \beta_1 \Box_{\mu\nu}(p) & = & 
	\decp{
		\nLV{\overline{\D}}{1}{\mu\nu}{a}(p) + \nLV{\overline{\D}}{1}{\mu\nu}{b}(p)
	}{\bullet}
\\
\nonumber
	-4 \beta_2 \Box_{\mu\nu}(p) & = &
	\decp{
		\nLV{\overline{\D}}{2}{\mu\nu}{a}(p)
		+ \nLV{\overline{\D}}{1}{\mu\rho}{a}(p)
		\Delta^{1\,1}_{\rho \sigma}(p) \nLV{\overline{\D}}{1}{\sigma\nu}{b}(p)	
		+ \nLV{\overline{\D}}{2}{\mu\nu}{b}(p)
	}{\bullet}
\\
\label{eq:beta2}
	& & \qquad + \gamma_1 \pder{}{\alpha} 
		\decp{
			\nLV{\overline{\D}}{1}{\mu\nu}{a}(p) + \nLV{\overline{\D}}{1}{\mu\nu}{b}(p)
		}{},
\eea
where we have used~\eqs{eq:EP-leading}{nLDa-factorize} and
have defined
\[
	\nLV{\overline{\D}}{n}{\mu\nu}{b}(p) \equiv
	\nLV{\D}{n}{\mu\nu}{b}(p)
	+
	\frac{1}{2}
	\sum_{n'=1}^{n-1} \nLV{\D}{n-n'}{\mu\rho}{b}(p)
	\Delta^{11}_{\rho\sigma}
	\nLV{\D}{n'}{\sigma\nu}{b}(p).
\]
Equations~\eqs{eq:beta1}{eq:beta2} reproduce the
results of~\cite{mgierg1,mgierg2,Thesis}, but written
in a remarkably compact form.

\section{Conclusion}
\label{sec:Conc}

We have constructed a manifestly gauge invariant
calculus for $SU(N)$ Yang-Mills, which can
be applied at any number of loops. The calculus, which
is entirely diagrammatic in nature,
has been comprehensively illustrated by demonstrating
that the $\beta$-function is independent of both the seed
action and the details of the covariantization of the cutoff,
to all orders in perturbation theory.

The inspiration for this methodology, the first
elements of which were developed
in~\cite{aprop}, is the immense freedom in the
construction of ERGs. Given
that we specialize not only to those ERGS that are
manifestly gauge invariant but to those which also allow 
convenient renormalization to any loop order,
there are still an infinite number with which
we can work. The differences between these
ERGs amount to non-universal details which
need never be exactly specified, instead satisfying
general constraints. In the computation of
universal quantities, these non-universal details
must cancel out; by leaving them unspecified,
we are thus guided towards a very constrained calculational
procedure. However, as recognized in~\cite{Thesis,Primer,RG2005},
such cancellations are embedded within the formalism
at a particularly deep level: at least
some of them occur in the calculation of quantities which
are not universal, such as $\beta$-function coefficients
beyond two loops. This observation formed the basis for developing
the diagrammatic calculus to the level
described in this paper.

Given a diagrammatic representation of the flow equation,
the calculus comprises an operator which implements 
the flow \ie $-\flow$,
a rule for implementing
the effects of \CC\ invariance, a set of primary
and secondary diagrammatic identities and the subtraction
techniques.

There are three types of primary diagrammatic identities
(see \sec{sec:Review}).
Those of the first type make no reference to
perturbation theory and reflect general
properties of the exact flow equation
and underlying theory. The single identity
of the second type is the effective propagator
relation, which arises as a solution to
the classical flow equation, given a 
convenient choice of
seed action we are free to make. 
The effective propagator relation states
that for each classical two-point vertex
(which cannot be consistently set to zero)
there exists an effective propagator, which
is the inverse of this vertex, up to a gauge remainder.
The primary
diagrammatic identities of the third type
follow directly from those of the first
and second types; they are stated,
nonetheless, due to the central role they play
in perturbative calculations.

The secondary diagrammatic identities, of which
there are two types, encode the equivalence
of distinct diagrammatic representations
of structures comprising
particular arrangements of components of
gauge remainders. Those of the second
type (see \sec{sec:D-ID-Secondary-II}) are applied only when we focus on
the component of a diagram which has been Taylor
expanded to zeroth order in its external momentum
(in this paper we have dealt only with diagrams
carrying a single external momentum);
those of the first type (see \sec{sec:D-ID-Secondary-I}) 
are more generally applicable.

The final element of the calculus is the subtraction
techniques. Working to some order in the external momentum
of a diagram, these techniques are used to isolate
those components which are of precisely the desired
order from those
which have an additional, non-polynomial contribution.

A crucial step which facilitates the practical
application of the calculus is the notation
introduced in \sec{sec:Notation}. As with
perturbative methods in general, the number
of diagrams contributing to some function grows
very rapidly with loop order, $n$. However, rather
than explicitly drawing each of these
diagrams, it was realized that we can construct
a function of both $n$ and the various diagrammatic
\emph{components} which can be used to generate the
correct set of diagrams, at any loop order.
Such expressions are very compact, with a beautifully
intuitive structure.

For the treatment of $\beta_n$,
we constructed
a set of diagrammatic functions 
$\nLV{\Ds}{n}{\mu\nu}{a}(p)$, $\nLV{\Ds}{n}{\mu\nu}{b}(p)$
and $\nLV{\Ds}{n}{\mu\nu}{c}(p)$
(see \sec{sec:Prelim-DiagFns})
which each include two external $A^1_\mu$s, carrying momenta $p_\mu$
and $-p_\nu$.  We proceeded by
focusing on the $\Op{2}$ 
components of these diagrammatic functions
(up to additional non-polynomial dependence on $p$)
 and
computed the flow of $\nLV{\Ds}{n}{\mu\nu}{a}(p)$. Amongst the terms generated
is a single instance of $\beta_n$, multiplied by
a universal coefficient. Of the remaining terms,
those involving either lower order $\beta$-function
coefficients or $\alpha$-derivatives were put to one side.
The remaining diagrams
were manipulated, wherever possible, using the 
following scheme.

The first stage
was to isolate any classical, two-point vertices
formed by the flow equation and, wherever
possible, to attach
either an external field or an effective propagator.
In the former case, the diagram possesses a manifestly $\Op{2}$
stub; these $\Op{2}$ terms were put to one side. 
In the latter case, the effective propagator
relation was applied. This
relation collapses all classical, two-point vertices
which are contracted into an effective propagator
down into a Kronecker~$\delta$ and a gauge remainder.
The gauge remainders were either be processed using
the primary diagrammatic identities or re-expressed
using the secondary diagrammatic identities. In
the former case, the whole procedure
was repeated: classical, two-point vertices were
isolated and decorated and the effective propagator
relation was applied {\it etc}. The
set of remaining diagrams now cancelled amongst
themselves, up to the $\Op{2}$ terms.

To process the $\Op{2}$ terms, the subtraction
techniques were employed, allowing the $\Op{2}$ terms
to be
effectively split into
those which have additional non-polynomial
dependence on $p$ and those which do not.
The latter terms were manipulated
using the primary and secondary diagrammatic
identities, yielding a set
of diagrams which, up to
further $\alpha$ and $\beta$-terms,
were precisely
cancelled by~\eq{eq:LdL-b-ba}.
The $\Op{2}$ diagrams containing additional,
non-polynomial, dependence on $p$ each factorized
into an un-manipulable component, $\decNTEs{\nLV{\Ds}{n-n'}{\mu\nu}{c}(p)}{}$,
and an $n'$-loop component of exactly the same form
as~\eq{eq:LdL-b-ba}. 
This allowed us to immediately write down an expression for
$\beta_n$  which is manifestly independent of
the seed action and details of the covariantization.
Finally, we neatened the expression up by processing
the $\alpha$ and $\beta$ terms, to yield~\eq{eq:bn-Complete}.

This expression was derived with the utmost rigour. In particular,
all components carrying non-polynomial dependence on $p$
have been retained, even though it is know that
they must cancel out. Indeed, it may well be
that these cancellations can be demonstrated diagrammatically
(this has already been done to some extent at two loops~\cite{mgierg2,Thesis}),
which would doubtless serve to considerably illuminate
underlying features of~\eq{eq:bn-Complete}. If one is
prepared to simply accept that these contributions do
indeed cancel, then~\eq{eq:bn-Complete} can be simplified
by focusing on the strictly $\Op{2}$ components 
of each diagram. This yields the considerably simpler
expression~\eq{eq:bn-Final}. 

From~\eq{eq:bn-Final}, the
diagrammatic expressions for $\beta_1$ and
$\beta_2$, \eqs{eq:beta1}{eq:beta2}, follow
directly. In the context of assessing the
practical usability of the calculus, the importance
of this can hardly be overstated. Recall
that the first time $\beta_2$ was computed in our
ERG framework~\cite{Thesis} the derivation of~\eq{eq:beta2}
involved the generation of $\order{10^4}$ diagrams (!)
at intermediate
stages of the calculation,
almost all of which cancelled to yield the final diagrammatic
expression.
Although the number of intermediate diagrams can
be hugely reduced using the refinements of~\cite{Primer},
the benefits of an expression from which the
diagrammatic expression can be read off at any loop
order is obvious.

The derivation of the new formula for $\beta_n$
provides insight into two subtle issues, both
of which relate to the regularization. First, although the
physical $SU(N)$ gauge theory is properly 
regularized, care must be taken with
the flow equation. In particular, the flow
equation generates (non-universal) diagrams in which
the kernel `bites its own tail' which 
are not properly regularized~\cite{ym,ymi,ymii,aprop,mgierg1}
(see the discussion around \fig{fig:Flow}).
Up until~\cite{Primer}, these diagrams were excluded,
via an imposed constraint on the covariantization.
However, as speculated in~\cite{Primer,Thesis},
all explicit instances
such non-universal objects should cancel out
in the reduction of $\beta$-function
coefficients to a form with no explicit dependence
on either the seed action or details of the
covariantization. This has been borne out here,
though it should be emphasised that implicit
dependence on these objects remains. It
thus seems very likely that, ultimately,
a suitable constraint on the covariantization
is required. 

Nonetheless, it is very interesting that
the basic structure of the flow equation
is sufficient to formally
remove certain non-universal objects
to $\beta_n$, even if these objects are
not strictly well defined. This very much
strengthens the case that
the diagrammatics is driving us towards a
new framework formulated directly
in terms of functions such as the $\Ds^i$,
with the seed action
and details of the covariantization of the
cutoff operating in the background.

The second issue pertaining to the regularization
that the new formula for $\beta_n$
sheds some light on is that of pre-regularization.
Recall that, in order the regularization
be properly defined, a pre-regulator must
be used in $D=4$~\cite{SU(N|N)}
to unambiguously define contributions
which are finite only by virtue of the 
Pauli-Villars regularization provided by
the massive regularizing fields. The
pre-regulator amounts to a prescription
for discarding otherwise non-vanishing
surface terms which can be generated by
shifting loop momenta. In this paper,
dimensional regularization was used.
However, the only place that the
pre-regulator is explicitly used in practise is to throw away diagrams
such as~\ref{nLV-c-MomDer} which possess a sum of momentum
derivative \wrt\ the loop momenta. Thus,
for the purposes of the calculation performed
in this paper,
we could have adopted the prescription
that we simply throw away all diagrams
of this form. Of course, the diagrammatic
functions contributing to $\beta_n$ require
pre-regularization themselves, as they
contain diagrams which are finite only
by virtue of the PV regularization. It is
not immediately obvious that the new prescription
suffices in this case but it will
be interesting to see in the future if
it can, in fact, be adopted 
in complete generality, thereby allowing
us to work directly in $D=4$.

Whilst the application of the calculus to
$\beta$-function coefficients was vital
in the development of the formalism,
it is very important to ask whether the techniques
of this paper can be applied in the computation of
other quantities. The most obvious
extension is to try to
calculate the expectation values
of gauge invariant variables in perturbation
theory; efforts in this direction are 
underway~\cite{FutureWork}.
Additionally, we should not lose
sight of the fact that the current
formulation was originally developed
with a view to non-perturbative 
applications~\cite{ymi,aprop}. This is
hardly surprising, since the ERG has proven
itself to be a flexible and powerful tool
for addressing such problems in a range of 
QFTs~\cite{Wilson:1974sk,Hasenfratz:1985dm,Wetterich:1992yh,Bergerhoff:1995zm,Bergerhoff:1995zq,Aoki:1996fh,Fisher:1998kv,Morris:1998da,Berges:2000ew,Bagnuls:2000ae,Polonyi:2001se,Salmhofer:2001tr,Pawlowski:2003hq,Pawlowski:2005xe,Reuter:1997gx,Bergerhoff:1997cv,Gies:2002af,Ellwanger:1996wy,Pawlowski:2003hq,Fischer:2004uk,Ellwanger:1997wv,Ellwanger:1999vc,Ellwanger:2002xa};
furthermore, an interesting link
has been recently made between
elements of the formalism presented
here and the
AdS / CFT correspondence~\cite{EMR}.

A very interesting question to ask from the
perspective of this paper is whether it is
possible to repeat the type of cancellations
seen in the treatment of $\beta_n$ in a non-perturbative context.
An immediate obstacle to this is that non-perturbative
ERG  treatments typically rely on truncations (in our case
of the Wilsonian effective action), after which
universality is lost. However, as stated already,
it seems that the diagrammatics are driving us
towards a more direct framework for
performing calculations in QFTs, whilst retaining
the advantages of an ERG approach. If this turns
out to be the case, then we can hope that the new
framework is defined non-perturbatively. Indeed, 
consider again the expression~\eq{eq:bn-Complete} for
$\beta_n$. We could imagine formally lifting
the diagrammatic expression on the \rhs\ to a
non-perturbative expression, such that its weak
coupling expansion just reproduces~\eq{eq:bn-Complete}.

Of course, even if this action is possible (and we
note that the presence of reduced vertices in~\eq{eq:bn-Complete}
complicates matters) we stress that
it would
represent only one step towards extracting non-perturbative
information. After all, our hypothetical non-perturbative
expression for the $\beta$-function would presumably
still require truncation before any numbers
could be computed. Nonetheless, the starting point
would surely be far more appealing with 
the explicit dependence on the seed action and
details of the covariantization removed from
the start.

\ack 
I would like to thank Daniel Litim for encouragement
and useful discussions.
I acknowledge financial support from PPARC.

\appendix

\section[]{The Diagrammatic Identities}
\label{app:D-ID}

\subsection*{The Primary Diagrammatic Identities}

\subsubsection*{The First Type}

\be
\label{eq:WID}
	\ensuremath{\begin{array}{c}\begin{picture}(0,0)%
\includegraphics{pstex/WID-contract-b.pstex}%
\end{picture}%
\setlength{\unitlength}{3947sp}%
\begingroup\makeatletter\ifx\SetFigFont\undefined%
\gdef\SetFigFont#1#2#3#4#5{%
  \reset@font\fontsize{#1}{#2pt}%
  \fontfamily{#3}\fontseries{#4}\fontshape{#5}%
  \selectfont}%
\fi\endgroup%
\begin{picture}(626,300)(2090,-1016)
\end{picture}%
 \end{array}} = \ensuremath{\begin{array}{c}\input{pstex/WID-PF.pstex_t} \end{array}} + \ensuremath{\begin{array}{c}\input{pstex/WID-PFb.pstex_t} \end{array}} - \ensuremath{\begin{array}{c}\input{pstex/WID-PB.pstex_t} \end{array}} - \ensuremath{\begin{array}{c}\input{pstex/WID-PBb.pstex_t} \end{array}} + \cdots
\ee

\bea
\label{eq:LdL-GRk-Pert}
	\hspace{0.8em} \stackrel{\bullet}{\GRk} & =  & 0
\\
\label{eq:dalpha-GRk}
	\begin{array}{c}
		\Dal
	\\[-1.5ex]
		\GRk
	\end{array} & = & 0
\eea

\be
\label{eq:D-ID-GR-TP}
	\ensuremath{\begin{array}{c}\begin{picture}(0,0)%
\includegraphics{pstex/GR-TP.pstex}%
\end{picture}%
\setlength{\unitlength}{3947sp}%
\begingroup\makeatletter\ifx\SetFigFont\undefined%
\gdef\SetFigFont#1#2#3#4#5{%
  \reset@font\fontsize{#1}{#2pt}%
  \fontfamily{#3}\fontseries{#4}\fontshape{#5}%
  \selectfont}%
\fi\endgroup%
\begin{picture}(757,318)(1880,-963)
\put(2278,-861){\makebox(0,0)[lb]{\smash{{\SetFigFont{11}{13.2}{\rmdefault}{\mddefault}{\updefault}{\color[rgb]{0,0,0}$S$}%
}}}}
\end{picture}%
 \end{array}} = 0
\ee

\be
\label{eq:Taylor}
	\ensuremath{\begin{array}{c}\begin{picture}(0,0)%
\includegraphics{pstex/Taylor-Parent-b.pstex}%
\end{picture}%
\setlength{\unitlength}{3947sp}%
\begingroup\makeatletter\ifx\SetFigFont\undefined%
\gdef\SetFigFont#1#2#3#4#5{%
  \reset@font\fontsize{#1}{#2pt}%
  \fontfamily{#3}\fontseries{#4}\fontshape{#5}%
  \selectfont}%
\fi\endgroup%
\begin{picture}(626,459)(2090,-1016)
\put(2520,-641){\makebox(0,0)[lb]{\smash{{\SetFigFont{8}{9.6}{\rmdefault}{\mddefault}{\updefault}{\color[rgb]{0,0,0}$0$}%
}}}}
\end{picture}%
 \end{array}} = \cdeps{Taylor-PFa} + \cdeps{Taylor-PFb} - \cdeps{Taylor-PBa} - \cdeps{Taylor-PBb} +\cdots
\ee

\subsubsection*{The Second Type}

\be
	\ensuremath{\begin{array}{c}\begin{picture}(0,0)%
\includegraphics{pstex/EffPropReln-b.pstex}%
\end{picture}%
\setlength{\unitlength}{3947sp}%
\begingroup\makeatletter\ifx\SetFigFont\undefined%
\gdef\SetFigFont#1#2#3#4#5{%
  \reset@font\fontsize{#1}{#2pt}%
  \fontfamily{#3}\fontseries{#4}\fontshape{#5}%
  \selectfont}%
\fi\endgroup%
\begin{picture}(1102,395)(2306,-1147)
\put(2576,-993){\makebox(0,0)[lb]{\smash{{\SetFigFont{11}{13.2}{\rmdefault}{\mddefault}{\updefault}{\color[rgb]{0,0,0}0}%
}}}}
\end{picture}%
 \end{array}}	= \ensuremath{\begin{array}{c}\begin{picture}(0,0)%
\includegraphics{pstex/K-Delta-b.pstex}%
\end{picture}%
\setlength{\unitlength}{3947sp}%
\begingroup\makeatletter\ifx\SetFigFont\undefined%
\gdef\SetFigFont#1#2#3#4#5{%
  \reset@font\fontsize{#1}{#2pt}%
  \fontfamily{#3}\fontseries{#4}\fontshape{#5}%
  \selectfont}%
\fi\endgroup%
\begin{picture}(224,395)(1941,-1006)
\end{picture}%
 \end{array}} - \ensuremath{\begin{array}{c}\begin{picture}(0,0)%
\includegraphics{pstex/FullGaugeRemainder-b.pstex}%
\end{picture}%
\setlength{\unitlength}{3947sp}%
\begingroup\makeatletter\ifx\SetFigFont\undefined%
\gdef\SetFigFont#1#2#3#4#5{%
  \reset@font\fontsize{#1}{#2pt}%
  \fontfamily{#3}\fontseries{#4}\fontshape{#5}%
  \selectfont}%
\fi\endgroup%
\begin{picture}(238,395)(2239,-930)
\end{picture}%
 \end{array}}
							= \ensuremath{\begin{array}{c}\begin{picture}(0,0)%
\includegraphics{pstex/K-Delta-b.pstex}%
\end{picture}%
\setlength{\unitlength}{3947sp}%
\begingroup\makeatletter\ifx\SetFigFont\undefined%
\gdef\SetFigFont#1#2#3#4#5{%
  \reset@font\fontsize{#1}{#2pt}%
  \fontfamily{#3}\fontseries{#4}\fontshape{#5}%
  \selectfont}%
\fi\endgroup%
\begin{picture}(224,395)(1941,-1006)
\end{picture}%
 \end{array}} - \ensuremath{\begin{array}{c}\begin{picture}(0,0)%
\includegraphics{pstex/DecomposedGR-b.pstex}%
\end{picture}%
\setlength{\unitlength}{3947sp}%
\begingroup\makeatletter\ifx\SetFigFont\undefined%
\gdef\SetFigFont#1#2#3#4#5{%
  \reset@font\fontsize{#1}{#2pt}%
  \fontfamily{#3}\fontseries{#4}\fontshape{#5}%
  \selectfont}%
\fi\endgroup%
\begin{picture}(402,395)(2074,-925)
\end{picture}%
 \end{array}}
	\label{eq:EPReln}
\ee

\subsubsection*{The Third Type}

\be
	\ensuremath{\begin{array}{c}\begin{picture}(0,0)%
\includegraphics{pstex/GR-TLTP.pstex}%
\end{picture}%
\setlength{\unitlength}{3947sp}%
\begingroup\makeatletter\ifx\SetFigFont\undefined%
\gdef\SetFigFont#1#2#3#4#5{%
  \reset@font\fontsize{#1}{#2pt}%
  \fontfamily{#3}\fontseries{#4}\fontshape{#5}%
  \selectfont}%
\fi\endgroup%
\begin{picture}(757,318)(1880,-963)
\put(2296,-857){\makebox(0,0)[lb]{\smash{{\SetFigFont{11}{13.2}{\rmdefault}{\mddefault}{\updefault}{\color[rgb]{0,0,0}$0$}%
}}}}
\end{picture}%
 \end{array}} = 0
\label{eq:GR-TLTP}
\ee

\be
	\ensuremath{\begin{array}{c}\begin{picture}(0,0)%
\includegraphics{pstex/GR-relation.pstex}%
\end{picture}%
\setlength{\unitlength}{3947sp}%
\begingroup\makeatletter\ifx\SetFigFont\undefined%
\gdef\SetFigFont#1#2#3#4#5{%
  \reset@font\fontsize{#1}{#2pt}%
  \fontfamily{#3}\fontseries{#4}\fontshape{#5}%
  \selectfont}%
\fi\endgroup%
\begin{picture}(314,174)(2239,-821)
\end{picture}
 \end{array}} = 1
\label{eq:GR-relation}
\ee

\be
	\ensuremath{\begin{array}{c}\begin{picture}(0,0)%
\includegraphics{pstex/EP-GRpr.pstex}%
\end{picture}%
\setlength{\unitlength}{3947sp}%
\begingroup\makeatletter\ifx\SetFigFont\undefined%
\gdef\SetFigFont#1#2#3#4#5{%
  \reset@font\fontsize{#1}{#2pt}%
  \fontfamily{#3}\fontseries{#4}\fontshape{#5}%
  \selectfont}%
\fi\endgroup%
\begin{picture}(628,174)(1926,-821)
\end{picture}
 \end{array}} \equiv \ensuremath{\begin{array}{c}\begin{picture}(0,0)%
\includegraphics{pstex/GR-PEP.pstex}%
\end{picture}%
\setlength{\unitlength}{3947sp}%
\begingroup\makeatletter\ifx\SetFigFont\undefined%
\gdef\SetFigFont#1#2#3#4#5{%
  \reset@font\fontsize{#1}{#2pt}%
  \fontfamily{#3}\fontseries{#4}\fontshape{#5}%
  \selectfont}%
\fi\endgroup%
\begin{picture}(786,174)(2089,-821)
\end{picture}
 \end{array}}
\label{eq:PseudoEP}
\ee

\subsection*{The Secondary Diagrammatic Identities}

\subsubsection*{The First Type}

\be
\label{D-ID-Trivial}
	\cdeps{GR-hook} = 0
\ee

\be
\label{eq:D-ID-Bitten-hook}
	\cdeps{Bitten-hook} - \cdeps{Bitten-hook-R} \equiv 0
\ee

\bea
	\nonumber
		\sum_{m'=1}^{m-1} \frac{1}{m'!} \ensuremath{\begin{array}{c}\input{pstex/TLTP-EP-ArbGRs.pstex_t} \end{array}} \hspace{2em}
		& = &  \sum_{m'=1}^{m-1} \frac{1}{(m'+1)!} \ensuremath{\begin{array}{c}\begin{picture}(0,0)%
\includegraphics{pstex/GR-ArbGRs.pstex}%
\end{picture}%
\setlength{\unitlength}{3947sp}%
\begingroup\makeatletter\ifx\SetFigFont\undefined%
\gdef\SetFigFont#1#2#3#4#5{%
  \reset@font\fontsize{#1}{#2pt}%
  \fontfamily{#3}\fontseries{#4}\fontshape{#5}%
  \selectfont}%
\fi\endgroup%
\begin{picture}(164,296)(785,791)
\put(785,955){\makebox(0,0)[lb]{\smash{{\SetFigFont{11}{13.2}{\rmdefault}{\mddefault}{\updefault}{\color[rgb]{0,0,0}$\decGR{ \ }{\GRkpr^{m'+1}}$}%
}}}}
\end{picture}%
 \end{array}} \hspace{2em}
	\\ 
	\label{eq:D-ID-G}
		&& - \sum_{m'=2}^{m-1} \sum_{m''=0}^{m'-2} \frac{1}{(m'-m'')!m''!} 
		\ensuremath{\begin{array}{c}\input{pstex/ArbGRs-TLTP-EP-ArbGRs.pstex_t} \end{array}}
\eea

\clearpage

\subsubsection*{The Second Type}

\be
\label{eq:D-ID-dGRk-GT-ring}
	2m(m-1) \ensuremath{\begin{array}{c}\begin{picture}(0,0)%
\includegraphics{pstex/GRs-dGRk.pstex}%
\end{picture}%
\setlength{\unitlength}{3947sp}%
\begingroup\makeatletter\ifx\SetFigFont\undefined%
\gdef\SetFigFont#1#2#3#4#5{%
  \reset@font\fontsize{#1}{#2pt}%
  \fontfamily{#3}\fontseries{#4}\fontshape{#5}%
  \selectfont}%
\fi\endgroup%
\begin{picture}(506,344)(1446,2019)
\put(1446,2069){\makebox(0,0)[lb]{\smash{{\SetFigFont{11}{13.2}{\rmdefault}{\mddefault}{\updefault}{\color[rgb]{0,0,0}$\decGR{ \hspace{3em} }{\GRkpr^{m-2}}$}%
}}}}
\end{picture}%
 \end{array}} \hspace{3em} = - \decGR{ \ }{\socket \GRkpr^m}.
\ee

\bea
	\lefteqn{
		\sum_{m'=0}^m 
		\left[
			\ensuremath{\begin{array}{c}\input{pstex/CTP-E-GRs-Combo-GRk-GR.pstex_t} \end{array}} 
			-\ensuremath{\begin{array}{c}\input{pstex/CTP-E-GRs-Combo-GR.pstex_t} \end{array}}
		\right]	
	}
	\nonumber
	\\ & &
		- \sum_{m'=0}^m \!\! \nCr{m'}{m''} \!\! \sum_{m''=0}^{m'} 
		\left[
			\ensuremath{\begin{array}{c}\input{pstex/CTP-E-GRs-Combo-CTP-EP-GR.pstex_t} \end{array}}
			\hspace{1.8em}
			-\ensuremath{\begin{array}{c}\input{pstex/CTP-E-GRs-Combo-GRs-CTP-EP.pstex_t} \end{array}}
			\hspace{1.8em}
		\right]
		= 0
	\label{eq:D-ID-Op2-G}
\eea

\clearpage

\section[]{Diagrammatic Expressions}

\subsection*{Decoration of diagram~\ref{j+2-DEP}}

\bcf[h]
	\[
	\begin{array}{l}
	\vspace{2.5ex}
		\left[
		\begin{array}{c}
			\ds
			\sum_{s=1}^n \sum_{m=0}^{2s} \sum_{j=-1}^{n+s-m-3}\frac{\norm_{j+s+1,j+1}}{m!}
		\\
			\ds
			\dec{
				\LO{\ensuremath{\begin{array}{c}\input{pstex/Dumbbell-vj_+kR-DEP-vkR.pstex_t} \end{array}} \!\! \Vertex{n_s,j}}{D-vj_+kR-DEP-vkR}
			}{11\Delta^{j+s+1} \GRkpr^m}
		\end{array}
		\right]
		-
		\left[
		\begin{array}{c}
			\ds
			\sum_{s=1}^n \sum_{m=0}^{2s-1} \sum_{j=-1}^{n+s-m-2}\frac{\norm_{j+s,j+1}}{m!}
		\\
			\dec{
			\sco[1]{
				\LDi{vj_+R-DEP}{vj_+R-DEP}
			}{\SumVertex}
			}{11\Delta^{j+s} \GRkpr^m}
		\end{array}
		\right]
	\\
	\vspace{2.5ex}
		-2
		\left[
		\begin{array}{c}
			\ds
			\sum_{s=1}^n \sum_{m=1}^{2s} \sum_{j=-1}^{n+s-m-2}\frac{\norm_{j+s,j+1}}{m!}
			\sum_{m'=1}^{m} \nCr{m}{m'}
		\\
			\ds
			\dec{
				\sco[1]{
					\LDi[1]{v_j+R-DEP-GRs}{v_j+R-DEP-GRs}
				}{\SumVertex}
			}{11\Delta^{j+s} \GRkpr^{m-m'}}
		\end{array}
		\right]
	\\
	\vspace{2.5ex}
		+
		\left[
			\begin{array}{c}
				\ds
				\sum_{s=1}^n \sum_{m=1}^{2s} \sum_{j=-2}^{n+s-m-2}\frac{\norm_{j+s,j+2}}{m!}
				\sum_{m'=1}^{m} \nCr{m}{m'} 
			\\
				\dec{
				\begin{array}{c}
				\vspace{1ex}
					\ds \nCr{m'}{m''}\sum_{m''=1}^{m'-1} 
					\LDi{GRs-DEP-GRs}{GRs-DEP-GRs-MV} \hspace{2em}
					+ \LDi{Ubend-DEP}{Ubend-MV} \hspace{0.5em}
				\\
				\vspace{1ex}
					\TopVertex
				\\
					\SumVertex
				\end{array}
			}{11\Delta^{j+s} \GRkpr^{m-m'}}
			\end{array}
		\right]
	\end{array}
	\]
\caption{Decoration of diagram~\ref{j+2-DEP} with
the differentiated effective propagator.}
\label{fig:app:DEP}
\ecf

\clearpage

\subsection*{Trapped gauge remainders}
\label{app:Trapped}

\bcf[h]
	\[
	\begin{array}{l}
	\vspace{2ex}
		-
		\bca{
			\sum_{s=1}^n \sum_{m=0}^{2s-2}  \sum_{j=-1}^{n+s-m-4} 
			\frac{\norm_{j+s,j+1}}{m!} \sum_{m'=0}^{m}
		}
			{
			\dec{
				\nCr{m}{m'} \!\!
				\sco[1]{
					\LDi{RV-RW-GR-B-RW}{RV-RW-GR-B-RW} \ 
				}{\SumVertex}
			}{11\Delta^{j+s}\GRkpr^{m-m'}}
		}	
	\\
	\vspace{2ex}
		-
		\bca{
			\sum_{s=1}^n \sum_{m=0}^{2s-2} \sum_{j=-2}^{n+s-m-4} 
			\frac{\norm_{j+s+1,j+2}}{m!}  \sum_{m'=0}^{m} \!\! \nCr{m}{m'} \!\!
		}
			{
			\dec{
				\Tower{\LDi{TLTP-RW-GR-B-RW}{TLTP-RW-GR-B-RW}}
			}{11\Delta^{j+s+1}\GRkpr^{m-m'}}
		}
	\\
		+
		\bca{
			\sum_{s=1}^n \sum_{m=0}^{2s-2} \sum_{j=-2}^{n+s-m-4}
			\frac{\norm_{j+s,j+2}}{m!}  \sum_{m'=0}^{m} \!\! \nCr{m}{m'} \!\!
		}
			{
			\dec{
				\Tower{\LDi{Trapped-KBK}{Trapped-KBK}}
			}{11\Delta^{j+s} \GRkpr^{m-m'}}
		}
	\end{array}
	\]
\caption{Terms spawned by diagrams~\ref{WGRx2} 
and~\ref{P-S_vj_+R-DW-GRx2}
which possess a trapped gauge remainder.}
\label{fig:GRsx2-P}
\ecf

\subsection*{Diagrams spawned by~\eq{eq:LdL-b-ba} in which a kernel
bites its own tail}

\bcf[h]
	\[
	\begin{array}{l}
	\vspace{2ex}
			2
		\bca{
			\sum_{s=2}^n \sum_{r=1}^{s-1} \sum_{m=0}^{2(s-r)-2} \sum_{j=-2}^{n+s-2r-m-3}
				\frac{\norm_{j+s+2-r,j+2}}{m!(r-1)!} \sum_{m'=0}^m \!\! \nCr{m}{m'} \!\! \sum_{m''=0}^{m'} \!\! \nCr{m'}{m''} \!\!
		}
			{
			\dec{
				\decp{
					\LDi{CTP-E-WBT-GR-GRs-dGR-GRs}{CTP-E-WBT-GR-GRs-dGR-GRs} 
					\hspace{1.8em}
					-\LDi{CTP-E-WBT-GR-GRs-GR-dGR-GRs}{CTP-E-WBT-GR-GRs-GR-dGR-GRs}
					\hspace{1.8em}
				}{\combo^{r-1}}
				\sco[1]{
					\TopVertex
				}{\SumVertex}
			}{1\Delta^{j+s+2-r}  \GRkpr^{m-m'}}
		}
	\\
	\vspace{2ex}
		+2
		\bca{
			\sum_{s=2}^n \sum_{r=1}^{s-1} \sum_{m=0}^{2(s-r)-1} \sum_{j=-2}^{n+s-2r-m-3}
				\frac{\norm_{j+s+2-r,j+2}}{m!(r-1)!} \sum_{m'=0}^m \!\! \nCr{m}{m'} \!\!
		}
			{
			\dec{
				\decp{
					\LDi{CTP-E-WBT-dGR-GRs}{CTP-E-WBT-dGR-GRs}
					-\LDi{CTP-E-WBT-GR-dGR-GRs}{CTP-E-WBT-GR-dGR-GRs}
				}{\combo^{r-1}}
				\sco[1]{
					\TopVertex
				}{\SumVertex}
			}{1\Delta^{j+s+2-r}  \GRkpr^{m-m'}}
		}
	\\
		+2
		\bca{
			\sum_{s=2}^n \sum_{r=1}^{s-1} \sum_{m=0}^{2(s-r)-1} \sum_{j=-2}^{n+s-2r-m-3}
				\frac{\norm_{j+s+1-r,j+2}}{m!(r-1)!} \sum_{m'=0}^m \!\! \nCr{m}{m'} \!\! \sum_{m''=0}^{m'} \!\! \nCr{m'}{m''} \!\!
		}
			{
			\dec{
				\decp{
					\LDi{CTP-E-GRs-Combo-WBT-GRs-CTP-EP}{CTP-E-GRs-Combo-WBT-GRs-CTP-EP}
					\hspace{1.8em}
				}{\combo^{r-1}}
				\sco[1]{
					\TopVertex
				}{\SumVertex}
			}{1\Delta^{j+s+1-r}  \GRkpr^{m-m'}}
		}
	\end{array}
	\]
\caption{Diagrams spawned by~\eq{eq:LdL-b-ba} in which a kernel
bites its own tail, part~1.}
\label{fig:nLDlb-KBT}
\ecf

\clearpage

\bcf
	\[
	\begin{array}{l}
		2
		\bca{
			\sum_{s=2}^n \sum_{r=1}^{s-1} \sum_{m=0}^{2(s-r)-1} \sum_{j=-2}^{n+s-2r-m-3}
				\frac{\norm_{j+s+1-r,j+2}}{m!(r-1)!} \sum_{m'=0}^m \!\! \nCr{m}{m'} \!\!
		}
			{
			\dec{
				\decp{
					\LDi{CTP-E-GRs-Combo-WBT}{CTP-E-GRs-Combo-WBT-b}
					-\LDi{CTP-E-GRs-Combo-WBT-GR}{CTP-E-GRs-Combo-WBT-GR}
				}{\combo^{r-1}}
				\sco[1]{
					\TopVertex
				}{\SumVertex}
			}{1\Delta^{j+s+1-r}  \GRkpr^{m-m'}}
		}	
	\end{array}
	\]
\caption{Diagrams spawned by~\eq{eq:LdL-b-ba} in which a kernel
bites its own tail, part~2.}
\label{fig:nLDlb-KBT-2}
\ecf

\subsection*{Subtractions and Additions}

\bcf[h]
	\[
	\begin{array}{l}
	\vspace{0ex}
		\mp 4
		\bca{
			\sum_{s=1}^n \sum_{m=0}^{2s-1} \sum_{j=-1}^{n+s-m-2}
			\frac{\norm_{j+s+1,j+1}}{m!} \sum_{m'=0}^m \!\! \nCr{m}{m'} \!\!
		}
			{
			\dec{
				\decp{
					\LLDi{dvh_j-DEP-GRs-Skt-CTP-E}{dvh_j-DEP-GRs-Skt-CTP-E-s}{dvh_j-DEP-GRs-Skt-CTP-E-a}
					-\LLDi{dPi_j+R-DW-GRs-Skt-CTP-E}{dPi_j+R-DW-GRs-Skt-CTP-E-s}{dPi_j+R-DW-GRs-Skt-CTP-E-a}
				}{}
				\SumVertex	
			}{1\Delta^{j+s+1}\GRkpr^{m-m'}}
		}
	\\
		\pm 4
		\bca{
			\sum_{s=1}^n \sum_{m=0}^{2s-1} \sum_{j=0}^{n+s-m-3}
			\frac{\norm_{j+s+2,j+1}}{m!} \sum_{m'=0}^m \!\! \nCr{m}{m'} \!\!
		}
			{
			\dec{
				\decp{
					\left[
						\LLDi{vh_k+R-DEP-GRs-Skt-CTP-E}{vh_kR-dvj+kh-s}{vh_kR-dvj+kh-a}
						-\LLDi{Pi_kR-DW-GRs-Skt-CTP-E}{Pi_j+R-dvj+kh-s}{Pi_j+R-dvj+kh-a}
					\right] \ensuremath{\begin{array}{c}\begin{picture}(0,0)%
\includegraphics{pstex/dvj+k.pstex}%
\end{picture}%
\setlength{\unitlength}{3947sp}%
\begingroup\makeatletter\ifx\SetFigFont\undefined%
\gdef\SetFigFont#1#2#3#4#5{%
  \reset@font\fontsize{#1}{#2pt}%
  \fontfamily{#3}\fontseries{#4}\fontshape{#5}%
  \selectfont}%
\fi\endgroup%
\begin{picture}(578,671)(654,1931)
\put(703,2105){\makebox(0,0)[lb]{\smash{{\SetFigFont{11}{13.2}{\rmdefault}{\mddefault}{\updefault}{\color[rgb]{0,0,0}$\nothing {v}^{j_+,k}$}%
}}}}
\end{picture}%
 \end{array}}
				}{}
				\SumVertex
			}{1\Delta^{j+s+2}\GRkpr^{m-m'}}
		}
	\end{array}
	\]
\caption{Subtractions and additions for diagrams~\ref{vh_j+R-DEP-GRs-Skt-CTP-E}
and~\ref{Pi_j+R-DW-GRs-Skt-CTP-E}, part~1.}
\label{fig:vh_j+R-DEP-GRs-Skt-CTP-E-Add}
\ecf

\bcf[!tbp]
	\[
	\begin{array}{l}
	\vspace{2ex}
		\mp 2
		\bca{
			\sum_{s=1}^n \sum_{m=1}^{2s-1} \sum_{j=-1}^{n+s-m-3}
			\frac{\norm_{j+s+1,j+1}}{m!} \sum_{m'=1}^m \!\! \nCr{m}{m'} \!\!
		}
			{
			\dec{
				\decp{
					\LLDi{vh_j-DEP-TEGRs-Skt-CTP-E}{vh_j-DEP-TEGRs-Skt-CTP-E-s}{vh_j-DEP-TEGRs-Skt-CTP-E-a}
					-\LLDi{Pi_j+R-DEP-TEGRs-Skt-CTP-E}{Pi_j+R-DEP-TEGRs-Skt-CTP-E-s}{Pi_j+R-DEP-TEGRs-Skt-CTP-E-a}
					\hspace{1em}
				}{}
				\SumVertex
			}{1\Delta^{j+s+1}\GRkpr^{m-m'}}
		}
	\\
	\vspace{2ex}
		\mp 2
		\bca{
			\sum_{s=1}^n \sum_{m=2}^{2s-1} \sum_{j=-1}^{n+s-m-3}
			\frac{\norm_{j+s+1,j+1}}{m!} \sum_{m'=2}^m \!\! \nCr{m}{m'} \!\!
			\sum_{m''=2}^{m'} \!\! \nCr{m'}{m''} \!\!
		}
			{
			\dec{
				\decp{
					\begin{array}{c}
						\LLDi{vh_j+R-DEP-GRs-Skt-CTP-E-b}{vh_j+R-DEP-GRs-Skt-CTP-E-b-s}{vh_j+R-DEP-GRs-Skt-CTP-E-a}
						\hspace{1.8em}
						-\LLDi{Pi_j+R-DW-GRs-Skt-CTP-E-b}{Pi_j+R-DW-GRs-Skt-CTP-E-b-s}{Pi_j+R-DW-GRs-Skt-CTP-E-b-a}
						\hspace{1.8em}
					\\
						\decGR{\ }{\socket \GRkpr^{m''}}	
					\end{array}
				}{}
				\SumVertex
			}{1\Delta^{j+s+1}\GRkpr^{m-m'}}
		}
	\\
		\pm 4
			\bca{
			\sum_{s=1}^n \sum_{m=0}^{2s-1} \sum_{j=-1}^{n+s-m-3}
			\frac{\norm_{j+s+1,j+1}}{m!} \sum_{m'=0}^m \!\! \nCr{m}{m'} \!\!
		}
			{
			\dec{	
				\decp{
					\LLDi{Pi_j+R-dK-GRs-Skt-CTP-E}{Pi_j+R-dK-GRs-Skt-CTP-E-s}{Pi_j+R-dK-GRs-Skt-CTP-E-a}
				}{}
				\SumVertex
			}{1\Delta^{j+s+1}\GRkpr^{m-m'}}
		}
	\end{array}
	\]
\caption{Subtractions and additions for diagrams~\ref{vh_j+R-DEP-GRs-Skt-CTP-E}
and~\ref{Pi_j+R-DW-GRs-Skt-CTP-E}, part~2.}
\label{fig:vh_j+R-DEP-GRs-Skt-CTP-E-Add-II}
\ecf

\bcf[!tbp]
	\[
	\begin{array}{l}
	\vspace{2ex}
		\mp 2 
		\bca{
			\sum_{s=1}^n \sum_{m=0}^{2s-1} \sum_{j=-1}^{n+s-m-2}
			\frac{\norm_{j+s+1,j+1}}{m!} \sum_{m'=0}^m \!\! \nCr{m}{m'} \!\!
		}
			{
			\dec{
				\begin{array}{c}
					\decp{
						\begin{array}{c}
							\LLDi{DW-GR-GRs-Skt-CTP-E}{DW-GR-GRs-Skt-CTP-E-dv-s}{DW-GR-GRs-Skt-CTP-E-dv-a}
							-2\LLDi{DW-GRs-Skt-CTP-E}{DW-GRs-Skt-CTP-E-dv-s}{DW-GRs-Skt-CTP-E-dv-a}
						\\
							\ensuremath{\begin{array}{c}\begin{picture}(0,0)%
\includegraphics{pstex/dvj+.pstex}%
\end{picture}%
\setlength{\unitlength}{3947sp}%
\begingroup\makeatletter\ifx\SetFigFont\undefined%
\gdef\SetFigFont#1#2#3#4#5{%
  \reset@font\fontsize{#1}{#2pt}%
  \fontfamily{#3}\fontseries{#4}\fontshape{#5}%
  \selectfont}%
\fi\endgroup%
\begin{picture}(540,594)(1127,1795)
\put(1203,1941){\makebox(0,0)[lb]{\smash{{\SetFigFont{11}{13.2}{\rmdefault}{\mddefault}{\updefault}{\color[rgb]{0,0,0}$\nothing {v}^{j_+}$}%
}}}}
\end{picture}%
 \end{array}}
						\end{array}
					}{}
				\\
					\SumVertex	
				\end{array}
			}{1\Delta^{j+s+1}\GRkpr^{m-m'}}
		}
	\\
		\pm 2 
		\bca{
			\sum_{s=1}^n \sum_{m=0}^{2s-1} \sum_{j=-2}^{n+s-m-3}
			\frac{\norm_{j+s+1,j+2}}{m!} \sum_{m'=0}^m \!\! \nCr{m}{m'} \!\!
		}
			{
			\dec{
				\begin{array}{c}
					\decp{
						\LLDi{dK-GR-GRs-Skt-CTP-E}{dK-GR-GRs-Skt-CTP-E-dv-s}{dK-GR-GRs-Skt-CTP-E-dv-a}
						-2\LLDi{dK-GRs-Skt-CTP-E}{dK-GRs-Skt-CTP-E-dv-s}{dK-GRs-Skt-CTP-E-dv-a}
					}{}
				\\
					\TopVertex \SumVertex	
				\end{array}
			}{1\Delta^{j+s+1}\GRkpr^{m-m'}}
		}
	\\
		\pm
		\bca{
			\sum_{s=1}^n \sum_{m=1}^{2s-1} \sum_{j=-2}^{n+s-m-3}
			\frac{\norm_{j+s+1,j+2}}{m!} \sum_{m'=1}^m \!\! \nCr{m}{m'} \!\!
		}
			{
			\dec{
				\sco[1]{
					\decp{
					\LLDi{DW-GR-TEGRs-Skt-CTP-E}{DW-GR-TEGRs-Skt-CTP-E-s}{DW-GR-TEGRs-Skt-CTP-E-a}
					-2\LLDi{DW-TEGRs-Skt-CTP-E}{DW-TEGRs-Skt-CTP-E-s}{DW-TEGRs-Skt-CTP-E-a}
					}{}
				}{\TopVertex \SumVertex}
			}{1\Delta^{j+s+1}\GRkpr^{m-m'}}
		}
	\end{array}
	\]
\caption{Subtractions and additions for 
diagrams~\ref{DW-GR-GRs-Skt-CTP-E} and~\ref{DW-GRs-Skt-CTP-E}, 
part~1.}
\label{fig:DW-GR-GRs-Skt-CTP-E-Add}
\ecf

\bcf[!tbp]
	\[
	\pm
		\bca{
			\sum_{s=1}^n \sum_{m=2}^{2s-1} \sum_{j=-2}^{n+s-m-3}
			\frac{\norm_{j+s+1,j+2}}{m!} \sum_{m'=2}^m \!\! \nCr{m}{m'} \!\!
			\sum_{m''=2}^{m'} \!\! \nCr{m'}{m''} \!\!
		}
			{
			\dec{
				\sco[1]{
					\decp{
					\begin{array}{c}
						\LLDi{DW-GR-GRs-Skt-CTP-E-b}{DW-GR-GRs-Skt-CTP-E-TEGR-s}{DW-GR-GRs-Skt-CTP-E-TEGR-a}
						\hspace{1.8em}
						-2\LLDi{DW-GRs-Skt-CTP-E-b}{DW-GRs-Skt-CTP-E-TEGR-s}{DW-GRs-Skt-CTP-E-TEGR-a}
						\hspace{1.8em}
					\\
						\decGR{\ }{\socket \GRkpr^{m''}}	
					\end{array}
					}{}
				}{\TopVertex \SumVertex}
			}{1\Delta^{j+s+1}\GRkpr^{m-m'}}
		}
	\]
\caption{Subtractions and additions for 
diagrams~\ref{DW-GR-GRs-Skt-CTP-E}
and~\ref{DW-GRs-Skt-CTP-E}, 
part~2.}
\label{fig:DW-GR-GRs-Skt-CTP-E-Add-II}
\ecf

\bcf[!tbp]
	\[
	\begin{array}{l}
	\vspace{2ex}
		\mp 4 
		\bca{
			\sum_{s=1}^n \sum_{m=2}^{2s-1} \sum_{j=-1}^{n+s-m-2}
			\frac{\norm_{j+s+1,j+1}}{m!} \sum_{m'=2}^m \!\! \nCr{m}{m'} \!\! \sum_{m''=2}^{m'} \!\! \nCr{m'}{m''} \!\!
		}
			{
			\dec{
				\begin{array}{c}
					\decp{
						\begin{array}{c}
							\LLDi{GRs-dEP-GR-Skt-CTP-E}{GRs-dEP-GR-Skt-CTP-E-dv-s}{GRs-dEP-GR-Skt-CTP-E-dv-a}
							\hspace{1.8em}
						\\
							\ensuremath{\begin{array}{c}\begin{picture}(0,0)%
\includegraphics{pstex/dvj+.pstex}%
\end{picture}%
\setlength{\unitlength}{3947sp}%
\begingroup\makeatletter\ifx\SetFigFont\undefined%
\gdef\SetFigFont#1#2#3#4#5{%
  \reset@font\fontsize{#1}{#2pt}%
  \fontfamily{#3}\fontseries{#4}\fontshape{#5}%
  \selectfont}%
\fi\endgroup%
\begin{picture}(540,594)(1127,1795)
\put(1203,1941){\makebox(0,0)[lb]{\smash{{\SetFigFont{11}{13.2}{\rmdefault}{\mddefault}{\updefault}{\color[rgb]{0,0,0}$\nothing {v}^{j_+}$}%
}}}}
\end{picture}%
 \end{array}}
						\end{array}
					}{}
				\\
					\SumVertex	
				\end{array}
			}{1\Delta^{j+s+1}\GRkpr^{m-m'}}
		}	
	\end{array}
	\]
\caption{Subtractions and additions for 
diagram~\ref{GRs-dEP-GR-Skt-CTP-E}, part~1.}
\label{fig:GRs-dEP-GR-Skt-CTP-E-Add}
\ecf

\bcf[!tbp]
	\[
	\begin{array}{l}
	\vspace{2ex}
		\pm 2
		\bca{
			\sum_{s=1}^n \sum_{m=2}^{2s-1} \sum_{j=-2}^{n+s-m-3}
			\frac{\norm_{j+s+1,j+2}}{m!} \sum_{m'=2}^m \!\! \nCr{m}{m'} \!\! \sum_{m''=2}^{m'} \!\! \nCr{m'}{m''} \!\!
		}
			{
			\dec{
				\sco[1]{
					\decp{
						\LLDi{TEGRs-dEP-GR-Skt-CTP-E}{TEGRs-dEP-GR-Skt-CTP-E-s}{TEGRs-dEP-GR-Skt-CTP-E-a}
						\hspace{1.8em}
						+\LLDi{GRs-dEP-GR-TEGRs-Skt-CTP-E}{GRs-dEP-GR-TEGRs-Skt-CTP-E-s}{GRs-dEP-GR-TEGRs-Skt-CTP-E-a}
						\hspace{1.8em}
					}{}
				}{\TopVertex \SumVertex}
			}{1\Delta^{j+s+1}\GRkpr^{m-m'}}
		}
	\\
		\pm 2
		\bca{
			\sum_{s=1}^n \sum_{m=2}^{2s-1} \sum_{j=-2}^{n+s-m-3}
			\frac{\norm_{j+s+1,j+2}}{m!} \sum_{m'=2}^m \!\! \nCr{m}{m'} \!\!
			\sum_{m''=2}^{m'} \!\! \nCr{m'}{m''} \!\! \sum_{m'''=2}^{m''} \!\! \nCr{m''}{m'''} \!\!
		}
			{
			\dec{
				\sco[1]{
					\decp{
					\begin{array}{c}
						\LLDi{GRs-dEP-GR-Skt-CTP-E-b}{GRs-dEP-GR-Skt-CTP-E-TEGRs-s}{GRs-dEP-GR-Skt-CTP-E-TEGRs-a}
						\hspace{2em}
					\\
						\decGR{\ }{\socket \GRkpr^{m'''}}	
					\end{array}
					}{}
				}{\TopVertex \SumVertex}
			}{1\Delta^{j+s+1}\GRkpr^{m-m'}}
		}
	\end{array}
	\]
\caption{Subtractions and additions for 
diagram~\ref{GRs-dEP-GR-Skt-CTP-E}, part~2.}
\label{fig:GRs-dEP-GR-Skt-CTP-E-Add-II}
\ecf

\end{document}